\documentclass[noline]{jfm}
\makeatletter

\gdef\@underjournal{}
\makeatother
\makeatletter

\let\oldps@titlepage\ps@titlepage
\def\ps@titlepage{\oldps@titlepage
	\def\@oddhead{\vbox{\vspace*{-1pc}\hbox to \textwidth{\@j@urnal \hfil\llap{\thepage}}\par}}%
}
\makeatother

\makeatletter
\def\absfooterflag{} 
\def\pagelimitfooter{} 
\makeatother

\usepackage{amsmath}
\usepackage{bm}
\usepackage{lipsum}
\usepackage{hyperref}
\usepackage{filecontents}
\usepackage{graphicx}
\usepackage{newtxtext}
\usepackage{newtxmath}
\usepackage{natbib}
\usepackage{hyperref}
\usepackage{subcaption}
\hypersetup{
    colorlinks = true, 
    urlcolor   = blue, 
    citecolor  = black, 
}

\newcommand{\RomanNumeralCaps}[1]
\linenumbers

\title{The extended gas-kinetic theory from Pullin equation: the relaxation rates, transport coefficients and model equation}

\author{Sha Liu\aff{1}\aff{2}\aff{3}\corresp{\email{shaliu@nwpu.edu.cn}}, 
  Ningchao Ding\aff{1}\corresp{\email{dingnc@mail.nwpu.edu.cn}}, Ming Fang\aff{4}, Hao Jin\aff{1}, Rui Zhang\aff{1}, Congshan Zhuo\aff{1}\aff{2}\aff{3} \and Chengwen Zhong\aff{1}\aff{2}\aff{3}}

\affiliation{\aff{1}School of Aeronautics, Northwestern Polytechnical University, Xi'an, Shaanxi 710072, China
	\aff{2} National Key Laboratory of Aircraft Configuration Design, Northwestern Polytechnical University, Xi'an, Shaanxi 710072, China
\aff{3} Institute of Extreme Mechanics, Northwestern Polytechnical University, Xi'an, Shaanxi 710072, China
\aff{4} China Aerodynamics Research and Development Center, Mianyang, Sichuan 621000, China}

\begin{document}
\maketitle

\begin{abstract}

The Borgnakke-Larsen model, widely used in rarefied flow predictions, serves as the mainstream energy-exchange kernel for polyatomic gases. However, it lacks integrability and does not guarantee detailed balance, limiting theoretical foundations for near-continuum relaxation mechanisms, transport coefficients, and relaxation model equations. In this work, we adopt the Pullin equation, which possesses an integrable collision kernel and satisfies detailed balance, to analyze near-continuum relaxation. Considering only translational and rotational degrees, we obtain explicit analytical expressions for the relaxation rates of macroscopic variables including stress, temperatures, and heat fluxes by approximating the distribution function in mixed Hermite and Laguerre spaces. Based on the same elementary moments, we derive transport coefficients via Chapman-Enskog expansion, rigorously confirming a long-standing speculation that thermal conductivity depends on the degree of thermal non-equilibrium; under equilibrium, the results reduce to those of Mason and Monchick. Using the correct relaxation rates, we propose a novel Rykov-type relaxation model that captures the coupled relaxation of translational and rotational heat fluxes, a mechanism ignored in the widely used Rykov equation. The model is validated against benchmark test cases.
\end{abstract}


\section{Introduction}
\label{sec:headings}
Gas kinetic theory, originally established by Maxwell and Boltzmann \citep{cercignani1971kinetic,  bird1970direct, chapman1990mathematical}, was developed to describe dilute gas flows under rarefied and micro-nano-scale conditions, and has since been extended to model a wide range of particle systems, including photons \citep{sun2015asymptotic, sun2017multidimensional, dreicer1964kinetic, li2020unified},  plasmas \citep{liu2017unified,pu2025unified}, phonons \citep{zhang2021transient, guo2016discrete, liu2024peridynamic, zhang2017unified},  and neutrons \citep{shuang2019parallel, zhou2020discrete, tan2020time}. The distribution function $f\left(\mathbf{x},\mathbf{c},t\right)$,  which is the fundamental field variable of the gas kinetic theory, represents the number density of molecules with velocity $\mathbf{c}$ at spatial coordinates $\mathbf{x}$ and time $t$. It can be viewed as dividing the macroscopic field variables into the specific molecules. For example, the macroscopic momentum is decomposed into the microscopic momentum of each individual molecule. By doing so, the fundamental property of rarefied and micro-nano flows, which can be expressed as the velocity distribution of molecules significantly deviates from the normal distribution (the Maxwellian distribution in gas kinetic theory), can be accurately captured. In other words, the gas kinetic theory can provide a comprehensive description of the non-normal (non-equilibrium) molecular system, along with a physically rigorous framework, which should be adopted in studying the non-equilibrium phenomena in rarefied and micro-nano flows. This theoretical foundation has yielded profound insights into a series of complex flow phenomena \citep{cercignani2000rarefied},  including the Knudsen layer \citep{guo2008analysis, gusarov2002gas, zhang2006capturing} and the Knudsen paradox \citep{akhlaghi2023comprehensive}. To extend the scope of gas kinetic theory to broader scale ranges, primarily in the context of numerical prediction methods, modern scale-adaptive strategy, such as a series of unified modeling approaches \citep{xu2010unified, guo2013discrete, chen2019conserved, yang2019improved,gallis2011investigation, fei2020unified,  zhang2019particle}, asymptotic preserving Monte Carlo \citep{ren2014asymptotic},  the general synthetic iterative scheme \citep{wu2014solving}, and time-relaxed Monte Carlo \citep{pareschi2001time, fei2023time} has been successful in accommodating the near-continuum flows and even the entire flow scale (regime) spanning from rarefied to continuum flows. On the other hand, in the context of theoretical analysis, rigorous connections between gas kinetic theory and continuum aerodynamics have been derived through the Chapman-Enskog (C-E) expansion \citep{chapman1990mathematical},  along with the moment methods \citep{grad1958principles, torrilhon2016modeling, struchtrup2003regularization, cai2010numerical, jiang2019computation},  yielding a system of constitutive relations and corresponding transport coefficients (such as thermal conductivity and viscosity) \citep{kremer2010introduction}.

However, significant challenges arise when accounting for the internal degrees of freedom within molecule structures (e.g., rotational, vibrational and electronic motions). Due to the lack of exact and analytically integrable differential collision cross-section, which is the pivot of a governing equation, formulating an accurate extended Boltzmann equation (EBE) becomes considerably difficult \citep{nagnibeda2009non, gamba2023cauchy, pavic2022kinetic}. Consequently, the theoretical derivation of the corresponding transport coefficients, which should be based on a sound master equation, also becomes difficult. The master equation corresponds to rarefied and micro-nano-flows, while the transport coefficients are for both continuum flows (in constitutive relations for diffusion, stress and heat flux) and rarefied flows (in constructing model equation for EBE). So far, almost all experimental data for transport coefficients, such as those published by National Institute of Standards and Technology \citep{NIST},  are predominantly confined to the thermal equilibrium condition where temperatures for all degrees of freedom are equal. This situation renders the acquisition of reliable transport coefficients particularly challenging under the common thermal non-equilibrium condition. These limitations in both the master equation and the corresponding transport coefficients hampers both the understanding of fundamental physics and the numerical simulations of flow fields, especially in scenarios of aerospace engineering and micro-nano-manufacturing, where thermal non-equilibrium are pervasive due to the much less molecular collisions. Therefore, experimental works in the electric arc shock tube (EAST) \citep{cruden2014absolute,park1989nonequilibrium} are conducted, which primarily focus on the acquisition of the relaxation time for internal energy and the chemical reaction rate \citep{nagnibeda2009non},  but without involving the transport coefficients. On the other hand, in scenarios of numerical prediction, the development of scale-adaptive methods for simulating the entire flow regime often relies on the model equation \citep{holway1966new,rykov1975model,gorji2013fokker} which is a mathematical simplification of EBE,  and its modeling constrains include these very transport coefficients.

To address these challenges, a possible route involves the following steps: (1) finding or constructing an analytically integrable EBE; (2) rigorously deriving the non-equilibrium relaxation rate based on EBE, and obtaining the transport coefficient using the same methodology; (3) constructing and validating the relaxation model equation for the EBE. The subsequent paragraphs in this section review the background relevant to these three objectives.
To ensure clarity throughout the derivation process and facilitate understanding of the underlying mechanisms, this study focuses exclusively on the rotational degrees of freedom. Other internal degrees of freedom can be treated analogously.

The key to constructing EBE lies in the accurate definition of the collision kernel, which determines the post-collision states of a particle pair based on their pre-collision states. This mapping relationship is precisely realized by the mechanism of the differential collision cross-section in the EBE framework. The construction of EBE, which entails defining a specific differential collision cross-section, primarily follows two routes: the state-to-state (StS) and the phenomenological approaches.
 
For StS-based EBE, the critical differential collision cross-sections are fitted or constructed based on molecular dynamics (MD) experimental results \citep{muller1997simple,  hu2025thermal} as well as potential energy surfaces (PES) obtained from quantum mechanical \citep{szabo2012modern} or simplified computational methods \citep{ebner1976density, te2020classical}. However, the construction of the StS database is constrained by the limited accuracy and high computational cost. Considerable efforts have been devoted to addressing these challenges. Addressing the challenge of improving the accuracy of state-to-state simulations while maintaining computational efficiency, a series of quantum-classical hybrid methods have been developed, including the mixed quantum-classical \citep{billing2003quantum} method and the quasi-classical trajectory method \citep{karplus1965exchange}. Currently, the PES established for the corresponding EBE exhibit fluctuations in precision due to the fact that no single method has been proven to possess optimal accuracy \citep{hong2020inelastic}. Furthermore, both MD simulations and PES calculations require highly detailed pre-collision molecular information (including the relative positions of internal atoms, vibrational velocities along bond directions, rotational velocities about three axes, etc.), leading to immense computational and experimental demands due to the high dimensionality and large number of discrete states involved. In response to this limitation, both traditional techniques (e.g., spline interpolation and nonlinear fitting methods) \citep{schatz1989analytical} and recent approaches driven by neural network fitting and artificial intelligence \citep{hong2023improved,li2020many} aim to reduce computational load and improve database efficiency while ensuring the accuracy of potential energy surface (PES) calculations. Although research on StS simulation has advanced rapidly in recent years, improving computational accuracy and reducing costs, the establishment of a comprehensive StS database still faces significant challenges. It follows that once StS differential collision cross-sections are obtained, they can be incorporated using the framework of the Wang Chang-Uhlenbeck (WCU) equations \citep{chang1951transport}. This framework effectively treats different internal energy states as distinct species. In fact, the WCU formalism represents an early exploration of constructing the EBE under the StS model, although accurate StS differential collision cross-sections were unavailable at that time, thus requiring approximation schemes to derive macroscopic transport properties, specifically viscosity, diffusion, and thermal conductivity coefficients \citep{chang1951transport}. The growing availability of established StS databases enhances the feasibility of this framework, as its application relies on detailed databases covering various temperature ranges and different degrees of thermal non-equilibrium generated by the aforementioned work \citep{hong2020inelastic, jiang2024global}. On the other hand, the WCU framework assumes by default that the internal energy states of molecules do not affect their translational motion. This treatment is also widely used in the model construction in Direct Simulation Monte Carlo (DSMC) methods\citep{borgnakke1975statistical}. However, some studies based on modeling practices have proposed that the influence of internal states on molecular collisional cross-sections (particularly the scattering angle) should be taken into account \citep{gamba2023cauchy}. Nevertheless, there is currently a relative lack of experimental or computational support to substantiate such work.
 
 Phenomenological EBE, in contrast, simplifies the detailed pre-collision state information by considering only translational velocity and  internal energy (e.g., rotational energy, vibrational energy), without involving the initial states of the rotational and vibrational modes. Ordinary with the decoupled influence of translational and internal energy on the collision model, the translation of a collision pair is modeled by the traditional model such as the variable hard-sphere (VHS) model, while energy exchange mechanisms are modeled separately. This phenomenological modeling is implemented either at the theoretical level in the mathematical construction of EBE \citep{pullin1978kinetic} or at the particle simulation level via specific procedures \citep{borgnakke1975statistical} to construct the stochastic collision process. Throughout the modeling process, strict compliance with the detailed balance principle must be maintained \citep{cercignani2000rarefied}. Regarding energy exchange mechanisms, phenomenological models generally employ two approaches: either probabilistically selecting between fully elastic and fully inelastic collisions \citep{bird1978monte, bird1976molecular},  or directly distributing energy according to predefined probability parameters \citep{pullin1978kinetic}.
 
 It is noteworthy that research on collision processes involving internal energy exchange for rarefied flow predictions has historically been predominantly conducted using the DSMC method. For modeling collisions of polyatomic gas molecules, the early energy sink model \citep{bird1976molecular} was used, with subsequent widespread adoption of the Borgnakke-Larsen (BL) model in later studies \citep{borgnakke1975statistical}. The energy sink model checks the equilibrium relationship between translational and internal energy for every colliding pair and relaxes all imbalance. However, the energy sink model distorts the high-velocity tail of the translational velocity equilibrium distribution. This issue was resolved by Borgnakke in the BL model \citep{bird1978monte}. The BL model simulates energy exchange probabilistically, assuming that only a fraction of colliding pairs undergo completely inelastic collisions, while the remainder collide elastically. For the completely inelastic collisions, the BL model samples both the post-collision translational energy (and thus the relative translational velocity) and the internal energy from their respective local equilibrium distributions (statistically determined from all molecules within a computational cell), while rigorously conserving total energy. Crucially, each sampling step must satisfy the constraint of the residual energy remaining from the previous step. However, it has not been conclusively proven that the BL model satisfies detailed balance principle. The BL model significantly broadened the applicability of DSMC, cementing its role as the primary model and method for simulating rarefied flows \citep{boyd1993temperature, valentini2023near}. However, it is essential to emphasize that the BL model was constructed specifically for direct simulation of the collision process. Consequently, its collision cross-section lacks an explicit integrable functional form. This limitation makes it difficult to derive analytical expressions for transport coefficients (e.g., viscosity, thermal conductivity) from the BL model, hindering further theoretical and completeness investigations.
 
Another seminal work in phenomenological EBE is Pullin's model \citep{pullin1978kinetic}. The phenomenological energy exchange component of its collision model employs a beta distribution to redistribute energy between colliding pairs, explicitly ensuring adherence to the important principle of detailed balance. Simultaneously, a governing equation with an analytically tractable form was constructed. Specifically, Pullin leveraged the mathematical property of the beta distribution enabling the decomposition of a normal distribution into multiple normal distributions \citep{pullin1978kinetic}, thereby preserving analytical integrability. Furthermore, the additional free parameter in the beta function allows for ensuring correct energy partitioning and matching macroscopic transport coefficients. Critically, incorporating the principle of detailed balance for the collision process guarantees collisional reversibility and the theoretical completeness of the equation within the framework of gas kinetic theory. Building upon similar energy redistribution concepts, Pavic et al. \citep{gamba2023cauchy} also proposed a collision model accounting for the increased collision likelihood of particles possessing high translational and high internal energy, constructing an EBE tailored for polyatomic gases.

 It is noteworthy that in the field of numerical computation, directly solving the Boltzmann equation poses significant challenges due to its mathematical complexity (specifically, the nonlinear multiple integrals in the collision term) and its physical stiffness. Although several representative efforts have addressed these difficulties \citep{aristov2001direct, morris2011monte, clarke2012novel},  both StS EBE and phenomenological EBE remain susceptible to the curse of dimensionality. To address this limitation, a series of model equations approximating the full Boltzmann transport equation and extended Boltzmann transport equation (EBTE) have been developed. These primarily fall into two categories: BGK-type model equations (also termed relaxation-type models) \citep{bhatnagar1954model,  morse1964kinetic, shakhov1968approximate, rykov1978macroscopic} and Fokker-Planck-type model equations \citep{lebowitz1960nonequilibrium, pawula1967approximation, yano2009analytical, gorji2011fokker}. Focusing on relaxation models for EBTE incorporating rotational energy, key contributions include the Rykov model \citep{rykov1975model} and ES-BGK model \citep{holway1966new}. The Rykov model adjusts Hermite polynomial coefficients to recover the correct thermal conductivity \citep{rykov1975model}. However, it assumes independent relaxation of translational and rotational heat fluxes, a feature inconsistent with the actual physical relaxation process \citep{wu2015kinetic}. While the ES-BGK model is grounded in the principle of maximum entropy, it suffers from similar unphysical limitations as the Rykov model. Furthermore, its reliance on an anisotropic Gaussian distribution restricts its ability to achieve exact relaxation for higher-order moments. Regarding Fokker-Planck models, the cubic-FP model has also been extended to include rotational energy \citep{gorji2011fokker}.
 
 Historically, constructing relaxation-type models (e.g., the Rykov model), particularly concerning their asymptotic preserving (AP) properties, centered on the guiding principle of recovering correct transport coefficients via C-E expansion. However, achieving the correct relaxation rates is a broader and more fundamental requirement. When accurate relaxation rates are attained, the correct macroscopic transport coefficients are naturally recovered. This critical point will be elaborated upon in detail in Section \ref{chapter2} of this work. Moreover, within both the C-E expansion and relaxation rate frameworks, the moment integration of the collision term constitutes a fundamental element. This integration exhibits significant universality across theoretical studies, particularly in deriving transport coefficients. Eucken analytically demonstrated that under thermal non-equilibrium conditions (where translational and rotational temperatures differ), the thermal conductivity coefficients contributing to the total heat flux differ for gradients in translational temperature versus rotational temperature, quantified by the Eucken factor \citep{eucken1913warmeleitvermogen}. In pursuit of its accurate value, Mason and Monchick \citep{mason1962heat} started from the WCU equations and employed the rigid sphere model to perform a series of mathematical simplifications, including the assumptions of perfect elastic collisions. They subsequently applied a conceptual equivalence treatment to the transport coefficients deferred in the original WCU formulation. This allowed them to derive expressions for the thermal relaxation coefficient and the Eucken factor for molecules with rotational energy via a simplified C-E expansion. Conversely, Wu et al. numerically explored and attempted to calibrate the near-equilibrium relaxation processes of translational and rotational heat fluxes using DSMC-BL simulations on benchmark problems, providing a contrasting perspective to theoretical results of Mason \citep{li2021uncertainty}. This numerical calibration formed the basis for constructing a relaxation-type model equation. However, it is crucial to recognize that numerical calibration outcomes can exhibit significant variations depending on the specific physical problem and the resulting form of the distribution function. Therefore, an analytical relaxation rates obtained from the EBE is essential for constructing model equations capable of accurately capturing relaxation processes.

 Hence, the present work aims to derive analytical relaxation rates and transport coefficients for molecular gases through theoretical formulation and, based on these results, to analyze thermal non-equilibrium relaxation processes and construct a kinetic model that simultaneously recovers both the relaxation rates and transport coefficients. The remainder of this paper is organized as follows. In $\S$ \ref{chapter2}, an approximate distribution function constructed via Hermite and Laguerre polynomial expansions is employed to approximate the near-equilibrium state. This enables the theoretical derivation of relaxation rates based on Pullin's energy exchange model, along with an optimization of the energy partitioning parameter. Using the derived relaxation rates, the thermal non-equilibrium relaxation process is analyzed, and the relationship between relaxation rates and macroscopic transport coefficients is elucidated and investigated. In $\S$ \ref{chapter3},  a Rykov-type kinetic model is developed using the previously derived relaxation rates, and the connection between relaxation rates and transport coefficients in the kinetic model is discussed. In $\S$ \ref{chapter4},  the proposed kinetic model is validated through DSMC simulations in both zero-dimensional relaxation and typical rarefied gas flows (e.g., normal shock wave, planar Couette flow, lid-driven cavity flow, and hypersonic flow past cylinder) with comparisons made against the Rykov model. Finally, conclusions are presented in $\S$ \ref{chapter5}.

\section {Relaxation rate}\label{chapter2}
 The relaxation process about a non-equilibrium gas system towards its equilibrium state is determined by the momentum and energy exchange between collision pairs, which is  characterized by the relaxation rates of non-equilibrium variables from macroscopic point of view. Analytical expressions for these relaxation rates can be obtained by calculating the moments of collision operator in EBE. In this work, the phenomenological Pullin equation \citep{pullin1978kinetic} is adopted, which starts from a extended polyatomic Boltzmann equation and utilizes the beta function to partition energy between collision pairs. Compared to the widely used and also phenomenological BL model for DSMC simulations, the Pullin's collision model satisfies the crucial detailed balance principle. Furthermore, when the Pullin's model is employed as the collision kernel for a master equation, the moments (relaxation rates) can be analytically obtained. Therefore, the Pullin equation can be used for both numerical prediction and theoretical analysis, and the later one is important for the strictness of the extension works in gas kinetic theory.
 
 The following of this section is structured as follows: the original Pullin equation is outlined in Subsection \ref{sec:2.1}, along with a proof of its detailed balance property. Subsection \ref{sec:2.2} is a derivation of relaxation rates, where combined Hermite and Laguerre expansions are used for approximating the non-equilibrium distribution, and extra elementary integral terms are introduced. Subsection \ref{sec:2.3} presents an analysis of the derived macroscopic relaxation rates and establishes the links between the relaxation rates and the transport coefficients through the C-E expansion.
 
\subsection{Pullin equation and detailed balance}
\label{sec:2.1}
To characterize a molecular system via the distribution function within the framework of gas kinetic theory, the conventional phase (spanned by time $t$, spatial coordinates $\mathbf{x}$, and molecular velocity $\mathbf{c}$) must be extended to account for internal motions of molecules. This extension introduces an additional independent variable, which may be selected as the internal energy $\epsilon$, the discrete quantum states, or a generalized energy coordinate (in terms of momentum) $\mathbf{q}=\{\mathbf{q}_1,\mathbf{q}_2...\mathbf{q}_{\nu}\}$, where $\nu$ denotes the number of rotational degrees of freedom considered in this work. When the generalized energy coordinate $\mathbf{q}$ is adopted, the resulting distribution function takes the form $F(t,\mathbf{x},\mathbf{c},\mathbf{q})$. In the absence of external forces and heat sources, the corresponding extended Boltzmann equation \citep{chapman1990mathematical} is given by:
\begin{equation}
	\frac{\partial {{F}_{{1}}}}{\partial t}+{{\mathbf{c}}_{1}}\cdot \frac{\partial {{F}_{1}}}{\partial \mathbf{x}}=\iint{\!\!\ldots \!\!\int{\!\!\left( {{F}_{1}}^{\prime }{{F}_{2}}^{\prime }-{{F}_{1}}{{F}_{2}} \right)g\hat{\sigma }d\mathbf{{e}'}d{{\mathbf{c}}_{2}}d{{\mathbf{q}}_{2}}}},
	\label{eq:general_boltzmann}
\end{equation}
where the left hand side of this equation represents the free-transport term, while the right hand side corresponds to the binary collision term. The subscripts ``1'' and ``2'' distinguish the two molecules in a collision pair, and the prime symbol denotes post-collision quantities. Within the collision integral, $g$ is the relative speed of the colliding pair, $d\mathbf{{e}'}=\sin \chi \,d\chi \,d\theta$ is the solid angle element for the post-collision relative velocity $g^{\prime}$ (where $\chi$ is the deflection angle and $\theta$ is the azimuth angle), and $\hat{\sigma }=\hat{\sigma }(g,{{\mathbf{q}}_{1}},{{\mathbf{q}}_{2}},\chi ,\theta )$ is the differential cross-section, which serves as the kernel of the collision operator.

Based rigorously on the extended Boltzmann equation, the phenomenological Pullin equation~\citep{pullin1978kinetic} is formulated as follows:
\begin{equation}
	\frac{\partial {{f}_{1}}}{\partial t}+{{\mathbf{c}}_{1}}\cdot \frac{\partial {{f}_{1}}}{\partial \mathbf{x}}=\iint{\!\!\ldots \!\!\int{\!\!\left[ {{{{f}}}_{1}}'{{{{f}}}_{2}}'\mathcal{J}-{{f}_{1}}{{f}_{2}} \right]}}\,g{\sigma }{R\left(\left.\bm{\epsilon}\right|{\bm{\epsilon}}'\right)}d\mathbf{{e}'}d{{\epsilon}^{\prime}}d{{\mathbf{c}}_{2}}d{{\epsilon }_{2}}.
	\label{eq:pullin_middle}
\end{equation}
In the Pullin equation, the detailed description of the generalized coordinates $\mathbf{q}$ is reduced to the internal energy $\epsilon$, yielding a reduced distribution function $f\left(t, \mathbf{x},\mathbf{c},\epsilon\right)$. Correspondingly, the differential cross-section $\hat{\sigma }(g,{{\mathbf{q}}_{1}},{{\mathbf{q}}_{2}},\chi ,\theta )$ in the extended Boltzmann equation is reduced to the product form ${\sigma }(g,\chi,\theta)R\left(\left.\bm{\epsilon}\right|{\bm{\epsilon}}'\right)$, where $R\left(\left.\bm{\epsilon}\right|{\bm{\epsilon}}'\right)$ denotes the transition probability from the pre-collision energy state $\bm{\epsilon}=\left({\epsilon}_{1},{\epsilon}_{2},{\epsilon}_{\rm{t}}\right)$ to the post-collision state $\bm{\epsilon}'$. Here, ${\epsilon}_{i}$ represents the internal energy of a molecule and ${\epsilon}_{\rm{t}}$ the translational energy of the collision pair. The term ${\sigma }(g,\chi,\theta)$ is identical to the traditional elastic collision model, such as the inverse-power potential or its equivalent VHS model in DSMC. That implies that velocity deflection (scattering) and internal energy transfer are independent, which is consistent with mainstream DSMC practice and also serves to simplify the analytical procedure. As a result of this decoupling, internal motion (particularly its directional aspect) is disregarded in the scattering process, rendering the azimuth angle $\theta$ in ${\sigma }(g,\chi,\theta)$ a dummy variable. The factor $\mathcal{J}=\left({\epsilon}_{1}{\epsilon}_{2}/{{\epsilon}'_{1}{\epsilon}'_{2}}\right)^{\zeta-1}$ originates from the degeneracy of generalized energy coordinates, as embodied by the relation $f\left(t, \mathbf{x},\mathbf{c},\epsilon\right) \propto {\epsilon}^{\zeta-1}F(t,\mathbf{x},\mathbf{c},\mathbf{q})$~\citep{pullin1978kinetic}. For notational brevity in subsequent expressions, $\zeta=\nu/2$ is defined. Since the differential cross-section ${\sigma }(g,\chi,\theta)$ is the well-established elastic scattering cross-section, the central task in constructing the Pullin model reduces to identifying a suitable energy transition probability $R\left(\left.\bm{\epsilon}\right|{\bm{\epsilon}}'\right)$ that satisfies the fundamental requirements of a differential cross-section: (1) energy conservation $\bm{\epsilon}=\bm{\epsilon}'$; (2) nonnegativity $R(\left.\bm{\epsilon}  \right|{\bm{\epsilon} }')>0$; (3) normalization $\int{R(\left.\bm{\epsilon}  \right|\bm{{\epsilon }}')}d{\bm{\epsilon}'}=1$. These basic properties are proved in \citep{pullin1978kinetic}. In the present work, we provide a detailed examination of an important and more advanced property of the Pullin equation, the detailed balance property, as a supplement to the original derivation.

It is noteworthy that the physical process governed by $R(\left.\bm{\epsilon}  \right|{\bm{\epsilon} }')$ corresponds to a redistribution of internal energy between the two molecules during a collision. In the Pullin model, this redistribution is controlled by a set of random variables $\mathbf{s} = \left(s_{1}, s_{2}, s_{3}, s_{4}, s_{5}\right)$, where each $s_{i}\in\left(0,1\right)$ follows a beta distribution of the form:
\begin{equation}
	\beta \left\langle s_{i}\left| {b_{1}},{b_{2}} \right.\right\rangle =\frac{1}{B\left( {{b_{1}}},{{b_{2}}} \right)}{{s_{i}}^{{b_{1}}-1}}{{\left( 1-s_{i} \right)}^{{{b_{2}}}-1}},
	\label{eq:beta_distribution}
\end{equation}
where $B\left( {b_{1}},{b_{2}} \right)$ denotes the standard beta function, and the shape parameters $b_{1}$ and $b_{2}$ vary for each $s_{i}$. The energy transition during a forward collision proceeds in two steps. First, the active energy ${{\epsilon }_{\rm{a}}}$ of the colliding pair is drawn from the pre-collision energies $\epsilon_1$, $\epsilon_2$ and $\epsilon_{\rm{t}}$ according to the following expression:
\begin{equation}
	{{\epsilon }_{\rm{a}}}={{s}_{1}}{{\epsilon }_{1}}+{{s}_{2}}{{\epsilon }_{2}}+{{s}_{3}}{{\epsilon }_{\rm{t}}},
	\label{eq:maintain_active}
\end{equation}
where $s_1$, $s_2$ and $s_3$ follow the beta distributions ${\beta \left\langle \left.{{s}_{1}} \right|\phi \zeta ,(1-\phi )\zeta  \right\rangle }$, ${\beta \left\langle \left.{{s}_{2}} \right|\phi \zeta ,(1-\phi )\zeta  \right\rangle}$ and $\beta \left\langle \left.{{s}_{3}} \right|\psi \eta ,(1-\psi )\eta  \right\rangle$, respectively. Here, $\phi$ and $\psi$ are the two key parameters in the Pullin model, which will be determined in Section~\ref{sec:2.2} based on the rotational collision number $Z_{\rm{rot}}$ and the energy equal-partition principle. The parameter $\eta=2-2/{\alpha}$ is an external parameter from the elastic collision model, with ${\alpha}$ denoting the index of the inverse-power potential. Then, the active energy is partitioned among the post-collision internal energy ($\epsilon_{1}'$ and $\epsilon_{2}'$) and the post-collision translational energy $\epsilon_{\rm{t}}'$ as follows:
\begin{equation}
	\begin{aligned}
		& \epsilon_{1}' = (1-s_{1})\epsilon_{1} + s_{5}s_{4}\epsilon_{\rm{a}},\\ 
		& \epsilon_{2}' = (1-s_{2})\epsilon_{2} + s_{5}(1-s_{4})\epsilon_{\rm{a}},\\ 
		& \epsilon_{\rm{t}}' = (1-s_{3})\epsilon_{\rm{t}} + (1-s_{5})\epsilon_{\rm{a}},\\ 
	\end{aligned}
	\label{eq:redistribution}
\end{equation}
where $s_4$ and $s_5$ follow the beta distributions $\beta \left\langle \left.{{s}_{4}} \right|\phi \zeta ,\phi \zeta  \right\rangle$ and $\beta \left\langle \left.{{s}_{5}} \right|2\phi \zeta ,\psi \eta  \right\rangle$, respectively.

On the other hand, for the inverse collision process, to ensure an exactly reversed transition from $\mathbf{\epsilon}'$ to $\mathbf{\epsilon}$, $\mathbf{s}'$ must satisfy the following relations.
\begin{equation}
	\begin{aligned}
		&s_{1}'=\epsilon^{({\rm{r}})}_{1}/\left(\epsilon^{({\rm{r}})}_{1}+\epsilon^{({\rm{i}})}_{1}\right),\\
		&s_{2}'=\epsilon^{({\rm{r}})}_{2}/\left(\epsilon^{({\rm{r}})}_{2}+\epsilon^{({\rm{i}})}_{2}\right),\\
		&s_{3}'=\epsilon^{({\rm{r}})}_{\rm{t}}/\left(\epsilon^{({\rm{r}})}_{\rm{t}}+\epsilon^{({\rm{i}})}_{\rm{t}}\right),\\
		&s_{4}'=\epsilon^{({\rm{a}})}_{1}/\left(\epsilon^{({\rm{a}})}_{1}+\epsilon^{({\rm{a}})}_{2}\right),\\
		&s_{5}'=\left(\epsilon^{({\rm{a}})}_{1}+\epsilon^{({\rm{a}})}_{2}\right)/\epsilon_{{\rm{a}}},\\
	\end{aligned}
	\label{eq:inverse_s}
\end{equation}
where the superscripts (a), (i), (r) denote the active, inert and redistribution energy components, respectively. For clarity, the corresponding energies are defined as follows:
\begin{equation}
	\begin{aligned}
		\epsilon^{({\rm{a}})}_{1}   &= s_{1}\epsilon_{1},       & \epsilon^{({\rm{a}})}_{2}   &= s_{2}\epsilon_{2},       & \epsilon^{({\rm{a}})}_{\rm{t}} &= s_{3}\epsilon_{\rm{t}}, \\
		\epsilon^{({\rm{i}})}_{1}   &= (1-s_{1})\epsilon_{1},       & \epsilon^{({\rm{i}})}_{2}   &= (1-s_{2})\epsilon_{2},       & \epsilon^{({\rm{i}})}_{\rm{t}} &= (1-s_{3})\epsilon_{\rm{t}}, \\
		\epsilon^{({\rm{r}})}_{1}   &= s_{4}s_{5}\epsilon_{{\rm{a}}},  & \epsilon^{({\rm{r}})}_{2}   &= (1-s_{4})s_{5}\epsilon_{{\rm{a}}}, & \epsilon^{({\rm{r}})}_{\rm{t}} &= (1-s_{5})\epsilon_{{\rm{a}}}.
	\end{aligned}
	\label{eq:s_middle}
\end{equation}
Here, $s'_{1}$, $s'_{2}$ and $s'_{3}$ serve to decompose a post-collision energy into the inert and redistribution parts in forward collision, while $s'_4$ and $s'_5$ function to partition the total active energy into $\epsilon^{({\rm{a}})}_1$, $\epsilon^{({\rm{a}})}_2$ and $\epsilon^{({\rm{a}})}_{\rm{t}}$. This configuration of $\mathbf{s}'$ ensures an exactly reversible process and will subsequently guarantee the detailed balance property derived below. Then, the inverse collision parameters $\mathbf{s}'$ can be reformulated in terms of the forward collision parameters as:
\begin{equation}
	\begin{aligned}
		&s_{1}'=s_{4}s_{5}{\epsilon}_{{\rm{a}}}/{\epsilon}'_{1},\\
		&s_{2}'=(1-s_{4})s_{5}{\epsilon}_{{\rm{a}}}/{\epsilon}'_{2},\\
		&s_{3}'=(1-s_{5}){\epsilon}_{{\rm{a}}}/{\epsilon}'_{\rm{t}},\\
		&s_{4}'=s_{1}{\epsilon}_{1}/(s_{1}{\epsilon}_{1}+s_{2}{\epsilon}_{2}),\\
		&s_{5}'=(s_{1}{\epsilon}_{1}+s_{1}{\epsilon}_{2})/{\epsilon}_{{\rm{a}}}.\\
	\end{aligned}
	\label{eq:inverse_s_final}
\end{equation}
Note that the active energy and the deflection angle remain invariant under this transformation, i.e., ${\epsilon}_{{\rm{a}}}'={\epsilon}_{{\rm{a}}}$ and ${\chi}'={\chi}$. Furthermore, the relative speed after the collision is given by $g'=\sqrt{4{\epsilon}'_{\rm{t}}/m}$ (where $m$ is the molecular mass), while it is not equal to the pre-collision relative speed ($g'\neq g$). Accordingly, the impact parameters for the inverse collision are fully determined. Therefore, the transition probability $R(\left.\bm{\epsilon}\right|{\bm{\epsilon}}')$ in the Pullin equation can be constructed as:
\begin{equation}
	R(\left.\bm{\epsilon}\right|{\bm{\epsilon}}')d{\bm{\epsilon}}'=\int{h(\mathbf{s})d\mathbf{s}},
	\label{eq:R_define}
\end{equation}
where $h(\mathbf{s})$ denotes the joint probability density of the forward variables, given by the product of beta distributions:
\begin{equation}
	\begin{aligned}
		h(\mathbf{s}) = &{\beta \left\langle \left.{{s}_{1}} \right|\phi \zeta ,(1-\phi )\zeta  \right\rangle }{\beta \left\langle \left.{{s}_{2}} \right|\phi \zeta ,(1-\phi )\zeta  \right\rangle }\beta \left\langle \left.{{s}_{3}} \right|\psi \eta ,(1-\psi )\eta  \right\rangle\\
		{\cdot}&\beta \left\langle \left.{{s}_{4}} \right|\phi \zeta ,\phi \zeta  \right\rangle \beta \left\langle \left.{{s}_{5}} \right|2\phi \zeta ,\psi \eta  \right\rangle.
	\end{aligned}
	\label{hs}
\end{equation}
Finally, the specific form of Pullin equation can be expressed as:
\begin{equation}
	\frac{\partial {{f}_{1}}}{\partial t}+{{\mathbf{c}}_{1}}\cdot \frac{\partial {{f}_{1}}}{\partial \mathbf{x}}=\int{\!\!\ldots \!\!\int{\left[ {{{{f}}}_{1}}'{{{{f}}}_{2}}'\mathcal{J}-{{f}_{1}}{{f}_{2}} \right]g\sigma d\mathbf{{e}'}\,h(\mathbf{s})d\mathbf{s}\,d{{\mathbf{c}}_{2}}}d{{\epsilon }_{2}}},
	\label{eq:pullin2}
\end{equation}
which is now ready for verification of the detailed balance property. 

It is noteworthy that the physical interpretation of the detailed balance property is that the probability of an inverse collision is identical to its corresponding forward collision. For Equation~(\ref{eq:pullin2}), the following equality must be established:
\begin{equation}
	\sigma {{g}}\epsilon _{1}^{\zeta -1}\epsilon _{2}^{\zeta -1}h\left(\mathbf{s}\right)d{\mathbf{e'}}d{\mathbf{s}}d{\mathbf{c}_{1}}d{\mathbf{c}_{2}}d{\epsilon}_{1}d{\epsilon}_{2}
	= {\sigma }'{{g}'}{\epsilon}_{1}'^{\zeta -1}{\epsilon }_{2}'^{\zeta -1}h\left(\mathbf{s}'\right)d{\mathbf{e}}d{\mathbf{s}'}d{\mathbf{c}'_{1}}d{\mathbf{c}'_{2}}d{\epsilon}'_{1}d{\epsilon}'_{2}.
	\label{eq:SimplifiyDB1}
\end{equation}
In the current form, however, the functional dependence between the integration variables and the integrand is not explicitly revealed. To facilitate the proof, it is therefore necessary to reformulate the proposition. Noting that the molecular velocities $\mathbf{c}_1$ and $\mathbf{c}_2$ can be expressed in the terms of relative velocity $\mathbf{g}$ and the center-of-mass velocity $\mathbf{G}$ (there is ${d\mathbf{c}_1}{d\mathbf{c}_2}=d\mathbf{g}d\mathbf{G}$), and the same holds for the inverse collision. This transformation is essential because the integrand explicitly depends on $g=\left|\mathbf{g}\right|$. Furthermore, momentum conservation in a binary collision implies $\mathbf{G}=\mathbf{G}'$. Since $d\mathbf{G}$ and $d\mathbf{G}'$ are irrelevant to the integrand, they can be omitted in subsequent steps. On the other hand, the differential $d\mathbf{g}$ can be written as $g^{2}\sin(\theta_{g})dgd{\theta}_{g}d{\phi}_{g}$. Neither ${\theta}_{g}$ nor ${\phi}_{g}$ appears in the integrand, and $\sin(\theta_{g})$ is eliminated by integrating $\theta_{g}$ over $(0,\pi)$. Moreover, since $\chi'=\chi$, and $d\mathbf{{e}'}=\sin \chi \,d\chi \,d\theta$, the solid angle $d\mathbf{{e}'}$ also cancels out. Consequently, the detailed balance condition to be verified reduces to the simplified form:
\begin{equation}
	\sigma {{g^3}}\epsilon _{1}^{\zeta -1}\epsilon _{2}^{\zeta -1}h\left(\mathbf{s}\right)d{\mathbf{s}}d{g}d{\epsilon}_{1}d{\epsilon}_{2}
	= {\sigma }'{{g}'^3}{\epsilon }_{1}'^{\zeta -1}{\epsilon }_{2}'^{\zeta -1}h\left(\mathbf{s}'\right)d{\mathbf{s}'}d{g'}d{\epsilon}'_{1}d{\epsilon}'_{2}.
	\label{eq:SimplifiyDB2}
\end{equation}
Using the relations ${\sigma }'{{{g}'}^{{4}/{\alpha }}}=\sigma {{g}^{{4}/{\alpha }}}$ and $(g/2m)dg= d{\epsilon}_{\rm{t}}$, the expression can be further simplified to:
\begin{equation}
	\epsilon_{\rm{t}}^{\eta -1} \epsilon_{1}^{\zeta -1} \epsilon_{2}^{\zeta -1}h\left(\mathbf{s}\right)d{\mathbf{s}}d{\epsilon}_{\rm{t}}d{\epsilon}_{1}d{\epsilon}_{2}
	= {\epsilon }_{\rm{t}}'^{\eta -1} {\epsilon }_{1}'^{\zeta -1}{\epsilon }_{2}'^{\zeta -1}h\left(\mathbf{s}'\right)d{\mathbf{s}'}d{\epsilon}'_{\rm{t}}d{\epsilon}'_{1}d{\epsilon}'_{2}.
	\label{eq:SimplifiyDB3}
\end{equation}
Multiplying both sides of this equation by ${\exp}(-\epsilon_{\rm{t}}-\epsilon_{1}-\epsilon_{2})$ and introducing the gamma distribution defined as:
\begin{equation}
	\gamma \left\langle \left.x \right|a  \right\rangle =[{1}/{\Gamma (a )}\;]{{x}^{a -1}}{{e}^{-x}},
	\label{eq:GammaDefine}
\end{equation}
where $\Gamma(a)$ is the gamma function. Then Equation~(\ref{eq:SimplifiyDB1}) finally becomes:
\begin{equation}
	\gamma \left\langle \left.{{{{\epsilon}}}'_{1}} \right|\zeta  \right\rangle \gamma \left\langle \left.{{{{\epsilon}}}'_{2}} \right|\zeta  \right\rangle \gamma \left\langle \left.{{{{\epsilon}}}'_{\rm{t}}} \right|\eta  \right\rangle h(\mathbf{{s}'})d\bm{{\epsilon}}'\,d\mathbf{{s}'}=\gamma \left\langle \left.{{\epsilon }_{1}} \right|\zeta  \right\rangle \gamma \left\langle \left.{{\epsilon }_{2}} \right|\zeta  \right\rangle \gamma \left\langle \left.{{\epsilon }_{\rm{t}}} \right|\eta  \right\rangle h(\mathbf{s})d\bm{\epsilon }\,d\mathbf{s}.
	\label{eq:detailed balance}
\end{equation}

At this point, the propositional expression is sufficiently clear and concise. Then, the proof proceeds from the following expression, denoted by $R$:
\begin{equation}
	R=\gamma \left\langle \left.{{\epsilon }_{1}} \right|\zeta  \right\rangle \gamma \left\langle \left.{{\epsilon }_{2}} \right|\zeta  \right\rangle \gamma \left\langle \left.{{\epsilon }_{\rm{t}}} \right|\eta  \right\rangle \cfrac{h(\mathbf{s})ds_{1}ds_{2}ds_{3}ds_{4}ds_{5}d{\epsilon_{\rm{t}}}d{\epsilon_{1}}d{\epsilon_{2}}}{h(\mathbf{{s}'})ds'_{1}ds'_{2}ds'_{3}ds'_{4}ds'_{5}d{\epsilon'_{\rm{t}}}d{\epsilon'_{1}}d{\epsilon'_{2}}}.\\
	\label{eq:detailed balanceR}
\end{equation}
The relation between beta and gamma distributions enables the decomposition of one gamma distribution into a product of two gamma distributions as follows:
\begin{equation}
	\begin{aligned}
		&{{\gamma}\langle{x}|a_{1}+a_{2}\rangle}{{\beta}{\langle{z}|a_{1},a_{2}\rangle}}dxdz ={{\gamma}\langle{x_1}|a_{1}\rangle}{{\gamma}\langle{x_2}|a_{2}\rangle}dx_{1}dx_{2},\\
		&x_1=a_{1}x,{\quad}x_2=a_{2}x.
	\end{aligned}
	\label{eq:BetaGamma}
\end{equation}
Therefore, the gamma distributions in Equation~(\ref{eq:detailed balanceR}) are decomposed successively by the beta distributions in $h(\mathbf{s})$, which corresponds to the physical energy redistribution process and ensures mathematical consistency.

First, the terms associated with the forward collision process are calculated, in which $s_{1}$, $s_{2}$ and $s_{3}$ extract the active energy components from $\epsilon_{1}$, $\epsilon_{2}$ and $\epsilon_{\rm{t}}$. At this step, $R$ becomes:
\begin{equation}
	\begin{aligned}
		R &=\left\{\gamma \left\langle \left.{{\epsilon}}_{1}^{({\rm{a}})} \right|{\phi \zeta} \right\rangle 
		\gamma \left\langle \left.{{\epsilon}}_{2}^{({\rm{a}})} \right|{\phi \zeta} \right\rangle \gamma \left\langle \left.{{\epsilon}}_{\rm{t}}^{({\rm{a}})} \right|{\psi \eta} \right\rangle{\cfrac{d{\epsilon_{\rm{t}}^{({\rm{a}})}}d{\epsilon_{1}^{({\rm{a}})}}d{\epsilon_{2}^{({\rm{a}})}}}{ds'_{4}ds'_{5}}}\right\}\\
		&{\cdot}\gamma \left\langle \left.{{\epsilon}}_{1}^{({\rm{i}})} \right|{(1-\phi )\zeta} \right\rangle {\gamma \left\langle \left.{{\epsilon}}_{2}^{({\rm{i}})} \right|{(1-\phi )\zeta} \right\rangle  } {\gamma \left\langle \left.{{\epsilon}}_{\rm{t}}^{({\rm{i}})} \right|{(1-\psi )\eta} \right\rangle}\\
		&{\cdot}\cfrac{{\beta \left\langle \left.{{s}_{4}} \right|\phi \zeta ,\phi \zeta  \right\rangle \beta \left\langle \left.{{s}_{5}} \right|2\phi \zeta ,\psi \eta  \right\rangle}ds_{4}ds_{5}d{\epsilon_{\rm{t}}}^{({\rm{i}})}d{\epsilon_{1}^{({\rm{i}})}}d{\epsilon_{2}^{({\rm{i}})}}}{{h(\mathbf{{s}'})}ds'_{1}ds'_{2}ds'_{3}d{\epsilon'_{\rm{t}}}d{\epsilon'_{1}}d{\epsilon'_{2}}}.
	\end{aligned}
	\label{eq:proof1}
\end{equation}
Next, the terms enclosed in the curly brackets are considered. Given the presence of $ds'_{4}$ and $ds'_{5}$ in the expressions, the inverse collision process must be introduced. Based on Equation~(\ref{eq:inverse_s}), the reverse process governed by $s'_{4}$ and $s'_{5}$ can be expressed as follows:
\begin{equation}
	\begin{aligned}
		\gamma&  \left\langle \left.{{\epsilon}}_{\rm{a}} \right|{2\phi \zeta+{\psi}\eta} \right\rangle\beta \left\langle \left.{{s}'_{4}} \right|\phi \zeta ,\phi \zeta  \right\rangle \beta \left\langle \left.{{s}'_{5}} \right|2\phi \zeta ,\psi \eta  \right\rangle d{\epsilon}_{{\rm{a}}} ds'_{4} ds'_{5}\\
		&= \gamma \left\langle \left.{{\epsilon}}_{1}^{({\rm{a}})} \right|{\phi \zeta} \right\rangle \gamma \left\langle \left.{{\epsilon}}_{2}^{({\rm{a}})} \right|{\phi \zeta} \right\rangle \gamma \left\langle \left.{{\epsilon}}_{\rm{t}}^{({\rm{a}})} \right|{\psi \eta} \right\rangle d{\epsilon}^{({\rm{a}})}_{1} d{\epsilon}^{({\rm{a}})}_{2} d{\epsilon}^{({\rm{a}})}_{\rm{t}}.
	\end{aligned}
	\label{eq:invers4ps5p}
\end{equation}
Here, following Equation~(\ref{eq:invers4ps5p}), the active energy components $\epsilon_{1}^{({\rm{a}})}$, $\epsilon_{2}^{({\rm{a}})}$ and $\epsilon_{\rm{t}}^{({\rm{a}})}$ are recombined into the total active energy $\epsilon_{{\rm{a}}}$. Therefore, by inserting Equation~(\ref{eq:invers4ps5p}) into Equation~(\ref{eq:proof1}), $R$ further becomes:
\begin{equation}
	\begin{aligned}
		&R =\left\{{\gamma  \left\langle \left.{{\epsilon}}_{\rm{a}} \right|{2\phi \zeta+(1-\psi )\eta} \right\rangle}{\beta \left\langle \left.{{s}_{4}} \right|\phi \zeta ,\phi \zeta  \right\rangle \beta \left\langle \left.{{s}_{5}} \right|2\phi \zeta ,\psi \eta  \right\rangle}ds_{4}ds_{5}\right\}\\
		&\cdot\gamma \left\langle \left.{{\epsilon}}_{1}^{({\rm{i}})} \right|{(1-\phi )\zeta} \right\rangle
		\gamma \left\langle \left.{{\epsilon}}_{2}^{({\rm{i}})} \right|{(1-\phi )\zeta} \right\rangle  \gamma \left\langle \left.{{\epsilon}}_{\rm{t}}^{({\rm{i}})} \right|{(1-\psi )\eta} \right\rangle\\
		&{\cdot}\cfrac{{d{\epsilon}_{{\rm{a}}}d{\epsilon_{\rm{t}}^{({\rm{i}})}}d{\epsilon_{1}^{({\rm{i}})}}d{\epsilon_{2}^{({\rm{i}})}}}}{{\beta \left\langle \left.{{{s}_{1}}'} \right|\phi \zeta ,(1-\phi )\zeta  \right\rangle }{\beta \left\langle \left.{{{s}_{2}}'} \right|\phi \zeta ,(1-\phi )\zeta  \right\rangle }\beta \left\langle \left.{{{s}_{3}}'} \right|\psi \eta ,(1-\psi )\eta  \right\rangle{ds'_{1}ds'_{2}ds'_{3}d{\epsilon'_{\rm{t}}}d{\epsilon'_{1}}d{\epsilon'_{2}}}}.
	\end{aligned}
	\label{eq:proof2}
\end{equation}
Following the procedure, the forward process governed by $s_{4}$ and $s_{5}$ within the curly brackets of Equation~(\ref{eq:proof2}) is evaluated. After rearrangement, $R$ reduces to the following form:
\begin{equation}
	\begin{aligned}
		R =&\cfrac{{\gamma \left\langle \left.{{\epsilon}}_{1}^{({\rm{i}})} \right|{(1-\phi )\zeta} \right\rangle
				\gamma \left\langle \left.{{\epsilon}}_{2}^{({\rm{i}})} \right|{(1-\phi )\zeta} \right\rangle  \gamma \left\langle \left.{{\epsilon}}_{\rm{t}}^{({\rm{i}})} \right|{(1-\psi )\eta} \right\rangle}\gamma \left\langle \left.{{\epsilon}}_{1}^{({\rm{r}})} \right|{\phi \zeta} \right\rangle \gamma \left\langle \left.{{\epsilon}}_{2}^{({\rm{r}})} \right|{\phi \zeta} \right\rangle \gamma \left\langle \left.{{\epsilon}}_{\rm{t}}^{({\rm{r}})} \right|{\psi \eta} \right\rangle }{{\beta \left\langle \left.{{{s}'_{1}}} \right|\phi \zeta ,(1-\phi )\zeta  \right\rangle }{\beta \left\langle \left.{{{s}'_{2}}} \right|\phi \zeta ,(1-\phi )\zeta  \right\rangle }\beta \left\langle \left.{{{s}'_{3}}} \right|\psi \eta ,(1-\psi )\eta  \right\rangle}\\
		&{\cdot}\cfrac{d{\epsilon}^{({\rm{r}})}_{\rm{t}}d{\epsilon}^{({\rm{r}})}_{1}d{\epsilon}^{({\rm{r}})}_{2}d{\epsilon_{\rm{t}}^{({\rm{i}})}}d{\epsilon_{1}^{({\rm{i}})}}d{\epsilon_{2}^{({\rm{i}})}}}{ds'_{1}ds'_{2}ds'_{3}d{\epsilon'_{\rm{t}}}d{\epsilon'_{1}}d{\epsilon'_{2}}}.
	\end{aligned}
	\label{eq:proof3}
\end{equation}
Equation~(\ref{eq:proof3}) describes the reverse process governed by $s'_{1}$, $s'_{2}$ and $s'_{\rm{t}}$, expressed as:
\begin{equation}
	\begin{aligned}
		&{\gamma \left\langle \left.{{\epsilon}}_{1}^{({\rm{i}})} \right|{(1-\phi )\zeta} \right\rangle}{\gamma \left\langle \left.{{\epsilon}}_{1}^{({\rm{r}})} \right|{\phi \zeta} \right\rangle} {d{\epsilon}^{({\rm{i}})}_{1}}{d{\epsilon}^{({\rm{r}})}_{1}}= {\gamma \left\langle \left.{{\epsilon }'_{1}} \right|\zeta  \right\rangle}{\beta \left\langle \left.{{{s}'_{1}}} \right|\phi \zeta ,(1-\phi )\zeta  \right\rangle }{d{\epsilon}'_{1}}{ds'_1},\\
		&\gamma \left\langle \left.{{\epsilon}}_{2}^{({\rm{i}})} \right|{(1-\phi )\zeta} \right\rangle  {\gamma \left\langle \left.{{\epsilon}}_{2}^{({\rm{r}})} \right|{\phi \zeta} \right\rangle} {d{\epsilon}^{({\rm{i}})}_{2}}{d{\epsilon}^{({\rm{r}})}_{2}}= {\gamma \left\langle \left.{{\epsilon }'_{2}} \right|\zeta  \right\rangle}{\beta \left\langle \left.{{{s}'_{2}}} \right|\phi \zeta ,(1-\phi )\zeta  \right\rangle }{d{\epsilon}'_{2}}{ds'_2},\\
		&\gamma \left\langle \left.{{\epsilon}}_{\rm{t}}^{({\rm{i}})} \right|{(1-\psi )\eta} \right\rangle{\gamma \left\langle \left.{{\epsilon}}_{\rm{t}}^{({\rm{r}})} \right|{\psi \eta} \right\rangle }{d{\epsilon}^{({\rm{i}})}_{\rm{t}}}{d{\epsilon}^{({\rm{r}})}_{\rm{t}}}={\gamma \left\langle \left.{{\epsilon }'_{\rm{t}}} \right|\eta  \right\rangle}{\beta \left\langle \left.{{{s}'_{3}}} \right|\psi \eta ,(1-\psi )\eta  \right\rangle}{d{\epsilon}'_{\rm{t}}}{ds'_3},\\
	\end{aligned}
	\label{eq:invers1ps2ps3p}
\end{equation}
where the inert and redistribution energies recombine to form the post-collision energy. Substituting Equation~(\ref{eq:invers1ps2ps3p}) into Equation~(\ref{eq:proof3}), the final form of $R$ simplifies to:
\begin{equation}
	R= \gamma \left\langle \left.{{{{\epsilon }}}_{1}}' \right|\zeta  \right\rangle \gamma \left\langle \left.{{{{\epsilon }}}_{2}}' \right|\zeta  \right\rangle \gamma \left\langle \left.{{{{\epsilon }}}_{\rm{t}}}' \right|\eta  \right\rangle.
\end{equation}
With this, the proposition is proved. To briefly summarize, the detailed balance property of Pullin model replies on two key elements: (1) the use of the beta function to probabilistically redistribute energies and its mathematical property for decomposing gamma distributions, and (2) the consistent definition of the inverse collision parameters $\mathbf{s}'$. 

Similar to the BL model, the Pullin model was originally developed for DSMC simulations of polyatomic gases. Although it has not seen widespread adoption, primarily due to its moderately high computational cost, the Pullin model offers distinct advantages in terms of analytical tractability and strict adherence to detailed balance principle, making it particularly suitable for theoretical investigations.

\subsection{Analytical relaxation rate}
\label{sec:2.2}
Time relaxation rates of non-equilibrium variables are essential in constructing thermal non-equilibrium governing equations. These variables include pressure tensor $\mathbf{p}$, translational and rotational temperatures ($T_{\rm{t}}$, $T_{\rm{r}}$), translational and rotational heat fluxes (${\mathbf{q}}_{\rm{t}}$, ${\mathbf{q}}_{\rm{t}}$), corresponding to both flow non-equilibrium and thermal non-equilibrium. For example, they serve as source terms in the thermal Navier-Stokes equations for continuum flows and the moment equations for low-speed rarefied flows, and provide essential modeling constraints for BGK-type and Fokker-Planck-type models for Boltzmann equation, which are widely used in constructing numerical schemes for multi-scale flows. Also, these relaxation rates share all the elementary integrals with the C-E expansion that connects the microscopic gas-kinetic theory with the macroscopic aerodynamics, and are also necessary for characterizing entropy evolution in non-equilibrium thermodynamics \citep{desvillettes2005trend}.

To determine these relaxation rates, the phase-space integrals (moments) of a extended Boltzmann equation should be calculated. However, the collision term cannot be analytically integrated when adopting the widely used BL model as the collision kernel. This is because of the forced energy conservation in sampling post-collision energies, which must be enforced in the BL for correct physical modeling. But, this treatment leads to multi-dimensional bounded integrations over the Maxwell-Boltzmann distribution for sampling post-collision energies, which consequently leads to nested error functions that cannot be analytically integrated. Therefore, the Pullin equation is adopted in the analytical derivations in this work, which is both analytically integrable and satisfies the detailed balance condition.

\subsubsection{Approximate distribution function}
Prior to calculating the moments of collision term, a definitive analytical approximation for the non-equilibrium distribution function must be constructed first. Accordingly, the following quasi-equilibrium distribution function is introduced, which serves as the basis for the approximate distribution function in subsequent procedures: 
 \begin{equation}
	{{f}^{\left( 0 \right)}}=n{{(\frac{m}{2\pi k{{T}_{\rm{t}}}})}^{\frac{3}{2}}}{{e}^{-\frac{m{{C}^{2}}}{2k{{T}_{\rm{t}}}}}}\cdot \frac{{{\epsilon }^{\frac{\nu }{2}-1}}}{\Gamma \left( \frac{\nu }{2} \right){{\left( k{{T}_{\rm{r}}} \right)}^{\frac{\nu }{2}}}}{{e}^{-\frac{\epsilon }{k{{T}_{\rm{r}}}}}}.
	\label{equilibrium distribution}
\end{equation}
This quasi-equilibrium distribution is a joint Maxwell-Boltzmann distribution, employing a Maxwell distribution for translational velocity and a Boltzmann distribution for internal energy. It is noteworthy that translational temperature $T_{\rm{t}}$ is used in Maxwell distribution and rotational temperature $T_{\rm{r}}$ is used in Boltzmann distribution. This configuration is used in \citep{chang1951transport} to handle strong thermal non-equilibrium where $T_{\rm{t}}$ and $T_{\rm{r}}$ exhibit a large departure. In contrast, the approach, choosing the averaged temperature $T$ in equilibrium state Eq.~(\ref{equilibrium distribution}) and expressing the difference of $T_{\rm{t}}$ and $T_{\rm{r}}$ by a second-order perturbation, can only handle small departures in different temperatures (weak thermal non-equilibrium).

Owing to the distinct functional forms of the equilibrium distributions for particle velocity and internal energy, the approximate distribution function is expanded in a coupled orthogonal basis comprising Hermite and Laguerre polynomials. The equilibrium distribution function for translational velocity corresponds to the weight function of the Hermite polynomial, $\omega_{\rm{H}} (\mathbf{\xi })={{{e}^{-{{\xi }^{2}}/2}}}/{{{(2\pi )}^{\frac{3}{2}}}}$, where $\xi=C/\sqrt{m/2kT_{\rm{t}}}$ is the dimensionless thermal velocity. Correspondingly, the equilibrium distribution for internal energy corresponds to the weight function of the Laguerre polynomial, $\omega_{\rm{L}} (\mathbf{\varepsilon })={{\varepsilon }^{N}}{{e}^{-\varepsilon }}$, where $\varepsilon=\epsilon/(kT_{\rm{r}})$ is the dimensionless internal energy. Based on the weight functions, the Hermite and Laguerre polynomials can be generated as:

\begin{equation}
	{{H}_{{{i}_{1}}{{i}_{2}}\ldots {{i}_{N}}}}(\xi )=\frac{{{(-1)}^{N}}}{\omega_{\rm{H}}(\mathbf{\xi }) }\frac{{{\partial }^{N}}\omega_{\rm{H}}(\mathbf{\xi }) }{\partial {{\xi }_{{{i}_{1}}}}\partial {{\xi }_{{{i}_{2}}}}\ldots \partial {{\xi }_{{{i}_{N}}}}},
	\label{Hermite define}
\end{equation}
\begin{equation}
	{{L}_{N}}\left( \varepsilon  \right)=\frac{{{\varepsilon }^{N}}}{\omega_{\rm{L}}(\mathbf{\varepsilon })N!}\frac{{{d}^{N}}\omega_{\rm{L}}(\mathbf{\varepsilon })}{d{{\varepsilon }^{N}}}.
	\label{Laguerre define}
\end{equation}
The approximate distribution function is expanded in a series of Hermite polynomials Eq.~(\ref{Hermite define}) and Laguerre polynomials Eq.~(\ref{Laguerre define}) as follows:
\begin{equation}
	\begin{aligned}
			f={{f}^{\left( 0 \right)}}&\left( a+{{a}_{i}}{{H}_{i}}+\frac{1}{2}{{a}_\mathit{ij}}{{H}_\mathit{ij}}+\ldots+\frac{1}{N!}{{a}_{{{i}_{1}}{{i}_{2}}\ldots {{i}_{N}}}}{{H}_{{{i}_{1}}{{i}_{2}}\ldots {{i}_{N}}}}+\ldots \right)\\
			\cdot&\left( b_0+{{b}_{1}}{{L}_{1}}+\ldots+{{b}_{N}}{{L}_{N}}+\ldots \right),
	\end{aligned}
	\label{define approximate distribution function}
\end{equation}
where ${{a}_{{{i}_{1}}{{i}_{2}}\ldots {{i}_{N}}}}\left( \mathbf{x},t \right)$ and ${{b}_{N}}\left( \mathbf{x},t \right)$ denote tensor coefficients dependent on spatial coordinate $\mathbf{x}$ and time $t$. To accurately characterize non-equilibrium states, the constructed approximate distribution function should ensure the accuracy of key macroscopic quantities, including mass density $\rho (\mathbf{x},t)$, bulk velocity $\mathbf{v}(\mathbf{x},t)$, pressure deviator ${{p}_\mathit{\langle ij\rangle }}(\mathbf{x},t)$, translational temperature ${T}_{\rm{t}}(\mathbf{x},t)$, rotational temperature ${T}_{\rm{r}}(\mathbf{x},t)$, translational heat flux $\mathbf{q}_{\rm{t}}(\mathbf{x},t)$ and rotational heat flux $\mathbf{q}_{\rm{r}}(\mathbf{x},t)$. For simplicity, the $(\mathbf{x},t)$ notation of  subsequent macroscopic variables is omitted. These macroscopic quantities can be calculated from the moment of the distribution function:
\begin{equation}
	\begin{aligned}
		& \rho =\int{mf\text{d}\mathbf{c}}\text{d}\epsilon, \quad 0=\int{m{{C}_{i}}f\text{d}\mathbf{c}}\text{d}\epsilon,  \\ 
		& {{T}_{\rm{t}}}=\frac{1}{3nk }\int{m{{C}^{2}}f\text{d}\mathbf{c}}\text{d}\epsilon, \quad {{T}_{\rm{r}}}=\frac{2}{\nu nk }\int{\epsilon f\text{d}\mathbf{c}}\text{d}\epsilon,  \\ 
		& {{p}_\mathit{\langle ij\rangle }}=\int{m{{C}_{\langle i}}{{C}_{j\rangle }}f\text{d}\mathbf{c}}\text{d}\epsilon, \quad {{q}_{{\rm{t}},i}}=\int{\frac{1}{2}m{{C}^{2}}{{C}_{i}}f\text{d}\mathbf{c}}\text{d}\epsilon,  \\ 
		& {{q}_{{\rm{r}},i}}=\int{\epsilon {{C}_{i}}f\text{d}\mathbf{c}}\text{d}\epsilon.  \\ 
	\end{aligned}
	\label{moment}
\end{equation}
In Equation~(\ref{define approximate distribution function}), the order of the Hermite polynomials corresponds to the power of particle velocity, and the order of the Laguerre polynomials to internal energy. To recover these key macroscopic quantities in Eq.~(\ref{moment}), the approximate distribution function is constructed using Hermite polynomials up to third order and Laguerre polynomials up to first order. The explicit expressions of the Hermite and Laguerre polynomials used are given as follows:
\begin{equation}
	\left\{
	\begin{aligned}
		& H = 1, \\
		& H_{i} = \xi_{i}, \\
		& H_\mathit{ij} = \xi_{i}\xi_{j} - \delta_\mathit{ij}, \\
		& H_\mathit{ijj} = \xi_{i}\xi_{j}\xi_{j} - 5\xi_{i},
	\end{aligned}
	\right.
	\quad \quad \quad
	\left\{
	\begin{aligned}
		& L_{0} = 1, \\
		& L_{1} = -\varepsilon + 1.
	\end{aligned}
	\label{Hermite-Laguerre}
	\right.
\end{equation}
Thus, with the accurate recovery of key macroscopic quantities, the expansion of the approximate distribution function is truncated as follows:
\begin{equation}
	f={{f}^{\left( 0 \right)}}\left(\left( a+{{a}_{i}}{{H}_{i}}+\frac{1}{2}{{a}_\mathit{ij}}{{H}_\mathit{ij}}+\frac{1}{10}{{a}_\mathit{ijj}}{{H}_\mathit{ijj}} \right)\cdot {{b}_{0}}+\left( a+{{a}_{i}}{{H}_{i}} \right)\cdot  {{b}_{1}}{{L}_{1}} \right).
	\label{distribution function0}
\end{equation}
The coefficients ${{a}_{{{i}_{1}}{{i}_{2}}\ldots {{i}_{N}}}}$ and ${{b}_{N}}$ can be solved by substituting and integrating the approximate distribution function Eq.~(\ref{distribution function0}) into the macroscopic moment Eq.~(\ref{moment}). The specific expression for the approximate distribution function can be written as:
\begin{equation}
	f={{f}^{\left( 0 \right)}}\left( 1+\frac{{{p}_\mathit{\langle ij\rangle }}}{2nk{{T}_{\rm{t}}}}{{\xi }_{i}}{{\xi }_{j}}+\frac{{{q}_{{\rm{t}},i}}}{5mn}{{\left( \frac{m}{k{{T}_{\rm{t}}}} \right)}^{\frac{3}{2}}}\left( {{\xi }_{i}}{{\xi }^{2}}-5{{\xi }_{i}} \right)+\frac{2}{\nu }{{q}_{{\rm{r}},i}}\sqrt{\frac{m}{k{{T}_{\rm{t}}}}}\frac{1}{nk{{T}_{\rm{r}}}}\left( \varepsilon -\frac{\nu }{2} \right){{\xi }_{i}} \right).
	\label{linearized function}
\end{equation}
At this stage, the explicit expression of the approximate distribution function is derived. This expression is also commonly used within the moment methods where the non-equilibrium macroscopic quantities are typically expressed by constitutive relations derived via the C-E expansion.
\subsubsection{Integration and simplification of the relaxation rates}

The relaxation rates can be derived by taking moments of the collision term in the Pullin equation. Therefore, the moment balance equations of the Pullin equation must be provided. Multiplying both sides of Pullin equation Eq.~(\ref{eq:pullin2}) by an arbitrary function $\varphi =\varphi (\mathbf{x},\mathbf{c},\epsilon ,t)$ and integrating over phase space $d{{\mathbf{c}}_{1}}d{{\epsilon }_{1}}$ yields the moment balance equation:
\begin{equation}
	\int{\varphi\frac{\partial {{f}_{1}}}{\partial t} \text{d}{{\mathbf{c}}_{1}}\text{d}{{\epsilon }_{1}}}+\int{\varphi \frac{\partial {{f}_{1}}}{\partial \mathbf{x}}}\cdot{{\mathbf{c}}_{1}} \text{d}{{\mathbf{c}}_{1}}\text{d}{{\epsilon }_{1}}=\int{\!\!\ldots \!\!\int{\varphi \left[ {{{{f}}}_{1}'}{{{{f}}}_{2}'}\mathcal{J}-{{f}_{1}}{{f}_{2}} \right]g\sigma \text{d}\mathbf{{e}'}\,h(\mathbf{s})\text{d}\mathbf{s}\,\text{d}{{\mathbf{c}}_{1}}\text{d}{{\mathbf{c}}_{2}}\text{d}{{\epsilon }_{1}}\text{d}{{\epsilon }_{2}}}}.
	\label{moment equation}
\end{equation}
where $\mathcal{J}=\left({\epsilon}_{1}{\epsilon}_{2}/{{\epsilon}'_{1}{\epsilon}'_{2}}\right)^{\zeta-1}$. In frameworks that calculate moments of the collision term, such as the C-E expansion, it is typically necessary to simplify the collision term using the detailed balance relation. According to Pullin's detailed balance relation Eq.~(\ref{eq:SimplifiyDB1}), the differential operators satisfy the following transformation relation:
\begin{equation}
	g\sigma \text{d}\mathbf{{e}'}\,h(\mathbf{s})\text{d}\mathbf{s}\,\text{d}{{\mathbf{c}}_{1}}\text{d}{{\mathbf{c}}_{2}}\text{d}{{\epsilon }_{1}}\text{d}{{\epsilon }_{2}}= {\left(\frac{{{\epsilon }'_{1}}{{\epsilon }'_{2}}}{{{\epsilon }_{1}}{{\epsilon }_{2}}}\right)}^{\zeta-1} \cdot {g}'{\sigma }'\text{d}\mathbf{e}\,h(\mathbf{{s}'})\text{d}\mathbf{{s}'}\,\text{d}{{\mathbf{{c}}}_{1}'}\text{d}{{\mathbf{{c}}}_{2}'}\text{d}{{{\epsilon }}_{1}'}\text{d}{{{\epsilon }}_{2}'},
	\label{detailed balance2}
\end{equation}
Additionally, Liouville's theorem states the conservation of phase-space volume for an ensemble of systems with same Hamiltonian function. Specifically, if at time $t$ these occupy an infinitesimal volume of the phase space, at any other time they occupy an equal volume \citep{chapman1990mathematical}. The Hamiltonian of the Pullin model is well-established \citep{pullin1978kinetic}. Therefore, according to Liouville's theorem, the phase spaces of the pre-collision and post-collision states satisfy $\text{d}\mathbf{c}_1 \text{d}\epsilon_1\text{d}\mathbf{c}_2 \text{d}\epsilon_2 =\text{d}\mathbf{c}_1' \text{d}\epsilon_1'\text{d}\mathbf{c}_2' \text{d}\epsilon_2' $. Based on Liouville's theorem and the reversibility of collisions, the transformation $\left( {{\mathbf{c}}_{1}},{{\mathbf{c}}_{2}},{{\epsilon }_{1}},{{\epsilon }_{2}} \right)\leftrightarrow \left( {{{\mathbf{{c}}}}_{1}'},{{{\mathbf{{c}}}}_{2}'},{{{{\epsilon }}}_{1}'},{{{{\epsilon }}}_{2}'} \right)$ is applied to the first term ${{f}_{1}'}{{f}_{2}'}$ of the collision term within Eq.~(\ref{moment equation}). Subsequently, to unify the integration domain, the first term of the collision integral is further simplified via detailed balance relation Eq.~(\ref{detailed balance2}), specifically:
\begin{equation}
	\begin{aligned}
		& \int{\ldots \int{\varphi \left[ {{{{f}}}_{1}'}{{{{f}}}_{2}}'\mathcal{J}-{{f}_{1}}{{f}_{2}} \right]g\sigma \text{d}\mathbf{{e}'}\,h(\mathbf{s})\text{d}\mathbf{s}\,\text{d}{{\mathbf{c}}_{1}}\text{d}{{\mathbf{c}}_{2}}\text{d}{{\epsilon }_{1}}\text{d}{{\epsilon }_{2}}}} \\ 
		= & \int{\ldots \int{\varphi \left[ {{{{f}}}_{1}'}{{{{f}}}_{2}}'{\left(\frac{{{\epsilon }_{1}}{{\epsilon }_{2}}}{{{\epsilon }'_{1}}{{\epsilon }'_{2}}}\right)}^{\zeta-1}-{{f}_{1}}{{f}_{2}} \right]g\sigma \text{d}\mathbf{{e}'}\,h(\mathbf{s})\text{d}\mathbf{s}\,\text{d}{{\mathbf{c}}_{1}}\text{d}{{\mathbf{c}}_{2}}\text{d}{{\epsilon }_{1}}\text{d}{{\epsilon }_{2}}}} \\ 
		= & \int{\ldots \int{{\varphi }'{{f}_{1}}{{f}_{2}}{\left(\frac{{{\epsilon }'_{1}}{{\epsilon }'_{2}}}{{{\epsilon }_{1}}{{\epsilon }_{2}}}\right)}^{\zeta-1}{g}'{\sigma }'\text{d}\mathbf{e}\,h(\mathbf{{s}'})\text{d}\mathbf{{s}'}\,d{{{\mathbf{{c}}}}_{1}'}\text{d}{{{\mathbf{{c}}}}_{2}'}\text{d}{{{{\epsilon }}}_{1}'}d{{{{\epsilon }}}_{2}'}}}- \\ 
		& \int{\ldots \int{\varphi {{f}_{1}}{{f}_{2}}g\sigma \text{d}\mathbf{{e}'}\,h(\mathbf{s})\text{d}\mathbf{s}\,\text{d}{{\mathbf{c}}_{1}}\text{d}{{\mathbf{c}}_{2}}\text{d}{{\epsilon }_{1}}\text{d}{{\epsilon }_{2}}}}  \\
		= & \int{\ldots \int{{\varphi }'{{f}_{1}}{{f}_{2}}g\sigma \text{d}\mathbf{{e}'}\,h(\mathbf{s})\text{d}\mathbf{s}\,\text{d}{{\mathbf{c}}_{1}}\text{d}{{\mathbf{c}}_{2}}\text{d}{{\epsilon }_{1}}\text{d}{{\epsilon }_{2}}}}- \\ 
		& \int{\ldots \int{\varphi {{f}_{1}}{{f}_{2}}g\sigma \text{d}\mathbf{{e}'}\,h(\mathbf{s})\text{d}\mathbf{s}\,\text{d}{{\mathbf{c}}_{1}}\text{d}{{\mathbf{c}}_{2}}\text{d}{{\epsilon }_{1}}\text{d}{{\epsilon }_{2}}}}  \\
		= & \int{\ldots \int{\left( {\varphi }'-\varphi  \right){{f}_{1}}{{f}_{2}}g\sigma \text{d}\mathbf{{e}'}\,h(\mathbf{s})\text{d}\mathbf{s}\,\text{d}{{\mathbf{c}}_{1}}\text{d}{{\mathbf{c}}_{2}}\text{d}{{\epsilon }_{1}}\text{d}{{\epsilon }_{2}}}},
	\end{aligned}
	\label{moment equation2}
\end{equation}
where the analytical expression of ${{f}_{1}}{{f}_{2}}$ is essential for taking moments of the collision term. The expression of ${{f}_{1}}{{f}_{2}}$ in the collision term can be obtained by using the approximate distribution function Eq.~(\ref{linearized function}):
\begin{equation}
	\begin{aligned}
		{{f}_{1}}{{f}_{2}}=&f_{1}^{\left( 0 \right)}f_{2}^{\left( 0 \right)}\left(1+P_1\right)\left(1+P_2\right)\\
		=&f_{1}^{\left( 0 \right)}f_{2}^{\left( 0 \right)}\left(1+P_1+P_2+P_1 P_2\right),
	\end{aligned}
\end{equation}
where $P_1$ and $P_2$ are the expansion polynomials of the approximate distribution function. The cross term $P_1 P_2$ , due to its higher-order nonlinearity, introduces significant complexity to the integration process. However, as the contribution of cross term $P_1 P_2$ to key macroscopic quantities in typical flows is weak, the present study focuses on the linearized formulation. Therefore, neglecting the nonlinear terms in the pressure deviator ${{p}_\mathit{\langle ij\rangle }}$ and heat fluxes ${{q}_{{\rm{t}},i}}$, ${{q}_{{\rm{r}},i}}$, the distribution function product ${{f}_{1}}{{f}_{2}}$ can be written as:
\begin{equation}
	{{f}_{1}}{{f}_{2}}=f_{1}^{\left( 0 \right)}f_{2}^{\left( 0 \right)}\left( \begin{aligned}
		& 1+\frac{{{p}_\mathit{\langle ij\rangle }}}{2nk{{T}_{\rm{t}}}}{{\xi }_{1,i}}{{\xi }_{1,j}}+\frac{{{p}_\mathit{\langle ij\rangle }}}{2nk{{T}_{\rm{t}}}}{{\xi }_{2,i}}{{\xi }_{2,j}}+ \\ 
		& \frac{{{q}_{{\rm{t}},i}}}{5mn}{{\left( \frac{m}{k{{T}_{\rm{t}}}} \right)}^{\frac{3}{2}}}\left( {{\xi }_{1,i}}\xi _{1}^{2}-5{{\xi }_{1,i}} \right)+\frac{{{q}_{{\rm{t}},i}}}{5mn}{{\left( \frac{m}{k{{T}_{\rm{t}}}} \right)}^{\frac{3}{2}}}\left( {{\xi }_{2,i}}\xi _{2}^{2}-5{{\xi }_{2,i}} \right)+ \\ 
		& \frac{2}{\nu }{{q}_{{\rm{r}},i}}\sqrt{\frac{m}{k{{T}_{\rm{t}}}}}\frac{1}{nk{{T}_{\rm{r}}}}\left( {{\varepsilon }_{1}}-\frac{\nu }{2} \right){{\xi }_{1,i}}+\frac{2}{\nu }{{q}_{{\rm{r}},i}}\sqrt{\frac{m}{k{{T}_{\rm{t}}}}}\frac{1}{nk{{T}_{\rm{r}}}}\left( {{\varepsilon }_{2}}-\frac{\nu }{2} \right){{\xi }_{2,i}} \\ 
	\end{aligned} \right).
	\label{f1f2}
\end{equation}

With the collision term simplified and the distribution function expressed approximately, the derivation of the relaxation rates can now be formally undertaken. 
When $\varphi$ takes ${mC_{1}^{2}}/{3nk}$,  ${2{{\varepsilon }_{1}}}/{{\nu}nk}$, $m{{C}_{1,i}}{{C}_{1,j}}$, ${mC_{1}^{2}{{C}_{1,i}}}/{2}$, ${{\varepsilon }_{1}}{{C}_{1,i}}$, the corresponding non-equilibrium quantities: the translational temperature relaxation rate ${{\mathcal{T}}_{\rm{t}}}$,  rotational temperature relaxation rate ${{\mathcal{T}}_{\rm{r}}}$, pressure deviator relaxation rate ${{\mathcal{P}}_\mathit{ij}}$,  translational heat flux relaxation rate ${{\mathcal{Q}}_{\rm{t}}}$, and rotational heat flux relaxation rate ${{\mathcal{Q}}_{\rm{r}}}$ can be obtained from the simplified collision term Eq.~(\ref{moment equation2}). For a concise expression during the derivation, a set of non-dimensional quantities is defined:
\begin{equation}
	\begin{cases}
		\xi_i = \dfrac{C_i}{\sqrt{\dfrac{kT_{\text{t}}}{m}}}, \\[25pt]
		\varepsilon = \dfrac{\epsilon}{kT_{\text{r}}}, \\[10pt]
		G_i^{*} = \dfrac{G_i}{\sqrt{\dfrac{kT_{\text{t}}}{m}}}, \\[10pt]
		g_i^{*} = \dfrac{g_i}{\sqrt{\dfrac{kT_{\text{t}}}{m}}},
	\end{cases}
	\qquad\qquad
	\begin{cases}
	p_\mathit{\langle ij\rangle }^{*} = \dfrac{p_\mathit{\langle ij\rangle }}{nkT_{\text{t}}}, \\[15pt]
		q_{\text{t},i}^{*} = \dfrac{q_{\text{t},i}}{nm\Bigl(\dfrac{kT_{\text{t}}}{m}\Bigr)^{\!3/2}}, \\[25pt]
		q_{\text{r},i}^{*} = \dfrac{q_{\text{r},i}}{n\sqrt{\dfrac{kT_{\text{t}}}{m}}\,kT_{\text{r}}}.
	\end{cases}
\end{equation}
It is noteworthy that during the non-dimensionalization process, the translational and rotational physical quantities are scaled using the translational and rotational temperatures, respectively. 

 In this work, to express the relaxation rate integrals compactly, an integral operator ${{\mathcal{I}}_{1}}\left[ A,B \right]$ is defined as follows:
\begin{equation}
	{{\mathcal{I}}_{1}}\left[ A,B \right]={{\left( \frac{k{{T}_{\rm{t}}}}{m} \right)}^{\frac{7}{2}}}{{\left( k{{T}_{\rm{r}}} \right)}^{2}}\int{\left( {A}'-A \right)f_{1}^{\left( 0 \right)}f_{2}^{\left( 0 \right)}\left( {{B}} \right){{g}^{*}}\sigma \text{d}\mathbf{{e}'}h(\mathbf{s})\text{d}\mathbf{s}\text{d}{{\bm{\xi}_{1}}}\text{d}{{\bm{\xi}_{2}}}\text{d}{{\varepsilon }_{2}}\text{d}{{\varepsilon }_{1}}}.
	\label{integrate1}
\end{equation}
In the calculation of the relaxation rates, the integrals can be simplified according to the parity of the dimensionless thermal velocity ${{\mathbf{\xi}}_{1}}$, ${{\mathbf{\xi}}_{2}}$ to reduce the complexity of the derivation. Based on Eq.~(\ref{integrate1}), the relaxation rates can be explicitly expressed as a combination of integral operators:
\begin{equation}
	{{\mathcal{T}}_{\rm{t}}}=\frac{\partial {{T}_{\rm{t}}}}{\partial t} =\frac{k{{T}_{\rm{t}}}}{3nk}{{\mathcal{I}}_{1}}\left[ \xi _{1}^{2},1 \right],
	\label{Tt_int}
\end{equation}
\begin{equation}
	{{\mathcal{T}}_{\rm{r}}}=\frac{\partial {{T}_{\rm{r}}}}{\partial t}=\frac{2k{{T}_{\rm{r}}}}{{\nu}nk}{{\mathcal{I}}_{1}}\left[ {{\varepsilon }_{1}},1 \right],
\end{equation}
\begin{equation}
	{{\mathcal{P}}_\mathit{ij}}=\frac{p_\mathit{\langle ij\rangle }}{\partial t}=\frac{k{{T}_{\rm{t}}}}{2}{{\mathcal{I}}_{1}}\left[ {{\xi }_{\langle 1,i}}{{\xi }_{1,j\rangle }},{{\xi }_{\langle 1,k}}{{\xi }_{1,l\rangle }}+{{\xi }_{\langle 2,k}}{{\xi }_{2,l\rangle }} \right]p_\mathit{\langle kl\rangle }^{*},
\end{equation}
\begin{equation}
		{{\mathcal{Q}}_{\rm{t}}}=\frac{\partial {q_{{\rm{t}},i}}}{\partial t}={{\left( \frac{k{{T}_{\rm{t}}}}{m} \right)}^{\frac{3}{2}}}\frac{1}{2}m\left(
		\begin{aligned}
			&{{\mathcal{I}}_{1}}\left[ \xi _{1}^{2}{{\xi }_{1,i}},{{\xi }^{2}}{{\xi }_{1,k}}+{{\xi }^{2}}{{\xi }_{2,k}} \right]\frac{q_{{\rm{t}},k}^{*}}{5}
			+{{\mathcal{I}}_{1}}\left[ \xi _{1}^{2}{{\xi }_{1,i}},{{\varepsilon }_{1}}{{\xi }_{1,k}}+{{\varepsilon }_{2}} {{\xi }_{2,k}} \right]\frac{2q_{{\rm{r}},k}^{*}}{\nu }\\
			&-{{\mathcal{I}}_{1}}\left[ \xi _{1}^{2}{{\xi }_{1,i}},{{\xi }_{1,k}}+{{\xi }_{2,k}} \right]\left( q_{{\rm{t}},k}^{*}+q_{{\rm{r}},k}^{*} \right) 
		\end{aligned}
		 \right)  ,
\end{equation}
\begin{equation}
	{{\mathcal{Q}}_{\rm{r}}}=\frac{\partial {q_{{\rm{r}},i}}}{\partial t}={{\left( \frac{k{{T}_{\rm{t}}}}{m} \right)}^{\frac{1}{2}}}k{{T}_{\rm{r}}}\left(
	\begin{aligned}
		&{{\mathcal{I}}_{1}}\left[ {{\xi }_{1,i}}{{\varepsilon }_{1}},{{\xi }^{2}}{{\xi }_{1,k}}+{{\xi }^{2}}{{\xi}_{2,k}}\right]\frac{q_{{\rm{t}},k}^{*}}{5}+{{\mathcal{I}}_{1}}\left[ {{\xi }_{1,i}}{{\varepsilon }_{1}},{{\varepsilon }_{1}} {{\xi }_{1,k}}+{{\varepsilon }_{2}} {{\xi }_{2,k}} \right]\frac{2q_{{\rm{r}},k}^{*}}{\nu } \\
		&-{{\mathcal{I}}_{1}}\left[ {{\xi }_{1,i}}{{\varepsilon }_{1}},{{\xi }_{1,k}}+{{\xi }_{2,k}} \right]\left( q_{{\rm{t}},k}^{*}+q_{{\rm{r}},k}^{*} \right)
	\end{aligned}
	  \right).
	\label{Qr_int}
\end{equation}

 In subsequent derivations, only a set of integrals in the ${{\mathcal{I}}_{1}}\left[ A,B \right]$ form need to be calculated. To decouple the velocity and energy parts of the collision integral, the integration variables are first transformed from the individual particle thermal velocities $\mathbf{C}_1$, $\mathbf{C}_2$ to the relative velocity $\mathbf{g}$ and center-of-mass velocity $\mathbf{G}$. This transformation is standard, as the differential cross-section $\sigma$ depends only on $g=\left|\mathbf{g}\right|$ and deflection angle $\chi$. The pre-collision thermal velocities ${{\mathbf{C}}_{1}}$ and ${{\mathbf{C}}_{2}}$ and post-collision thermal velocities ${{\mathbf{{C}}}_{1}'}$ and ${{\mathbf{{C}}}_{2}'}$ are expressed as:
  \begin{equation}
  	{{\mathbf{C}}_{1}}=\mathbf{G}-\frac{1}{2}\mathbf{g},\quad \quad {{\mathbf{C}}_{2}}=\mathbf{G}+\frac{1}{2}\mathbf{g},\quad \quad {{\mathbf{{C}}}_{1}'}=\mathbf{G}'-\frac{1}{2}\mathbf{{g}'},\quad \quad {{\mathbf{{C}}}_{2}'}=\mathbf{G}'+\frac{1}{2}\mathbf{{g}'}.
  	\label{tranGg}
  \end{equation}
 The corresponding transformation of integration variables gives $\text{d}{\mathbf{C}_{1}}\text{d}{\mathbf{C}_{2}}=\text{d}{{\mathbf{G}}}\text{d}{{\mathbf{g}}}$. Noting momentum conservation implies $\mathbf{G}=\mathbf{{G}'}$, while energy exchange between translational and rotational modes leads to $\left|\mathbf{g}\right| \neq \left|\mathbf{g'}\right|$. Based on the assumption of decoupled velocity scattering and energy scattering in Pullin model, the integration over the relative velocity is separated into directional and magnitude components. The treatment of velocity scattering is first carried out. The differential cross-section can be expressed as $\sigma \text{d}\mathbf{{e}'}=b\text{d}b\text{d}\theta$, where $b$ is the miss distance and $\theta$ is the azimuthal angle. The relative velocity $\mathbf{g}$ is oriented along the $x_3$-axis, without loss of generality, and the post-collision relative velocity is expressed in spherical coordinates $\left( g,\chi ,\theta  \right)$:
\begin{equation}
	\begin{aligned}
		(\mathbf{g})=g\left( \begin{matrix}
			0  \\
			0  \\
			1  \\
		\end{matrix} \right),\quad \quad \quad \quad (\mathbf{{g}'})={g}'\left( \begin{matrix}
			\sin \chi \cos \theta   \\
			\sin \chi \sin \theta   \\
			\cos \chi   \\
		\end{matrix} \right),
	\end{aligned}
	\label{eq:directional scatter}
\end{equation}
where $\chi$ and $\theta$ are the deflection angle and the azimuthal angle respectively. The transformations in Eqs.~(\ref{tranGg}) and (\ref{eq:directional scatter}) are applied to the integral operator ${{\mathcal{I}}_{1}}\left[ A,B \right]$, followed by integrating the azimuthal angle $\theta$ over $\left(0,2\pi\right)$. The deflection angle $\chi$ will be treated collectively at the final stage. After this treatment, integral operator ${{\mathcal{I}}_{1}}\left[ A,B \right]$ can be simplified to integral operator ${{\mathcal{I}}_{2}}\left[ A \right]$:
\begin{equation}
	\begin{aligned}
		{{\mathcal{I}}_{2}}\left[ A \right]= \sqrt{\frac{k{{T}_{\rm{t}}}}{m}}
		\int & A\cdot {{n}^{2}}{{\left( \frac{1}{2\pi } \right)}^{3}}\frac{\varepsilon _{1}^{\frac{\nu }{2}-1}}{\Gamma \left( \frac{\nu }{2} \right)}\frac{\varepsilon _{2}^{\frac{\nu }{2}-1}}{\Gamma \left( \frac{\nu }{2} \right)}{{e}^{-{{G}^{{{*}^{2}}}}}}{{e}^{-\frac{{{g}^{{{*}^{2}}}}}{4}}}{{e}^{-{{\varepsilon }_{1}}}}{{e}^{-{{\varepsilon }_{2}}}} \\
		& \cdot {{g}^{*}} b\text{d}b\,h(\mathbf{s})\text{d}\mathbf{s}\,\text{d}{{\mathbf{G}}^{*}}\text{d}{{\mathbf{g}}^{*}}\text{d}{{\varepsilon }_{2}}\text{d}{{\varepsilon }_{1}}.
	\end{aligned}
	\label{integrate2}
\end{equation}
The correspondence between the specific integral operators ${{\mathcal{I}}_{1}}\left[ A,B \right]$ and ${{\mathcal{I}}_{2}}\left[ A \right]$ used in the relaxation rate integrals is provided in Appendix \ref{appendixC}. The center-of-mass velocity ${{\mathbf{G}}}$ and the solid angle of relative velocity ${{\mathbf{g}}}$ are subsequently integrated, while the magnitude of the relative velocity $g$ is maintained. Specifically, ${d{\mathbf{g}}}$ can be expressed in spherical coordinates as $g^{2}\sin(\theta_{g})dgd{\theta}_{g}d{\phi}_{g}$, with ${\theta}_{g}$ and ${\phi}_{g}$ integrated over $\left(0,\pi\right)$ and $\left(0,2\pi\right)$, respectively. With the treatment of velocity scattering largely complete here, Integral operator ${{\mathcal{I}}_{2}}\left[ A \right]$ can be further simplified to integral operator ${{\mathcal{I}}_{3}}\left[ A \right]$:
\begin{equation}
	{{\mathcal{I}}_{3}}\left[ A \right]=\sqrt{\frac{k{{T}_{\rm{t}}}}{m}}\int{A\cdot {{\left( \frac{1}{2\pi } \right)}^{3}}\frac{\varepsilon _{1}^{\frac{\nu }{2}-1}}{\Gamma \left( \frac{\nu }{2} \right)}\frac{\varepsilon _{2}^{\frac{\nu }{2}-1}}{\Gamma \left( \frac{\nu }{2} \right)}{{n}^{2}}{{e}^{-\frac{{{g}^{{{*}^{2}}}}}{4}}}{{e}^{-{{\varepsilon }_{1}}}}{{e}^{-{{\varepsilon }_{2}}}}{{g}^{*}}b\text{d}bh(\mathbf{s})\text{d}\mathbf{s}\text{d}{{g}^{*}}\text{d}{{\varepsilon }_{2}}\text{d}{{\varepsilon }_{1}}}.
	\label{integrate3}
\end{equation}
The explicit computational steps from integral operator ${{\mathcal{I}}_{1}}\left[ A,B \right]$ to ${{\mathcal{I}}_{2}}\left[ A \right]$ and  from ${{\mathcal{I}}_{2}}\left[ A,B \right]$ to ${{\mathcal{I}}_{3}}\left[ A \right]$ are provided in Appendix \ref{appendixC}. The subsequent treatment of energy scattering in the collision process is the core of this work and a key difference from previous studies on elastic collisions. In this integral step, calculating the post-collision translational and rotational energies is essential. According to the redistribution rule in Eq.~(\ref{eq:redistribution}), the post-collision rotational energies ${{\epsilon }_{1}}',{{\epsilon }_{2}}'$ and relative translational energy ${{{\epsilon }}_{\rm{t}}}'$ of the colliding particle pair can be explicitly expressed in terms of the pre-collision variables $({{\epsilon }_{1}},{{\epsilon }_{2}},{{{\epsilon }}_{\rm{t}}})$ and the partition variables $\mathbf{s} = \left(s_1, s_2, s_3, s_4, s_5\right)$. Additionally, the post-collision relative velocity can be expressed as:
\begin{equation}
	{g}'=\sqrt{\frac{4}{m}(1-{{s}_{3}}){{\epsilon }_{\rm{t}}}+(1-{{s}_{5}})\left( {{s}_{1}}{{\epsilon }_{1}}+{{s}_{2}}{{\epsilon }_{2}}+{{s}_{3}}{{\epsilon }_{\rm{t}}} \right)}.
	\label{after-g}
\end{equation}
Based on the energy-scattering rules in Eqs.~(\ref{eq:redistribution}) and (\ref{after-g}), the integration over the phase space $\text{d}{{\epsilon }_{1}}\text{d}{{\epsilon }_{2}}\text{d}\mathbf{s}$, spanned by rotational energies ${\epsilon }_{1},{\epsilon }_{2}$,  and partition variables $\mathbf{s}$, can be carried out. Noting that the partition variable $\mathbf{s}$ represents the fraction of energy redistributed during the scattering process, its integration is carried out over the interval $\left(0,1\right)$. 

After completing the integration for energy scattering, integral operator ${{\mathcal{I}}_{3}}\left[ A \right]$ can be written as functions of a series of element integrals like the use of $\Omega$ in traditional gas kinetic theory.
Once the molecular potential is specified, the values of the elementary integrals can be determined. Thus, the most compact form of the relaxation rates is presented first. At the end of this subsection, the specific expressions using the VHS model is provided. The integral over velocity scattering $b\text{d}b\text{d}g$, can be compactly expressed through the collision cross-section. The elementary collision cross-section integral ${{\Omega }_{l,r}}$ and ${{\hat{\Omega }}_{r}}$ are defined as follows:
\begin{equation}
	{{\Omega }_{l,r}}=\int{{{e}^{-\frac{{{g}^{*}}^{2}}{4}}}{{\left( {{g}^{*}} \right)}^{2r+3}}\left( 1-{{\cos }^{l}}\left[\chi\right]  \right)\,b\,\text{d}b\,\text{d}{{g}^{*}}},
\end{equation}

\begin{equation}
	{{\hat{\Omega }}_{r}}=\int{{{e}^{-\frac{{{g}^{*}}^{2}}{4}}}{{\left( {{g}^{*}} \right)}^{2r+3}} \,b\,\text{d}b\,\text{d}{{g}^{*}}},
\end{equation}
The definition of ${{\Omega }_{l,r}}$ is consistent with that of the collision cross-section integral in traditional theory, while ${{\hat{\Omega }}_{r}}$ serves as its supplement. These elementary collision cross-sections are analytically integrable. On the other hand, there are also the non-analytical integral terms in the heat flux relaxation rates, Eqs.~(\ref{I3_Omega+1}) and (\ref{I3_Omega+2}). 
For the conciseness of relaxation rates, these non-analytical terms are temporarily denoted as:
\begin{equation}
	\begin{aligned}
		\Omega _{1}^{+}=\int & \frac{{\pi }^{\frac{1}{2}}}{48}\cos \left[\chi\right] {{{{g}'}}^{*}}{{g}^{{{*}^{3}}}}\left( 10+{{g}^{{{*}^{2}}}} \right)\left( {{\varepsilon }_{1}}-{{\varepsilon }_{2}} \right)\\
		&\cdot\frac{\varepsilon _{1}^{\frac{\nu }{2}-1}}{\Gamma \left( \frac{\nu }{2} \right)}\frac{\varepsilon _{2}^{\frac{\nu }{2}-1}}{\Gamma \left( \frac{\nu }{2} \right)}{{e}^{-\frac{{{g}^{{{*}^{2}}}}}{4}}}{{e}^{-{{\varepsilon }_{1}}}}{{e}^{-{{\varepsilon }_{2}}}}{{g}^{*}}b\mathrm{d}b\,h(\mathbf{s})\mathrm{d}\mathbf{s}\,\mathrm{d}{{g}^{*}}\,\mathrm{d}{{\varepsilon }_{2}}\,\mathrm{d}{{\varepsilon }_{1}},
	\end{aligned}
	\label{omega1+}
\end{equation}

\begin{equation}
	\begin{aligned}
		\Omega _{2}^{+}=\int& -\frac{{{\pi }^{\frac{1}{2}}}}{12}\cos \left[\chi\right]{{{{g}'}}^{*}}{{g}^{{{*}^{3}}}}{{{{\varepsilon }}}_{1}}'\left( -{{\varepsilon }_{1}}+{{\varepsilon }_{2}} \right)\\
		&\cdot \frac{\varepsilon _{1}^{\frac{\nu }{2}-1}}{\Gamma \left( \frac{\nu }{2} \right)}\frac{\varepsilon _{2}^{\frac{\nu }{2}-1}}{\Gamma \left( \frac{\nu }{2} \right)}{{e}^{-\frac{{{g}^{{{*}^{2}}}}}{4}}}{{e}^{-{{\varepsilon }_{1}}}}{{e}^{-{{\varepsilon }_{2}}}}{{g}^{*}}b\mathrm{d}b\,h(\mathbf{s})\mathrm{d}\mathbf{s}\,\mathrm{d}{{g}^{*}}\,\mathrm{d}{{\varepsilon }_{2}}\,\mathrm{d}{{\varepsilon }_{1}}.
	\end{aligned}
	\label{omega2+}
\end{equation}
Once the adopted molecular potential is specified, the specific values of these non-analytical terms can be obtained via numerical integration. Using the Gaussian quadrature method, we evaluate Eqs.~(\ref{omega1+})–(\ref{omega2+}) under equilibrium temperature conditions ($T_{\rm{t}}=T_{\rm{r}}$). Table \ref{Omega numerical value} lists the numerical values of $\Omega _{1}^{+}$ and $\Omega _{2}^{+}$ for the hard sphere (HS) model, the variable hard sphere (VHS) model, and the variable soft sphere (VSS) models, all normalized by ${d^2_{\rm{ref}}}$. Evidently, when the HS or VHS model is employed, both $\Omega _{1}^{+}$ and $\Omega _{2}^{+}$ are numerically 0. This result can likewise be mathematically demonstrated by integrating $\Omega _{1}^{+}$ and $\Omega _{2}^{+}$ over the deflection angle $\chi$. Notably, for the VSS model, $\Omega _{1}^{+}$ is also zero while $\Omega _{2}^{+}$ takes a non-negligible value. The role of the non-analytical term $\Omega _{2}^{+}$ is reflected in the influence of the rotational heat flux on its own relaxation process.

\begin{table}
	\centering
	\renewcommand{\arraystretch}{1.7}   
	\setlength{\tabcolsep}{18pt}        
	
	\def~{\hphantom{0}}
	{\large  
		\begin{tabular}{lcc}
			       & ${\Omega _{1}^{+}}/{{d^2_{\rm{ref}}}}$ & ${\Omega _{2}^{+}}/{{d^2_{\rm{ref}}}}$\\[6pt]
			HS     & 8.11$\times 10^{-31}$ & 3.33$\times 10^{-15}$ \\[4pt]
			VHS    & 1.41$\times 10^{-30}$ & 4.83$\times 10^{-14}$ \\[4pt]
			VSS    & 1.82$\times 10^{-15}$ & 9.68 \\
	\end{tabular}}
	
	\caption{Numerical values of non-analytic terms $\Omega _{1}^{+}$ and $\Omega _{2}^{+}$.}
	\label{Omega numerical value}
\end{table}
 Based on the defined collision cross-section integral, the integral operators ${{\mathcal{I}}_{1}}\left[ A,B \right]$ that constitute the relaxation rates Eqs.~(\ref{Tt_int})-(\ref{Qr_int}) can be expressed in the following form:

\begin{equation}
	{{\mathcal{I}}_{1}}\left[ \xi _{1}^{2},1 \right]\text{=}\frac{{{n}^{2}}\sqrt{\pi }\sqrt{\frac{k{{T}_{\text{t}}}}{m}}}{4\left( v\phi +\eta \psi  \right)}\nu \phi \psi \left( \frac{{{T}_{\text{r}}}}{{{T}_{\text{t}}}}4\eta {{{\hat{\Omega }}}_{0}}-{{{\hat{\Omega }}}_{1}} \right),
\end{equation}

\begin{equation}
	{{\mathcal{I}}_{1}}\left[ {{\varepsilon }_{1}},1 \right]\text{=}\frac{{{n}^{2}}\sqrt{\pi }\sqrt{\frac{k{{T}_{\text{t}}}}{m}}}{8\left( v\phi +\eta \psi  \right)}\nu \phi \psi \left( -4\eta {{{\hat{\Omega }}}_{0}}+\frac{{{T}_{\text{t}}}}{{{T}_{\text{r}}}}{{{\hat{\Omega }}}_{1}} \right),
\end{equation}

\begin{equation}
	\begin{aligned}
		&{{\mathcal{I}}_{1}}\left[ {{\xi }_{\langle 1,i}}{{\xi }_{1,j\rangle }},{{\xi }_{\langle 1,k}}{{\xi }_{1,l\rangle }}+{{\xi }_{\langle 2,k}}{{\xi }_{2,l\rangle }} \right]\\
		=&\frac{{{n}^{2}}\sqrt{\pi }\sqrt{\frac{k{{T}_{\text{t}}}}{m}}\left( {{\delta }_{ik}}{{\delta }_{jl}}\text{+}{{\delta }_{il}}{{\delta }_{jk}} \right)}{240\left( v\phi +\eta \psi  \right)}\left( \begin{aligned}
			& 8\frac{{{T}_{\text{r}}}}{{{T}_{\text{t}}}}\nu \phi \psi \eta {{{\hat{\Omega }}}_{1}}-2\nu \phi \psi {{{\hat{\Omega }}}_{2}} \\ 
			& -12\frac{{{T}_{\text{r}}}}{{{T}_{\text{t}}}}\nu \phi \psi \eta {{\Omega }_{2,1}}-3\left( v\phi +\eta \psi -\nu \phi \psi  \right){{\Omega }_{2,2}} \\ 
		\end{aligned} \right),
	\end{aligned}
\end{equation}

\begin{equation}
	\begin{aligned}
		&{{\mathcal{I}}_{1}}\left[ \xi _{1}^{2}{{\xi }_{1,i}},\xi _{1}^{2}{{\xi }_{1,k}}+\xi _{2}^{2}{{\xi }_{2,k}} \right]\\
		=&\frac{{{n}^{2}}\sqrt{\pi }\sqrt{\frac{k{{T}_{\text{t}}}}{m}}{{\delta }_{ik}}}{48\left( v\phi +\eta \psi  \right)}\left( \begin{aligned}
			& 200\frac{{{T}_{\text{r}}}}{{{T}_{\text{t}}}}\nu \phi \psi \eta {{{\hat{\Omega }}}_{0}}+\left( -25+22\frac{{{T}_{\text{r}}}}{{{T}_{\text{t}}}}\eta  \right)2\nu \phi \psi {{{\hat{\Omega }}}_{1}}-11\nu \phi \psi {{{\hat{\Omega }}}_{2}} \\ 
			& -16\frac{{{T}_{\text{r}}}}{{{T}_{\text{t}}}}\nu \phi \psi \eta {{\Omega }_{2,1}}-4\left( v\phi +\eta \psi -\nu \phi \psi  \right){{\Omega }_{2,2}} \\ 
		\end{aligned} \right),
	\end{aligned}
\end{equation}

\begin{equation}
	{{\mathcal{I}}_{1}}\left[ \xi _{1}^{2}{{\xi }_{1,i}},{{\varepsilon }_{1}}{{\xi }_{1,k}}+{{\varepsilon }_{2}}{{\xi }_{2,k}} \right]\text{=}\frac{5{{n}^{2}}\sqrt{\pi }\sqrt{\frac{k{{T}_{\text{t}}}}{m}}{{\delta }_{ik}}}{24\left( v\phi +\eta \psi  \right)}\nu \phi \psi \left( \frac{{{T}_{\text{r}}}}{{{T}_{\text{t}}}}4\left( 1+\nu  \right)\eta {{{\hat{\Omega }}}_{0}}-\nu {{{\hat{\Omega }}}_{1}} \right)+{{{n}^{2}}\delta_{ik}\sqrt{\frac{k{{T}_{\text{t}}}}{m}}}\Omega _{1}^{+},
\end{equation}

\begin{equation}
	{{\mathcal{I}}_{1}}\left[ \xi _{1}^{2}{{\xi }_{1,i}},{{\xi }_{1,k}}+{{\xi }_{2,k}} \right]\text{=}-\frac{5{{n}^{2}}\sqrt{\pi }\sqrt{\frac{k{{T}_{\text{t}}}}{m}}{{\delta }_{ik}}}{12\left( v\phi +\eta \psi  \right)}\nu \phi \psi \left( -\frac{{{T}_{\text{r}}}}{{{T}_{\text{t}}}}4\eta {{{\hat{\Omega }}}_{0}}+{{{\hat{\Omega }}}_{1}} \right),
\end{equation}

\begin{equation}
	{{\mathcal{I}}_{1}}\left[ {{\xi }_{1,i}}{{\varepsilon }_{1}},\xi _{1}^{2}{{\xi }_{1,k}}+\xi _{2}^{2}{{\xi }_{2,k}} \right]\text{=}\frac{5{{n}^{2}}\sqrt{\pi }\sqrt{\frac{k{{T}_{\text{t}}}}{m}}{{\delta }_{ik}}}{96\left( v\phi +\eta \psi  \right)}\nu \phi \psi \left( -24\eta {{{\hat{\Omega }}}_{0}}+\left( 6\frac{{{T}_{\text{t}}}}{{{T}_{\text{r}}}}-4\eta  \right){{{\hat{\Omega }}}_{1}}+\frac{{{T}_{\text{t}}}}{{{T}_{\text{r}}}}{{{\hat{\Omega }}}_{2}} \right),
\end{equation}

\begin{equation}
	\begin{aligned}
		&{{\mathcal{I}}_{1}}\left[ {{\xi }_{1,i}}{{\varepsilon }_{1}},{{\varepsilon }_{1}}{{\xi }_{1,k}}+{{\varepsilon }_{2}}{{\xi }_{2,k}} \right]\text{ }\\
		=&\frac{{{n}^{2}}\sqrt{\pi }\sqrt{\frac{k{{T}_{\text{t}}}}{m}}{{\delta }_{ik}}}{48\left( v\phi +\eta \psi  \right)}\nu \left( -12\left( 1+\nu  \right)\eta \phi \psi {{{\hat{\Omega }}}_{0}}+\left( 3\frac{{{T}_{\text{t}}}}{{{T}_{\text{r}}}}\nu \phi \psi -2\left( v\phi +\eta \psi  \right) \right){{{\hat{\Omega }}}_{1}} \right)+{{{n}^{2}}\delta_{ik}\sqrt{\frac{k{{T}_{\text{t}}}}{m}}}\Omega _{2}^{+},
	\end{aligned}
\end{equation}

\begin{equation}
	{{\mathcal{I}}_{1}}\left[ {{\xi }_{1,i}}{{\varepsilon }_{1}},{{\xi }_{1,k}}+{{\xi }_{2,k}} \right]\text{=}\frac{{{n}^{2}}\sqrt{\pi }\sqrt{\frac{k{{T}_{\text{t}}}}{m}}{{\delta }_{ik}}}{8\left( v\phi +\eta \psi  \right)}\nu \phi \psi \left( -4\eta {{{\hat{\Omega }}}_{0}}+\frac{{{T}_{\text{t}}}}{{{T}_{\text{r}}}}{{{\hat{\Omega }}}_{1}} \right).
\end{equation}
The integral operator ${{\mathcal{I}}_{1}}\left[ A,B \right]$ in its simplest form is obtained through the above mathematical derivation and integration. By substituting this simplest form into the definition of the relaxation rates Eqs.~(\ref{Tt_int})-(\ref{Qr_int}), the relaxation rate can be analytically expressed as:
\begin{equation}
	{{\mathcal{T}}_{\rm{t}}}= -\frac{2}{3nk}\sqrt{\frac{k{{T}_{\rm{t}}}}{m}}\frac{\sqrt{\pi }{{n}^{2}}k{{T}_{\rm{t}}}{\nu}\phi \psi \left( {{{\hat{\Omega }}}_{1}}-4\eta \frac{{{T}_{\rm{r}}}}{{{T}_{\rm{t}}}}\,{{{\hat{\Omega }}}_{0}} \right)}{8\left( {\nu}\phi +\eta \psi  \right)},
	\label{pullin_Tt_relax}
\end{equation}
\begin{equation}
	{{\mathcal{T}}_{\rm{r}}}=\frac{2}{{\nu}nk}\sqrt{\frac{k{{T}_{\rm{t}}}}{m}}\frac{\sqrt{\pi }{{n}^{2}}k{{T}_{\rm{r}}}{\nu}\phi \psi \left( \frac{{{T}_{\rm{t}}}}{{{T}_{\rm{r}}}}{{{\hat{\Omega }}}_{1}}-4\eta \,{{{\hat{\Omega }}}_{0}} \right)}{8\left( {\nu}\phi +\eta \psi  \right)},
	\label{pullin_Tr_relax}
\end{equation}
\begin{equation}
	{{\mathcal{P}}_\mathit{ij}}=\frac{n\sqrt{\frac{\pi k{{T}_{\rm{r}}}}{m}}}{240\left( {\nu}\phi +\eta \psi  \right)}\left(\begin{aligned}
		&-12{\nu}\frac{{{T}_{\rm{r}}}}{{{T}_{\rm{t}}}}\eta \phi \psi {{\Omega }_{\left( 2,1 \right)}}-3\left( {\nu}\phi +\eta \psi -{\nu}\phi \psi  \right){{\Omega }_{\left( 2,2 \right)}}\\
		&+8{\nu}\frac{{{T}_{\rm{r}}}}{{{T}_{\rm{t}}}}\eta \phi \psi {{{\hat{\Omega }}}_{1}}-2{\nu}\phi \psi {{{\hat{\Omega }}}_{2}} 
	\end{aligned}\right)p_\mathit{\langle ij\rangle },
	\label{pullin_pij_relax}
\end{equation}
\begin{equation}
	\begin{aligned}
		{{\mathcal{Q}}_{{\rm{t}},i}}= &\frac{n\sqrt{\frac{\pi k{{T}_{\rm{r}}}}{m}}}{\left( {\nu}\phi +\eta \psi  \right)}\left(\begin{aligned}&\frac{1}{120}\left(  -4\frac{{{T}_{\rm{r}}}}{{{T}_{\rm{t}}}}\eta{\nu} \phi \psi {{\Omega }_{\left( 2,1 \right)}}-\left( {\nu}\phi +\eta \psi -{\nu}\phi \psi  \right){{\Omega }_{\left( 2,2 \right)}} \right)q_{{\rm{t}},i}\\
			&+\frac{1}{480}\left( -200\frac{{{T}_{\rm{r}}}}{{{T}_{\rm{t}}}}\eta{\nu} \phi \psi {{{\hat{\Omega }}}_{0}}+\left( 50+44\frac{{{T}_{\rm{r}}}}{{{T}_{\rm{t}}}}\eta  \right){\nu}\phi \psi {{{\hat{\Omega }}}_{1}}-11{\nu}\phi \psi {{{\hat{\Omega }}}_{2}} \right)q_{{\rm{r}},i}\\
			&+\frac{5}{6} {\nu}\eta \phi \psi {{{\hat{\Omega }}}_{0}}q_{{\rm{r}},i}
		\end{aligned}\right)\\
		&+\frac{{{T}_{\rm{t}}}}{{{T}_{\rm{r}}}}\frac{n\sqrt{\frac{k{{T}_{\mathrm{t}}}}{m}}}{\nu}\Omega _{1}^{+}q_{{\rm{r}},i},
	\end{aligned}
		\label{pullin_Qt_relax}
\end{equation}
\begin{equation}
	\begin{aligned}
		{{\mathcal{Q}}_{{\rm{r}},i}}= &\frac{n\sqrt{\frac{\pi k{{T}_{\rm{r}}}}{m}}}{\left( {\nu}\phi +\eta \psi  \right)}\left(\begin{aligned}&\frac{\nu \phi \psi}{96}\left( 24\frac{{{T}_{\rm{r}}}}{{{T}_{\rm{t}}}}\eta {{{\hat{\Omega }}}_{0}}-\left( 6+4\frac{{{T}_{\rm{r}}}}{{{T}_{\rm{t}}}}\eta  \right) {{{\hat{\Omega }}}_{1}}+ {{{\hat{\Omega }}}_{2}} \right)q_{{\rm{t}},i}\\
			&+\frac{1}{12}\left(6\eta \phi \psi {{{\hat{\Omega }}}_{0}}+\left( \nu\phi + \eta\psi  \right){{{\hat{\Omega }}}_{1}} \right)q_{{\rm{r}},i}
		\end{aligned}\right)\\
		&+\frac{2n\sqrt{\frac{k{{T}_{\mathrm{t}}}}{m}}}{\nu}\Omega _{2}^{+}q_{{\rm{r}},i}.
	\end{aligned}
		\label{pullin_Qr_relax}
\end{equation}
An analytical relaxation rate applicable to arbitrary molecular collision models is obtained through mathematical derivation. Consistent with conventional gas kinetic theory, the relaxation rates derived in this work depend on elementary collision cross-section integrals ${{\Omega }_{l,r}},{{\hat{\Omega }}_{r}}$ and macroscopic quantities like $n$, $m$, ${{T}_{\rm{t}}}$, ${{T}_{\rm{r}}}$, with free parameters $\phi$ and $\psi$ to be specified subsequently. 

\begin{figure}
	\centering
	\includegraphics[width=0.67\textwidth]{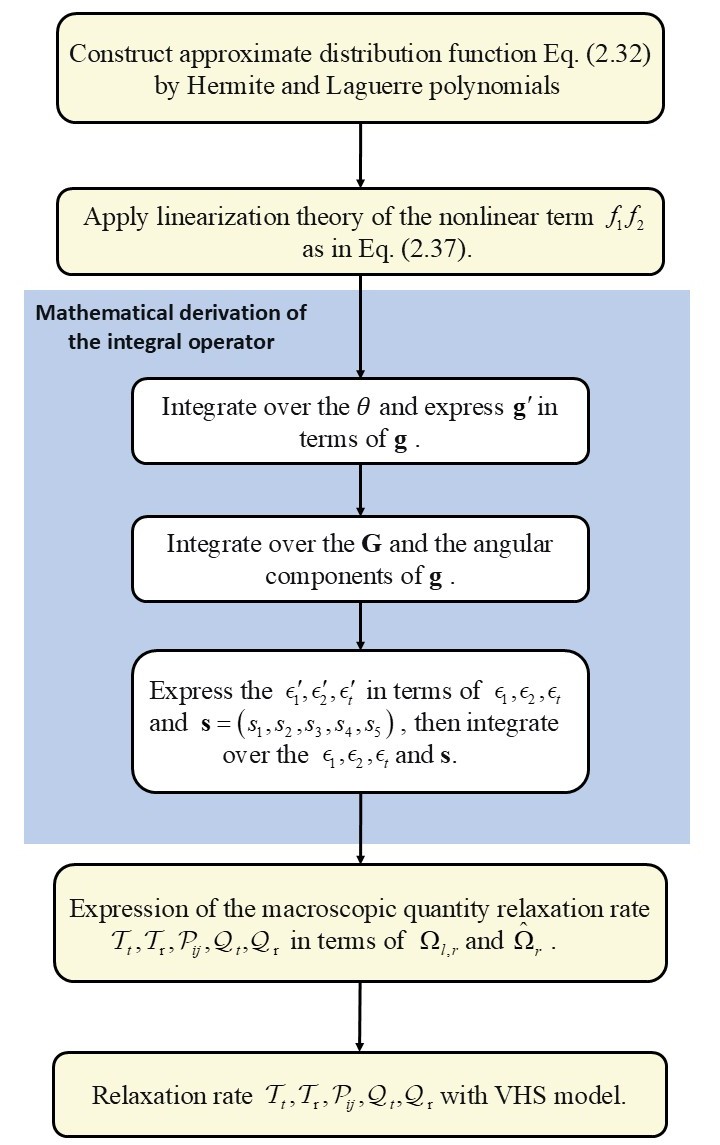}
	\caption{Mathematical derivation process of macroscopic quantity relaxation rate}
	\label{Mathematical Derivation Process}
\end{figure}

For the subsequent model validation, the relaxation rates implemented using the VHS model are provided. The collision cross-section integrals using the VHS model are given by:
\begin{equation}
	\begin{aligned}
		&\left\{
		\begin{matrix}
			\hat{\Omega}_{0}=4\left( \dfrac{2}{g_{\rm{ref}}} \right)^{1-2\omega} d_{\rm{ref}}^{2}\Gamma \left[ \dfrac{5}{2}-\omega  \right],  \\
			\hat{\Omega}_{1}=4\left( \dfrac{2}{g_{\rm{ref}}} \right)^{1-2\omega} d_{\rm{ref}}^{2}\Gamma \left[ \dfrac{7}{2}-\omega  \right],  \\
			\hat{\Omega}_{2}=4\left( \dfrac{2}{g_{\rm{ref}}} \right)^{1-2\omega} d_{\rm{ref}}^{2}\Gamma \left[ \dfrac{9}{2}-\omega  \right],  \\
		\end{matrix} \right. \\[12pt]
		&\left\{
		\begin{matrix}
			\Omega_{2,1}=\dfrac{8}{3}\left( \dfrac{2}{g_{\rm{ref}}} \right)^{1-2\omega} d_{\rm{ref}}^{2}\Gamma \left[ \dfrac{7}{2}-\omega  \right],  \\
			\Omega_{2,2}=\dfrac{8}{3}\left( \dfrac{2}{g_{\rm{ref}}} \right)^{1-2\omega} d_{\rm{ref}}^{2}\Gamma \left[ \dfrac{9}{2}-\omega  \right],  \\
		\end{matrix} \right. \\[12pt]
		&\left\{
		\begin{matrix}
			\Omega_{1}^{+}=0,  \\[8pt]
			\Omega_{2}^{+}=0.  \\
		\end{matrix} \right.
	\end{aligned}
\end{equation}
The collision cross-section integrals of the VHS model are substituted into the relaxation rate integrals Eqs.~(\ref{pullin_Tt_relax})–(\ref{pullin_Qr_relax}). The relaxation time ${\tau}_{\rm t}$ adopts the same definition as that in elastic collisions and can be evaluated from the ratio of the shear viscosity $\mu$ to the translational pressure ${p}_{\rm{t}}$. The relaxation time for the VHS model is given by:
 \begin{equation}
 	{\tau}_{\rm{t}} = \cfrac{15}{{{4}^{2-\omega }}d_{\rm{ref}}^{2}g_{\rm{ref}}^{2\omega -1}n{{\left( \cfrac{k{{T}_{\rm{t}}}}{m} \right)}^{1-\omega }}\sqrt{\pi }\Gamma \left[ \cfrac{9}{2}-\omega  \right]}.
 	\label{eq:tau}
 \end{equation}
Finally, the relaxation rates for the VHS model are given as follows: 
\begin{equation}
	{{\mathcal{T}}_{\rm{t}}}=\cfrac{\left( T-{{T}_{\rm{t}}} \right)}{{{Z}_{\rm{rot}}}{\tau}_{\rm{t}} },
	\label{relax_Tt_VHS}
\end{equation}
\begin{equation}
	{{\mathcal{T}}_{\rm{r}}}=\cfrac{\left( T-{{T}_{\rm{r}}} \right)}{{{Z}_{\rm{rot}}}{\tau}_{\rm{t}} },
\end{equation}
\begin{equation}
	{{\mathcal{P}}_\mathit{ij}}= -\cfrac{{{p}_\mathit{\langle ij\rangle }}}{{\tau}_{\rm{t}}},
\end{equation}
\begin{equation}
	{{\mathcal{Q}}_{{\rm{t}},i}}= -\left(\cfrac{2}{3}+ \frac{ 5\nu\left( \cfrac{{{T}_{\rm{r}}}}{{{T}_{\rm{t}}}}\left( \omega -1 \right)-\left( \omega -2 \right) \right)}{6\left(3+\nu\right){{Z}_{\rm{rot}}}} \right)\cfrac{{{q}_{\rm{t},i}}}{{\tau}_{\rm{t}}}+\cfrac{5}{2\left(3+\nu\right){{Z}_{\rm{rot}}}}\cfrac{{{q}_{\rm{r},i}}}{{\tau}_{\rm{t}}},
	\label{relax_Qt_VHS}
\end{equation}
\begin{equation}
	{{\mathcal{Q}}_{{\rm{r}},i}}= \cfrac{\nu\left( \cfrac{{{T}_{\rm{t}}}}{{{T}_{\rm{r}}}}\left( \omega -1 \right)-\left( \omega -2 \right) \right)}{2\left(3+\nu\right){{Z}_{\rm{rot}}}}\cfrac{{{q}_{\rm{t},i}}}{{\tau}_{\rm{t}}}-\left( \cfrac{5}{\left( 7-2\omega  \right)}+\cfrac{3}{2\left(3+\nu\right){{Z}_{\rm{rot}}}} \right)\cfrac{{{q}_{\rm{r},i}}}{{\tau}_{\rm{t}}},
	\label{relax_Qr_VHS}
\end{equation}
where ${{Z}_{\rm{rot}}}$ represents the rotational collision number, defined as:
\begin{equation}
	{{Z}_{\rm{rot}}}=\frac{{{\tau }_{\rm{r}}}}{{{\tau}_{\rm{t}}}}.
\end{equation}
The translational relaxation time ${{\tau }_{\rm{t}}}$ and rotational relaxation time ${{\tau }_{\rm{r}}}$ can be obtained from the previously derived temperature relaxation rate Eq.~(\ref{pullin_Tr_relax}) ${{\mathcal{T}}_{\rm{r}}}=\left({T-{{T}_{\rm{r}}}}\right)/{{{\tau }_{\rm{r}}}}$ and pressure deviator relaxation rate Eq.~(\ref{pullin_pij_relax}) ${\mathcal{P}_\mathit{ij}}=-{{{p}_\mathit{\langle ij\rangle }}}/{{{\tau}_{\rm{t}}}}$, respectively. Therefore, the rotational collision number can be explicitly expressed as:
\begin{equation}
	{{Z}_{\rm{rot}}}=\cfrac{\left( 7-2\omega  \right)\left( {\nu}\phi +\eta \psi  \right)}{5\left(3+\nu\right)\phi \psi }.
	\label{Zrot1}
\end{equation}
To clarify the derivation process of the relaxation rate, figure~\ref{Mathematical Derivation Process} presents the complete procedure in a flowchart form.
\subsubsection{Determination of free parameters}
In the Pullin model, $\phi $ and $\psi $ are key parameters that control the energy distribution ratio at the microscopic collision process. The selection of free parameters should ensure the accuracy of both microscopic processes and macroscopic parameters. Pullin derived the relationship between the rotational collision number and the parameters $\phi $ and $\psi $ through the C-E expansion \citep{pullin1978kinetic}. However, if the goal is to ensure only the accuracy of the rotational collision number, one of the free parameters is redundant. Accordingly, Pullin introduced the assumption $\phi = \psi$ under the hard-sphere model. The explicit expressions of $\phi $ and $\psi $ under this assumption are:
\begin{equation}
	\phi =\psi =\frac{8\left( 2+\nu  \right)}{5\pi }\frac{1}{{{Z}_{\rm{rot}}}}.
	\label{phi_Pullin}
\end{equation}
Pullin originally determined the free parameters by focusing solely on recovering the rotational collision number $Z_{\rm{rot}}$ and introduced the simplifying assumption $\phi = \psi$. While effective for matching $Z_{\rm{rot}}$, this approach leaves an extra degree of freedom unconstrained and does not explicitly consider the microscopic energy partitioning process. In this work, the extra parameter is used to give more reasonable microscopic process. According to the energy equal-partition principle, the energy between two colliding particles is redistributed based on their respective degrees of freedom. In the energy partitioning function Eq.~(\ref{hs}), ${\beta \left\langle \left.{{s}_{1}} \right|\phi \zeta ,(1-\phi )\zeta  \right\rangle }$ and $\beta \left\langle \left.{{s}_{3}} \right|\psi \eta ,(1-\psi )\eta  \right\rangle$ respectively control the energy distribution between rotational and translational energy during the collision process, where $\phi \zeta$ and $\psi \eta$ represent the allocation ratios of rotational energy and translational energy, respectively. According to the energy equal-partition principle, the relationship between the parameters $\phi \zeta$ and $\psi \eta$ in the energy partitioning function is given as follows:
\begin{equation}
	\frac{\phi \zeta }{\psi \eta }=\frac{\nu }{3}.
	\label{phi-psi}
\end{equation}
Meanwhile, to ensure the accuracy of the rotational collision number, the correspondence between the rotational collision number and the free parameters given by Eq.~(\ref{Zrot1}) must also be satisfied. Likewise, the free parameters $\phi$ and $\psi$ can be obtained from Eq.~(\ref{Zrot1}) and Eq.~(\ref{phi-psi}) as follows:
\begin{equation}
	\phi =\frac{2\left( 3+2{\nu} \right)\eta \left( 1+\eta  \right)}{15\left(3+\nu\right){{Z}_{\rm{rot}}}},
	\label{phi_present}
\end{equation}
\begin{equation}
	\psi =\frac{\left( 3+2{\nu} \right)\left( 1+\eta  \right)}{5\left(3+\nu\right){{Z}_{\rm{rot}}}}.
	\label{psi_present}
\end{equation}
It is noteworthy that the determination of free parameters via either Eq.~(\ref{phi_Pullin}) or Eqs.~(\ref{phi_present})-(\ref{psi_present}), does not affect the relaxation rates of the non-equilibrium macroscopic quantities. In contrast, Eqs.~(\ref{phi_present})-(\ref{psi_present}), which are derived from the equipartition theorem, provide a more physically reasonable description of the microscopic partitioning process. Moreover, the new determination of free parameters extends the applicability of Pullin model from hard sphere model to the variable hard sphere model.

 Thus far, the analytical relaxation rates for polyatomic gases considering rotational energy excitation have been derived by taking moments of the collision term in the Pullin model. The analytical expression for the relaxation rates elucidates the relaxation process of key macroscopic physical quantities, thereby providing a clear theoretical foundation and framework for analyzing thermally non-equilibrium relaxation behavior. Furthermore, the analytical relaxation rates implies that the non-equilibrium degree ${{T}_{\rm{t}}}/{{T}_{\rm{r}}}$ has a discernible influence on the relaxations of the heat flux. Meanwhile, these analytical relaxation rates can be directly linked to the macroscopic transport coefficients obtained from the C-E expansion. These points will be addressed in greater detail in the following subsections.
 
\subsection{Theoretical analysis of the obtained relaxation rate}
\label{sec:2.3}
Compared with monoatomic gases, polyatomic gases contain additional rotational energy, which introduces energy exchange between translational and rotational modes during non-equilibrium relaxation. As a result, the relaxation process of polyatomic gases becomes inherently more complex. In spatially homogeneous systems, the temperature relaxation behavior can be described by the Jeans–Landau equation:

 \begin{equation}
	{{\mathcal{T}}_{\rm{t}}}=\frac{p_{\rm{t}}}{\mu}\frac{T-{{T}_{\rm{t}}}}{Z_{\rm{rot}}},
	\label{Jeans_Tt}
\end{equation}
 \begin{equation}
	{{\mathcal{T}}_{\rm{r}}}=\frac{p_{\rm{t}}}{\mu}\frac{T-{{T}_{\rm{r}}}}{Z_{\rm{rot}}},
		\label{Jeans_Tr}
\end{equation}
where ${p}_{\rm{t}}$ is kinetic pressure, $\mu $ is shear viscosity. The temperature relaxation rate can be controlled by adjusting the rotational collision number ${Z_{\rm{rot}}}$. The relaxation processes of the translational and rotational heat fluxes can be described as follows \citep{mason1962heat,mccormack1968kinetic}:
\begin{equation}
	\begin{aligned}
		& {{\mathcal{Q}}_{\rm{t}}}=-\frac{p_{\rm{t}}}{\mu}\left( {{A}_{\rm{tt}}}{{{q}_{{\rm{t}},i}}}+{{A}_{\rm{tr}}}{{{q}_{{\rm{r}},i}}} \right), \\ 
		& {{\mathcal{Q}}_{\rm{r}}}=-\frac{p_{\rm{t}}}{\mu}\left( {{A}_{\rm{rt}}}{{{q}_{{\rm{t}},i}}}+{{A}_{\rm{rr}}}{{{q}_{{\rm{r}},i}}} \right). \\ 
	\end{aligned}
	\label{heatflux_relax}
\end{equation}
where ${{A}_{\rm{tt}}},{{A}_{\rm{tr}}},{{A}_{\rm{rt}}},{{A}_{\rm{rr}}}$ are heat flux relaxation coefficients, which characterize the influence of translational and rotational heat fluxes on the heat flux relaxation rates. The heat flux relaxation framework provided by Eq.~(\ref{heatflux_relax}) is derived from the classical Wang Chang-Uhlenbeck (WCU) equation. Due to the historical lack of comprehensive state-to-state collision cross-section databases, the heat flux relaxation coefficients within this framework could not be determined or could only be obtained by introducing specific assumptions. However, introducing assumptions such as elastic collisions and equilibrium temperatures leads to the loss of key physical mechanisms during the relaxation process. On the other hand, based on the heat flux relaxation framework in Eq.~(\ref{heatflux_relax}), the heat flux relaxation coefficients can be extracted via the least-squares method from the results of DSMC simulation \citep{li2021uncertainty} employing the Borgnakke–Larsen (BL) model. Through using DSMC to investigate heat flux relaxation rates, the numerical values of the heat flux relaxation coefficients under specific conditions are clarified, and the linear relationship between the heat flux relaxation coefficients and $1/{{Z}_{\rm{rot}}}$ is further identified. However, in the absence of theoretical derivation, this approach cannot yield explicit expressions for the heat flux relaxation rates, making it difficult to provide a more detailed and universally applicable description of the regularities governing the heat flux relaxation process. The determining factors and physical behavior of the coefficients $A_{ij}$ remain unclear, limiting an in-depth understanding of nonequilibrium transport mechanisms in polyatomic gases.

In this work, the specific and analytical relaxation rates derived from the Pullin equation provide a more complete and explicit theoretical foundation for analyzing relaxation processes. These relaxation rates exhibit good consistency in form with the classical relaxation framework given by Eqs.~(\ref{Jeans_Tt})-(\ref{relax_Qr_VHS}). In particular, the heat flux relaxation coefficients, which remain unspecified in the classical framework, can be explicitly expressed via these analytical results as:
\begin{equation}
	\begin{aligned}
		& {{A}_{\rm{tt}}}=\cfrac{2}{3}+ \frac{ 5\nu}{6\left(3+\nu\right){{Z}_{\rm{rot}}}}\left( \cfrac{{{T}_{\rm{r}}}}{{{T}_{\rm{t}}}}\left( \omega -1 \right)-\left( \omega -2 \right) \right), \\ 
		& {{A}_{\rm{tr}}}=-\cfrac{5}{2\left(3+\nu\right){{Z}_{\rm{rot}}}}, \\ 
		& {{A}_{\rm{rt}}}=-\cfrac{\nu}{2\left(3+\nu\right){{Z}_{\rm{rot}}}}\left( \cfrac{{{T}_{\rm{r}}}}{{{T}_{\rm{t}}}}\left( \omega -1 \right)-\left( \omega -2 \right) \right), \\ 
		& {{A}_{\rm{rr}}}=\cfrac{{5}}{(7 - 2{\omega})}+\cfrac{3}{2\left(3+\nu\right){{Z}_{\rm{rot}}}}.
	\end{aligned}
	\label{Aij}
\end{equation}
 According to the explicit expression in Eq.~(\ref{Aij}), the heat flux relaxation coefficients depend on the viscosity index $\omega$, the rotational collision number $Z_{\rm{rot}}$, and the translational-to-rotational temperature ratio $T_{\rm{t}}/T_{\rm{r}}$. The viscosity index $\omega $ and the rotational collision number ${{Z}_{\rm{rot}}}$ reflect the effects of shear viscosity and bulk viscosity on the thermal relaxation process respectively. This influence has a clear physical basis: exchanges of molecular momentum and energy between translational and rotational modes manifest macroscopically as shear viscosity Eq.~(\ref{pullin_shear}) and bulk viscosity Eq.~(\ref{pullin_bulk}), and these momentum and energy exchanges also influence higher-order heat flux. Specifically, the heat flux relaxation coefficients exhibit a linear relationship with $1/{{Z}_{\rm{rot}}}$. In the limit ${{Z}_{\rm{rot}}}\to \infty $, ${{A}_{\rm{tt}}}$ approaches ${2}/{3}$ (i.e., the Prandtl number), ${{A}_{\rm{rr}}}$ approaches the Schmidt number (Sc), while ${{A}_{\rm{tr}}}$ and ${{A}_{\rm{rt}}}$ tend to zero. This corresponds to the case of elastic collisions, where the relaxations of translational and rotational heat fluxes are decoupled. This behavior of the relaxation coefficients with respect to the rotational collision number is corroborated by DSMC-based investigations that extract heat flux relaxation rates \citep{li2021uncertainty}.

 Notably, the dependence on ${{T}_{\rm{t}}}/{{T}_{\rm{r}}}$, which quantifies the degree of thermodynamic non-equilibrium in the flow, indicates that heat flux relaxation is affected by the evolving non-equilibrium conditions. Specifically, the relaxation coefficients characterizing the influence of translational heat flux, ${{A}_{\rm{tt}}}$ and ${{A}_{\rm{rt}}}$, depend on ${{T}_{\rm{t}}}/{{T}_{\rm{r}}}$. Moreover, in terms of their specific numerical values, the self-influence coefficients ${{A}_{\rm{tt}}}$ and ${{A}_{\rm{rr}}}$ are positive. Because of the negative sign in Eq.~(\ref{heatflux_relax}), the translational and rotational heat fluxes effectively attenuate their own relaxation. Similarly, the cross-influence coefficients ${{A}_{\rm{tr}}}$ and ${{A}_{\rm{rt}}}$ are negative, which in practice leads to an reinforcing effect. For example, a large translational heat flux promotes an increase in the rotational heat flux via the coupled relaxation process. The coupling mechanism between heat flux relaxations of different energy modes exhibits similarities with the coupled relaxation process among different species in the gas mixtures. In the WCU equation, treating gas molecules of different energy modes as distinct species is also intended to capture the interactions between relaxation processes of different energy modes.
  
 The microscopic non-equilibrium relaxation behavior of gas molecules manifests macroscopically as constitutive relations in the Navier-Stokes (N-S) framework. This connection can be rigorously established through the C-E expansion (see Appendix~\ref{appendixA}). Within the N-S framework, the pressure tensor and the heat flux are described by the laws of Navier-Stokes and Fourier, respectively:
\begin{equation}
	{{p}_\mathit{ij}}=\left( p-{{\mu }_{\rm{b}}}\frac{\partial {{U}_{r}}}{\partial {{x}_{r}}} \right){{\delta }_\mathit{ij}}-2\mu \frac{\partial {{U}_{\langle i}}}{\partial {{x}_{j\rangle }}}
\end{equation}
\begin{equation}
	{{q}_{i}}=-\kappa \frac{\partial T}{\partial {{x}_{i}}}
\end{equation}
where $\mu$ is the shear viscosity, ${{\mu }_{\rm{b}}}$ is the bulk viscosity and $\kappa$ is the thermal conductivity. The particle momentum exchange during non-equilibrium relaxation is characterized by the pressure tensor relaxation rate. This exchange produces a shear stress during shear expansion or compression of the gas, which is quantified by the shear viscosity within the N-S framework. In the Pullin model, the shear viscosity can be written via the C-E expansion as:
\begin{equation}
	\mu =\tau_{\rm{t}}{{p}_{\rm{t}}} .
	\label{pullin_shear}
\end{equation}
 Compared to monatomic gases, polyatomic molecular gases exhibit an additional exchange between translational and internal (rotational) energy during non-equilibrium processes, which can be described by the temperature relaxation rate. During shear-free expansion or compression of a gas, the translational energy adjusts preferentially due to pressure work, whereas the adjustment of internal energy relies on inelastic collisions with a finite relaxation time and therefore typically lags behind the translational energy. This dissipative effect arising from the delayed relaxation of internal energy is phenomenologically represented by the bulk viscosity within the N-S framework. In the Pullin equation, the bulk viscosity can be written via the C-E expansion as:
\begin{equation}
	{{\mu }_{\rm{b}}}=\frac{2{\nu}}{3\left( 3+{\nu} \right)}{{Z}_{\rm{rot}}}\tau_{\rm{t}}{{p}_{\rm{t}}}.
	\label{pullin_bulk}
\end{equation}
As shown in Eqs.~(\ref{pullin_shear}) and (\ref{pullin_bulk}), the relaxation rates of momentum and energy exchange macroscopically correspond to shear and bulk viscosity dissipation; similarly, the relaxation rates of heat flux macroscopically correspond to heat transport. Via the C-E expansion, constitutive relation Eqs.~(\ref{Pullin_lambda_t}) and (\ref{Pullin_lambda_r}) can be extracted from the relaxation rates in the near-equilibrium relaxation. The correspondence between the thermal conductivity and the heat flux relaxation coefficient therein can be written as:
\begin{equation}
	\left[ \begin{matrix}
		{{\kappa}_{\rm{t}}}  \\
		{{\kappa}_{\rm{r}}}  \\
	\end{matrix} \right]=\frac{k\mu }{2m}{{\left[ \begin{matrix}
				{{A}_{\rm{tt}}} & {{A}_{\rm{tr}}}  \\
				{{A}_{\rm{rt}}} & {{A}_{\rm{rr}}}  \\
			\end{matrix} \right]}^{-1}}\left[ \begin{matrix}
		5  \\
		{\nu}  \\
	\end{matrix} \right].
	\label{thermal_define}
\end{equation}
Conventionally, to concisely express the relationship between thermal conductivity and shear viscosity, Eucken \citep{eucken1913warmeleitvermogen} defined dimensionless factors ${{f}_{\rm{eu}}}$, ${{f}_{\rm{t}}}$ and ${{f}_{\rm{r}}}$:
\begin{equation}
	{{f}_{\rm{eu}}}=\cfrac{3}{3+\nu}{{f}_{\rm{t}}}+\cfrac{\nu }{3+\nu}{{f}_{\rm{r}}}=\cfrac{2}{3+\nu}	\cfrac{\kappa m}{\mu k},
	\label{Kapa_define}
\end{equation}
\begin{equation}
	{{f}_{\rm{t}}}=\cfrac{2}{3}\cfrac{{{\kappa}_{\rm{t}}}m}{\mu k}, \;\;\; {{f}_{\rm{r}}}=\cfrac{2}{\nu}\cfrac{{{\kappa}_{\rm{r}}}m}{\mu k}.
\end{equation}
Under Eucken's framework, the total Eucken factor ${{f}_{\rm{eu}}}$ is a result of combining the translational Eucken factor ${{f}_{\rm{t}}}$ and rotational Eucken factor ${{f}_{\rm{r}}}$ at equilibrium temperature. Based on the thermal conductivities ${{\kappa}_{\rm{t}}}$ and ${{\kappa}_{\rm{r}}}$ derived from the C-E expansion, the translational and rotational Eucken factors can be directly expressed in terms of the heat flux relaxation coefficient $A_\mathit{ij}$ as:
\begin{equation}
	\left[ \begin{matrix}
		{{f}_{\rm{t}}}  \\
		{{f}_{\rm{r}}}  \\
	\end{matrix} \right]={{\left[ \begin{matrix}
				3{{A}_{\rm{tt}}} & {\nu}{{A}_{\rm{tr}}}  \\
				3{{A}_{\rm{rt}}} & {\nu}{{A}_{\rm{rr}}}  \\
			\end{matrix} \right]}^{-1}}\left[ \begin{matrix}
		5  \\
		{\nu}  \\
	\end{matrix} \right].
	\label{Eucken_define}
\end{equation}
The explicit expression of the Eucken factors can be derived from the heat flux relaxation coefficient $A_\mathit{ij}$ in Eq.~(\ref{Aij}):
\begin{equation}
	{{f}_{\rm{t}}}=\frac{\left(3+\nu\right)\left( {(7 - 2{\omega})} +{2}Z_{\rm{rot}}\right)}{{20(3+\nu)}Z_{\rm{rot}} +  6{(7 - 2{\omega})} +25\nu\left(\cfrac{{{T}_{\rm{r}}}}{{{T}_{\rm{t}}}} (\omega-1)-\left(\omega-2\right)\right)},
	\label{Eucken_ft}
\end{equation}
\begin{equation}
	{{f}_{\rm{r}}}=	\cfrac{\left(3+\nu\right){(7 - 2{\omega})}\left(4Z_{\text{rot}} + 5\left(\cfrac{{{T}_{\rm{r}}}}{{{T}_{\rm{t}}}} (\omega-1)-\left(\omega-2\right)\right)\right)}{{20(3+\nu)}Z_{\rm{rot}} +  6{(7 - 2{\omega})} +25\nu\left(\cfrac{{{T}_{\rm{r}}}}{{{T}_{\rm{t}}}} (\omega-1)-\left(\omega-2\right)\right)},
	\label{Eucken_fr}
\end{equation}
\begin{equation}
	{f}_{\rm{eu}}=\frac{(150+4\nu{(7 - 2{\omega})})Z_{\mathrm{rot}}+{(7 - 2{\omega})} \bigl(15+5\nu\left(\cfrac{{{T}_{\rm{r}}}}{{{T}_{\rm{t}}}} (\omega-1)-\left(\omega-2\right)\right)\bigr)}{{20(3+\nu)}Z_{\rm{rot}} +  6{(7 - 2{\omega})} +25\nu\left(\cfrac{{{T}_{\rm{r}}}}{{{T}_{\rm{t}}}} (\omega-1)-\left(\omega-2\right)\right)}.
	\label{Eucken_feu}
\end{equation}
Evidently, the dependence of the heat flux relaxation process on the non-equilibrium temperature  ${{T}_{\rm{t}}}/{{T}_{\rm{r}}}$ leads to the dependence of the macroscopic Eucken factor on the non-equilibrium temperature ${{T}_{\rm{t}}}/{{T}_{\rm{r}}}$, stemming from reliance on identical elementary integrals. The similar dependence of gas property on non-equilibrium degrees $T_{\rm{t}}/T_{\rm{r}}$ is also found in the classic research on rotation number $Z_{\rm{rot}}$ by Parker \citep{parker1959rotational} and Zhang \citep{zhang2014nonequilibrium}. Prior to the establishment of Eqs.~(\ref{Eucken_ft})-(\ref{Eucken_feu}), as research on transport coefficients was predominantly carried out under thermal equilibrium conditions \citep{mason1962heat}, the dependence on the non-equilibrium degree remained largely unrecognized. 

Mason and Monchick derived thermal conductivity and the Eucken factor from the WCU equation using a modified Chapman-Enskog expansion. Due to the technical limitations of calculating state-to-state collision cross-sections in early work, a series of assumptions (including thermal equilibrium, elastic collisions, and the rough-sphere model) were introduced to calculate the complex collision integrals. However, in strongly non-equilibrium flows such as shock waves and expansion waves, the introduction of these specific assumptions inevitably leads to the loss of information such as the influence of non-equilibrium temperature $T_{\rm{t}}/T_{\rm{r}}$. In contrast, Eqs.~(\ref{Eucken_ft})-(\ref{Eucken_feu}) derived in this work captures the missing information by being based on a more comprehensive theoretical framework that circumvents the need for state-to-state collision cross-sections. Additionally, by reintroducing the thermal equilibrium assumption, Eqs.~(\ref{Eucken_ft})-(\ref{Eucken_fr}) can be simplified and systematically compared with earlier results. For comparison with results of \citep{mason1962heat}, Eqs.~(\ref{Eucken_ft})-(\ref{Eucken_fr}) are first simplified under the conditions of thermal equilibrium ${{T}_{\rm{t}}}/{{T}_{\rm{r}}}=1$ and the relation ${\rho D}/{\mu } = {(7 - 2{\omega})}/{5}$, yielding:

\begin{equation}
	{{f}_{\rm{t}}}=\frac{5}{2}\left[ 1-\frac{5\nu}{4(3+\nu)}\cdot \dfrac{4\left(3+\nu\right)\left(1-\dfrac{2\rho D}{5\mu }\right) }{4(3+\nu){Z_{\rm{rot}}}+ \left\{6\dfrac{\rho D}{\mu }+5\nu\right\} }  \right],
	\label{eucken_ft_D}
\end{equation}

\begin{equation}
	{{f}_{\rm{r}}}=\frac{\rho D}{\mu }\left[ 1+\frac{15}{4(3+\nu)}\cdot \dfrac{4\left(3+\nu\right)\left(1-\dfrac{2\rho D}{5\mu }\right) }{4(3+\nu){Z_{\rm{rot}}}+ \left\{6\dfrac{\rho D}{\mu }+5\nu\right\} }  \right],
	\label{eucken_fr_D}
\end{equation}
where $D$ denotes the diffusion coefficient. A further simplification of the Eucken factor is achieved by adopting Mason's approach, keeping only first-order terms of $1/Z_{\rm{rot}}$. This is equivalent to neglecting the terms enclosed in curly braces, leading to:
\begin{equation}
	\begin{aligned}
		& {{f}_{\rm{t}}}=\frac{5}{2}\left[ 1-\frac{5\nu}{4(3+\nu){{Z}_{\rm{rot}}}}\left( 1-\frac{2\rho D}{5\mu } \right) \right], \\ 
		& {{f}_{\rm{r}}}=\frac{\rho D}{\mu }\left[ 1+\frac{15}{4(3+\nu){{Z}_{\rm{rot}}}}\left( 1-\frac{2\rho D}{5\mu } \right) \right].
		\label{Eucken_eq}
	\end{aligned}
\end{equation}
Clearly, the results derived in this work reduce to the classical results of \citep{mason1962heat} within the equilibrium temperature framework. Equations (\ref{Eucken_ft})–(\ref{Eucken_feu}) provide a more comprehensive expression for the Eucken factor within a thermal non-equilibrium framework, whereas result of Mason represents first-order approximation of it under the condition $T_t/T_r=1$.

\begin{figure}
	\centering
	\includegraphics[width=0.6\linewidth]{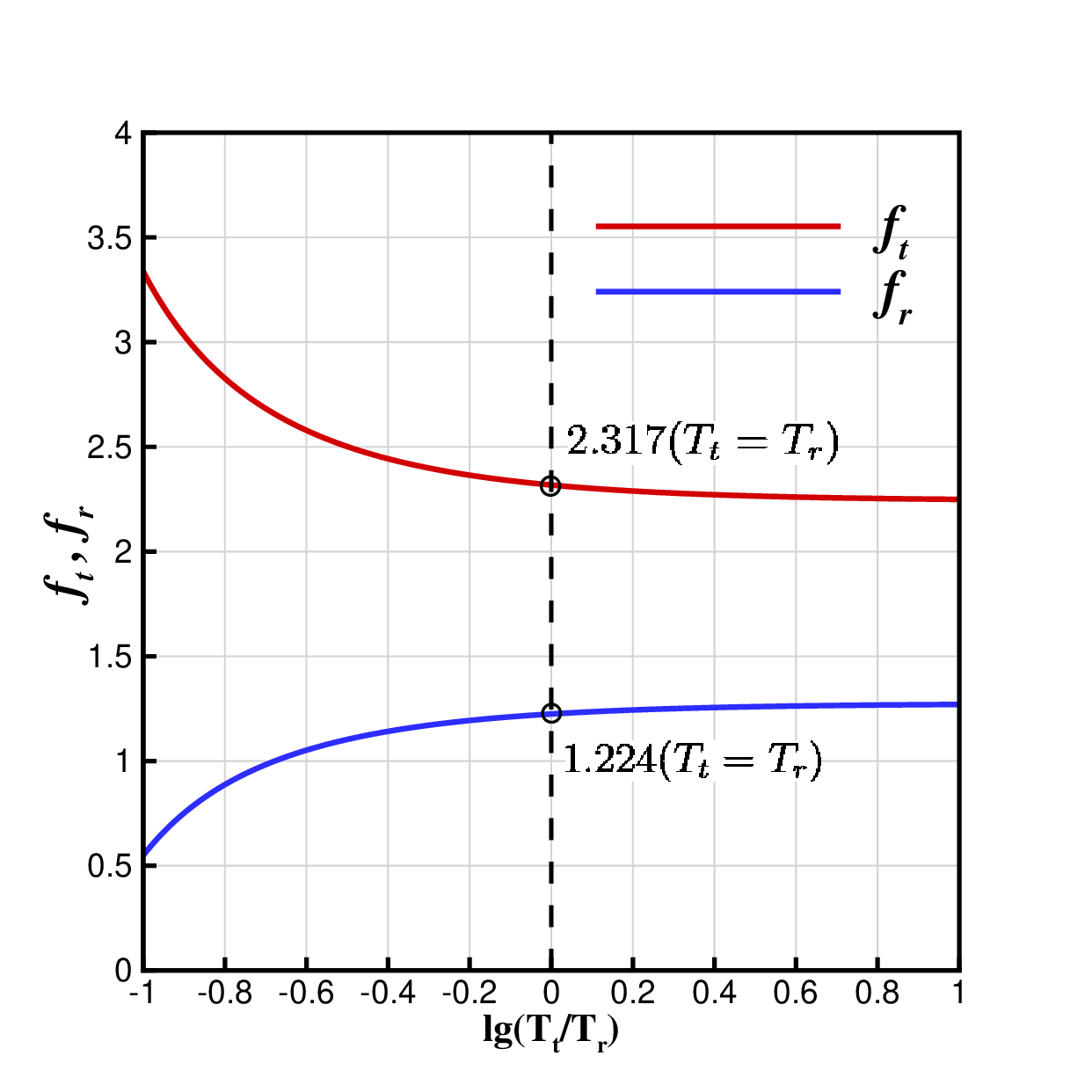}
	\caption{Variation of the Eucken factor with the degree of non-equilibrium}
	\label{Eucken_TrTt}
\end{figure}

 Moreover, to provide a more intuitive analysis of the influence of non-equilibrium temperature on the heat flux, figure~\ref{Eucken_TrTt} illustrates the relationship between the Eucken factor and the ratio of translational-to-rotational temperature ${{T}_{\rm{t}}}/{{T}_{\rm{r}}}$. The curves are constructed for nitrogen molecules, with the rotational degrees of freedom set as $\nu = 2$,  the rotational collision number as $Z_{\rm{rot}} = 3$,  and the viscosity index as $\omega = 0.74$. The solid red and blue lines show the variation trends of the translational and rotational Eucken factors, respectively. The black dashed line indicates the equilibrium condition $T_{\rm{t}}/T_{\rm{r}} = 1$, with the corresponding Eucken factor values ${{f}_{\rm{t}}}=2.317$ and ${{f}_{\rm{r}}}=1.224$ annotated. As $\log(T_{\rm{t}} / T_{\rm{r}})$ increases, the translational Eucken factor ${{f}_{\rm{t}}}$ decreases while the rotational Eucken factor ${{f}_{\rm{r}}}$ increases. Furthermore, ${{f}_{\rm{t}}}$ and ${{f}_{\rm{r}}}$ approach asymptotic values of 2.242 and 1.275, respectively, in the limit $\log(T_{\rm{t}} / T_{\rm{r}})\to 0$ (i.e., the limiting state approaching absolute zero with rotational modes unexcited). When $T_{\rm{t}}>T_{\rm{r}}$, the non-equilibrium temperature ${{T}_{\rm{t}}}/{{T}_{\rm{r}}}$ exerts limited influence on the Eucken factor. This corresponds to compressive flows, such as those encountered in aerospace re-entry problems, where macroscopic kinetic energy is first converted into translational thermal energy. Conversely, when $T_{\rm{r}}>T_{\rm{t}}$, the ratio ${{T}_{\rm{t}}}/{{T}_{\rm{r}}}$ exerts a significant influence on the Eucken factor. This corresponds to expansion or jet flows in applications such as aerospace propulsion or micro/nanofabrication, where ${{T}_{\rm{t}}}$ decreases along the jet.

\begin{figure}
	\centering
	\begin{subfigure}[b]{0.45\textwidth}
		\includegraphics[width=\linewidth]{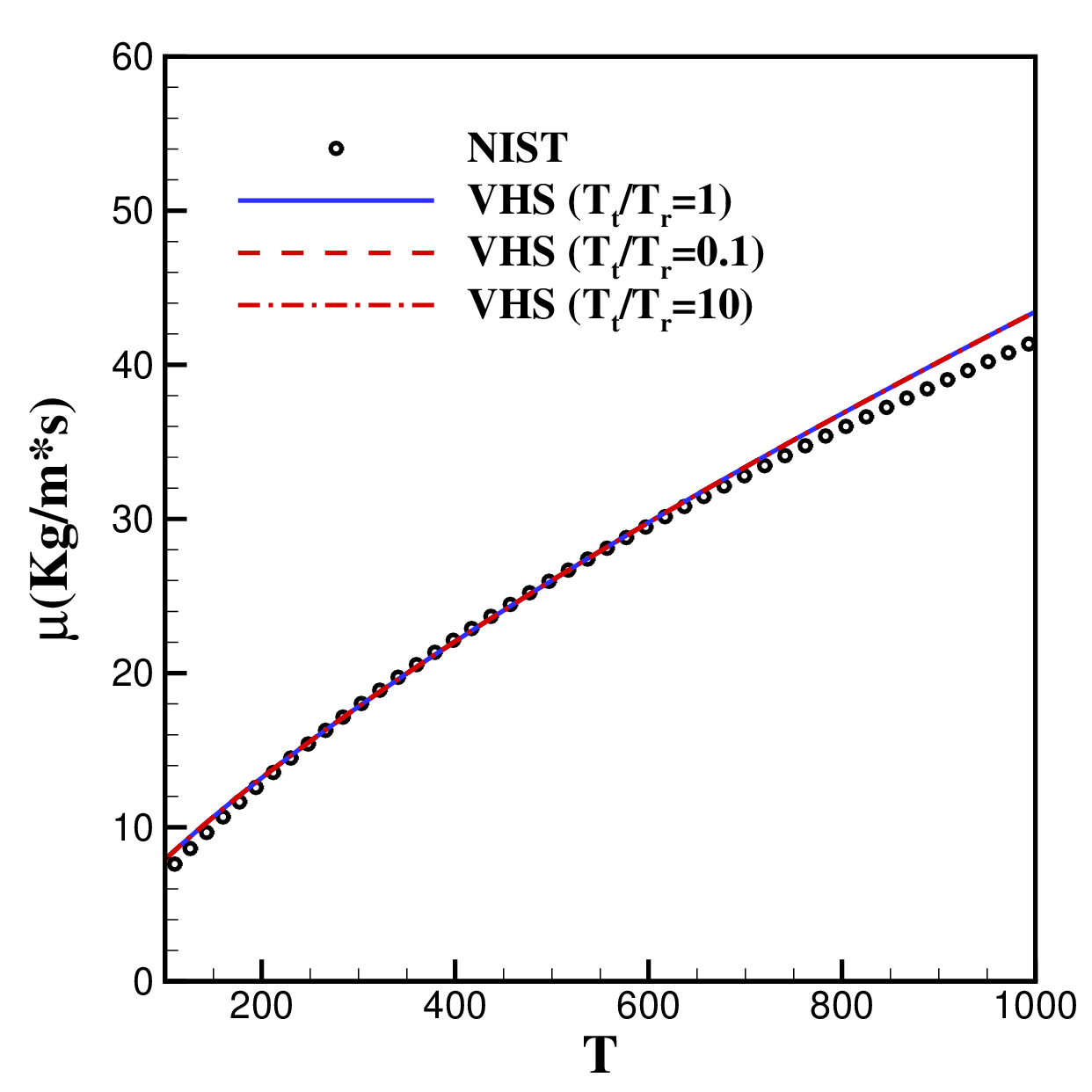}
		\caption{shear viscosity}
		\label{cmp_NIST_viscosity}
	\end{subfigure}
	\hfill
	\begin{subfigure}[b]{0.45\textwidth}
		\includegraphics[width=\linewidth]{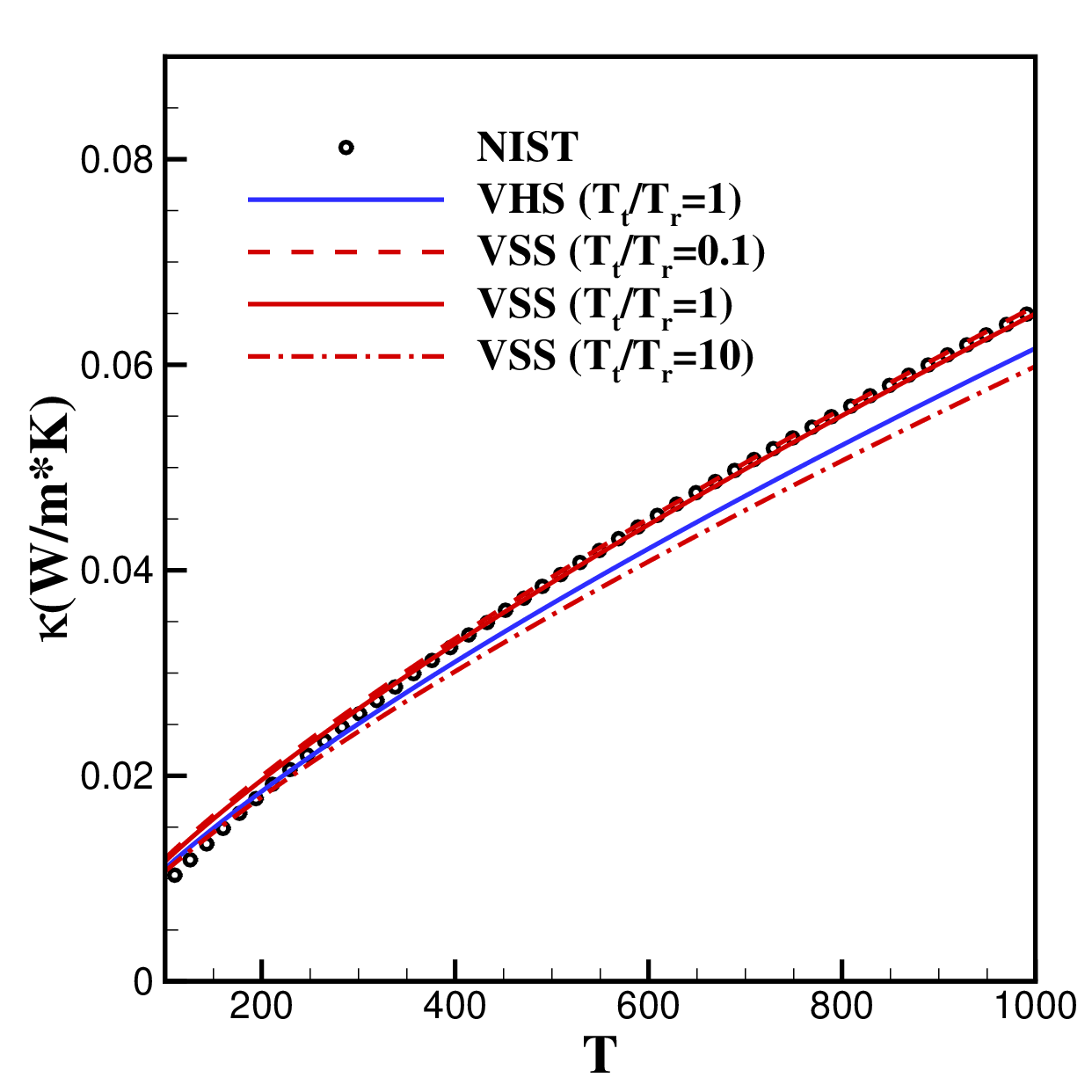}
		\caption{thermal conductivity}
		\label{cmp_NIST_thermal}
	\end{subfigure}
	\caption{Comparison of calculated macroscopic transport coefficients with the NIST database}
	\label{cmp_NIST}
\end{figure}

 To further check the derived result in this work, the shear viscosity $\mu$ and thermal conductivity $\kappa$ of nitrogen gas are compared with the experimentally fitted data from the National Institute of Standards and Technology (NIST) \citep{NIST} over the temperature range of 100–1000~K. The NIST data include both physical experiments and their fitted values. It is noteworthy that most measurements of viscosity and thermal conductivity are conducted under thermal equilibrium conditions. The comparative results of viscosity using the VHS model ($\omega=0.74$) are presented in figure~\ref{cmp_NIST_viscosity}, where the blue solid line represents results obtained under the equilibrium temperature assumption and black circles represent NIST data. The red dashed and dotted-dashed lines represent the viscosity under different non-equilibrium temperatures, which remains identical to that at equilibrium temperature. This stems from the decoupling of velocity and energy collisions in the Pullin model, and thus the viscosity is unaffected by thermal non-equilibrium temperature. Comparison with the NIST data shows that shear viscosity achieve good agreement. In the thermal conductivity calculation, the rotational collision number adopts the fitting results of Larina and Rykov \citep{larina1976,rykov2007numerical}:
 \begin{equation}
 	{{Z}_{\rm{rot}}}=\frac{3}{4}\pi \frac{\Phi \left( t \right)}{{{t}^{{1}/{6}\;}}}\frac{9t}{t+8}\left( \frac{{{T}_{\rm{r}}}}{{{T}_{\rm{t}}}} \right)\left[ 0.461+0.5581\left( \frac{{{T}_{\rm{r}}}}{{{T}_{\rm{t}}}} \right)+0.0358{{\left( \frac{{{T}_{\rm{r}}}}{{{T}_{\rm{t}}}} \right)}^{2}} \right],
 \end{equation}
 \begin{equation}
 	\Phi \left( t \right)=0.767+0.233{{t}^{-\frac{1}{6}}}\exp \left[ -1.17\left( t-1 \right) \right],
 \end{equation}
 \begin{equation}
 	t=\frac{{{T}_{\rm{t}}}}{{\tilde{T}}},
 \end{equation}
where ${\tilde{T}}$ is the reduced temperature; for nitrogen ${\tilde{T}}=91.5K$. The thermal conductivity using the VHS model ($\omega=0.74$) under the equilibrium temperature ($T_{\rm{t}}/T_{\rm{r}}=1$) is presented as blue solid line in figure~\ref{cmp_NIST_thermal}. Furthermore, by substituting the self-diffusion coefficient $D$ of the VSS model ($\alpha=1.36$) into Eqs.~(\ref{eucken_ft_D})-(\ref{eucken_fr_D}), the corresponding thermal conductivity under the equilibrium temperature ($T_{\rm{t}}/T_{\rm{r}}=1$) and different non-equilibrium temperature conditions ($T_{\rm{t}}/T_{\rm{r}}=0.1, 10$) is represented by the red lines. The comparison shows that, under equilibrium temperature conditions, the results obtained using the VSS model demonstrate better accuracy than using the VHS model. Moreover, as $T_{\rm{t}}/T_{\rm{r}}$ decreases, the total thermal conductivity exhibits an increasing trend, which becomes more pronounced when $T_{\rm{t}}/T_{\rm{r}}<1$.

\newpage

\section {Kinetic Model}\label{chapter3}
Due to the high computational cost of directly solving the extended Boltzmann equation, there is a need to develop simplified models that are both efficient and accurate. The establishment of a simplified model requires explicit and accurate relaxation rates to recover the correct relaxation process. In the past, when relaxation rates were unavailable, some studies resorted to the C-E expansion to calibrate transport coefficients for model construction such as relaxation-type (BGK-type) model equations and Fokker-Planck-type model equations. However, due to insufficient understanding of the relaxation process, modeling conducted this way is often incomplete.

 To yield accurate relaxation rates and transport coefficients simultaneously, a new relaxation-type model equation is constructed based on the accurate analytical relaxation rates provided in Eqs.~(\ref{relax_Tt_VHS})–(\ref{relax_Qr_VHS}). In reviewing previous work, the basic framework of the Rykov model was sound. However, due to the absence of relaxation rates as a basis for modeling, its description of near-equilibrium relaxation remains incomplete. Therefore, constructing an appropriate equilibrium state to recover the relaxation rates is key to modeling. The subsequent sections are organized as follows. In subsection \ref{subsection:rykov model}, a brief review of the Rykov model is provided to highlight its key assumptions and limitations. In subsection \ref{subsection:new kinetic model}, the model construction procedure based on macroscopic relaxation rates is presented. The relaxation behavior and the recovery of transport coefficients in the proposed kinetic model are then examined.

\subsection{The Rykov kinetic model}\label{subsection:rykov model}
The original Rykov model was developed for diatomic gases and later extended to polyatomic gases. Its construction adopts the approach of treating elastic and full-inelastic collisions separately. Without external forces, the evolution of the molecular distribution function is described by:
\begin{equation}
	\frac{\partial f}{\partial t}+\mathbf{c}\cdot \frac{\partial f}{\partial \mathbf{x}}=\frac{{{g}_{\rm{t,Rykov}}}-f}{{{\tau}_{\rm{t}}}}+\frac{{{g}_{\rm{r,Rykov}}}-{{g}_{\rm{t,Rykov}}}}{{{Z}_{\rm{rot}}}{{\tau}_{\rm{t}}}}.
	\label{kinetic_equation}
\end{equation}
where the left hand side of this equation represents the free-transport term, the right-hand side represents elastic and inelastic collisions terms respectively. All molecules first relax to the equilibrium state $g_{\rm{t}}$ via elastic collisions, which mediate the exchange of momentum between molecules, with a concurrent redistribution of translational energy occurring across different spatial directions. Then, a fraction ${1}/{Z_{\rm{rot}}}$ of the molecules undergo further inelastic collisions, relaxing to the equilibrium state $g_{\rm{r}}$, during which translational and rotational energy are exchanged. Analogous to the approach of constructing approximate distribution functions in theoretical analysis, ${{g}_{\rm{t,Rykov}}}$ and ${{g}_{\rm{r,Rykov}}}$ are similarly expressed as the product of perturbation polynomial and equilibrium state functions:

\begin{equation}
	{{g}_{\rm{t,Rykov}}}= {{E}_{\rm{t}}}\left( {T}_{\rm{t}} \right) {{E}_{\rm{r}}}\left( {T}_{\rm{r}} \right) 
	\left[ 1+\frac{2m{{q}_{\rm{t},i}}{{C}_{i}}}{15k{{T}_{\rm{t}}}{{p}_{\rm{t}}}}\left( \frac{m{{C}^{2}}}{2k{{T}_{\rm{t}}}}-\frac{5}{2} \right)+\frac{2\left( 1-\delta  \right)}{\nu }\frac{m{{q}_{\rm{r},i}}{{C}_{i}}}{k{{T}_{\rm{t}}}{{p}_{\rm{r}}}}\left( \frac{\epsilon }{k{{T}_{\rm{r}}}}-\frac{\nu }{2} \right) \right],
\end{equation}
\begin{equation}
	{{g}_{\rm{r,Rykov}}}= {{E}_{\rm{t}}}\left( T \right) {{E}_{\rm{r}}}\left( {T} \right)  
	\left[ 1+{{\omega }_{0}}\frac{2m{{q}_{\rm{t},i}}{{C}_{i}}}{15kTp}\left( \frac{m{{C}^{2}}}{2kT}-\frac{5}{2} \right)+\frac{2{{\omega }_{1}}\left( 1-\delta  \right)}{\nu }\frac{m{{q}_{\rm{r},i}}{{C}_{i}}}{kTp}\left( \frac{\epsilon }{kT}-\frac{\nu }{2} \right) \right],
\end{equation}
where ${{E}_{\rm{t}}}\left( T \right)$ and ${{E}_{\rm{r}}}\left( T \right)$ are the equilibrium state functions as:
\begin{equation}
	{{E}_{\rm{t}}}\left( T \right)=n{{(\frac{m}{2\pi kT})}^{\frac{3}{2}}}\exp \left( -\frac{m{{C}^{2}}}{2k{{T}_{\rm{t}}}} \right),
\end{equation}
\begin{equation}
	{{E}_{\rm{r}}}\left( T \right)=\frac{{{\epsilon }^{\frac{{\nu}}{2}-1}}}{\Gamma \left( \frac{\nu }{2} \right){{\left( kT \right)}^{\frac{\nu }{2}}}}\exp \left( -\frac{\epsilon }{kT} \right).
\end{equation}
The Rykov model obtains its transport coefficients through the C-E expansion and recovers the correct bulk viscosity, shear viscosity, and thermal conductivity by adjusting its free parameters. Specifically, the Rykov model adjusts the collision number $Z_{\rm{rot}}$ to control the proportion of inelastic collisions, thereby recovering the correct bulk viscosity. For the shear viscosity, the model sets the relaxation time for elastic collisions as $\tau_{\rm{t}} ={\mu }/{{{p}_{\rm{t}}}}$ to achieve its recovery. Regarding thermal conductivity, the model matches the correct thermal conductivity values by tuning the free parameters ${{\omega }_{0}}$ and ${{\omega }_{1}}$ in the heat flux correction term (where $\delta ={\mu }/{\rho D}$ characterizes diffusion properties). 

More significantly, the relaxation behavior of the Rykov model is analyzed. By taking the third-order moment of the collision term, the heat flux relaxation rates for the Rykov model can be obtained. Following the heat flux relaxation relationship Eq.~(\ref{heatflux_relax}), the corresponding heat flux relaxation coefficients can be extracted for analysis:

\begin{equation}
	\begin{aligned}
		A_{\mathrm{tt}} &= \frac{2}{3} + \frac{1 - \omega_0}{3 Z_{\mathrm{rot}}}, &\quad A_{\mathrm{tr}} &= 0, \\
		A_{\mathrm{rt}} &= 0, &\quad A_{\mathrm{rr}} &= \cfrac{\mu}{\rho D} + \frac{\left(1 - \cfrac{\mu}{\rho D}\right)(1 - \omega_1)}{Z_{\mathrm{rot}}}.
	\end{aligned}
\end{equation}
Notably, the heat flux relaxation coefficients ${A}_{\rm{tr}}$ and ${A}_{\rm{rt}}$ are set to zero. This implies the model assumes independent relaxation between translational and rotational heat fluxes without mutual influence. While this simplification allows the model to recover the correct transport coefficients (viscosity and thermal conductivity) via the C-E expansion, it fails to capture the coupled relaxation dynamics between ${\mathbf{q}}_{\rm{t}}$ and ${\mathbf{q}}_{\rm{r}}$. This discrepancy highlights a key limitation: matching macroscopic transport coefficients in the continuum limit does not guarantee the accuracy of the underlying microscopic relaxation processes. A more complete kinetic model should be grounded in the correct relaxation rates.

\subsection{A kinetic model based on relaxation rates}\label{subsection:new kinetic model}
In Section \ref{chapter2}, we derived analytical expressions for relaxation rates of key macroscopic quantities (i.e., temperature, pressure tensor and heat flux) within Pullin model framework. These theoretical results provide complete and explicit information on the constraints for constructing relaxation-type model equations that can recover accurate relaxation rates. The new kinetic model adopts the same form as Rykov's model:
\begin{equation}
	\frac{\partial f}{\partial t}+\mathbf{c}\cdot \frac{\partial f}{\partial \mathbf{x}}=\frac{{{g}_{\rm{t,new}}}-f}{{{\tau}_{\rm{t}}}}+\frac{{{g}_{\rm{r,new}}}-{{g}_{\rm{t,new}}}}{{{Z}_{\rm{rot}}}{{\tau}_{\rm{t}}}}.
	\label{new_kinetic_equation}
\end{equation}
The construction of this new kinetic model follows a systematic approach centered on relaxation rates. The overarching objective is to ensure that taking moments of the collision term on the right-hand side of Eq.~(\ref{new_kinetic_equation}) exactly recovers the analytical relaxation rates derived in Section \ref{chapter2}. This is achieved by designing the reference distribution functions $g_{\rm{t,new}}$ and $g_{\rm{r,new}}$. As the relaxation rates are now more clearly expressed, the perturbative polynomials within the equilibrium states ${{g}_{\rm{t,new}}}$ and ${{g}_{\rm{r,new}}}$ can be constructed in a simpler and more compact form:
\begin{equation}
	{{g}_{\rm{t,new}}}={{E}_{\rm{t}}}\left( {T}_{\rm{t}} \right){{E}_{\rm{r}}}\left( {T}_{\rm{r}} \right)\left[ 1+\frac{2}{5nm}{{\left( \frac{m}{k{{T}_{\rm{t}}}} \right)}^{2}}{{q}_{0,i}}{{C}_{i}}\left( \frac{m}{2k{{T}_{\rm{t}}}}{{C}^{2}}-\frac{5}{2} \right) \right],
\end{equation}
\begin{equation}
	{{g}_{\rm{r,new}}}={{E}_{\rm{t}}}\left( T \right){{E}_{\rm{r}}}\left( T \right)\left[ 1+\frac{2}{\nu nm}{{\left( \frac{m}{kT} \right)}^{2}}{{q}_{1,i}}\sqrt{\frac{m}{kT}}{{C}_{i}}\left( \frac{1}{kT}\epsilon -\frac{{\nu}}{2} \right) \right],
\end{equation}
where ${{q}_{0,i}}$ and ${{q}_{1,i}}$ are linear combinations of the translational heat flux ${{q}_{\rm{t},i}}$ and rotational heat flux ${{q}_{\rm{r},i}}$ designed specifically to recover heat flux relaxation rates:
\begin{equation}
		\begin{aligned}
			& {{q}_{0,i}}=\frac{{{Z}_{\rm{rot}}}}{{{Z}_{\rm{rot}}}-1}\left(
			\left( \frac{1}{3}-\frac{\cfrac{{{T}_{\rm{r}}}}{{{T}_{\rm{t}}}}\left( \omega -1 \right)-\left( \omega -2 \right)}{3{{Z}_{\rm{rot}}}} \right){{q}_{\rm{t},i}}+ \frac{\left( \cfrac{{{T}_{\rm{r}}}}{{{T}_{\rm{t}}}}\left( \nu -1 \right)-\left( \nu -2 \right) \right)}{2{{Z}_{\rm{rot}}}}{{q}_{\rm{r},i}}\right), \\ 
			& {{q}_{1,i}}={{Z}_{\rm{rot}}}\left( \frac{\cfrac{{{T}_{\rm{t}}}}{{{T}_{\rm{r}}}}\left( \omega -1 \right)-\left( \omega -2 \right)}{5{{Z}_{\rm{rot}}}} {{q}_{\rm{t},i}}+\left( 1- \frac{5\nu }{2\left( 7-2\omega  \right)}-\frac{3\left( \left( \nu -1 \right)-\left( \nu -2 \right)\cfrac{{{T}_{\rm{t}}}}{{{T}_{\rm{r}}}} \right)}{10{{Z}_{\rm{rot}}}} \right){{q}_{\rm{r},i}}\right). \\ 
		\end{aligned}
\end{equation}

The new reference distribution function is constructed around the equilibrium distribution and orthogonal polynomials in thermal velocity ${{C}_{i}}$, rotational energy ${{\epsilon }}$,  and their corresponding moments ${{q}_{0,i}}$, ${{q}_{1,i}}$. The parameters in these combinations, including $\tau_{\rm{t}}$, $Z_{\rm{rot}}$, $q_{0,i}$, $q_{1,i}$, are respectively designed to recover the correct relaxation rates of stress, temperature, and heat flux. To achieve practical application, we now detail the determination of these parameters and examine both the microscopic relaxation processes and the macroscopic transport coefficients of the new kinetic model equation.
\subsubsection{Relaxation of pressure deviator}
Regarding pressure and the pressure tensor relaxations, multiplying the collision term in the kinetic equation Eq.~(\ref{new_kinetic_equation}) by $m{{C}^{2}}/3$ and $m{{C}_{i}}{{C}_{j}}$, and integrating over the thermal velocity $\mathbf{C}$ and the rotational energy $\epsilon$ yields:
\begin{equation}
	\frac{\partial {{p}_{\rm{t}}}}{\partial t}=\frac{p-{{p}_{\rm{t}}}}{{{Z}_{\rm{rot}}}\tau_{\rm{t}} },
	\label{model_p_relax}
\end{equation}
\begin{equation}
	\frac{\partial {{p}_\mathit{ij}}}{\partial t}=\frac{{{p}_{\rm{t}}}{{\delta }_\mathit{ij}}-{{p}_\mathit{ij}}}{\tau_{\rm{t}} }+\frac{p{{\delta }_\mathit{ij}}-{{p}_{\rm{t}}}{{\delta }_\mathit{ij}}}{{{Z}_{\rm{rot}}}\tau_{\rm{t}} }.
	\label{model_pij_relax}
\end{equation}
The relaxation rates for the pressure deviator ${\partial {{p}_{\mathit{\langle ij\rangle }}}}={{p}_\mathit{ij}}-p\delta_\mathit{ij}$ can be derived from Eqs.~(\ref{model_p_relax}) and (\ref{model_pij_relax}) as follows:
\begin{equation}
	\frac{\partial {{p}_\mathit{\langle ij\rangle }}}{\partial t}=-\frac{{{p}_\mathit{\langle ij\rangle }}}{{{\tau_{\rm{t}} }}}.
\end{equation}
To recover accurate relaxation rates for the pressure deviator, the relaxation time $\tau_{\rm{t}}$ employed in the kinetic model equations should be determined according to the pressure deviator relaxation Eq.~(\ref{pullin_pij_relax}) derived previously. Specifically, taking the results of the VHS model as an example, the relaxation time $\tau_{\rm{t}}$ is given by:
 \begin{equation}
	{\tau}_{\rm{t}} = \cfrac{15}{{{4}^{2-\omega }}d_{\rm{ref}}^{2}g_{\rm{ref}}^{2\omega -1}n{{\left( \cfrac{k{{T}_{\rm{t}}}}{m} \right)}^{1-\omega }}\sqrt{\pi }\Gamma \left[ \cfrac{9}{2}-\omega  \right]}.
\end{equation}
Moreover, it is essential to verify the accuracy of the shear viscosity while ensuring the correct recovery of the pressure deviator relaxation rate. The shear viscosity for the kinetic model equation can be derived via the C-E expansion as(see Appendix~\ref{appendixB}):
\begin{equation}
	\mu ={{p}_{\rm{t}}}\tau_{\rm{t}} .
	\label{model_mu}
\end{equation}
Comparison with the result derived from the Pullin model, Eq.~(\ref{pullin_shear}), demonstrates that correctly recovering the pressure tensor relaxation rate ensures the accuracy of the model's shear viscosity. 

\subsubsection{Relaxation of energy}
The relaxation process of energy can be described by the temperature relaxation rate. The translational and rotational temperature relaxation rates can be obtained by multiplying the collision term in the kinetic equation Eq.~(\ref{new_kinetic_equation}) by $m{{C}^{2}}/3nk$ and $2\epsilon/\nu nk$, and integrating over the thermal velocity $\mathbf{C}$ and the rotational energy $\epsilon$:
\begin{equation}
	\frac{\partial {{T}_{\rm{t}}}}{\partial t}=\frac{T-{{T}_{\rm{t}}}}{Z_{\rm{rot}}\tau_{\rm{t}} },
\end{equation}
\begin{equation}
	\frac{\partial {{T}_{\rm{r}}}}{\partial t}=\frac{T-{{T}_{\rm{r}}}}{Z_{\rm{rot}}\tau_{\rm{t}} }.
\end{equation}
Based on the previously derived temperature relaxation rates Eq.~(\ref{pullin_Tt_relax}) and Eq.~(\ref{pullin_Tr_relax}), adjusting the rotational collision number $Z_{\rm{rot}}$ enables the kinetic model to accurately describe the relaxation processes of translational and rotational energies. Specifically, the rotational collision number under the VHS model is adopted as follows:
\begin{equation}
	{{Z}_{\rm{rot}}}=\cfrac{\left( 7-2\omega  \right)\left( {\nu}\phi +\eta \psi  \right)}{5\left(3+\nu\right)\phi \psi }.
\end{equation}
Furthermore, since the bulk viscosity coefficient $\mu_{\rm{b}}$ fundamentally originates from dissipative effects during translational-rotational energy exchange, the correctly recovered energy relaxation rates naturally should ensure the accuracy of the bulk viscosity coefficient. This correspondence is rigorously established through C-E expansion (see Appendix \ref{appendixB}). The bulk viscosity of the kinetic model equation can be expressed as:
 \begin{equation}
	{{\mu }_{\rm{b}}}=\frac{2{\nu}}{3\left( 3+{\nu} \right)}{{Z}_{\rm{rot}}}\mu.
\end{equation}
By comparison with the bulk viscosity derived from the Pullin model Eq.~(\ref{pullin_bulk}), this demonstrates that the kinetic model equation can recover the correct bulk viscosity provided accurate temperature relaxation is maintained.

\subsubsection{Relaxation of heat flux}
For the relaxation of heat flux in the kinetic model equation, multiplying the collision term in the kinetic equation Eq.~(\ref{new_kinetic_equation}) by $m{{C}^{2}}{{C}_{i}}/2$ and $\epsilon{C}_{i}$, then integrating over thermal velocity $C_i$ and rotational energy $\epsilon$ to obtain translational and rotational heat flux relaxation rates:
\begin{equation}
	\frac{\partial {{q}_{\rm{t},i}}}{\partial t}=\frac{1}{\tau_{\rm{t}} }\left( \frac{{{Z}_{\rm{rot}}}-1}{{{Z}_{\rm{rot}}}}{{q}_{0,i}}-{{q}_{\rm{t},i}} \right),
	\label{model_qt_relax}
\end{equation}

\begin{equation}
	\frac{\partial {{q}_{\rm{r},i}}}{\partial t}=\frac{1}{\tau_{\rm{t}} }\left( \frac{1}{{{Z}_{\rm{rot}}}}{{q}_{1,i}}-{{q}_{\rm{r},i}} \right).
	\label{model_qr_relax}
\end{equation}
where $q_{0,i}$ and $q_{1,i}$ are the free parameters that serve to restore the heat flux relaxation rate and correct thermal conductivity. To ensure that the relaxation rates in Eqs.~(\ref{model_qt_relax})-(\ref{model_qr_relax}) are consistent with the derived results given by Eqs.~(\ref{pullin_Qt_relax})-(\ref{pullin_Qr_relax}), a system of linear equations for $q_{0,i}$ and $q_{1,i}$ in terms of $q_{{\rm{t}},i}$, $q_{{\rm{r}},i}$, and the heat flux relaxation coefficients $A_{ij}$ is established. Within the heat flux relaxation framework of Eq.~(\ref{heatflux_relax}), the free parameters $q_{0,i}$ and $q_{1,i}$ can be solved as follows:
\begin{equation}
	\left[ \begin{matrix}
		{{q}_{0,i}}  \\
		{{q}_{1,i}}  \\
	\end{matrix} \right]=\left[ \begin{matrix}
		\cfrac{{{Z}_{\rm{rot}}}}{{{Z}_{\rm{rot}}}-1}\left( 1-{{A}_{\rm{tt}}} \right) & -\cfrac{{{Z}_{\rm{rot}}}}{{{Z}_{\rm{rot}}}-1}{{A}_{\rm{tr}}}  \\
		-{{Z}_{\rm{rot}}}{{A}_{\rm{rt}}} & {{Z}_{\rm{rot}}}\left( 1-{{A}_{\rm{rr}}} \right)  \\
	\end{matrix} \right]\left[ \begin{matrix}
		{{q}_{\rm{t},i}}  \\
		{{q}_{\rm{r},i}}  \\
	\end{matrix} \right].
	\label{eq:q0,q1}
\end{equation}
Specifically, under the VHS model, the heat flux relaxation coefficients ${{A}_\mathit{ij}}$ in Eq.~(\ref{eq:q0,q1}) adopt the result from Eq.~(\ref{Aij}). The corresponding free parameters can be written as:
\begin{equation}
	\begin{aligned}
		& {{q}_{0,i}}=\frac{{{Z}_{\rm{rot}}}}{{{Z}_{\rm{rot}}}-1}\left(
		\left( \frac{1}{3}-\frac{\cfrac{{{T}_{\rm{r}}}}{{{T}_{\rm{t}}}}\left( \omega -1 \right)-\left( \omega -2 \right)}{3{{Z}_{\rm{rot}}}} \right){{q}_{\rm{t},i}}+ \frac{\left( \cfrac{{{T}_{\rm{r}}}}{{{T}_{\rm{t}}}}\left( \nu -1 \right)-\left( \nu -2 \right) \right)}{2{{Z}_{\rm{rot}}}}{{q}_{\rm{r},i}}\right), \\ 
		& {{q}_{1,i}}={{Z}_{\rm{rot}}}\left( \frac{\cfrac{{{T}_{\rm{t}}}}{{{T}_{\rm{r}}}}\left( \omega -1 \right)-\left( \omega -2 \right)}{5{{Z}_{\rm{rot}}}} {{q}_{\rm{t},i}}+\left( 1- \frac{5\nu }{2\left( 7-2\omega  \right)}-\frac{3\left( \left( \nu -1 \right)-\left( \nu -2 \right)\cfrac{{{T}_{\rm{t}}}}{{{T}_{\rm{r}}}} \right)}{10{{Z}_{\rm{rot}}}} \right){{q}_{\rm{r},i}}\right). \\ 
	\end{aligned}
\end{equation}
Furthermore, to simultaneously verify the accuracy of thermal conductivity, we derived the translational thermal conductivity ${\kappa}_{\rm{t}}$ and rotational thermal conductivity ${\kappa}_{\rm{r}}$ from the kinetic model equation via the C-E expansion in the continuous limit:
 \begin{equation}
	\left[ \begin{matrix}
		{{\kappa}_{\rm{t}}}  \\
		{{\kappa}_{\rm{r}}}  \\
	\end{matrix} \right]=\frac{k\mu }{2m}{{\left[ \begin{matrix}
				{{A}_{\rm{tt}}} & {{A}_{\rm{tr}}}  \\
				{{A}_{\rm{rt}}} & {{A}_{\rm{rr}}}  \\
			\end{matrix} \right]}^{-1}}\left[ \begin{matrix}
		5  \\
		{\nu}  \\
	\end{matrix} \right].
\end{equation}
Result confirms that the ${\kappa}_{\rm{t}}$ and ${\kappa}_{\rm{r}}$ obtained from C-E expansion are identical to the thermal conductivity results theoretically derived from Pullin model. This conclusively demonstrates that when the model accurately describes the heat flux relaxation process (i.e., with correct relaxation coefficients ${{A}_\mathit{ij}}$),  the accuracy of thermal conductivities is ensured.

At this stage, all free parameters in the new kinetic model are determined. The examination of the relaxation rates and transport coefficients shows that, while ensuring the accuracy of the relaxation rates, the transport coefficients naturally match those in Pullin model. This demonstrates that kinetic modeling based on relaxation rates is more complete. The new kinetic model is founded on theoretically derived relaxation rates, thereby achieving more accurate relaxation processes compared to Rykov's model. Moreover, by employing relaxation-rate-based kinetic modeling, the theoretically derived relaxation rates not only serve for theoretical analysis but are also applied to the simulation and prediction of actual flows.

\section{Validation of the kinetic model}\label{chapter4}

In the preceding section, we constructed a new kinetic model based on the analytically derived relaxation rates from the Pullin equation. This model is designed to capture the correct relaxation process, a capability absent in traditional models like Rykov model. In this section, we solve the newly developed kinetic model (hereafter referred to as ``the kinetic model'') using the unified gas kinetic scheme (UGKS) method, which employs a discrete velocity space to capture the non-equilibrium distribution function, and we perform numerical simulations for classical test cases. 
To validate the kinetic model, numerical results are evaluated against DSMC data for accuracy and compared with the Rykov model, thereby revealing the differences from traditional models established via the C-E expansion. Notably, in DSMC simulations employing the VHS model, the temperature relaxation process is described by:
\begin{equation}
	\frac{\partial {{T}_{\rm{r}}}}{\partial t}=\frac{30\left( T-{{T}_{\rm{r}}} \right)}{\left( 7-2\omega  \right)\left( 5-2\omega  \right){{Z}_{\rm{r,DSMC}}}}.
\end{equation}
While the kinetic model strictly follows the Jeans-Landau relaxation equation, for the convenience of model validation, it is necessary to clarify the relationship between the rotational collision numbers in DSMC and the kinetic model:
\begin{equation}
	{{Z}_{\rm{rot,DSMC}}}=\frac{30}{\left( 7-2\omega  \right)\left( 5-2\omega  \right)}Z_{\rm{rot}}.
	\label{Zrot_DSMC}
\end{equation}

\subsection{Unified gas-kinetic scheme}
\subsubsection{General framework of unified gas-kinetic scheme}
Int this work, the UGKS procedure used to solve the kinetic model equation follows that of \citep{xu2010unified}, with a simplified multiscale flux adopted \citep{zhang2024conservative}. A brief overview of the solution procedure is provided here. The gas-kinetic relaxation model equation adopted in the UGKS can be expressed in the following form:
\begin{equation}
	\frac{\partial f}{\partial t}+\mathbf{c}\cdot \frac{\partial f}{\partial \mathbf{x}}= \frac{g-f}{\tau_{\rm{t}}},
	\label{eq:relaxequation}
\end{equation}
where $g$ is the equilibrium distribution function. The collision term in the new kinetic model can be recast into the general form on the right-hand side of Eq.~(\ref{eq:relaxequation}), where $g = {{g}_{\rm{t,new}}} + {{g}_{\rm{r,new}}}$. The construction of UGKS for diatomic gas is based on reduced model equations in the classical finite volume framework. By discretizing the model equation Eq.~(\ref{eq:relaxequation}) at time $t_{n+1}$ by a backward Euler method, we can obtain the discrete governing equation:
\begin{equation}
	f_{i}^{n+1}-f_{i}^{n}+\frac{\Delta t}{\left| {{V}_{i}} \right|}F_{i}^{n+1/2}=\frac{\Delta t}{2}\left( \frac{g_{i}^{n+1}-f_{i}^{n+1}}{\tau_{{\rm{t}},i}^{n+1}} + \frac{g_{i}^{n}-f_{i}^{n}}{\tau_{{\rm{t}},i}^{n}} \right).
	\label{eq:discrete equation}
\end{equation}
The ${{V}_{i}}$ is the volume of cell i, and $\mathbf{c}_{k}$ is the discrete particle velocity space. $\Delta t=t_{n+1}-t_{n}$ is the time step. The microscopic flux $F_{i}^{n+1/2}$ can be calculated as follows:
\begin{equation}
	F_{i}^{n+1/2}=\sum\limits_{j}{\mathbf{c}_{k}\cdot \mathbf{A}_{i}^{j}f\left( \mathbf{x}_{i}^{j},\mathbf{c}_{k},{{t}_{n+1/2}} \right)},
	\label{micro-flux}
\end{equation}
where $\mathbf{A}_{i}^{j}$ is the outward normal vector of the $j$th face of cell $i$ with an area of $\left|\mathbf{A}_{i}^{j}\right|$, and $\mathbf{x}_{i}^{j}$ is the center of this face.  In this work, a simplified multi-scale flux is adopted \citep{zhang2024conservative}. By integrating relaxation model equation Eq.~(\ref{eq:relaxequation}) along the characteristic line (in the direction of particle velocity) from $t_n$ to $t_{n+1/2}$, the interface distribution function $f\left(\mathbf{x}_{i}^{j},\mathbf{c}_{k},{{t}_{n+1/2}} \right)$ can be expressed as:
\begin{equation}
	f\left( \mathbf{x}_{i}^{j},\mathbf{c}_{k},{{t}_{n+1/2}} \right)=\frac{2{{\tau }^{n+1/2}}}{2{{\tau }^{n+1/2}}+\Delta t}f\left( \mathbf{x}_{i}^{j}-\mathbf{c}_{k}\frac{\Delta t}{2},\mathbf{c}_{k},{{t}_{n}} \right)+\frac{\Delta t}{2{{\tau }^{n+1/2}}+\Delta t}g\left( \mathbf{x}_{i}^{j},\mathbf{c}_{k},{{t}_{n+1/2}} \right)
	\label{eq:interface distribution function}
\end{equation}
where the free transport distribution (representing the microscopic mechanism) and equilibrium distribution (representing the macroscopic mechanism) are coupled rationally, resulting in the multi-scale property of the present UGKS. Take the moment of Eq~(\ref{eq:interface distribution function}) and we can directly obtain the macroscopic flow variable $\mathbf{W}\left( \mathbf{x}_{i}^{j},\mathbf{c}_{k},{{t}_{n+1/2}} \right)$, which is used to evaluate the equilibrium state $g\left( \mathbf{x}_{i}^{j},\mathbf{c}_{k},{{t}_{n+1/2}} \right)$.

To perform the microscopic evolution for obtaining $f_{i}^{n+1}$, the equilibrium state $g_{i}^{n+1}$ is required. The equilibrium distribution $g$ depends on the macroscopic variables $\mathbf{W} = (\rho, \rho \mathbf{U}, \rho E, \rho E_{\rm{t}}, \rho E_{\rm{r}})^{T}$, where $\rho$, $\rho \mathbf{U}$, $\rho E$, $\rho E_{\rm{t}}$, and $\rho E_{\rm {r}}$ are the density, momentum, total energy, translational energy, and rotational energy, respectively. Therefore, by taking the moments of Eq.~(\ref{eq:discrete equation}), the macroscopic evolution equation can be obtained:
\begin{equation}
	\mathbf{W}_{i}^{n+1}=\mathbf{W}_{i}^{n}-\frac{\Delta t}{\left| {{V}_{i}} \right|}\int{\bm{\psi }F_{i}^{n+1/2}d\mathbf{c}_{k}}+\frac{1}{2}\left( \bm{S}_{i}^{n+1}+\bm{S}_{i}^{n} \right),
	\label{eq:ugks1}
\end{equation}
where $\int{{\bm{\psi }}F_{i}^{n+1/2}d\mathbf{c}_{k}}$ is the macroscopic flux, $\bm{\psi }$ is the collision invariant, and $\bm{S}$ is the source term. Once the macroscopic variables are updated, the implicit equation Eq.~(\ref{eq:discrete equation}) can be explicitly expressed as:
\begin{equation}
	f_{i}^{n+1}={{\left( 1+\frac{\Delta t}{2\tau _{i}^{n+1}} \right)}^{-1}}\left[ f_{i}^{n}-\frac{\Delta t}{\left| {{V}_{i}} \right|}F_{i}^{n+1/2}+\frac{\Delta t}{2}\left( \frac{g_{i}^{n+1}}{\tau _{i}^{n+1}}+\frac{g_{i}^{n}-f_{i}^{n}}{\tau _{i}^{n}} \right) \right].
	\label{eq:ugks2}
\end{equation}
In summary, the calculation procedure of the UGKS with simplified multi-scale numerical flux from time level $t_n$ to
$t_{n+1}$ is summarized in the following steps:\\
\textbf{Step~1.}~Given the initial macroscopic flow variables $\mathbf{W}_{i}^{0}$ and calculate the equilibrium distribution functions $g_{i}^{n+1}$.\\
\textbf{Step~2.}~Compute the micro-flux $F_{i}^{n+1/2}$ across the cell interface of control volumes.
 
 \textbf{(a)}~Calculate the macroscopic flow variables $\mathbf{W}\left( \mathbf{x}_{i}^{j},\mathbf{c}_{k},{{t}_{n+1/2}} \right)$ by taking the moment of Eq.~(\ref{eq:interface distribution function}).
 
 \textbf{(b)}~Calculate the equilibrium distribution functions $g\left( \mathbf{x}_{i}^{j},\mathbf{c}_{k},{{t}_{n+1/2}} \right)$  from the macroscopic flow variables $\mathbf{W}\left( \mathbf{x}_{i}^{j},\mathbf{c}_{k},{{t}_{n+1/2}} \right)$.
 
 \textbf{(c)}~Calculate the distribution functions $f\left( \mathbf{x}_{i}^{j},\mathbf{c}_{k},{{t}_{n+1/2}} \right)$ at the cell interface using Eq.~(\ref{eq:interface distribution function}).
 
 \textbf{(d)}~Calculate the micro-flux $F_{i}^{n+1/2}$ using Eq.~(\ref{micro-flux}).\\
\textbf{Step~3.}~Update the macroscopic flow variables $\mathbf{W}_{i}^{n+1}$ in each cell i according to Eq.~(\ref{eq:ugks1}).\\
\textbf{Step~4.}~Update the distribution function $f_{i}^{n+1}$ in each cell i according to Eq.~(\ref{eq:ugks2}).

Therefore, in the UGKS, the gas-kinetic relaxation model equations can be solved by sequentially updating the macroscopic variables and distribution functions respectively.
\subsubsection{Reduced distribution function}

 Because of the quasi-linear nature common to BGK-type relaxation model equations, their dimensionality can often be reduced when energy is treated continuously, achieving lower computational cost. For the reader's convenience, the following reduced distribution functions are employed: ${{R}_{1}}=\int{f\left( t,\mathbf{x},\mathbf{c},{{\epsilon}} \right)d{{\epsilon}}},{{R}_{2}}=\int{{{\epsilon}}f\left( t,\mathbf{x},\mathbf{c},{{\epsilon}} \right)d{{\epsilon}}}$. This effectively reduces solution complexity. Eliminating the rotational energy variable decreases computational memory requirements and costs. Multiplying Eq.~(\ref{new_kinetic_equation}) by 1 and $\epsilon$, then integrating over $\epsilon$ from $\left( 0,\infty \right)$,  yields equations for the reduced distribution functions $R_1$ and $R_2$:
 \begin{equation}
 	\frac{\partial {{R}_{1}}}{\partial {t}}+\mathbf{c}\cdot \frac{\partial {{R}_{1}}}{\partial \mathbf{x}}=\frac{{{R}_{1,\rm{t}}}-{{R}_{1}}}{\tau_{\rm{t}}}+\frac{{{R}_{1,\rm{r}}}-{{R}_{1,\rm{t}}}}{{{Z}_{\rm{rot}}}\tau_{\rm{t}} },
 \end{equation}
 
 \begin{equation}
 	\frac{\partial {{R}_{2}}}{\partial t}+\mathbf{c}\cdot \frac{\partial {{R}_{2}}}{\partial \mathbf{x}}=\frac{{{R}_{2,\rm{t}}}-{{R}_{2}}}{\tau_{\rm{t}} }+\frac{{{R}_{2,\rm{r}}}-{{R}_{2,\rm{t}}}}{{{Z}_{\rm{rot}}}\tau_{\rm{t}} },
 \end{equation}
 where the reference reduced distribution functions ${R}_{1,\rm{t}}$, ${R}_{1,\rm{r}}$,${R}_{2,\rm{t}}$ and ${R}_{2,\rm{r}}$ are given by:
 \begin{equation}
 	{{R}_{1,\rm{t}}}\left( t,\mathbf{x},\mathbf{c} \right)={{E}_{\rm{t}}}\left( {T}_{\rm{t}} \right)\left[ 1+\frac{2}{5nm}{{\left( \frac{m}{k{{T}_{\rm{t}}}} \right)}^{2}}{{q}_{0,i}}{{C}_{i}}\left( \frac{m}{2k{{T}_{\rm{t}}}}{{C}^{2}}-\frac{5}{2} \right) \right],
 \end{equation}
 
 \begin{equation}
 	{{R}_{1,\rm{r}}}\left( t,\mathbf{x},\mathbf{c} \right)={{E}_{\rm{t}}}\left( {T}_{\rm{t}} \right) ,
 \end{equation}
 
 \begin{equation}
 	{{R}_{2,\rm{t}}}\left( t,\mathbf{x},\mathbf{c} \right)={{E}_{\rm{t}}}\left( {T}_{\rm{t}} \right) \frac{{\nu}k{{T}_{\rm{r}}}}{2}\left[ 1+\frac{2}{5nm}{{\left( \frac{m}{k{{T}_{\rm{t}}}} \right)}^{2}}{{q}_{0,i}}{{C}_{i}}\left( \frac{m}{2k{{T}_{\rm{t}}}}{{C}^{2}}-\frac{5}{2} \right) \right],
 \end{equation}
 
 \begin{equation}
 	{{R}_{2,\rm{r}}}\left( t,\mathbf{x},\mathbf{c} \right)={{E}_{\rm{t}}}\left( {T}_{\rm{t}} \right) \frac{{\nu}kT}{2}\left[ 1+\frac{2}{{\nu}nm}{{\left( \frac{m}{kT} \right)}^{2}}{{q}_{1,i}}{{C}_{i}} \right].
 \end{equation}
 Finally, the macroscopic quantities are calculated as:
 \begin{equation}
 	\begin{aligned}
 		& n=\int{{{R}_{1}}d\mathbf{c}},\quad \mathbf{U}=\int{{{R}_{1}}\frac{\mathbf{c}}{n}d\mathbf{c}},\quad {{p}_\mathit{ij}}=\int{{{R}_{1}}m{{C}_{i}}{{C}_{j}}d\mathbf{c}},\quad {{T}_{\rm{t}}}=\int{{{R}_{1}}\frac{m{{C}^{2}}}{3nk}d\mathbf{c}}, \\ 
 		& {{T}_{\rm{r}}}=\int{{{R}_{2}}\frac{2}{{\nu}nk}d\mathbf{c}},\quad {{\mathbf{q}}_{\rm{t}}}=\int{{{R}_{1}}\frac{1}{2}m{{C}^{2}}\mathbf{C}d\mathbf{c}},\quad {{\mathbf{q}}_{\rm{r}}}=\int{{{R}_{2}}\mathbf{C}d\mathbf{c}}. \\ 
 	\end{aligned}
 \end{equation}
 \subsection{Dimensionless analysis}
 In the calculations, the dimensionless quantities normalized  by the reference length, density, temperature, and velocity are introduced as follows:
 \begin{equation}
 	L_{\rm{ref}}=L_{\rm{c}},\quad \rho_{\rm{ref}}=\rho_{\rm{\infty}},\quad T_{\rm{ref}}=T_{\rm{\infty}},\quad U_{\rm{ref}}=\sqrt{2RT_{\rm{ref}}},
 \end{equation}
where $L_{\rm{c}}$ is the characteristic length scale of the flow, $\rho_{\rm{\infty}}$, $T_{\rm{\infty}}$ are the density and temperature of the free-stream, respectively. Then the following dimensionless quantities can be obtained:
\begin{equation}
	\begin{aligned}
		& \hat{L}=\frac{L}{{{L}_{\rm{ref}}}},\quad \hat{\rho }=\frac{\rho }{{{\rho }_{\rm{ref}}}},\quad \hat{T}=\frac{T}{{{T}_{\rm{ref}}}},\quad \hat{U}=\frac{U}{{{U}_{\rm{ref}}}}, \\ 
		& \hat{t}=\frac{t}{{{L}_{\rm{ref}}}U_{\rm{ref}}^{-1}},\quad \hat{n}=\frac{n}{L_{\rm{ref}}^{-3}},\quad \hat{m}=\frac{m}{{{\rho }_{\rm{ref}}}L_{\rm{ref}}^{3}},\quad \hat{\mu }=\frac{\mu }{{{\rho }_{\rm{ref}}}{{U}_{\rm{ref}}}{{L}_{\rm{ref}}}}, \\ 
		& \hat{p}=\frac{p}{{{\rho }_{\rm{ref}}}U_{\rm{ref}}^{2}},\quad \mathbf{\hat{q}}=\frac{\mathbf{q}}{{{\rho }_{\rm{ref}}}U_{\rm{ref}}^{3}},\quad \hat{\tau }=\frac{\tau}{{{L}_{\rm{ref}}}U_{\rm{ref}}^{-1}}. \\ 
	\end{aligned}
\end{equation}
Finally, a complete dimensionless system is obtained. In the following, all variables without the ``hat'' are nondimensionalized by default, unless stated otherwise.

\subsection{ 0-Dimensional homogenous relaxation}
In the zero-dimensional homogeneous relaxation case, the flow field is spatially uniform, allowing us to neglect the transport terms on the left-hand side of the model equation and focus exclusively on the effects of intermolecular collisions. This configuration is particularly suitable for investigating non-equilibrium relaxation processes in flow fields. Given the spatial homogeneity, the governing equation simplifies to:
\begin{equation}
	\frac{\partial f}{\partial t}=\frac{{{g}_{\rm{t}}}-f}{{{\tau}_{\rm{t}}}}+\frac{{{g}_{\rm{r}}}-{{g}_{\rm{t}}}}{{{Z}_{\rm{rot}}}{{\tau}_{\rm{t}}}}.
\end{equation}

\begin{figure}
	\centering
	
	\begin{subfigure}[b]{0.32\textwidth}
		\includegraphics[width=\linewidth]{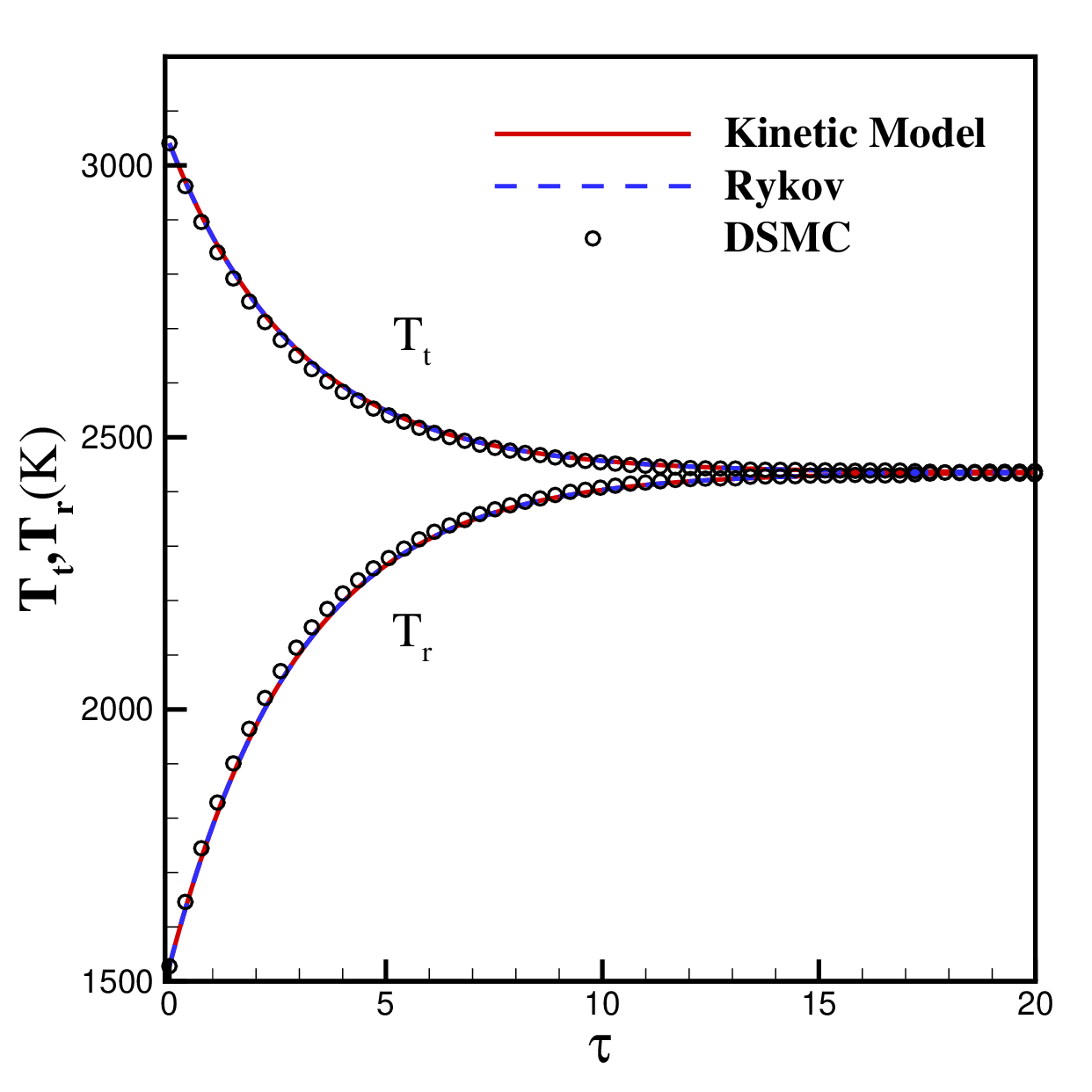}
		\caption{}
	\end{subfigure}
	\hfill
	\begin{subfigure}[b]{0.32\textwidth}
		\includegraphics[width=\linewidth]{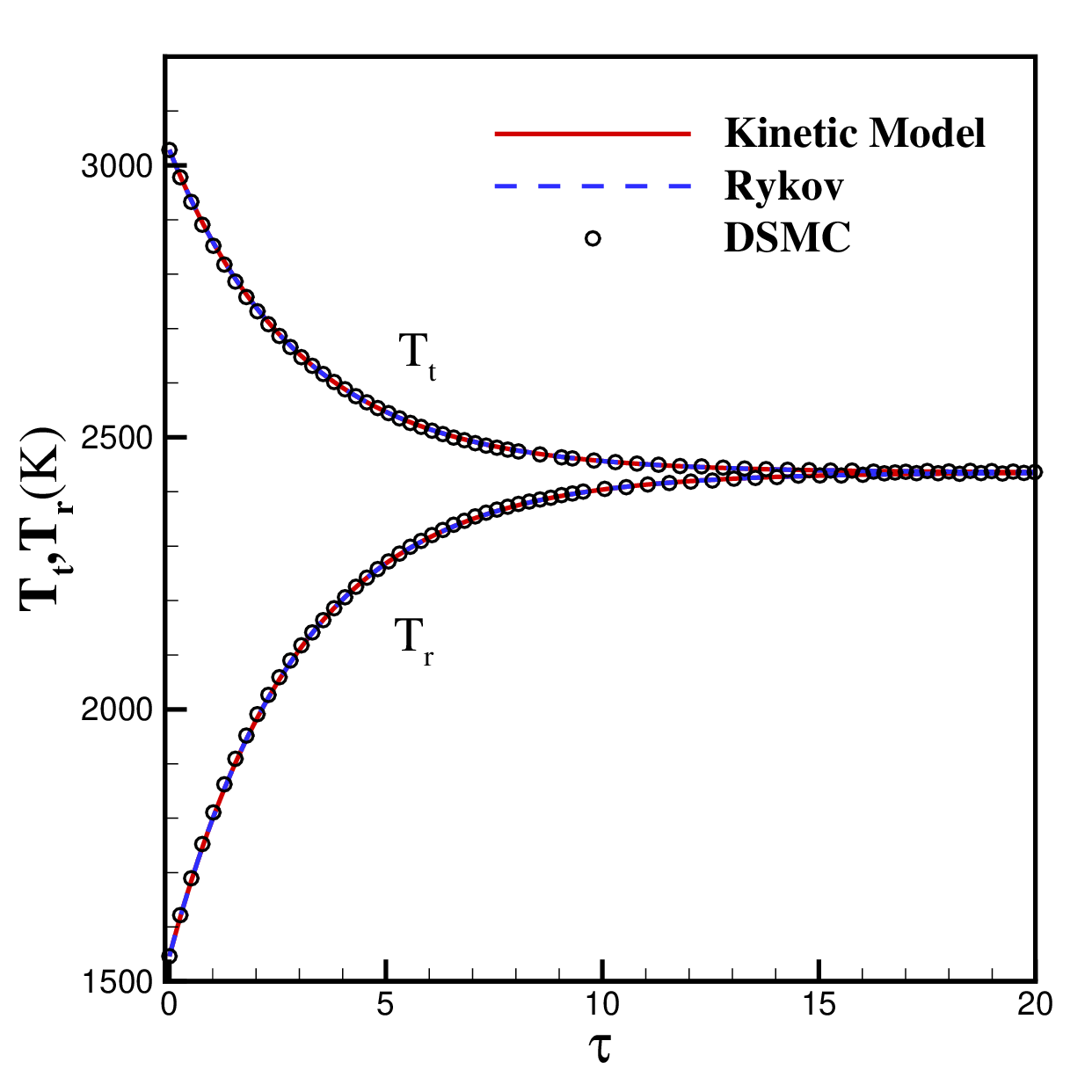}
		\caption{}
	\end{subfigure}
	\hfill
	\begin{subfigure}[b]{0.32\textwidth}
		\includegraphics[width=\linewidth]{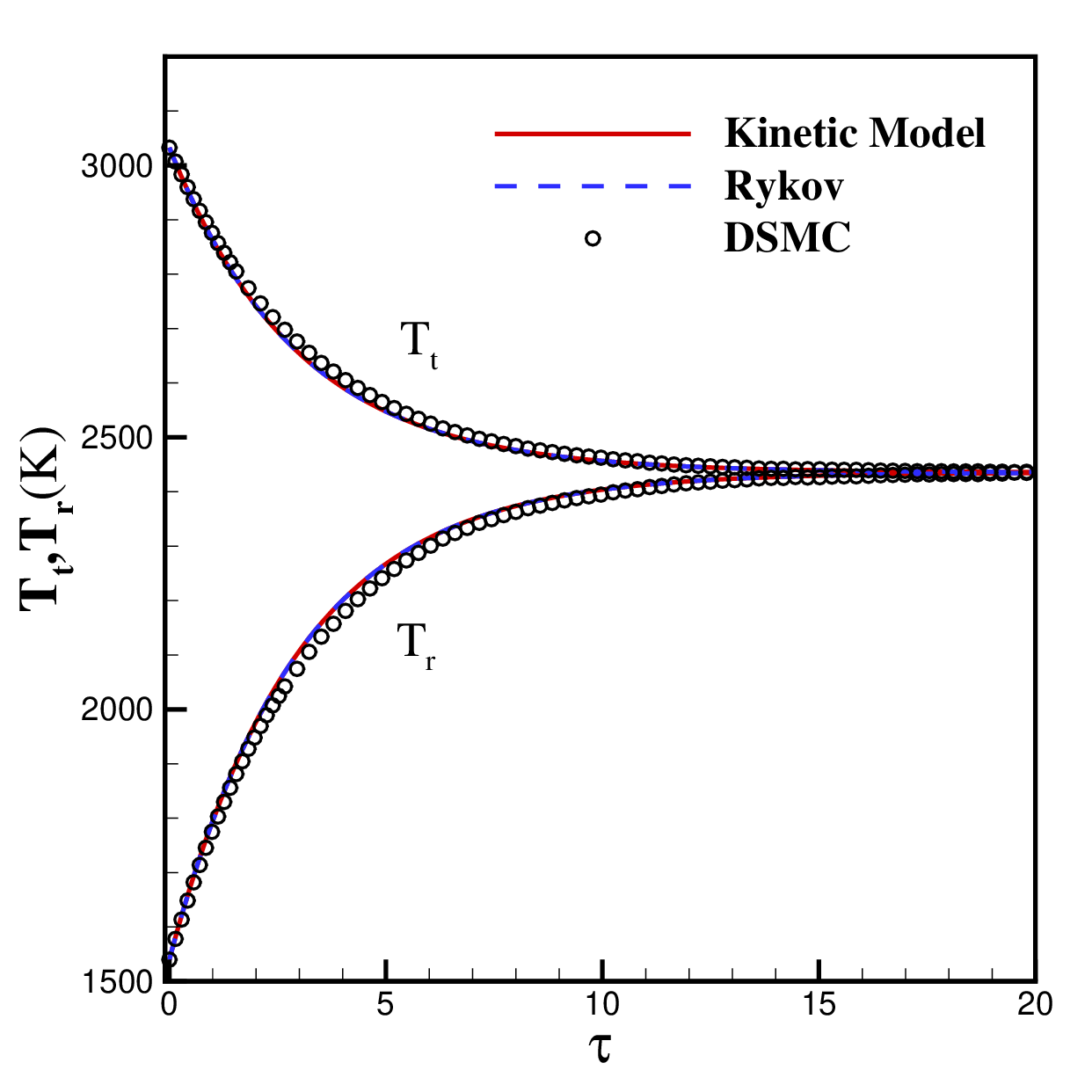}
		\caption{}
	\end{subfigure}	
	
	\begin{subfigure}[b]{0.32\textwidth}
		\includegraphics[width=\linewidth]{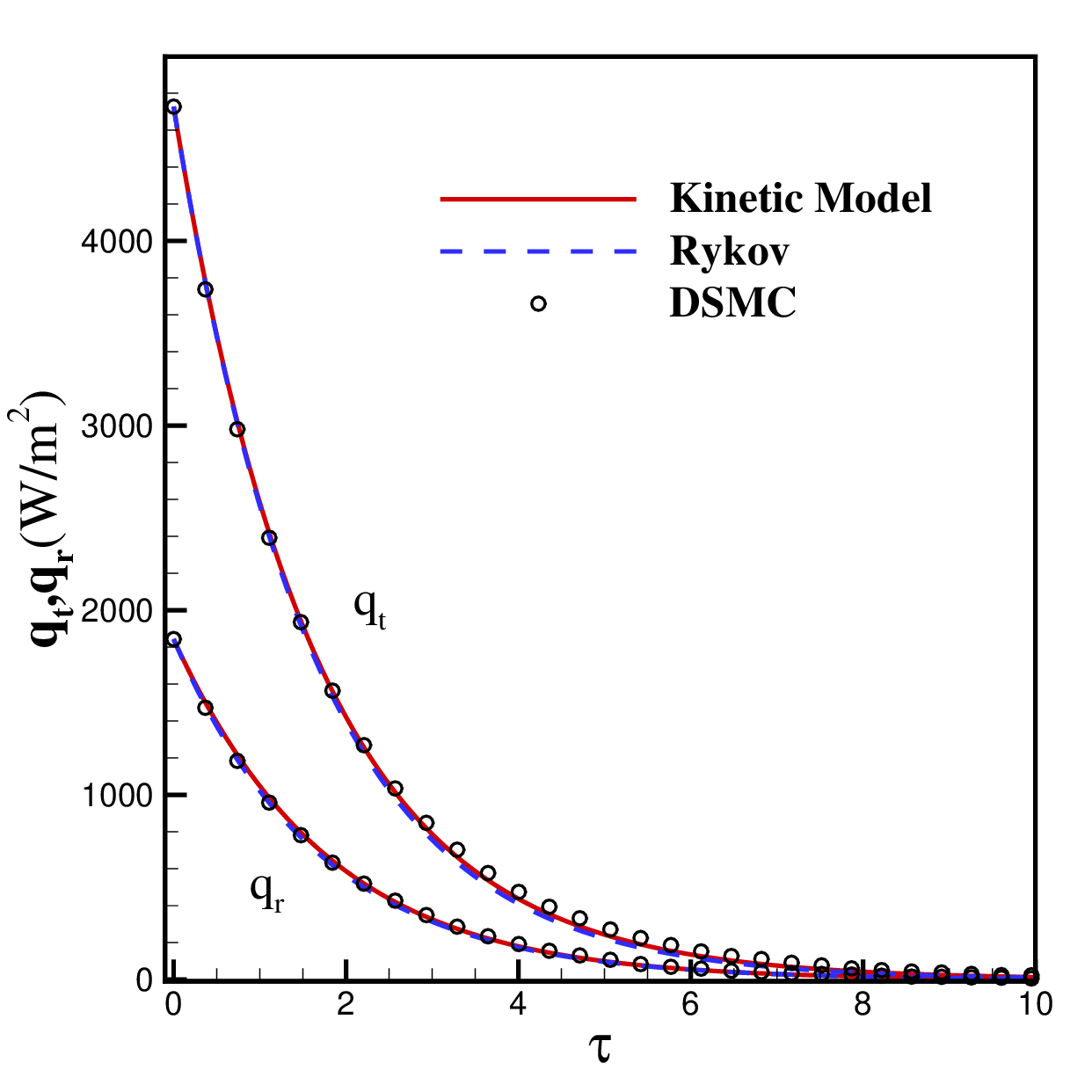}
		\caption{}
	\end{subfigure}
	\hfill
	\begin{subfigure}[b]{0.32\textwidth}
		\includegraphics[width=\linewidth]{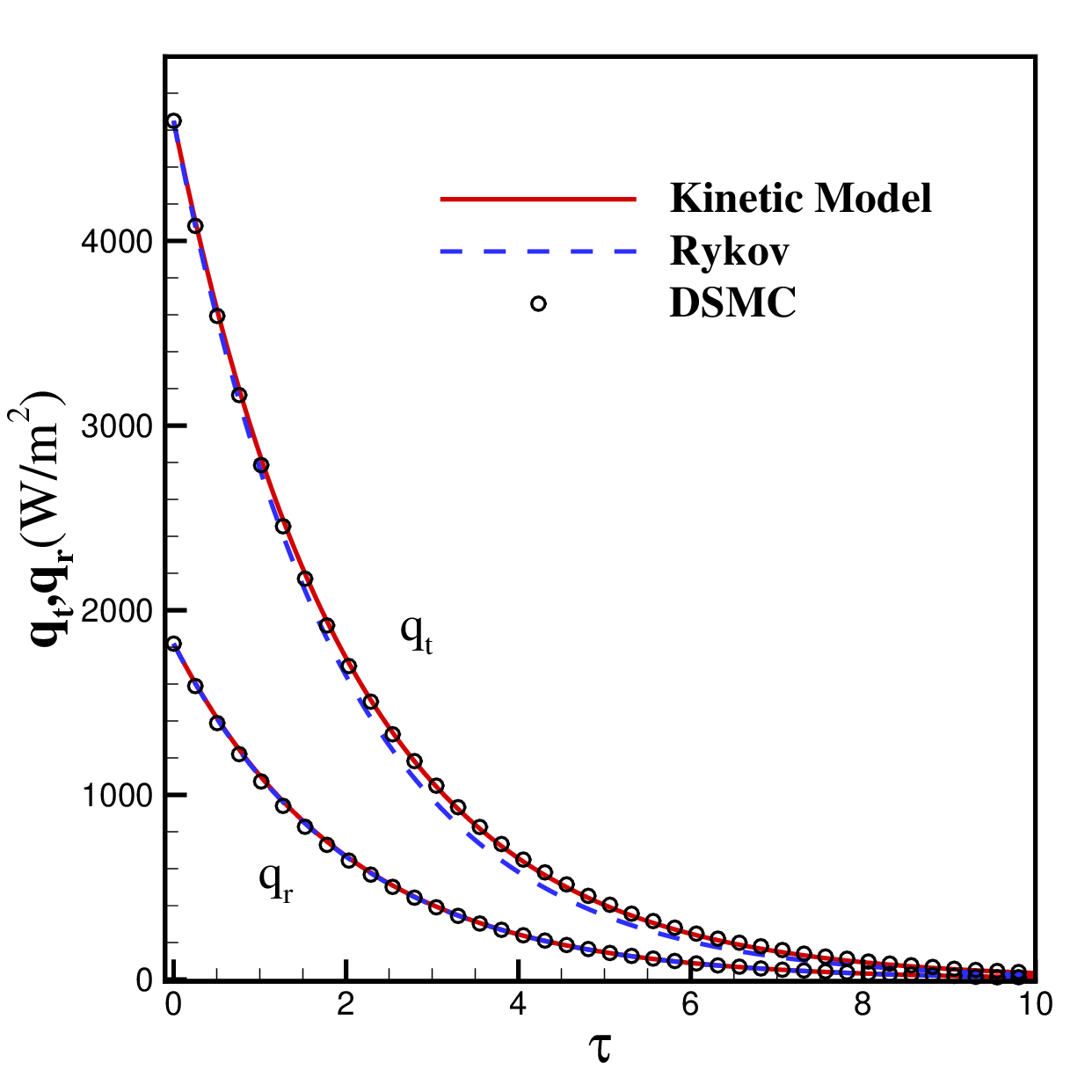}
		\caption{}
	\end{subfigure}
	\hfill
	\begin{subfigure}[b]{0.32\textwidth}
		\includegraphics[width=\linewidth]{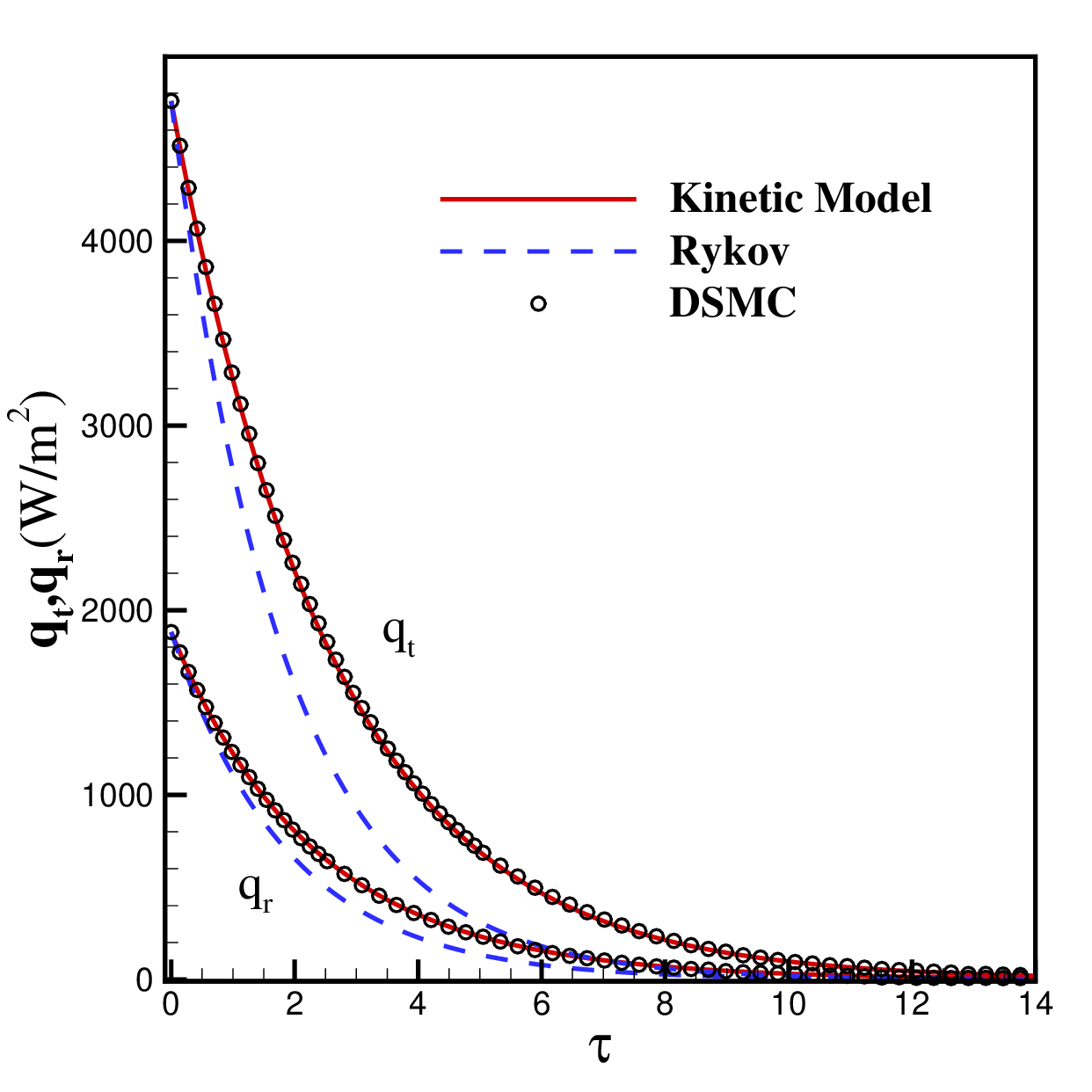}
		\caption{}
	\end{subfigure}
	
	\caption{Time evolution of temperature and heat flux during zero-dimensional homogeneous relaxation from a strongly non-equilibrium initial state corresponding to a Ma=8 normal shock. (a)-(c) are the temperature profiles with HS, VHS and Maxwell molecules. (d)-(f) are the heat flux profiles with HS, VHS and Maxwell molecules.}
	\label{0dim_relax_Ma8}
\end{figure}

In this zero-dimensional case, all particles are considered to reside within a single computational cell. To facilitate observation of the relaxation process, the flow field is initialized in a non-equilibrium state, evolving toward equilibrium through collisional relaxation. We initialize the particle distribution using two distinct groups following Maxwellian distributions, mimicking the strongly non-equilibrium velocity distribution functions found in shock waves. The macroscopic variables for these two initial Maxwellian distributions, corresponding to Mach numbers $\rm{Ma=8}$, are specified as follows:

\begin{equation}
	Ma=8\quad \left\{ \begin{matrix}
		{{\rho }_{\rm{A}}}=4.65\times {{10}^{-6}}{\rm{kg/m^{3}}},{{U}_{\rm{A}}}=2695.2{\rm{m/s}},{{T}_{\rm{A}}}=273{\rm{K}}  \\
		{{\rho }_{\rm{B}}}=4.65\times {{10}^{-6}}{\rm{kg/m^{3}}},{{U}_{\rm{B}}}=336.9{\rm{m/s}},{{T}_{\rm{B}}}=2730{\rm{K}}  \\
	\end{matrix} \right.
\end{equation}
In this case study, we solve both the kinetic model and Rykov model with rotational collision number fixed at ${{Z}_{\rm{rot,DSMC}}}=3$. To examine the impact of collision models on relaxation processes, we implement three distinct molecular interaction models: hard sphere molecules ($\omega=0.5$), nitrogen molecules ($\omega=0.74$),  Maxwell molecules ($\omega=1$). Parallel DSMC simulations are performed with $10^5$ simulator particles to minimize statistical fluctuations. Crucially, the collision number of the kinetic model is determined through Eq.~(\ref{Zrot_DSMC}). Figure~\ref{0dim_relax_Ma8} presents zero-dimensional homogeneous relaxation simulations within normal shock wave at $\rm{Ma=8}$, comparing results from the new kinetic model, Rykov model and DSMC simulations. For temperature relaxation, the kinetic model exhibits excellent agreement with DSMC data, confirming its accuracy in energy exchange processes; regarding heat flux relaxation, the kinetic model maintains consistent alignment with DSMC across all three collision models (hard sphere, nitrogen, Maxwell), while Rykov model shows varying degrees of deviation attributable to its neglect of translational-rotational heat flux coupling. The new model overcomes this limitation by explicitly incorporating cross-coupling effects, non-equilibrium intensity dependence, and molecular collision model sensitivity into heat flux relaxation dynamics.

\subsection{normal shock wave}
Normal shock wave represents prevalent phenomena in high-speed flows and constitute a focal point in non-equilibrium flow research. The normal shock wave manifests as a strong discontinuity, with its thickness typically spanning approximately 20 molecular mean free paths. Gas compression within confined space induces severe non-equilibrium effects, particularly intense energy exchange processes for molecular gases within the normal shock wave. Consequently, this scenario serves as both a benchmark and challenging test case for numerical methods predicting rarefied and multiscale flows. The Rankine-Hugoniot relations prescribe the inlet and outlet boundary conditions as follows:
\begin{equation}
	\begin{aligned}
		& \frac{{{\rho }_{2}}}{{{\rho }_{1}}}=\frac{\left( \gamma +1 \right){{\rm{Ma}}^{2}}}{\left( \gamma -1 \right){{\rm{Ma}}^{2}}+2}, \\ 
		& \frac{{{T}_{2}}}{{{T}_{1}}}=\frac{\left( 1+\dfrac{\gamma -1}{2}{{\rm{Ma}}^{2}} \right)\left( \dfrac{2\gamma }{\gamma -1}{{\rm{Ma}}^{2}}-1 \right)}{{{\rm{Ma}}^{2}}\left( \dfrac{2\gamma }{\gamma -1}+\dfrac{\gamma -1}{2} \right)}, \\ 
	\end{aligned}
\end{equation}

The subscripts ``1'' and ``2'' denote variables upstream and downstream of the normal shock wave, respectively, with specific heat ratio $\gamma=5/3$ and inlet Mach number $\rm{Ma}$. Three sets of numerical simulations were conducted on nitrogen normal shock wave at Mach numbers 1.53, 4.0, and 7.0. Viscosity is calculated using the VHS model. For all cases, the pre-shock density ${{\rho}_{1}}=1.7413\times{{10}^{-2}}{\rm{kg/m^{3}}}$ and temperature ${{T}_{1}}=226.149\rm{K}$  remain identical. Dimensionless quantities are adopted in computations. Reference values are defined as: ${{L}_{\rm{ref}}}={{\lambda }_{1}},{{\rho }_{\rm{ref}}}={{\rho }_{1}},{{T}_{\rm{ref}}}={{T}_{1}},{{U}_{\rm{ref}}}=\sqrt{{2k{{T}_{\rm{ref}}}}/{m}}$ where the mean free path $\lambda_1$ is computed by:
\begin{equation}
	\lambda =\frac{2\mu \left( 7-2\omega  \right)\left( 5-2\omega  \right)}{15n{{\left( 2\pi mkT \right)}^{{1}/{2}\;}}}.
\end{equation}
The computational domain spans $[-100\lambda_1, 100\lambda_1]$ discretized with a uniform grid of 500 cells ($\Delta x=0.5\lambda_1$). Velocity space is discretized into 401 uniformly distributed points within $[-18, 18]$, with numerical integration carried out via the Newton-Cotes quadrature method.

\begin{figure}
 	\centering
 	
 	\begin{subfigure}[b]{0.32\textwidth}
 		\includegraphics[width=\linewidth]{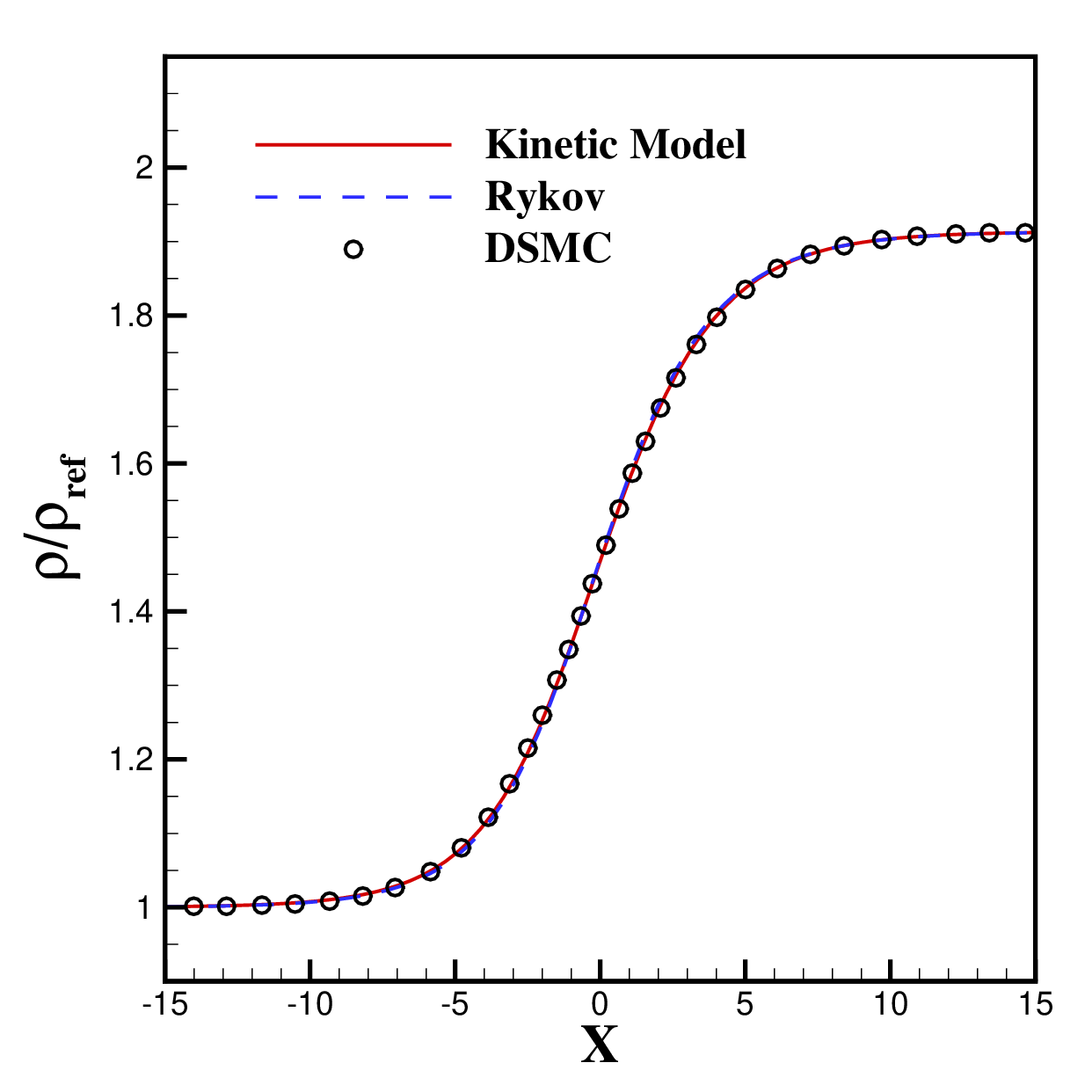}
 		\caption{Density\\(Ma=1.53)}
 	\end{subfigure}
 	\hfill
 	\begin{subfigure}[b]{0.32\textwidth}
 		\includegraphics[width=\linewidth]{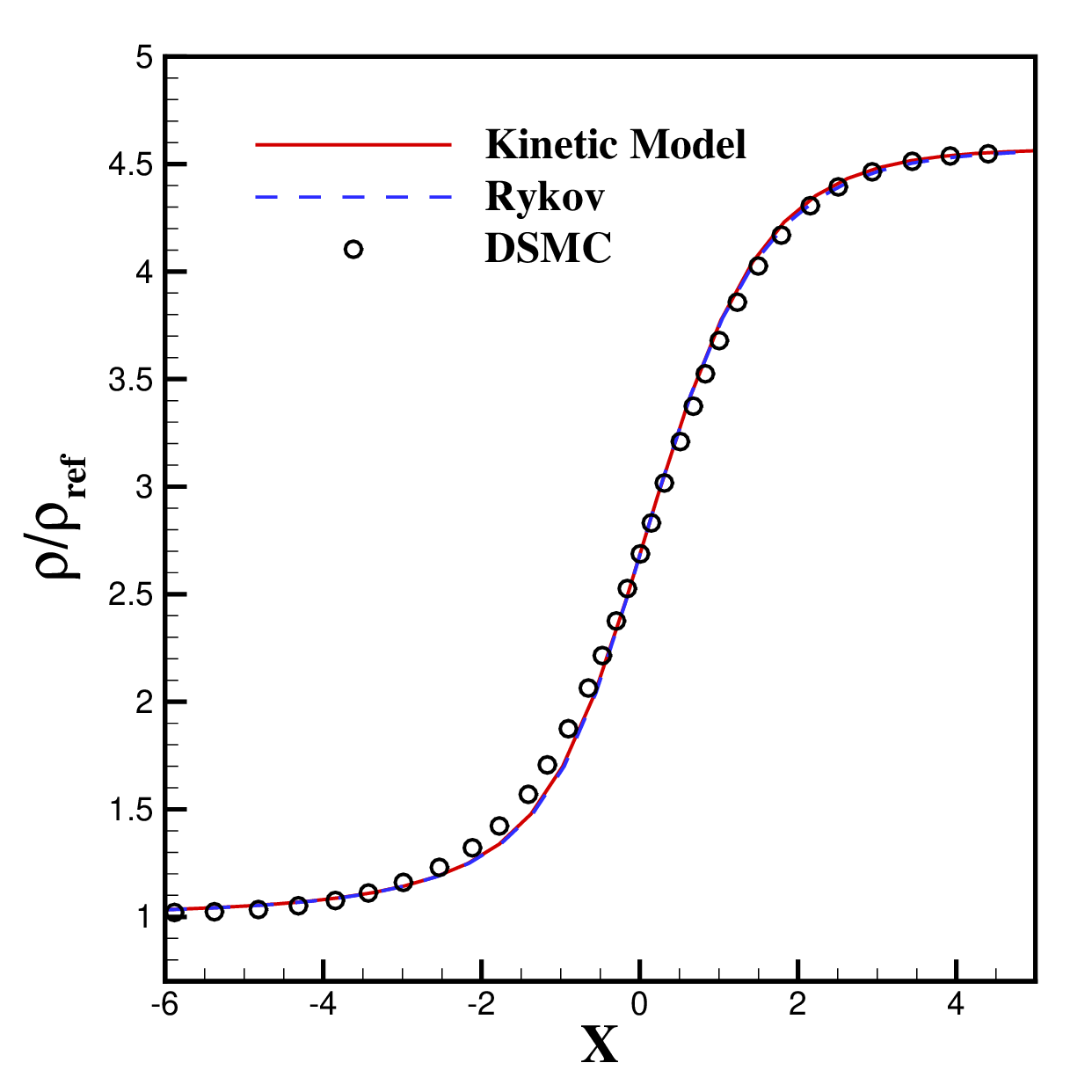}
 		\caption{Density\\(Ma=4.0)}
 	\end{subfigure}
 	\hfill
 	\begin{subfigure}[b]{0.32\textwidth}
 		\includegraphics[width=\linewidth]{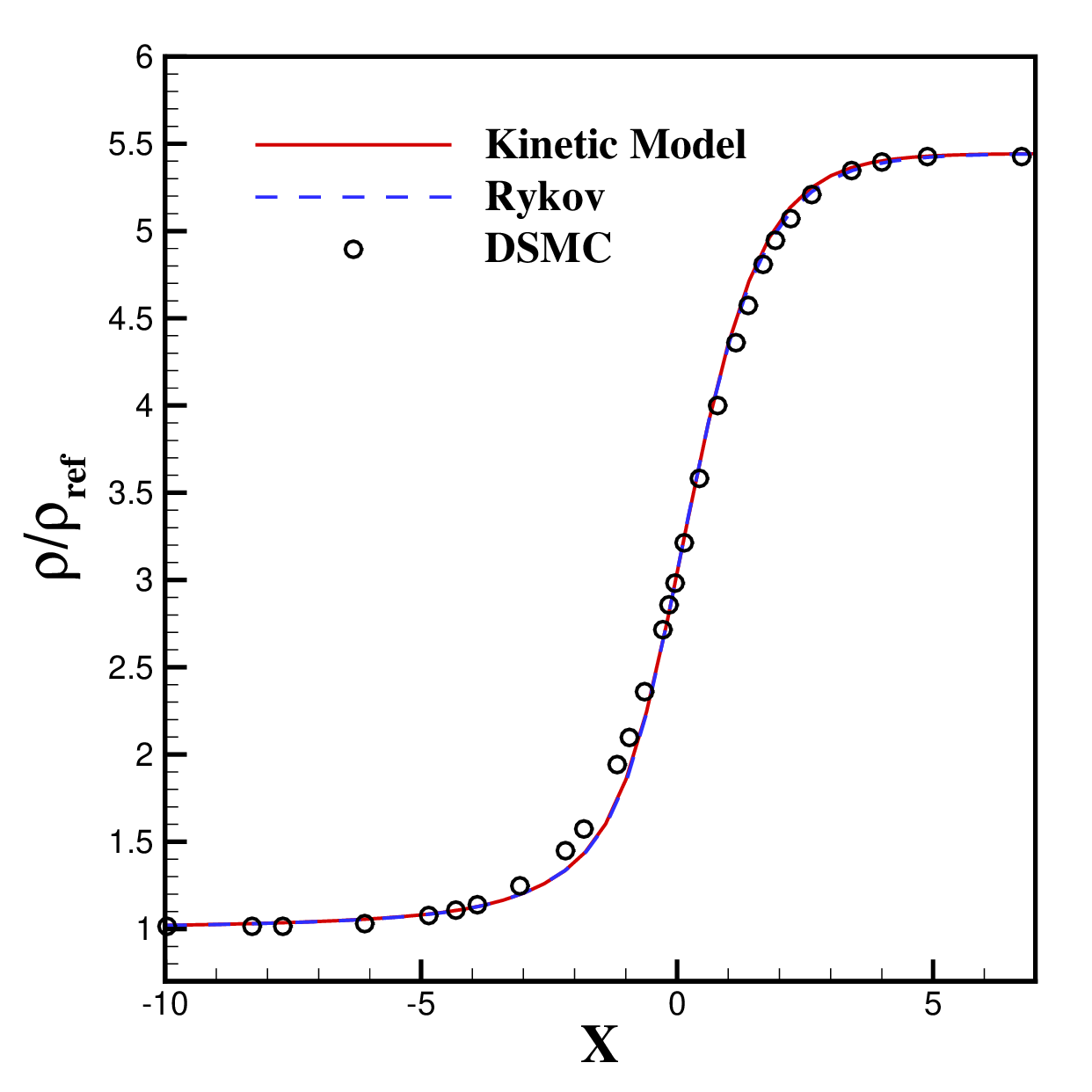}
 		\caption{Density\\(Ma=7.0)}
 	\end{subfigure}
 	
 	\vspace{0.25cm}
 	
 	\begin{subfigure}[b]{0.32\textwidth}
 		\includegraphics[width=\linewidth]{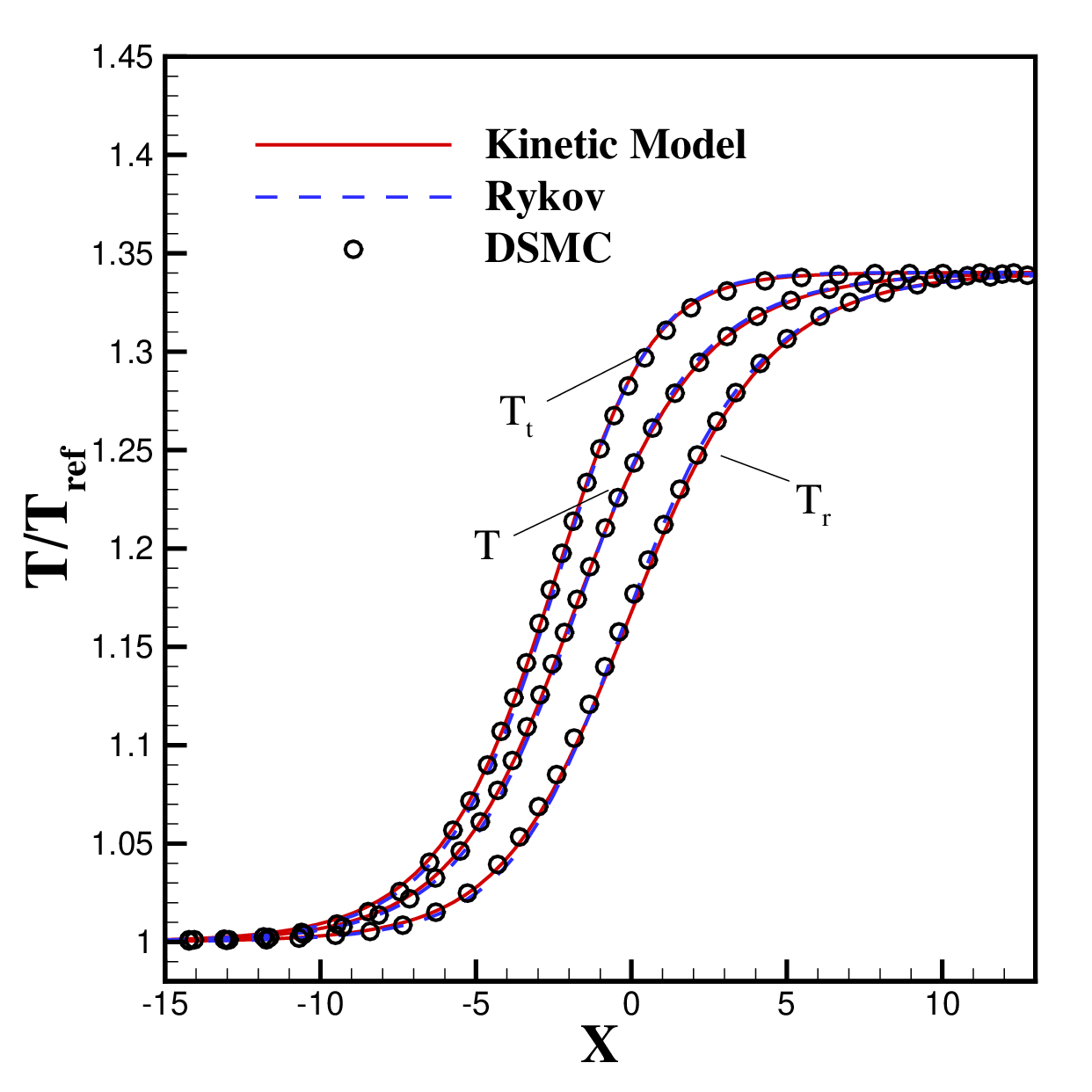}
 		\caption{Temperature\\(Ma=1.53)}
 	\end{subfigure}
 	\hfill
 	\begin{subfigure}[b]{0.32\textwidth}
 		\includegraphics[width=\linewidth]{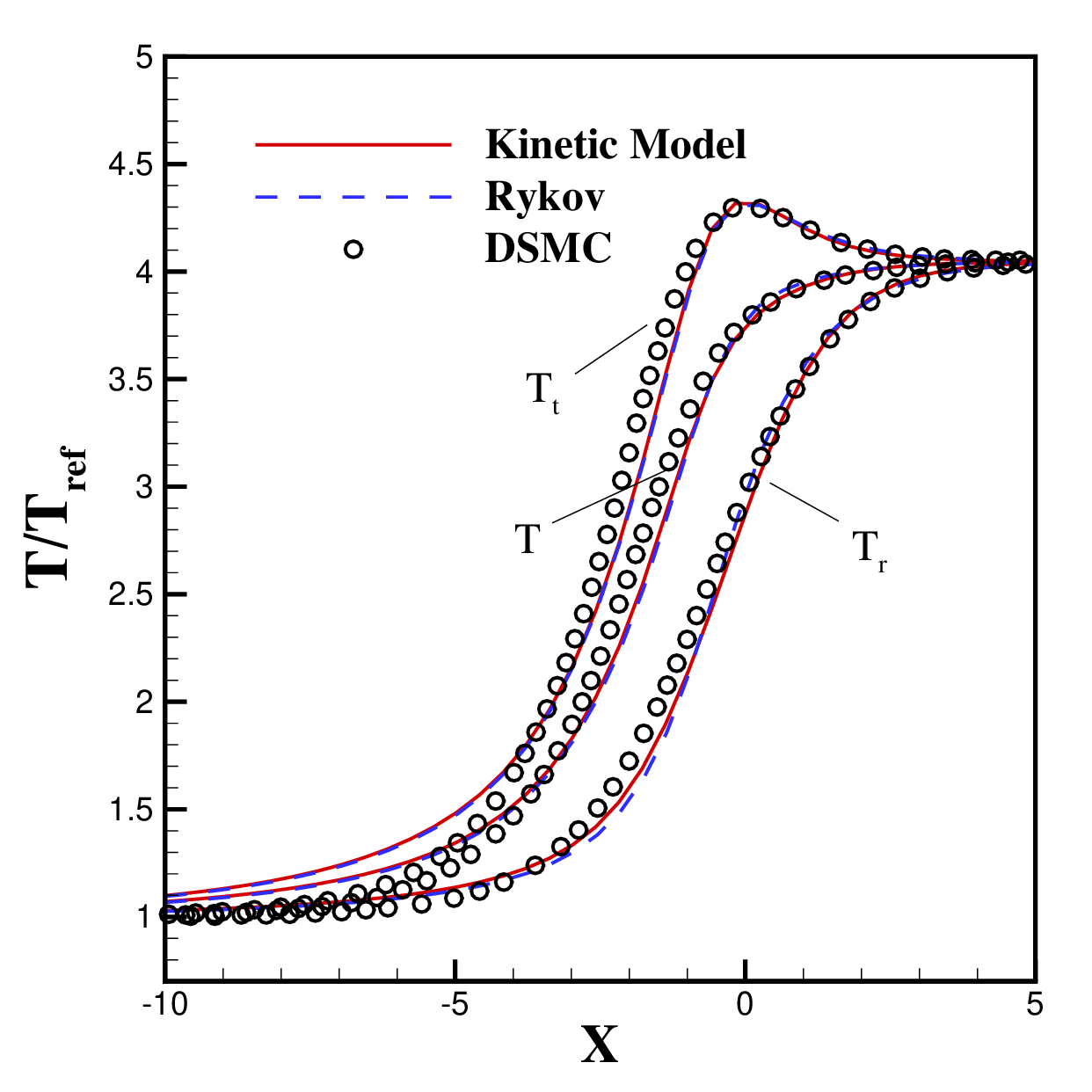}
 		\caption{Temperature\\(Ma=4.0)}
 	\end{subfigure}
 	\hfill
 	\begin{subfigure}[b]{0.32\textwidth}
 		\includegraphics[width=\linewidth]{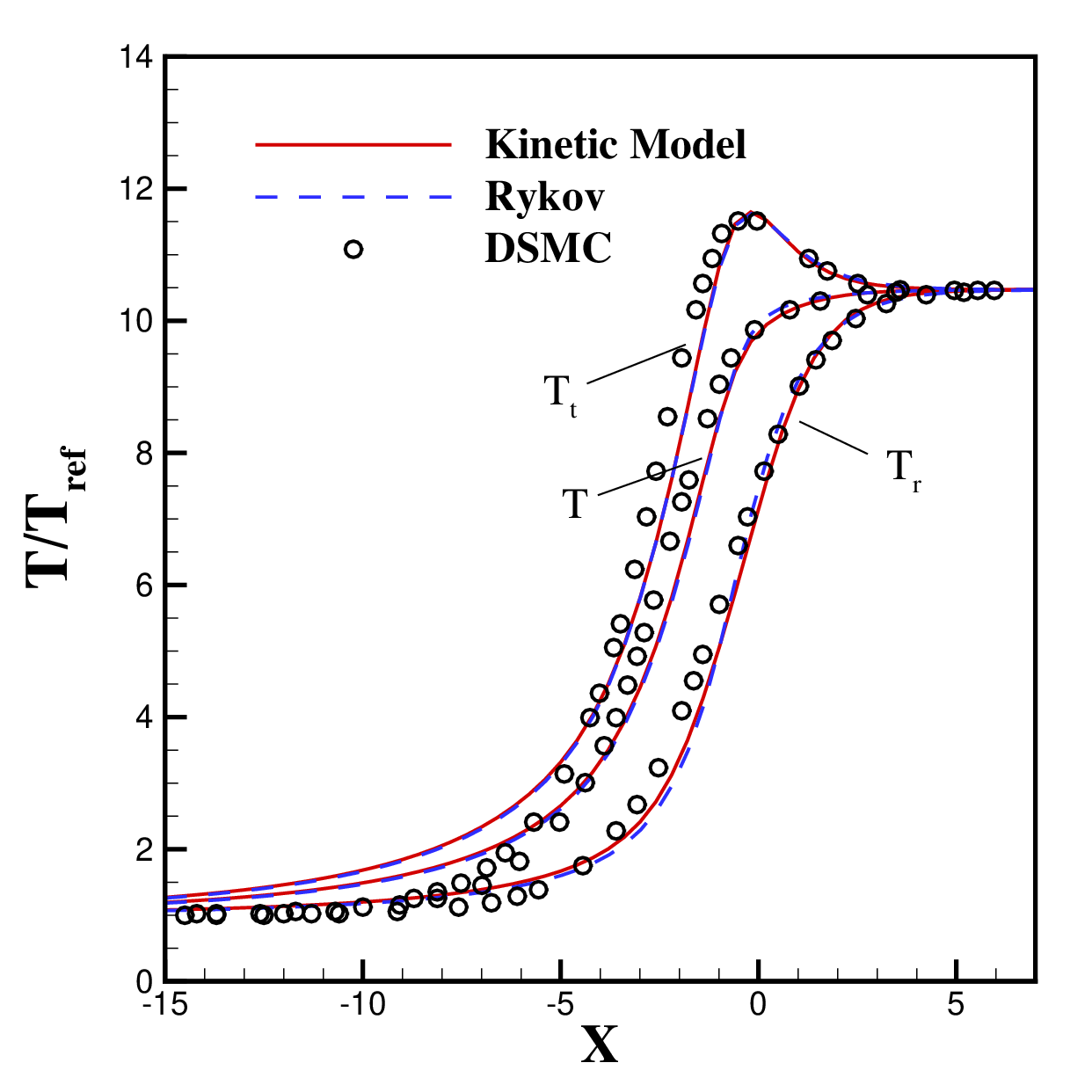}
 		\caption{Temperature\\(Ma=7.0)}
 	\end{subfigure}
 	
 	\caption{Shock structure profiles for nitrogen at Ma = 1.53, 4.0, 7.0. (a)-(c) Density distributions. (d)-(f) Temperature distributions.}
 	\label{shock_rho_T}
 \end{figure}

\begin{figure}
 	\centering
 	\includegraphics[width=0.32\textwidth]{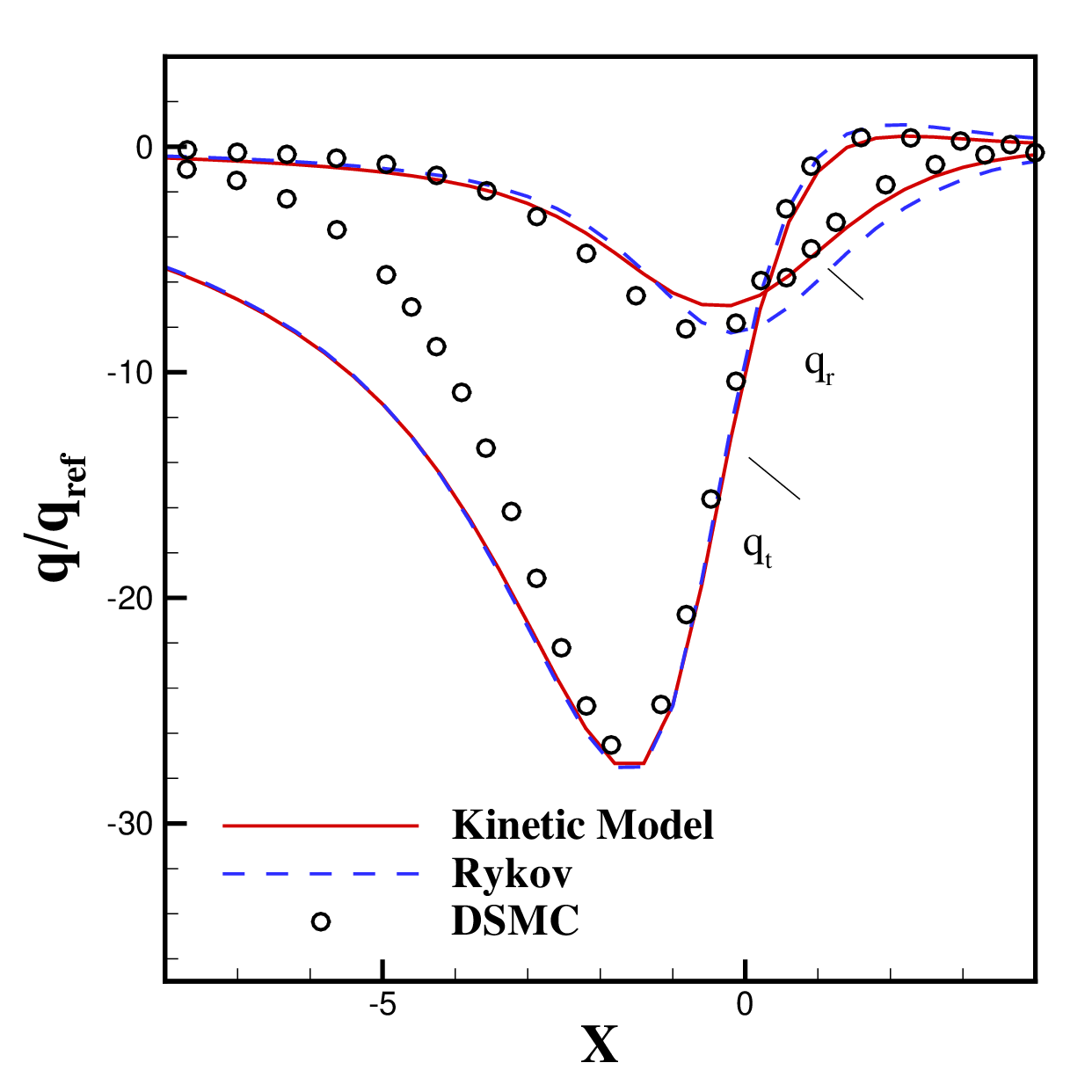}
 	\caption{Translational and rotational heat flux distributions for the nitrogen shock structure at Ma = 7.0.}
 	\label{shock_q}
 \end{figure}

In this case study, we solve both the kinetic model and the Rykov model, setting the viscosity index $\omega=0.72$ and rotational collision number $Z_{\rm{rot}}=2.4$ to benchmark against DSMC results from \citep{liu2014unified}. Figure~\ref{shock_rho_T} compares density and temperature distributions, revealing excellent agreement between the kinetic model and DSMC for density profiles across Mach numbers. While temperature distributions align well with DSMC at low Mach numbers, both kinetic models exhibit premature upstream temperature rise at high Mach numbers. This issue also appears in the heat flux profile, causing the same premature upstream rise (figure~\ref{shock_q}). This is a well-known limitation of BGK-type models, which use a single relaxation time $\tau_{\rm{t}}$ for all particles. However, in actual physical processes, high-speed particles tend to have shorter relaxation times. This limitation has been resolved in the modeling process by assigning shorter relaxation times to high-speed particles \citep{xu2021modeling}, while this refinement is out of scope of this work. Subsequently, we adjust parameters to $\omega=0.74$ and $Z_{\rm{rot}}=2.6671$ to compare heat flux distributions with \citep{jianan2022kinetic}. As shown in figure~\ref{shock_q}, in the high-temperature region downstream of the shock wave, the translational and rotational heat fluxes predicted by the kinetic model perform better overall than those from the Rykov model, indicating the importance of the coupled heat flux relaxation mechanism and its application in constructing model equations.

 \subsection{Planar Couette flow}
\begin{figure}
 	\centering
 	
 	\begin{subfigure}[b]{0.32\textwidth}
 		\includegraphics[width=\linewidth]{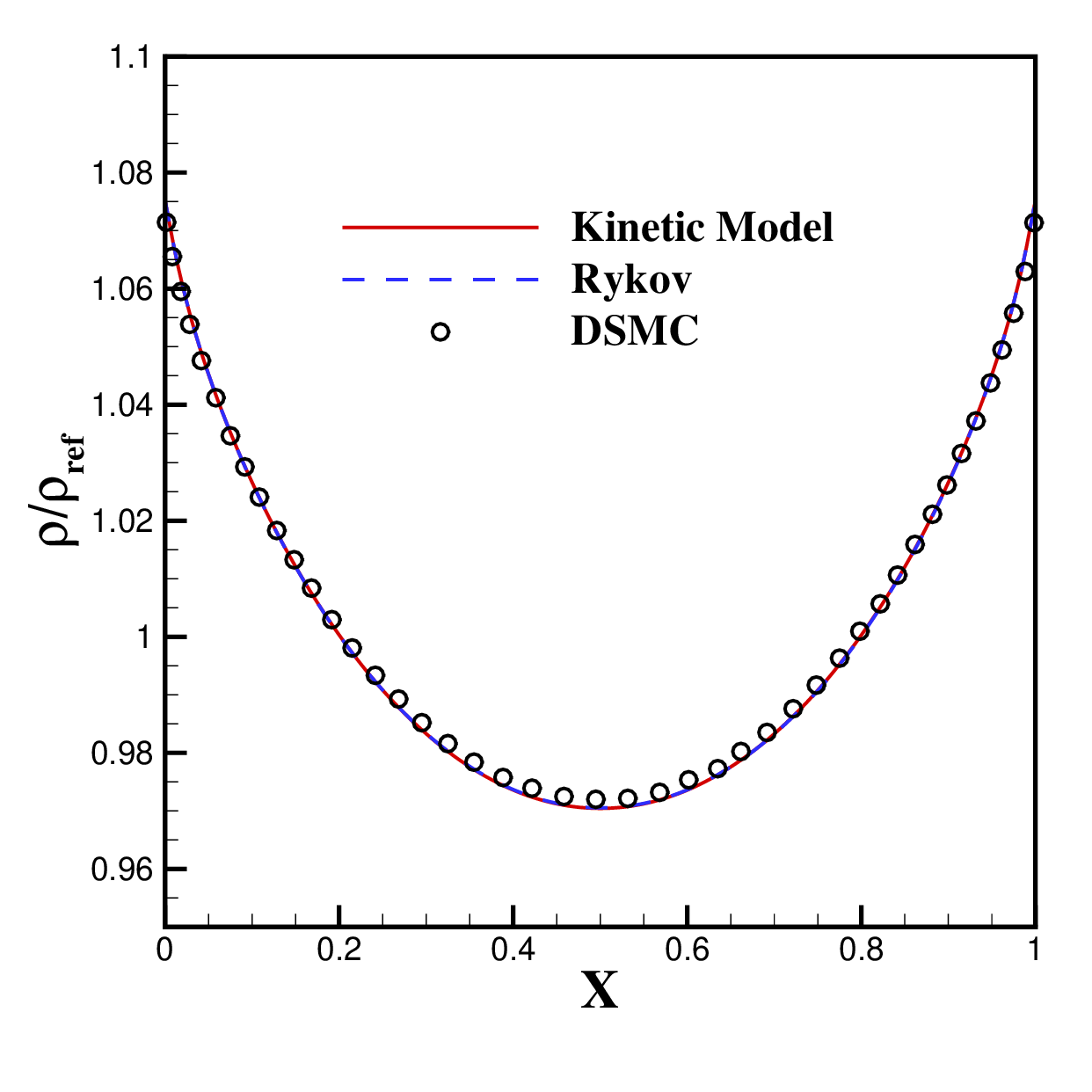}
 		\caption{}
 	\end{subfigure}
 	\hfill
 	\begin{subfigure}[b]{0.32\textwidth}
 		\includegraphics[width=\linewidth]{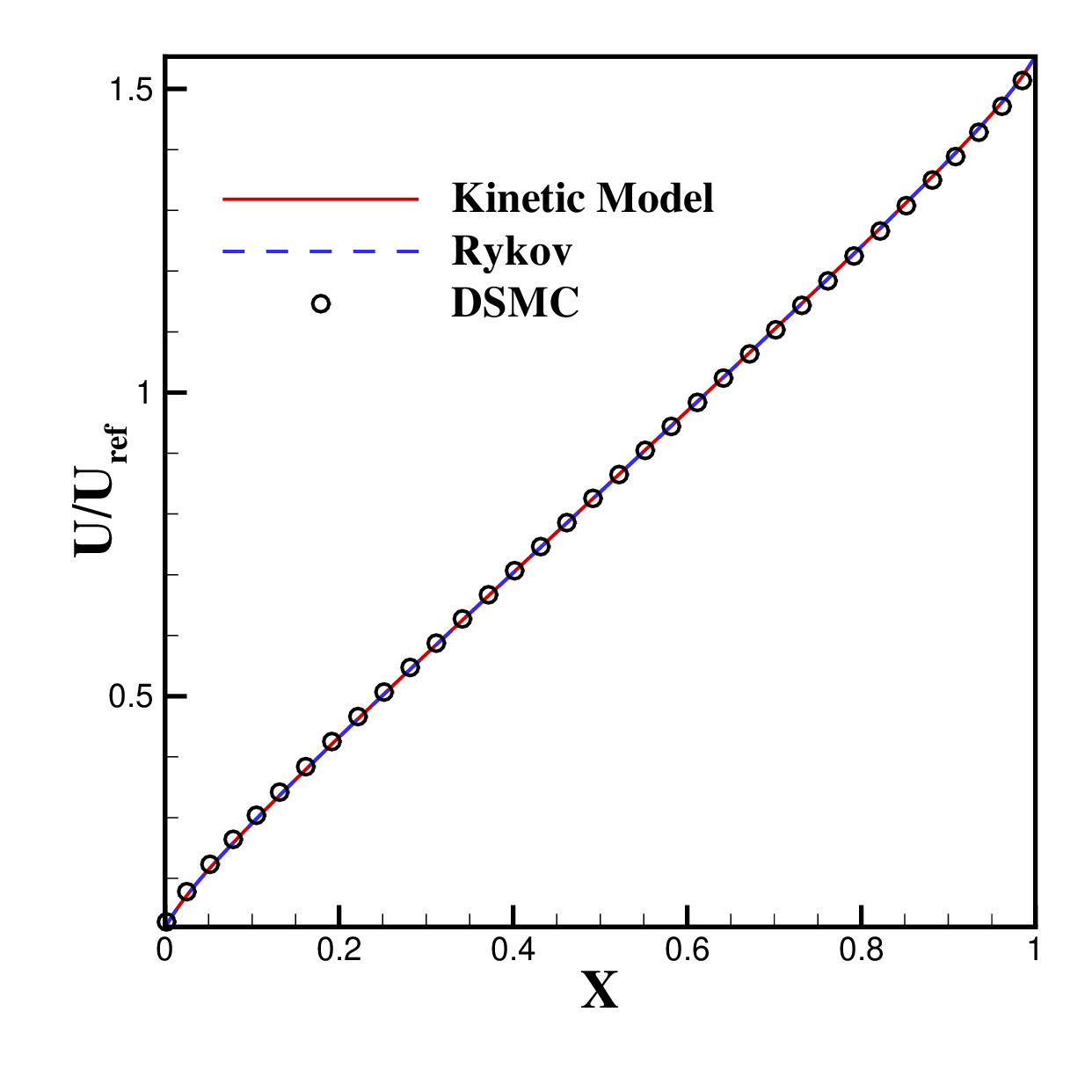}
 		\caption{}
 	\end{subfigure}
 	\hfill
 	\begin{subfigure}[b]{0.32\textwidth}
 		\includegraphics[width=\linewidth]{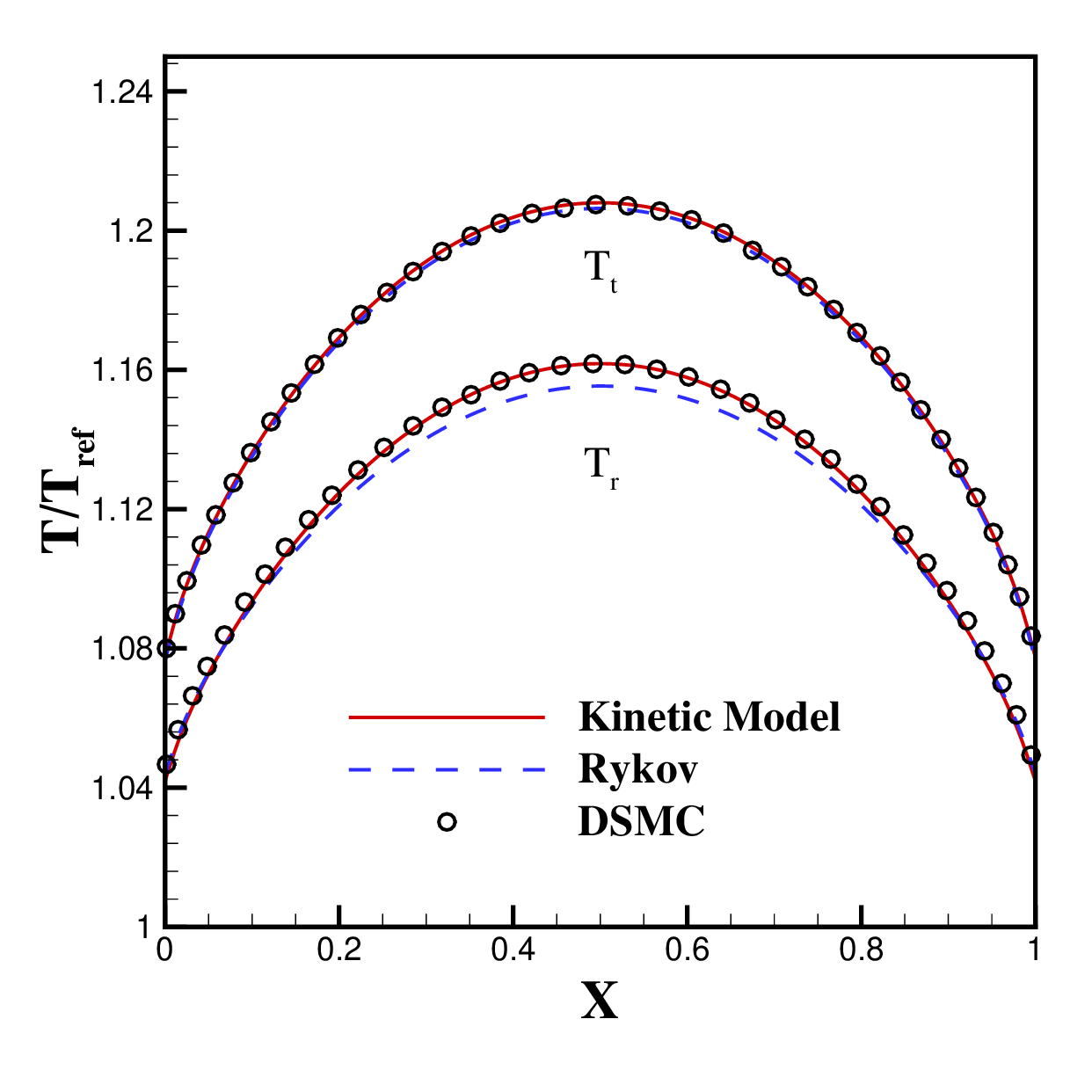}
 		\caption{}
 	\end{subfigure}
 	
 	\caption{Profiles for planar Couette flows of nitrogen at Kn=0.1. (a) Density, (b) velocity and (c) temperature.}
 	\label{Couette}
 \end{figure}

 This case examines planar Couette flow between two parallel plates separated by distance L, both maintained at temperature $T_0$. The lower plate moves with velocity $U=-V_{\rm{m}}$, while the upper plate moves in the opposite direction, with diffuse reflection boundary conditions applied. Nitrogen gas is simulated using the VHS model with viscosity index $\omega=0.74$. For the $\rm{Kn=0.1}$ condition, the rotational collision number is set to $Z_{\rm{rot}}=3.2384$ in the kinetic and Rykov model solutions, and to $Z_{\rm{rot}}=5$ in DSMC simulations according to Eq.~(\ref{Zrot_DSMC}). figure~\ref{Couette} compares density, velocity, and temperature distributions between the plates. Results demonstrate excellent agreement between the kinetic model predictions and DSMC for all macroscopic quantities. The Rykov model shows nearly identical density and velocity profiles to the kinetic model but underpredicts rotational temperature distributions near the centerline. This discrepancy may arise because the translational heat flux slightly exceeds the rotational heat flux in this case, whereas the Rykov model's neglect of the coupling between these heat fluxes leads to an underprediction of the rotational heat flux, consequently resulting in a lower peak rotational temperature.
 
 \subsection{Lid-driven cavity ﬂow}
 The lid-driven cavity flow benchmark serves as an excellent test case for validating the capability of proposed model to accurately simulate viscous effects. This study simulates cavity flows across a wide range of Knudsen numbers (0.075, 1, 10) to verify the model's effectiveness. For nitrogen under standard conditions, the reference lengths $L=8.27\times10^{-7}{\rm{m}},6.2\times10^{-8}{\rm{m}},6.2\times10^{-9}{\rm{m}}$ corresponding to Knudsen numbers of $\rm{Kn=0.075, 1, 10}$ respectively. The cavity features walls of length L, with the top wall moving rightward at constant velocity $U_{\rm{w}}$ while other walls remain stationary, all employing diffuse reflection boundary conditions. Nitrogen gas is simulated using a VHS molecular model with viscosity index $\omega=0.74$ and rotational collision number $Z_{\rm{rot}}=5$. All walls and the initial flow field maintain 273 K temperature, with the lid velocity set at 54.7 m/s and speed of sound at 336.9 m/s. 
 
\begin{figure}
 	\centering
 	
 	\begin{subfigure}[b]{0.32\textwidth}
 		\includegraphics[width=\linewidth]{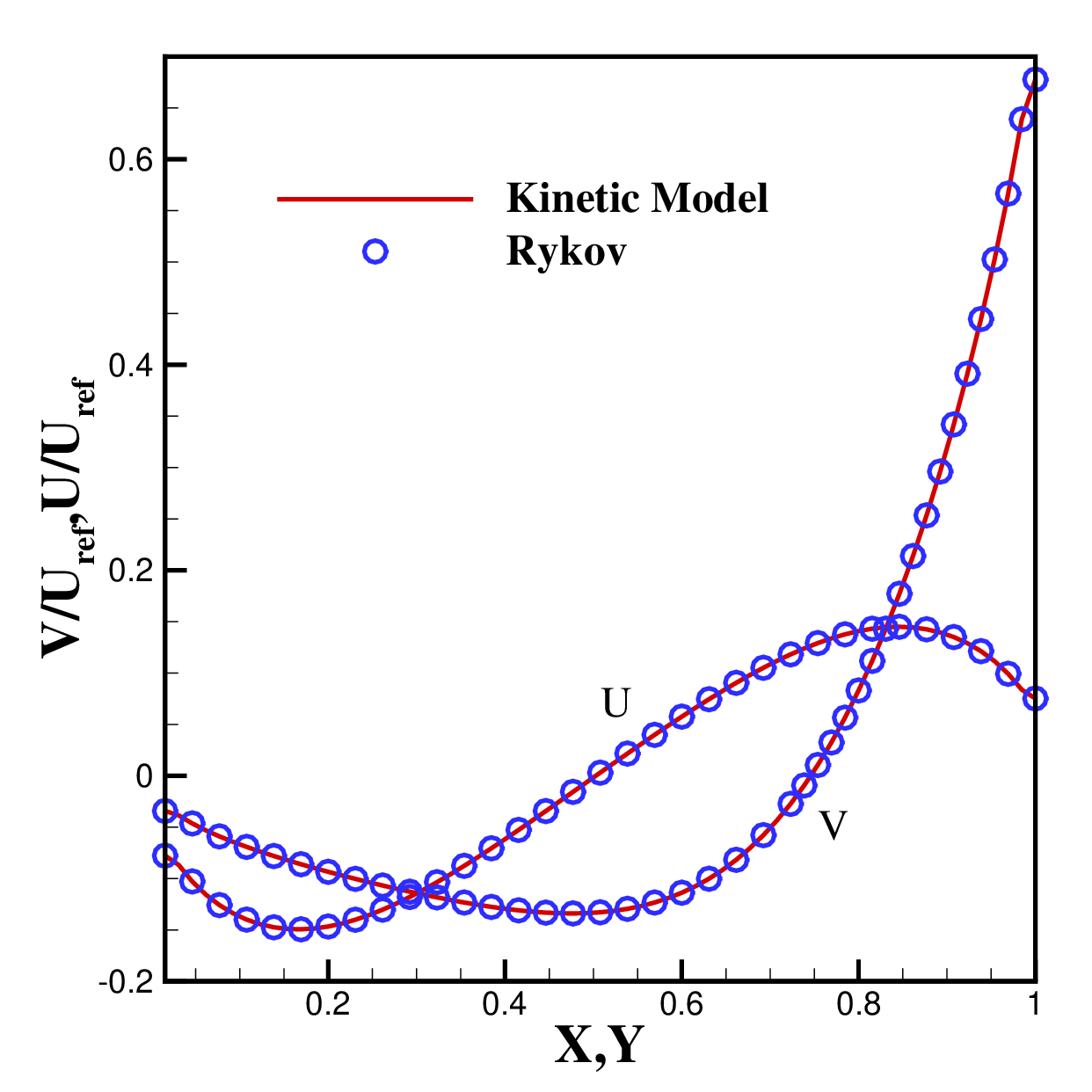}
 		\caption{Kn=0.075}
 	\end{subfigure}
 	\hfill
 	\begin{subfigure}[b]{0.32\textwidth}
 		\includegraphics[width=\linewidth]{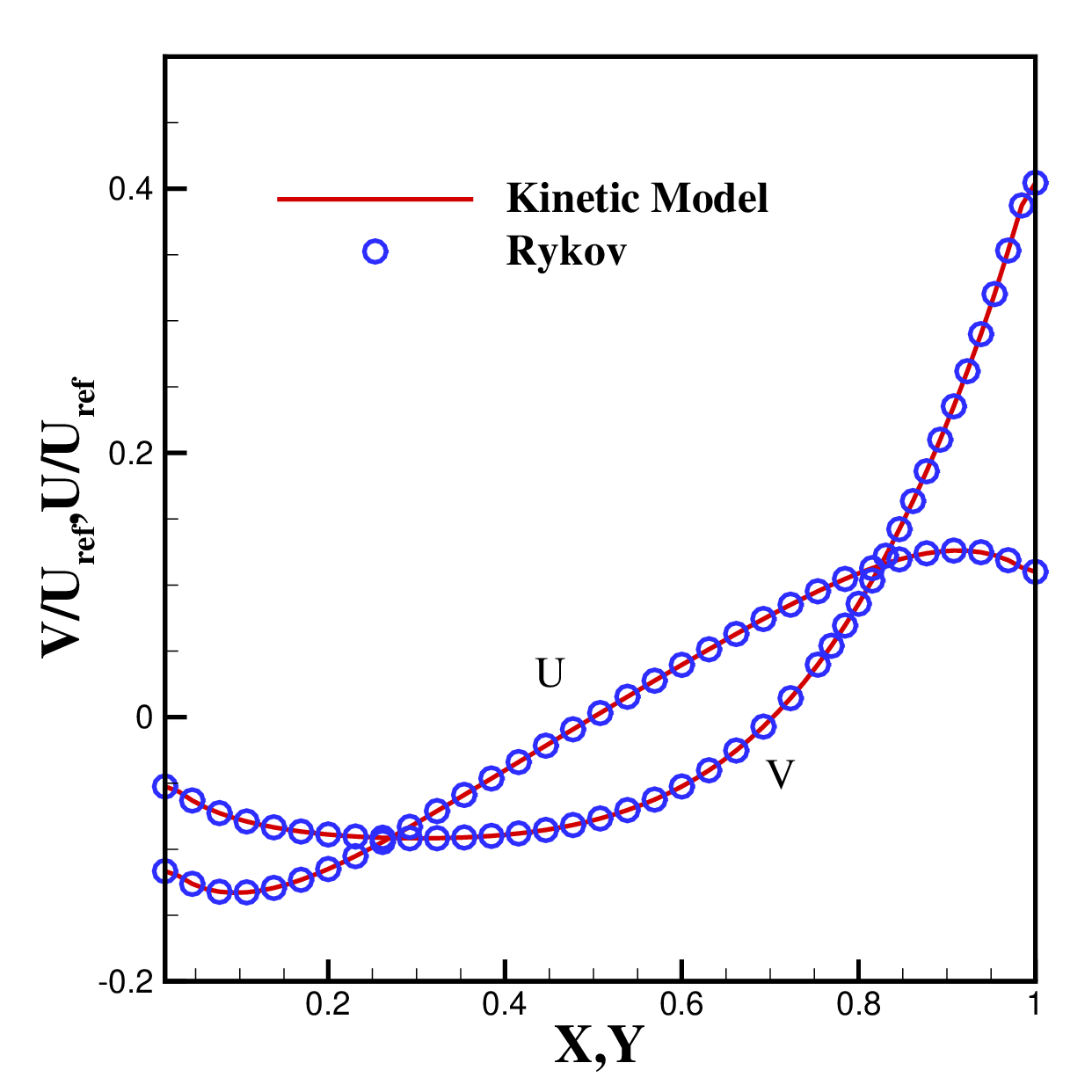}
 		\caption{Kn=1.0}
 	\end{subfigure}
 	\hfill
 	\begin{subfigure}[b]{0.32\textwidth}
 		\includegraphics[width=\linewidth]{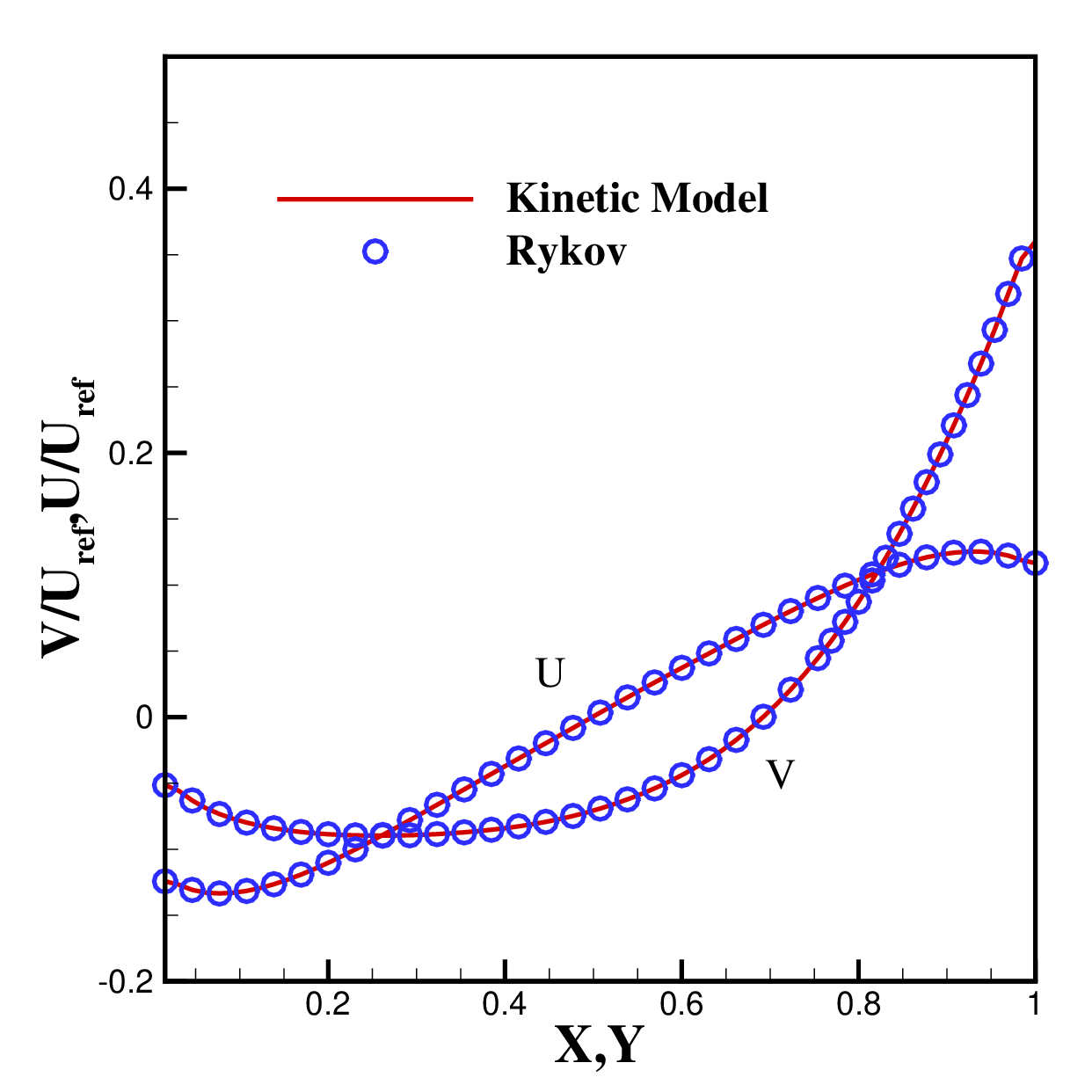}
 		\caption{Kn=10}
 	\end{subfigure}
 	
 	\caption{Velocity profiles along the vertical and horizontal centerlines in lid-driven cavity flows at Kn = 0.075, 1.0, and 10.}
 	\label{cavity_UV}
 \end{figure}

\begin{figure}
 	\centering
 	
 	\begin{subfigure}[b]{0.32\textwidth}
 		\includegraphics[width=\linewidth]{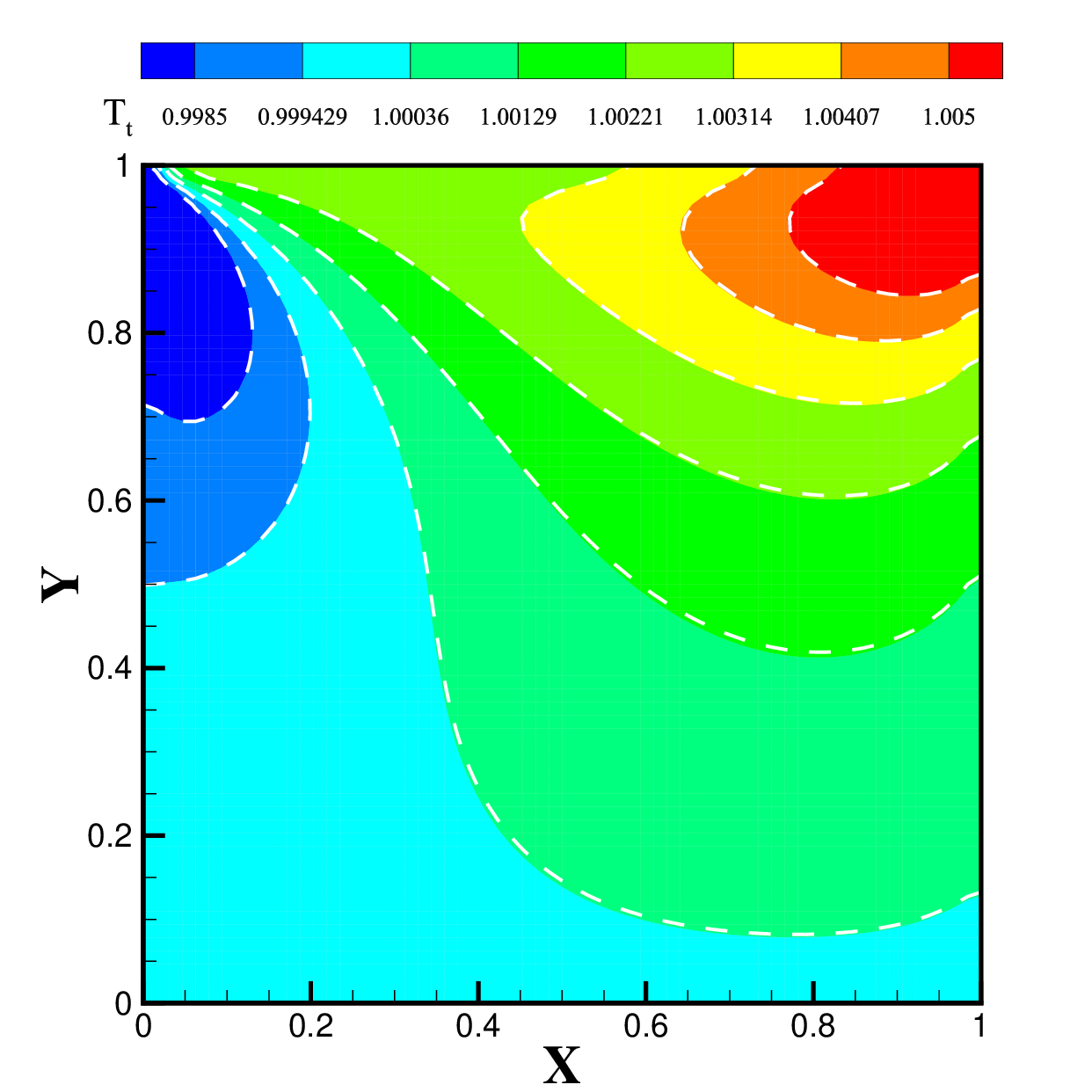}
 		\caption{Kn=0.075}
 	\end{subfigure}
 	\hfill
 	\begin{subfigure}[b]{0.32\textwidth}
 		\includegraphics[width=\linewidth]{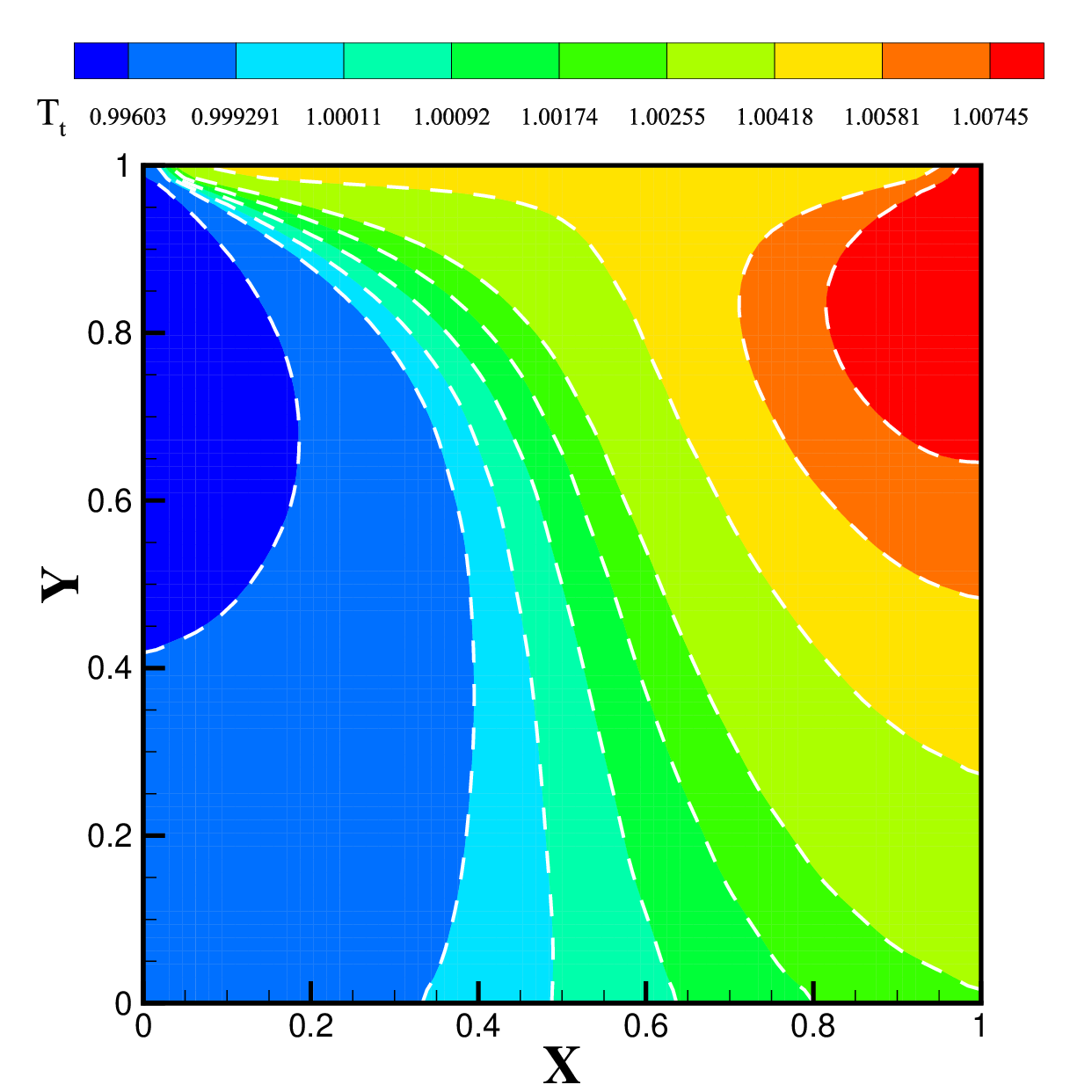}
 		\caption{Kn=1.0}
 	\end{subfigure}
 	\hfill
 	\begin{subfigure}[b]{0.32\textwidth}
 		\includegraphics[width=\linewidth]{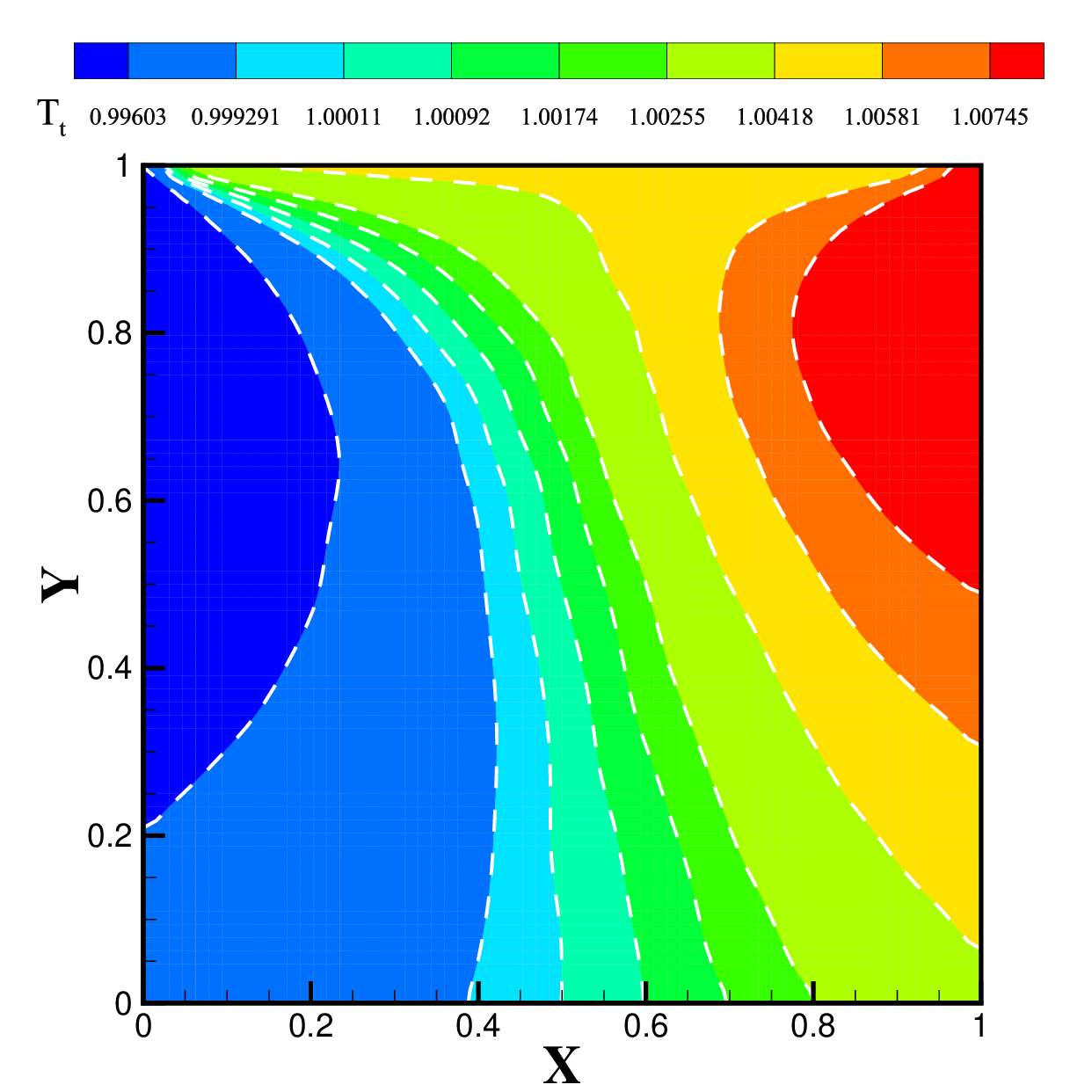}
 		\caption{Kn=10}
 	\end{subfigure}

 	\begin{subfigure}[b]{0.32\textwidth}
 		\includegraphics[width=\linewidth]{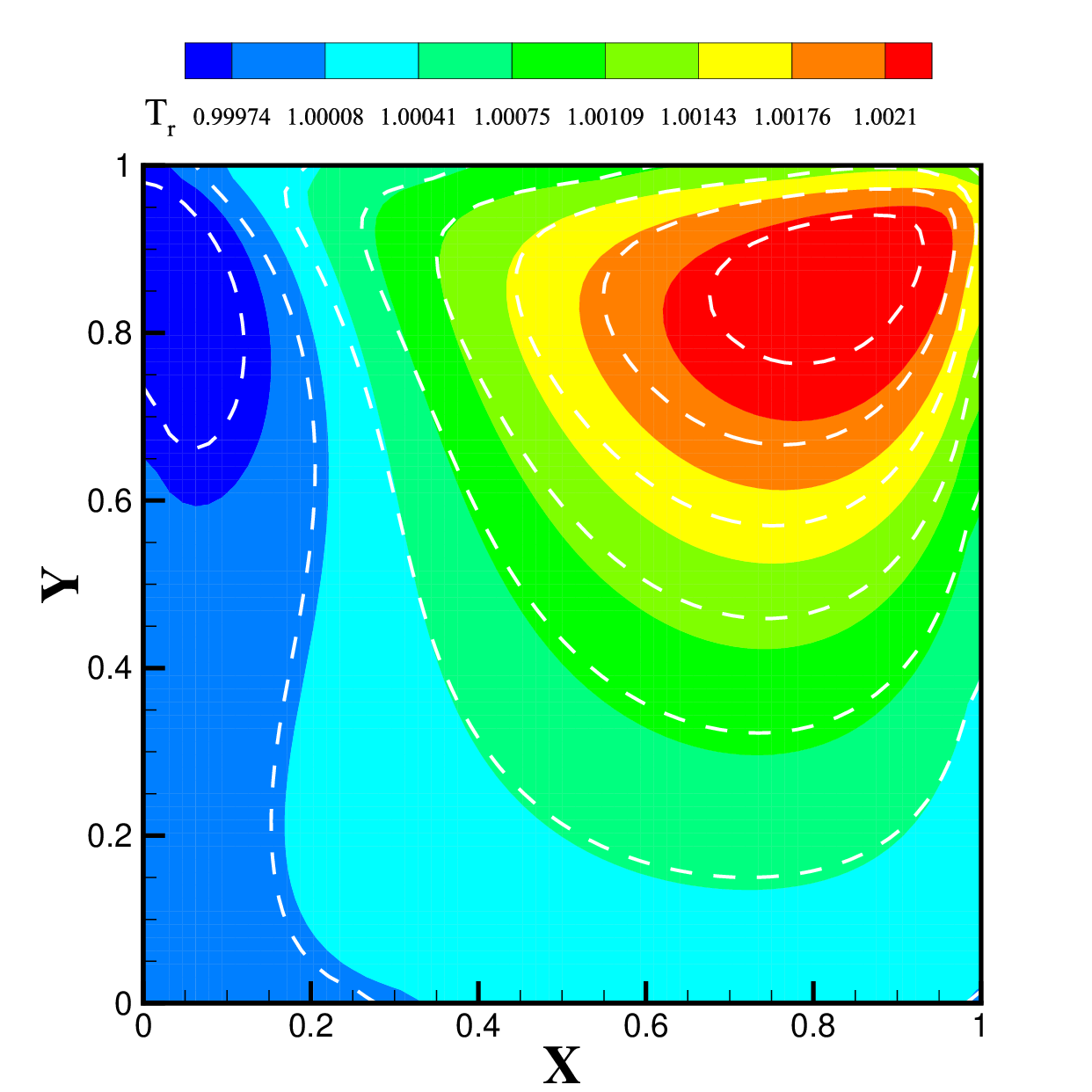}
 		\caption{Kn=0.075}
 	\end{subfigure}
 	\hfill
 	\begin{subfigure}[b]{0.32\textwidth}
 		\includegraphics[width=\linewidth]{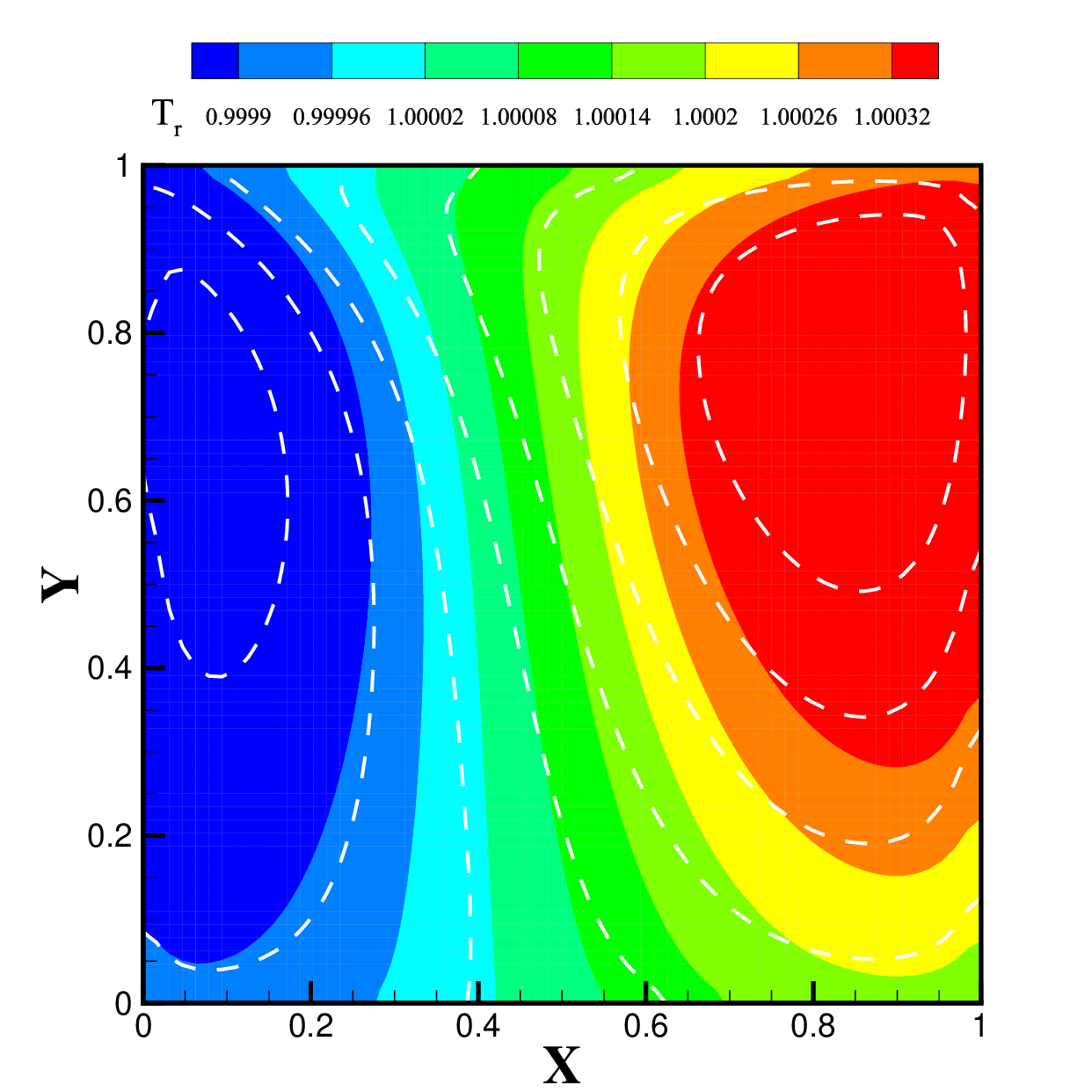}
 		\caption{Kn=1.0}
 	\end{subfigure}
 	\hfill
 	\begin{subfigure}[b]{0.32\textwidth}
 		\includegraphics[width=\linewidth]{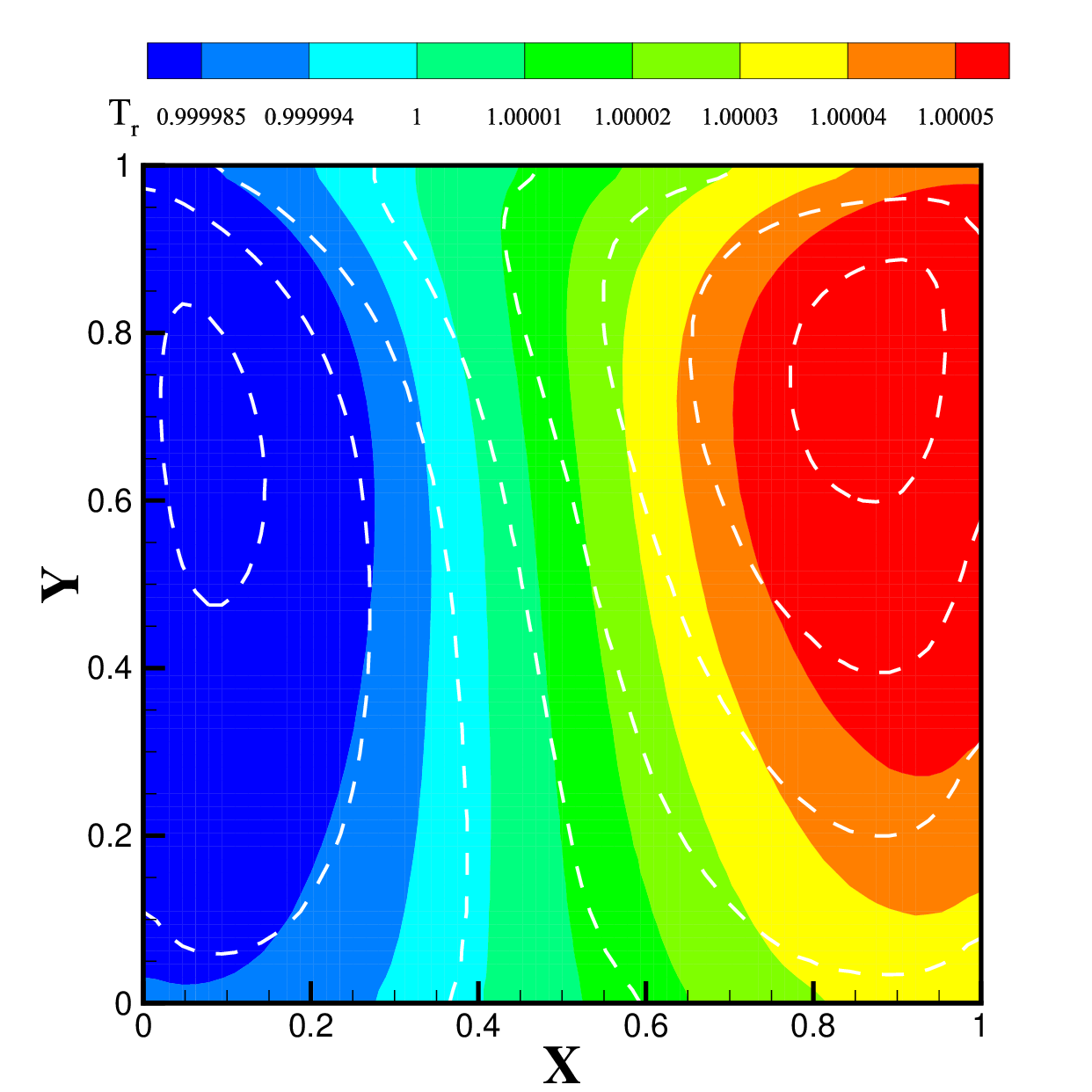}
 		\caption{Kn=10}
 	\end{subfigure}
 	
 	\caption{Temperature contours in lid-driven cavity flows at Kn = 0.075, 1.0, and 10. Results from the new kinetic model are shown as colored contours, while those from the Rykov model are overlaid as white dashed lines. (a)-(c) Translational temperature. (d)-(f) Rotational temperature.}
 	\label{cavity_T}
 \end{figure}

\begin{figure}
 	\centering
 	
 	\begin{subfigure}[b]{0.49\textwidth}
 		\includegraphics[width=\linewidth]{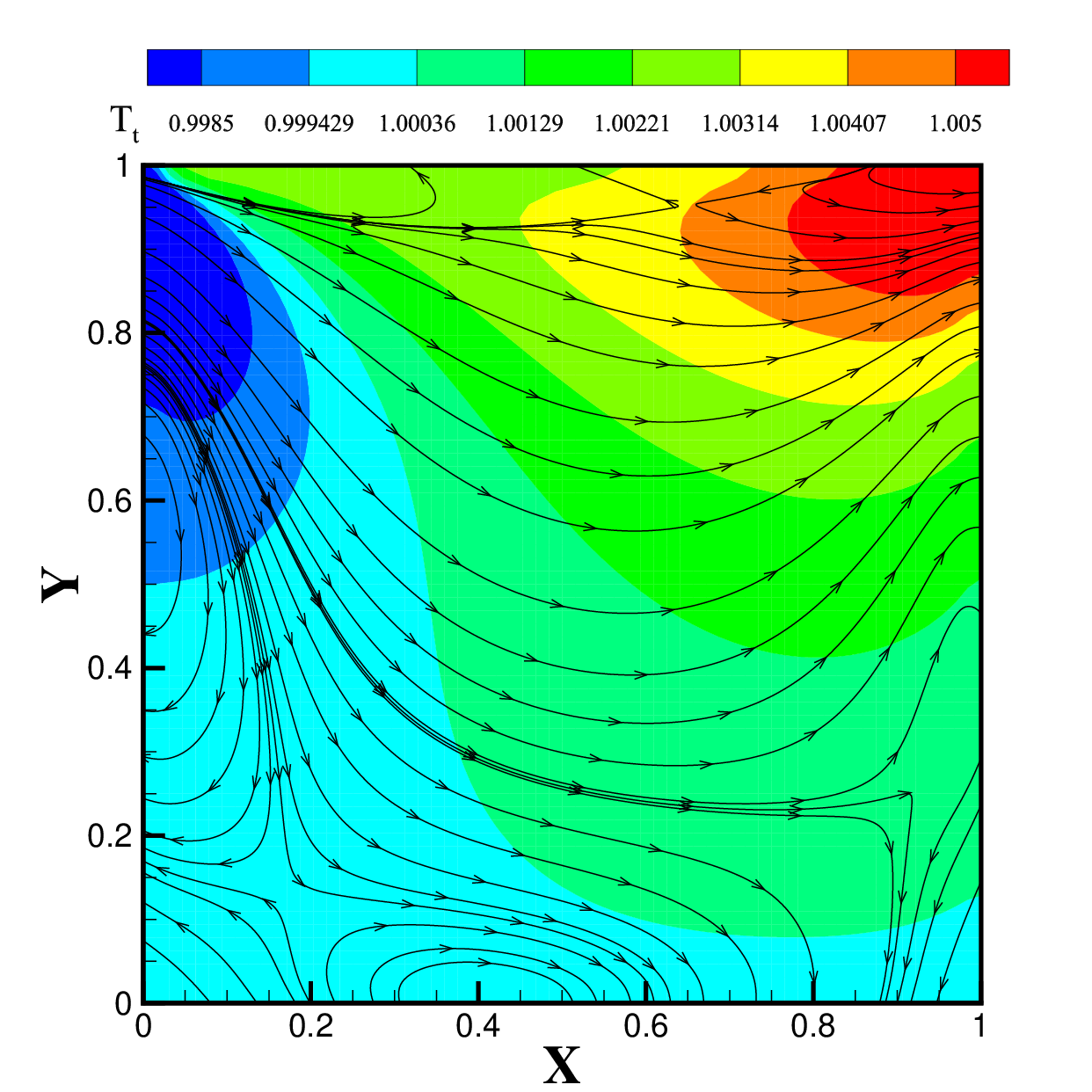}
 		\caption{$T_{\rm{t}}$ and $\mathbf{q}_{\rm{t}}$ by new kinetic model}
 	\end{subfigure}
 	\hfill
 	\begin{subfigure}[b]{0.49\textwidth}
 		\includegraphics[width=\linewidth]{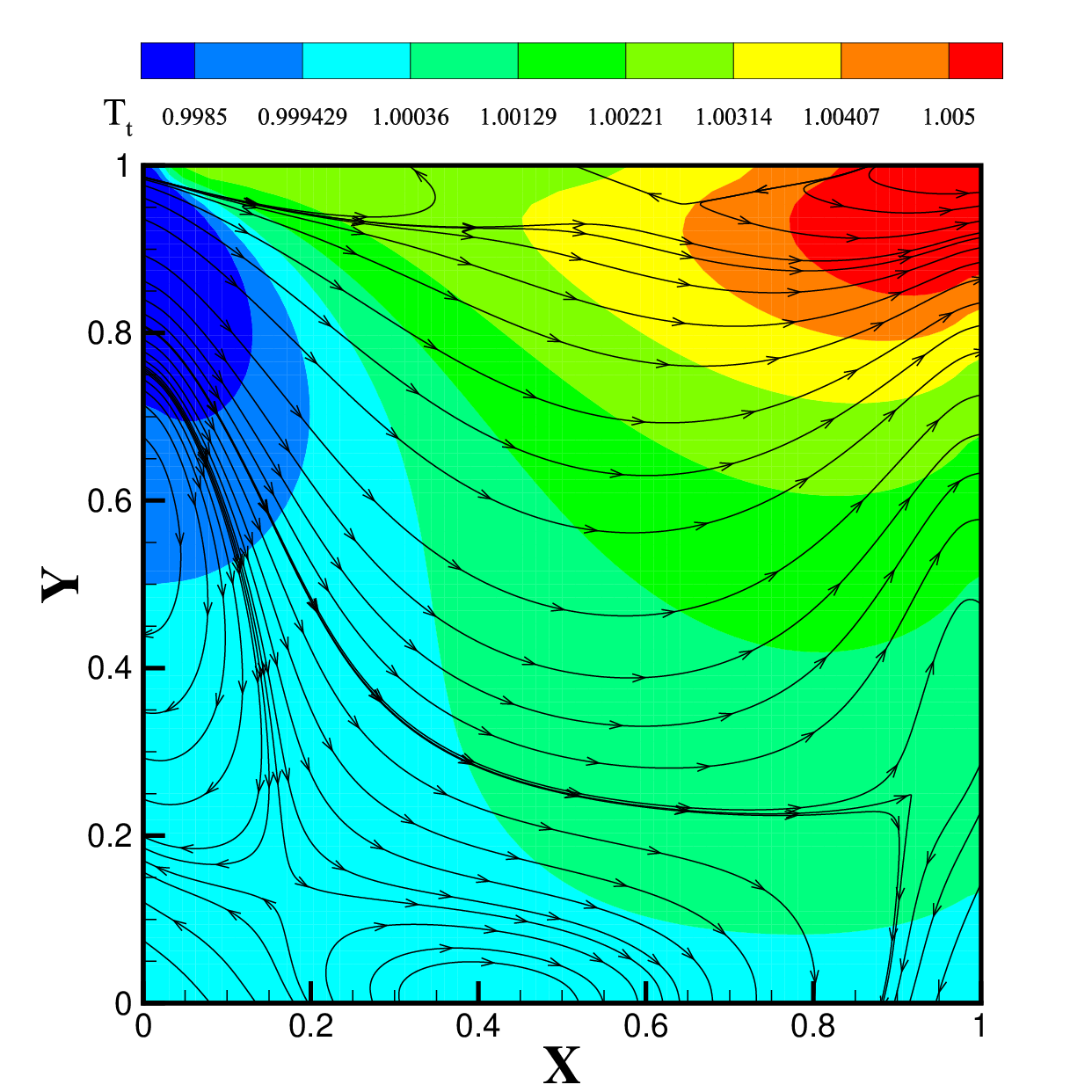}
 		\caption{$T_{\rm{t}}$ and $\mathbf{q}_{\rm{t}}$ by Rykov model}
 	\end{subfigure}

 	\begin{subfigure}[b]{0.49\textwidth}
 		\includegraphics[width=\linewidth]{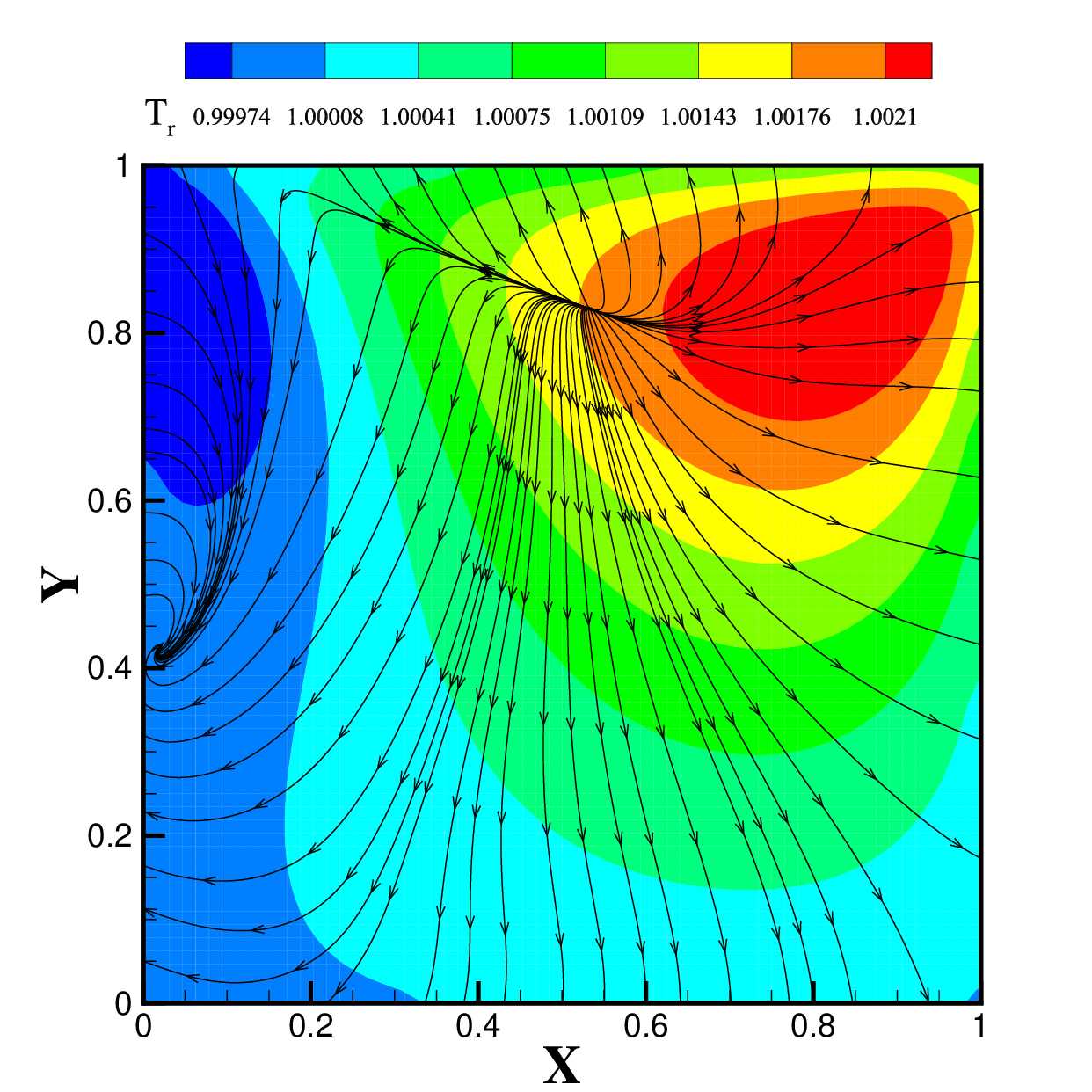}
 		\caption{$T_{\rm{r}}$ and $\mathbf{q}_{\rm{r}}$ by new kinetic model}
 	\end{subfigure}
 	\hfill
 	\begin{subfigure}[b]{0.49\textwidth}
 		\includegraphics[width=\linewidth]{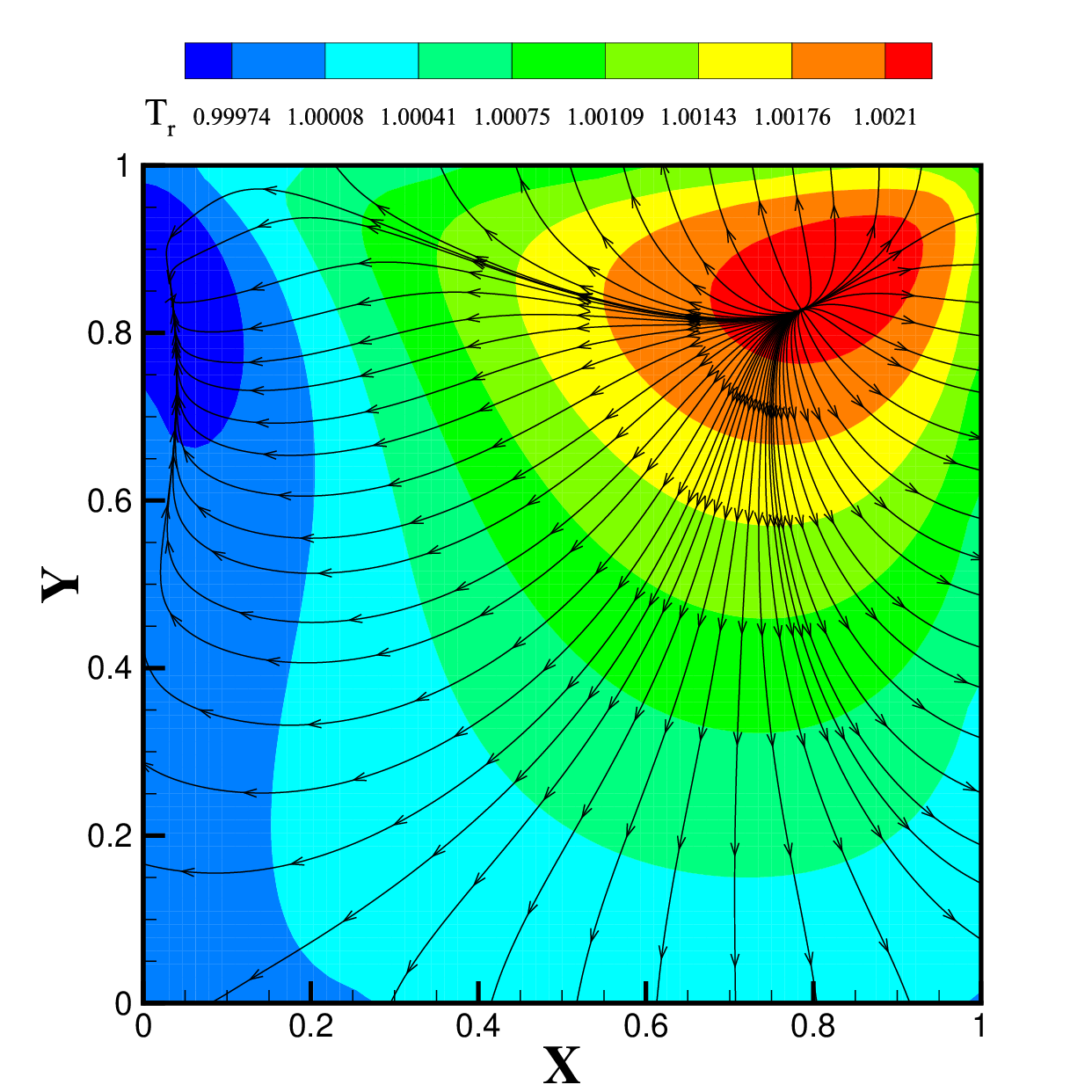}
 		\caption{$T_{\rm{r}}$ and $\mathbf{q}_{\rm{r}}$ by Rykov model}
 	\end{subfigure}

 	\caption{Comparison of the temperature fields (colored fields) and heat flux vectors (black solid lines) between the kinetic model and the Rykov model in lid-driven cavity flow at Kn = 0.075. (a)-(b) Translational temperature and heat flux. (c)-(d) rotational temperature and heat flux.}
 	\label{cavity_q_Kn0.075}
 \end{figure}

\begin{figure}
 	\centering
 	
 	\begin{subfigure}[b]{0.49\textwidth}
 		\includegraphics[width=\linewidth]{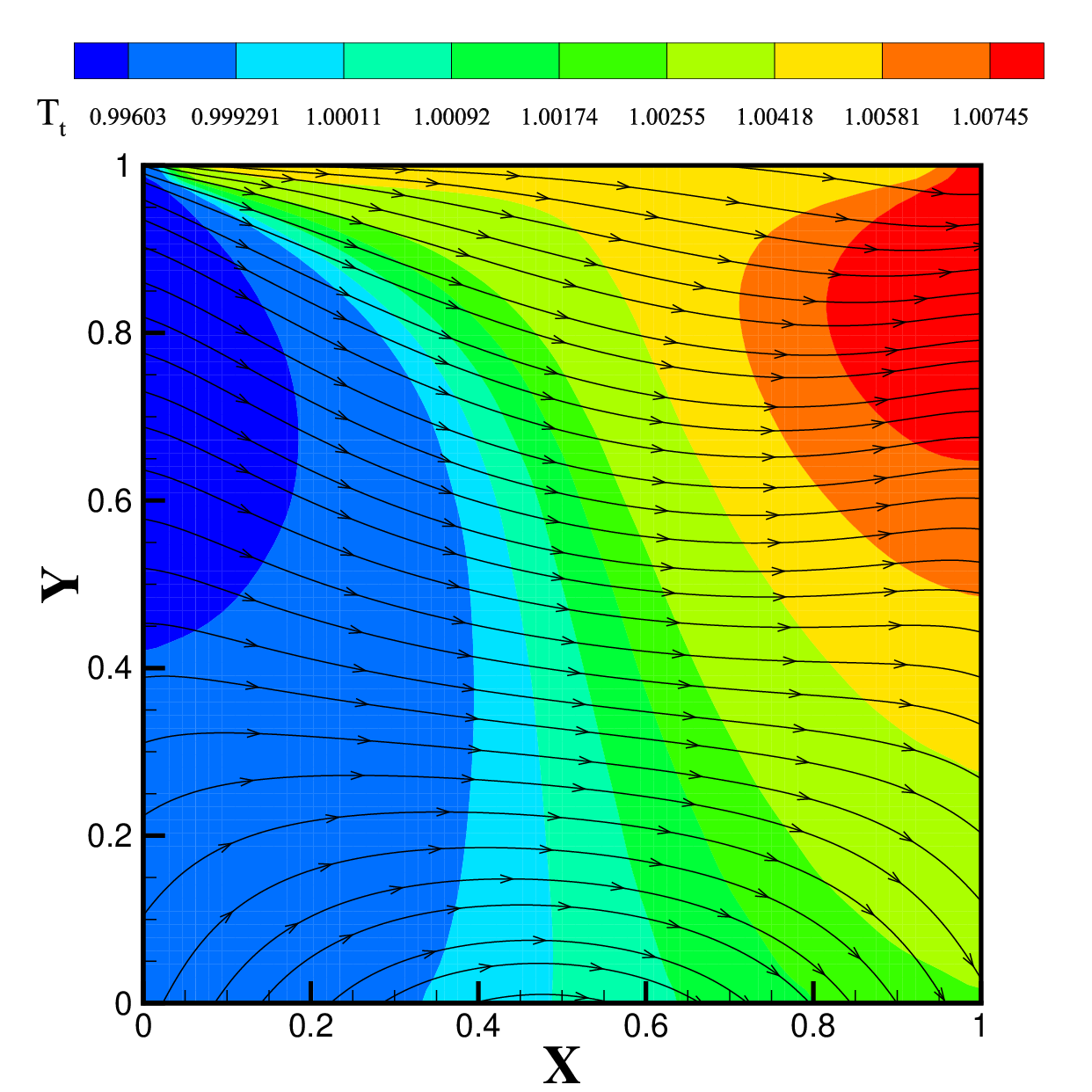}
 		\caption{$T_{\rm{t}}$ and $\mathbf{q}_{\rm{t}}$ by new kinetic model}
 	\end{subfigure}
 	\hfill
 	\begin{subfigure}[b]{0.49\textwidth}
 		\includegraphics[width=\linewidth]{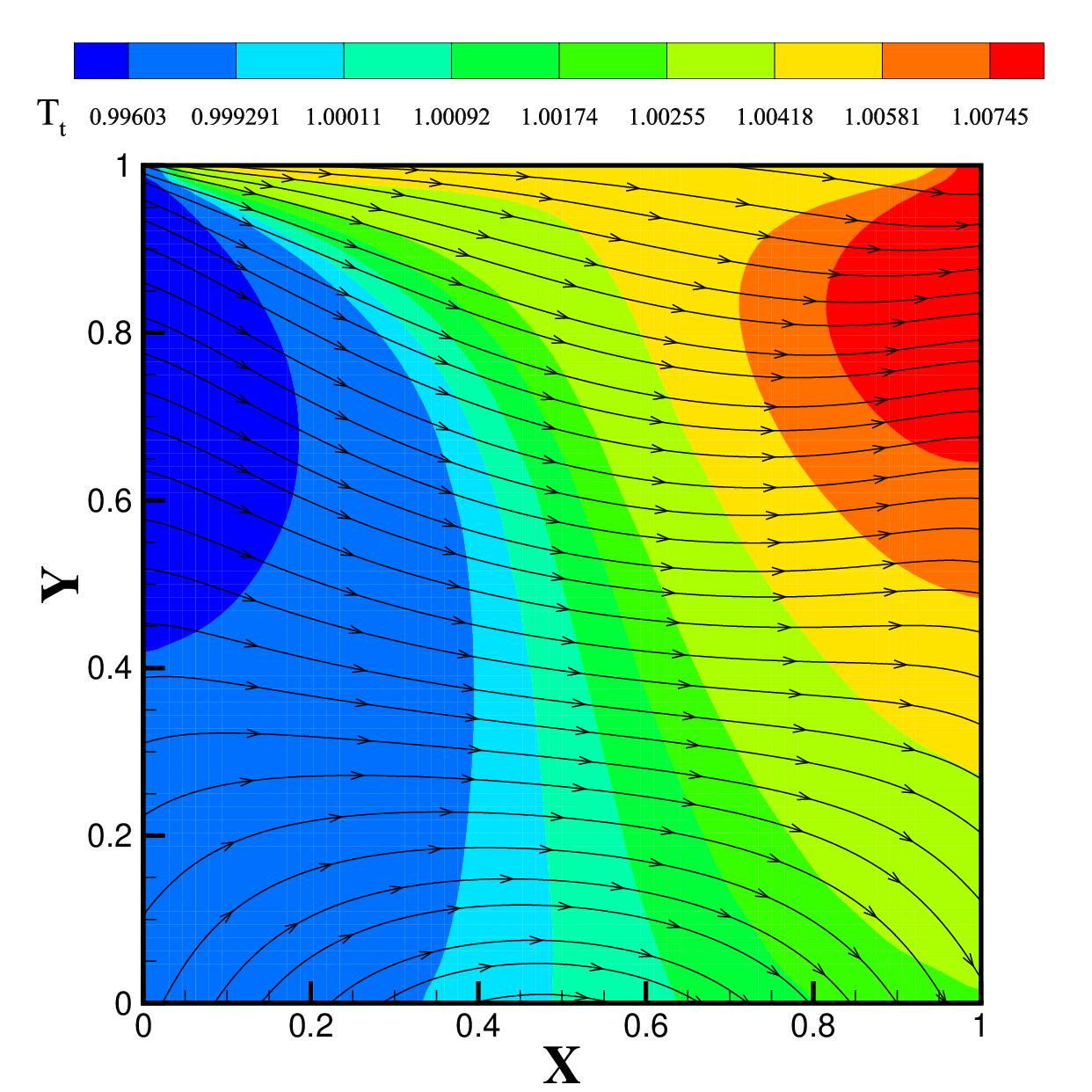}
 		\caption{$T_{\rm{t}}$ and $\mathbf{q}_{\rm{t}}$ by Rykov model}
 	\end{subfigure}
 	
 	\begin{subfigure}[b]{0.49\textwidth}
 		\includegraphics[width=\linewidth]{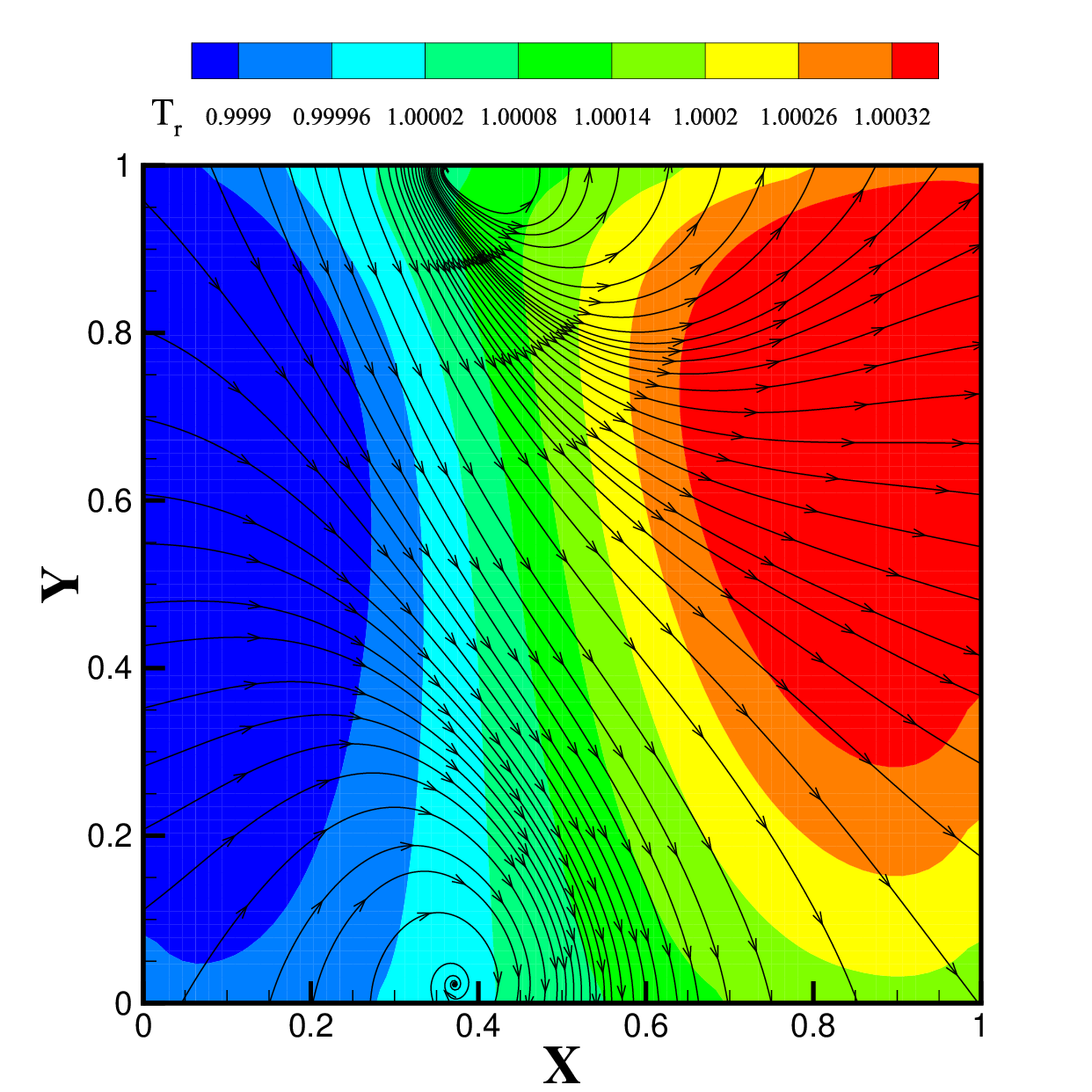}
 		\caption{$T_{\rm{r}}$ and $\mathbf{q}_{\rm{r}}$ by new kinetic model}
 	\end{subfigure}
 	\hfill
 	\begin{subfigure}[b]{0.49\textwidth}
 		\includegraphics[width=\linewidth]{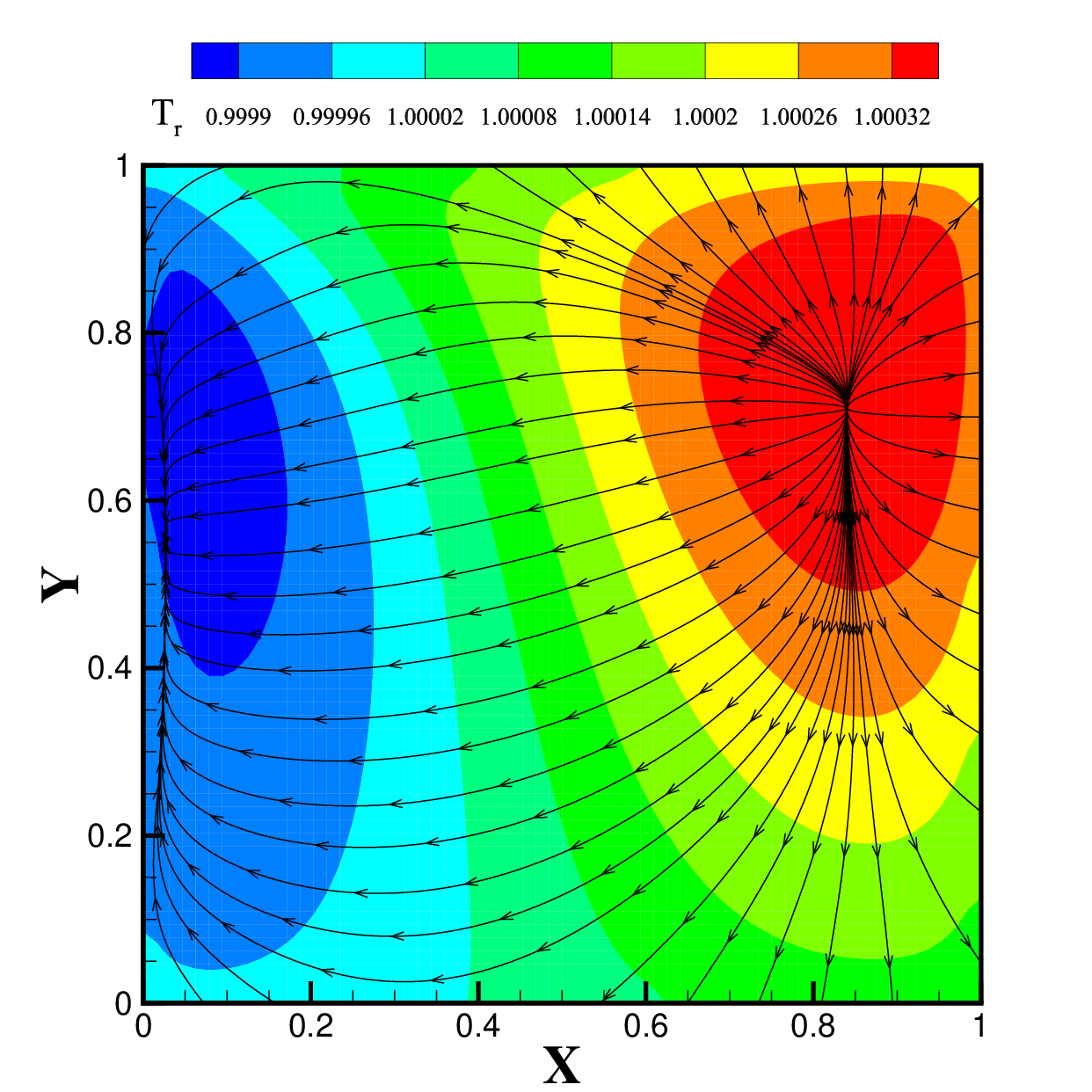}
 		\caption{$T_{\rm{r}}$ and $\mathbf{q}_{\rm{r}}$ by Rykov model}
 	\end{subfigure}

 	\caption{Comparison of the temperature fields (colored fields) and heat flux vectors (black solid lines) between the kinetic model and the Rykov model in lid-driven cavity flow at Kn = 1.0. (a)-(b) Translational temperature and heat flux. (c)-(d) rotational temperature and heat flux.}
 	\label{cavity_q_Kn1.0}
 \end{figure}

\begin{figure}
 	\centering
 	
 	\begin{subfigure}[b]{0.49\textwidth}
 		\includegraphics[width=\linewidth]{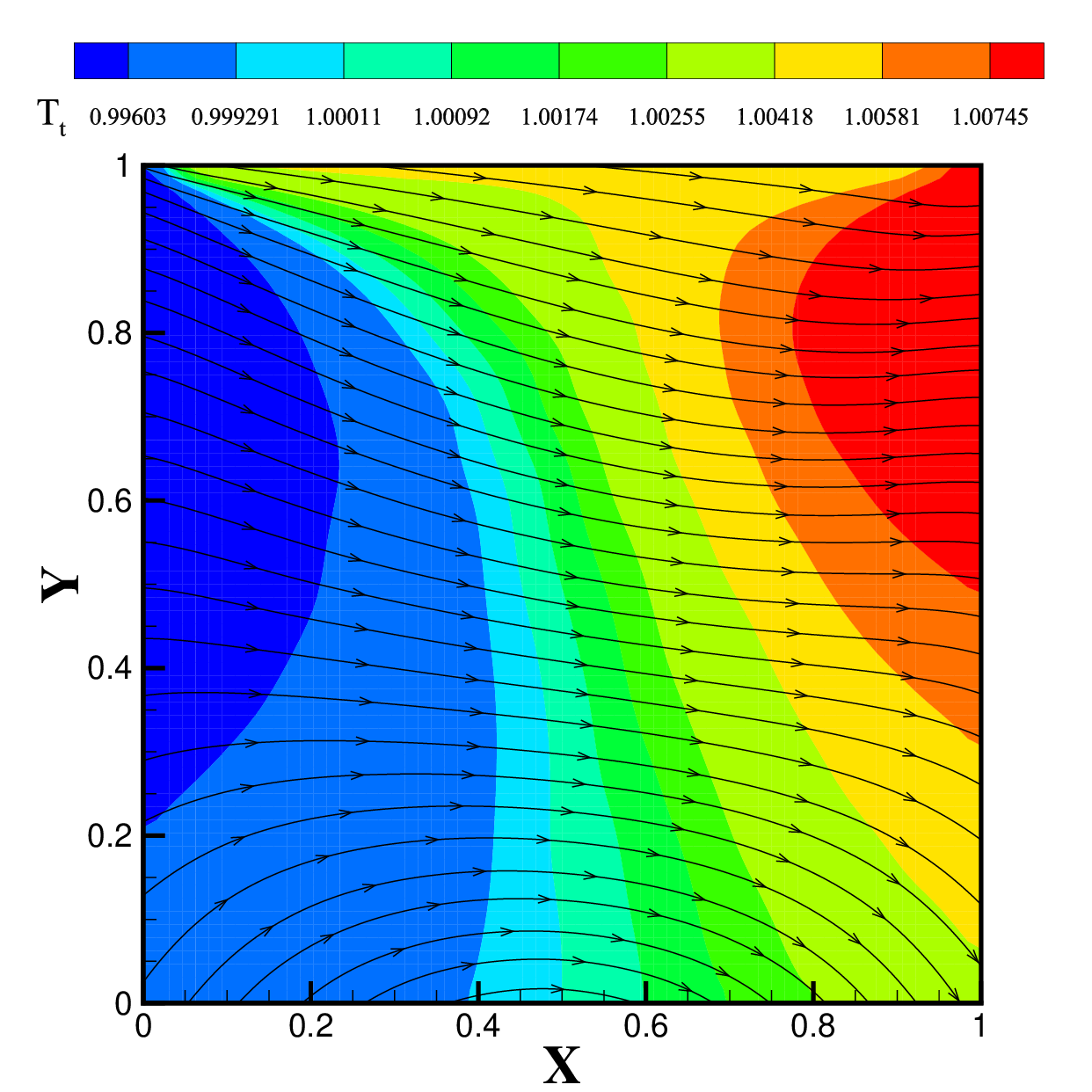}
 		\caption{$T_{\rm{t}}$ and $\mathbf{q}_{\rm{t}}$ by new kinetic model}
 	\end{subfigure}
 	\hfill
 	\begin{subfigure}[b]{0.49\textwidth}
 		\includegraphics[width=\linewidth]{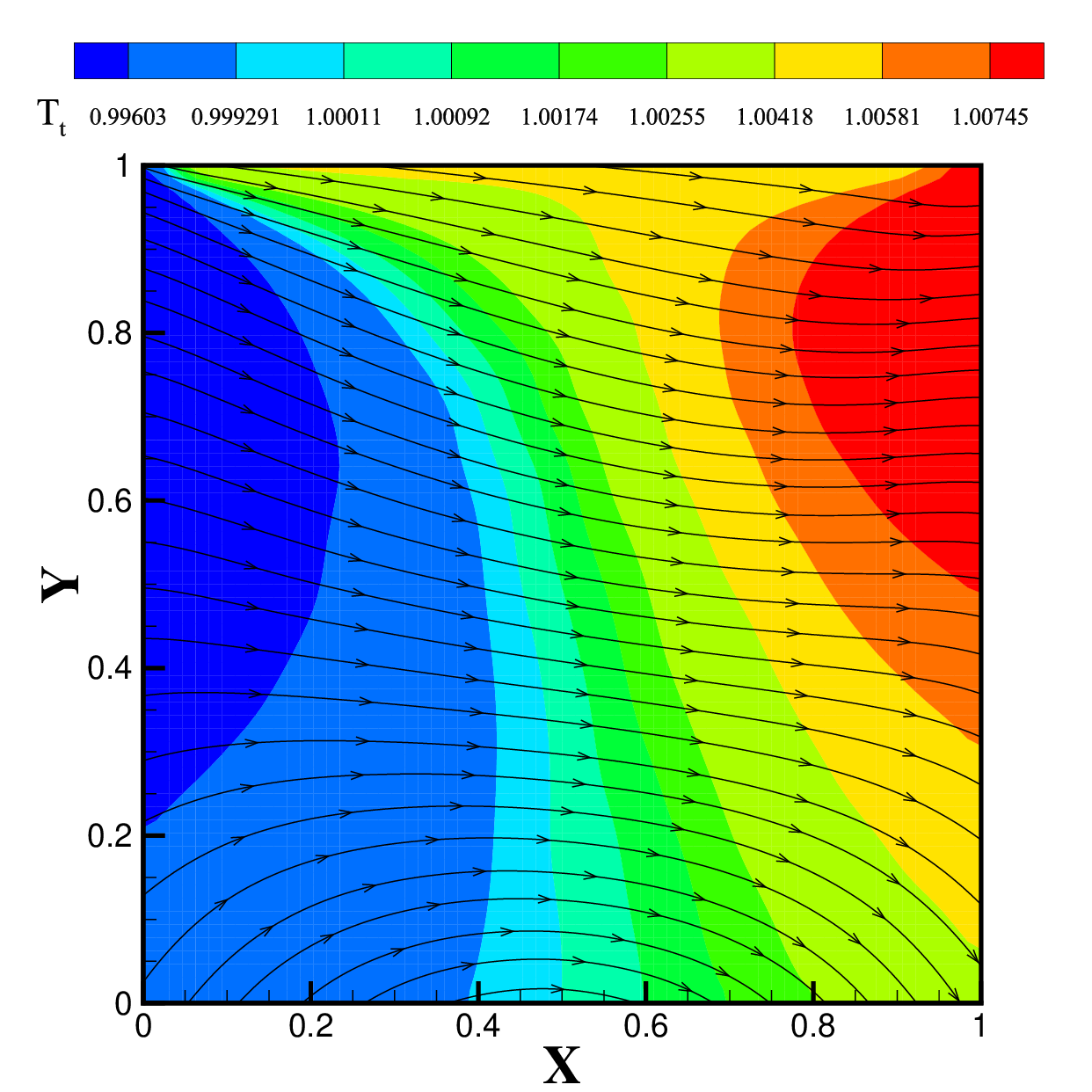}
 		\caption{$T_{\rm{t}}$ and $\mathbf{q}_{\rm{t}}$ by Rykov model}
 	\end{subfigure}
 	
 	\begin{subfigure}[b]{0.49\textwidth}
 		\includegraphics[width=\linewidth]{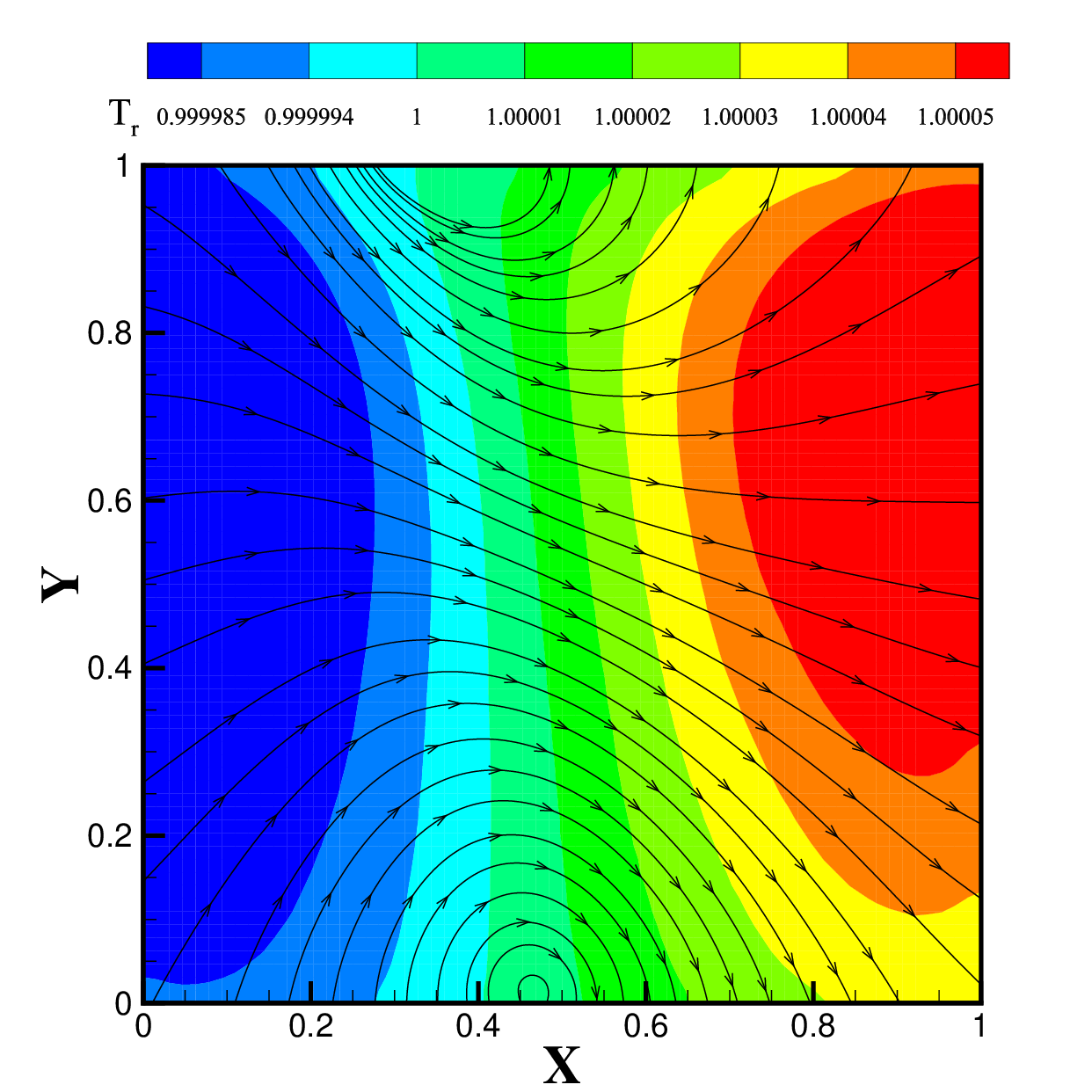}
 		\caption{$T_{\rm{r}}$ and $\mathbf{q}_{\rm{r}}$ by new kinetic model}
 	\end{subfigure}
 	\hfill
 	\begin{subfigure}[b]{0.49\textwidth}
 		\includegraphics[width=\linewidth]{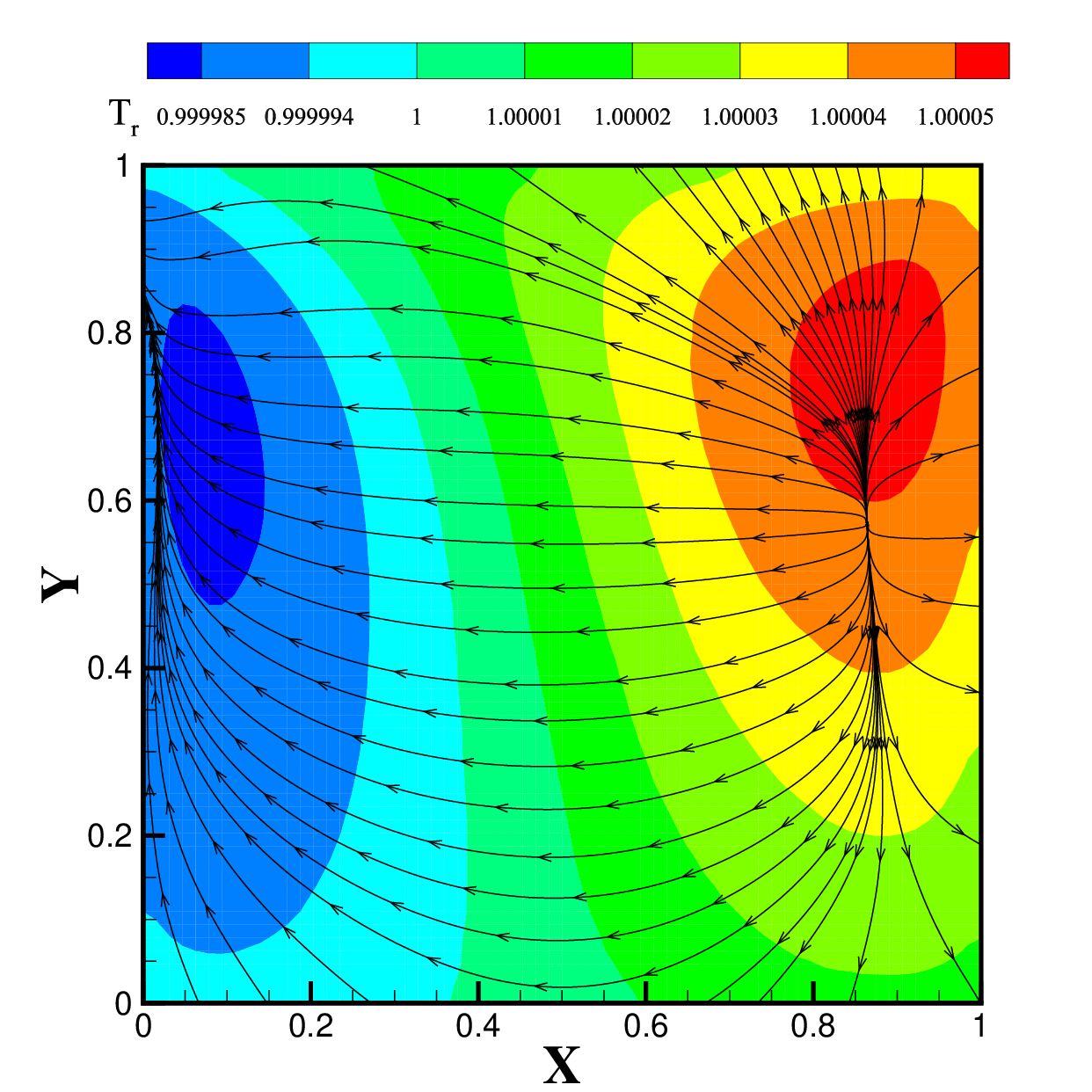}
 		\caption{$T_{\rm{r}}$ and $\mathbf{q}_{\rm{r}}$ by Rykov model}
 	\end{subfigure}

 	\caption{Comparison of the temperature fields (colored fields) and heat flux vectors (black solid lines) between the kinetic model and the Rykov model in lid-driven cavity flow at Kn = 10. (a)-(b) Translational temperature and heat flux. (c)-(d) rotational temperature and heat flux.}
 	\label{cavity_q_Kn10}
 \end{figure}

 We solve both the current kinetic model and Rykov model using the UGKS method for Knudsen numbers 0.075, 1, and 10. Figure~\ref{cavity_UV} compares centerline velocity profiles, demonstrating the kinetic model's accuracy in predicting flow velocities across rarefaction regimes, confirming proper recovery of viscous effects. Figure~\ref{cavity_T} presents the translational and rotational temperature distributions, with the kinetic model's results shown as colored fields and the Rykov model's as white dashed lines. Comparative results show that while the translational temperature distributions essentially match Rykov's, the rotational temperatures exhibit discrepancies, with the kinetic model demonstrating more concentrated rotational energy. This concentration is analyzed by comparing the translational and rotational heat fluxes distributions in figures~\ref{cavity_q_Kn0.075}-\ref{cavity_q_Kn10}. While both models agree well on the translational heat flux, a key difference emerges in the rotational component: the Rykov model's rotational heat flux always opposes the temperature gradient (defined as pointing from low to high temperature), whereas the kinetic model's displays a direction reversal that intensifies with the Knudsen number. This reversal occurs because the kinetic model incorporates coupling between the heat flux. The rotational heat flux is approximately two orders of magnitude smaller than the translational heat flux. Therefore, the influence of the rotational heat flux on the translational component is negligible. However, as the Knudsen number rises, the translational heat flux increases while the rotational heat flux decreases. According to the coupling mechanism outlined by the heat flux relaxation Equations~(\ref{relax_Qt_VHS})-(\ref{relax_Qr_VHS}), the translational heat flux exerts a stronger influence on the rotational component in more rarefied flows. Consequently, the rotational heat flux is pulled toward the direction of the counter-gradient translational heat flux. This mechanism enables more rotational heat flux in the kinetic model to flow from low- to high-temperature regions, leading to the more concentrated rotational temperature distribution shown in figure~\ref{cavity_T}. This difference underscores the importance of the heat flux coupling relaxation mechanism for accurate thermal flow prediction.
 
 \subsection{hypersonic flow past cylinder}
 In high-speed molecular gas flows past objects within rarefied environments, thermodynamic non-equilibrium arises not only from translational non-equilibrium (manifested as bimodal distributions and significant velocity slip) but also from disparities in translational and rotational temperatures within shock layers, boundary layers, and wake regions. While earlier studies emphasized shock and boundary layers, the wake region is in fact equally important and requires models to possess the ability to capture detailed flow information. To validate the kinetic model's reliability for high-speed flows, we solve both the current kinetic model and Rykov model using the UGKS method, comparing results against each other and against DSMC simulations. This study simulates hypersonic nitrogen flow ($\rm{Ma = 5}$) past a cylinder at Knudsen numbers ($\rm{Kn = 0.01, 0.1, 1}$). The cylinder radius is ${{R}_{\rm{c}}}=0.5{\rm{m}}$ with freestream temperature ${{T}_{\infty }}=273{\rm{K}}$. VHS molecular model ($\omega=0.74$) is employed under isothermal wall conditions ($T_{\rm{w}}=273{\rm{K}}$). Rotational collision numbers are set to ${{Z}_{\rm{rot}}}=3.5$ for the kinetic model and ${{Z}_{\rm{rot}}}=5.404$ for DSMC according to Eq.~(\ref{Zrot_DSMC}). Boundaries employ fully accommodated diffuse reflection \citep{cercignani1969mathematical} at walls, inlet and outlet.
 
\begin{figure}
 	\centering
 	
 	\begin{subfigure}[b]{0.32\textwidth}
 		\includegraphics[width=\linewidth]{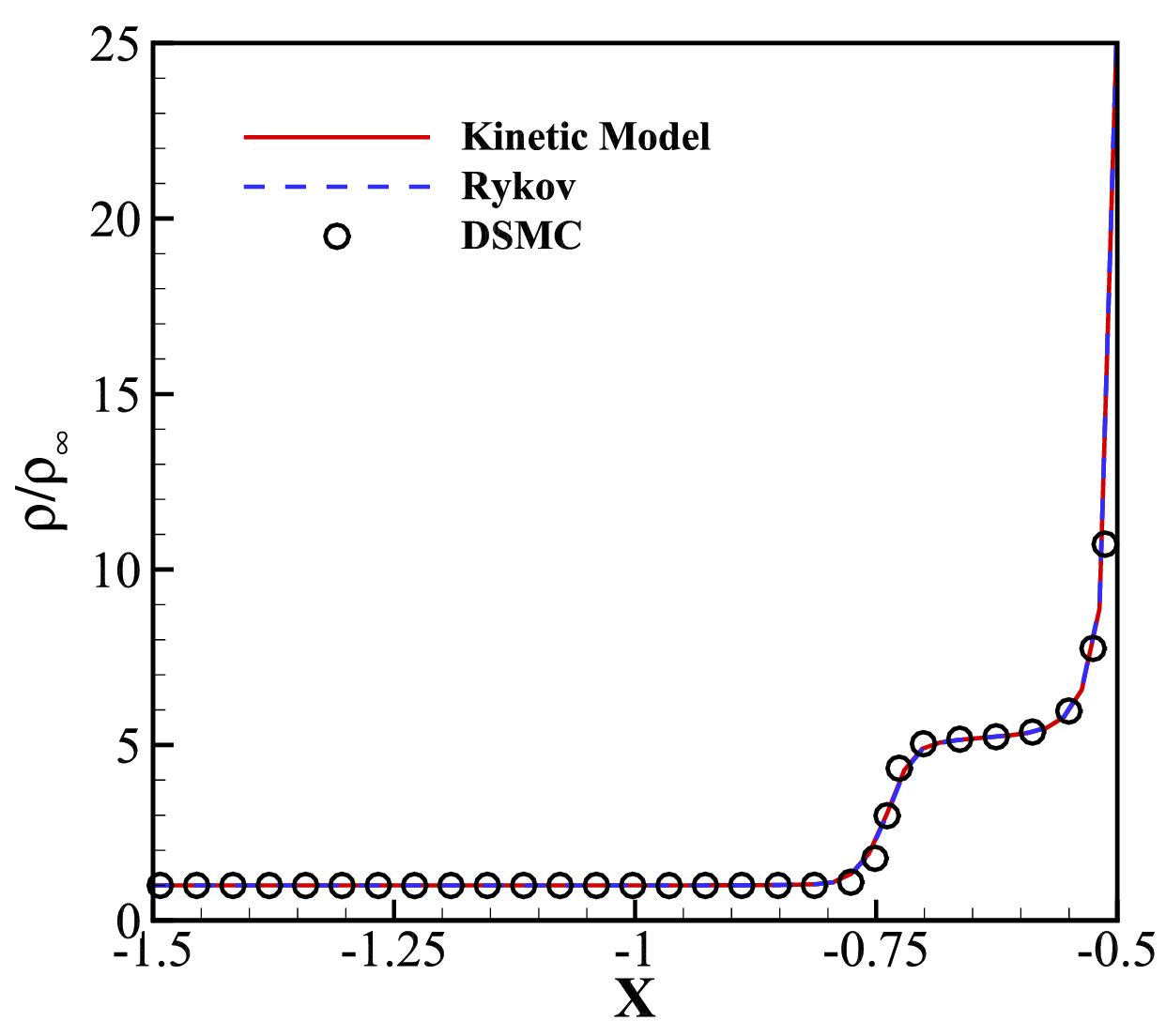}
 		\caption{}
 	\end{subfigure}
 	\hfill
 	\begin{subfigure}[b]{0.32\textwidth}
 		\includegraphics[width=\linewidth]{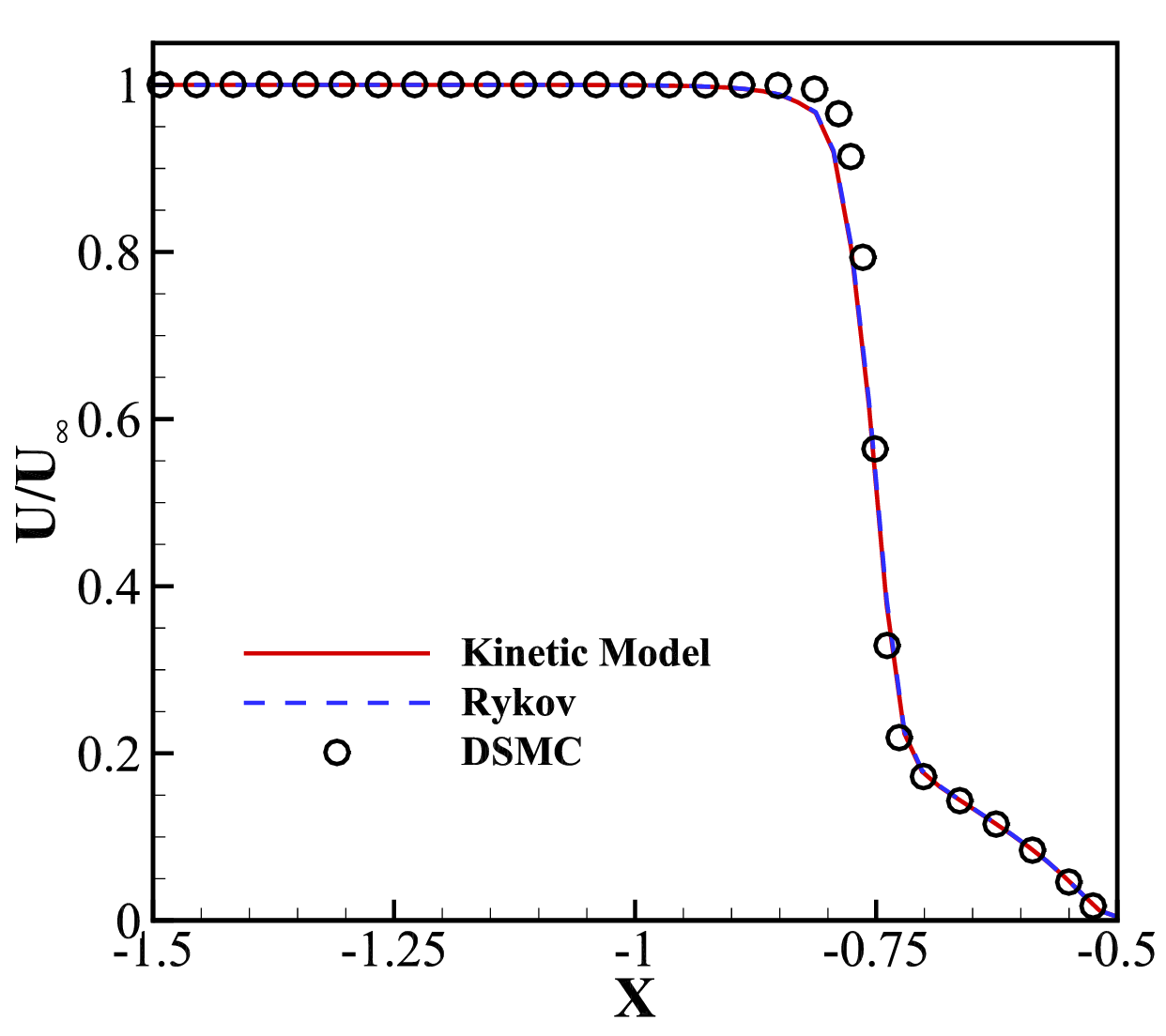}
 		\caption{}
 	\end{subfigure}
 	\hfill
 	\begin{subfigure}[b]{0.32\textwidth}
 		\includegraphics[width=\linewidth]{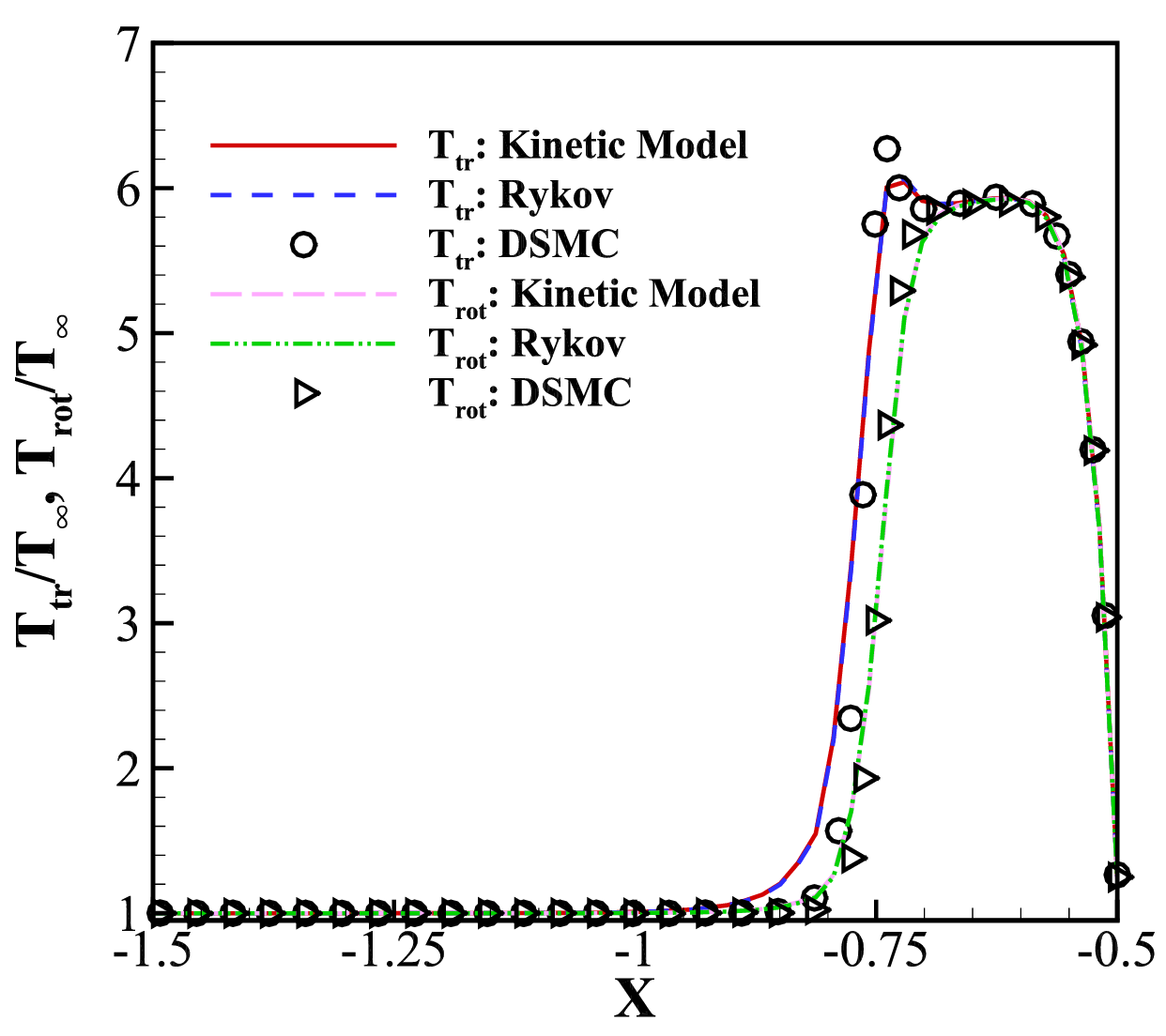}
 		\caption{}
 	\end{subfigure}
 	
 	\caption{Comparison of flow quantities along the stagnation line for hypersonic flow past a cylinder at Kn = 0.01. (a) Density, (b) velocity, and (c) translational and rotational temperatures.}
 	\label{cylinder_Kn0.01_mac}
 \end{figure}

\begin{figure}
 	\centering
 	
 	\begin{subfigure}[b]{0.32\textwidth}
 		\includegraphics[width=\linewidth]{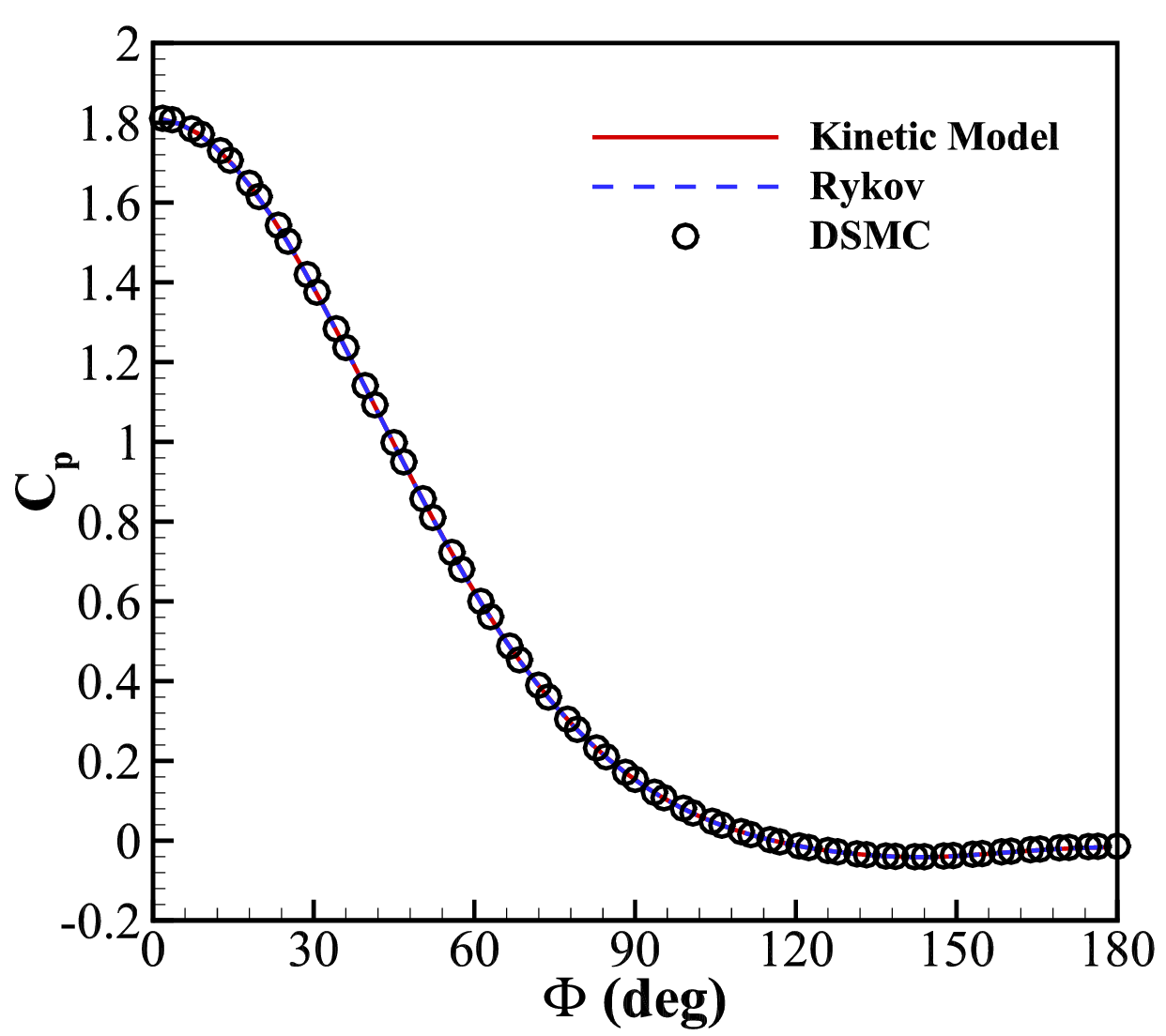}
 		\caption{}
 	\end{subfigure}
 	\hfill
 	\begin{subfigure}[b]{0.32\textwidth}
 		\includegraphics[width=\linewidth]{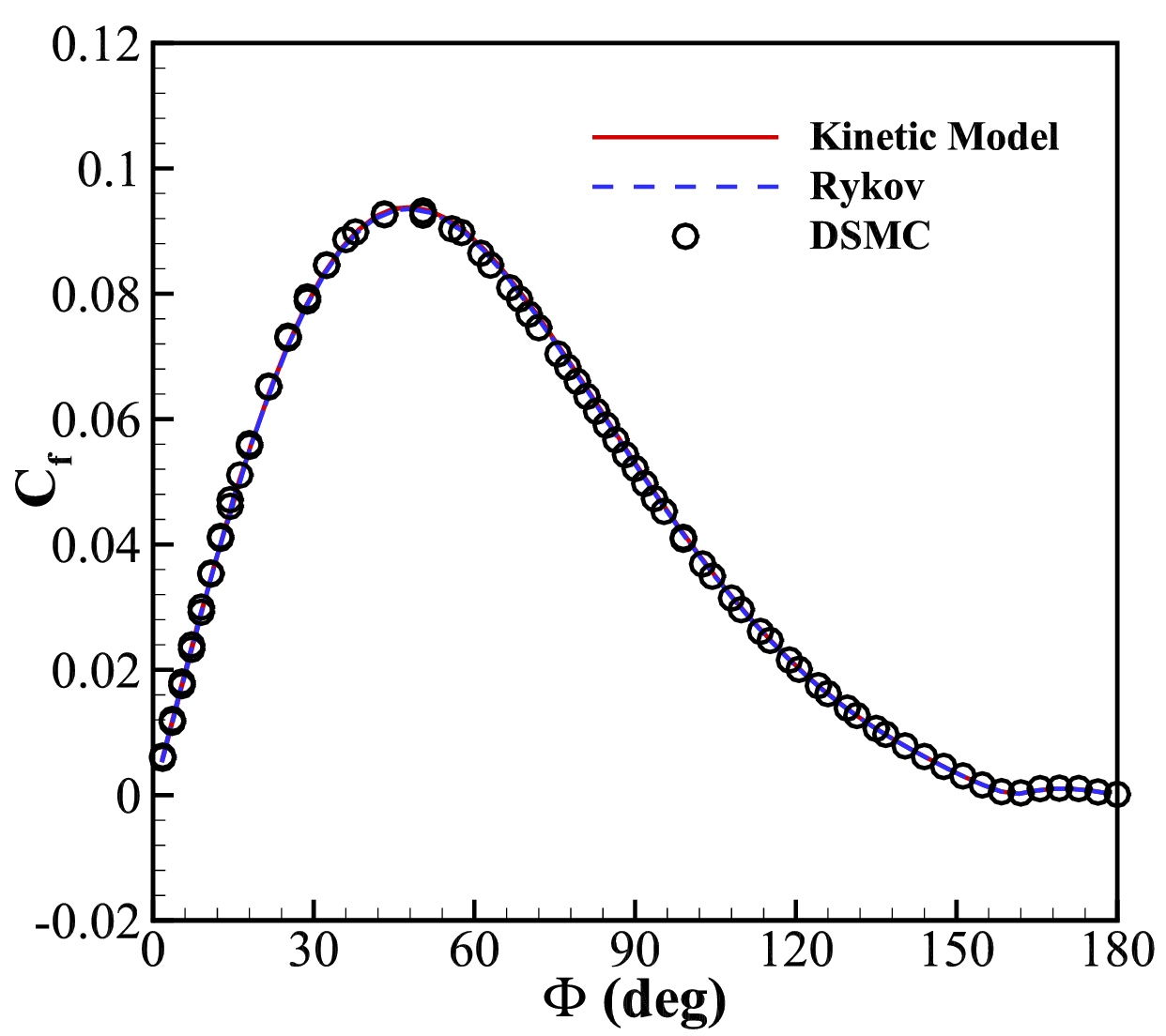}
 		\caption{}
 	\end{subfigure}
 	\hfill
 	\begin{subfigure}[b]{0.32\textwidth}
 		\includegraphics[width=\linewidth]{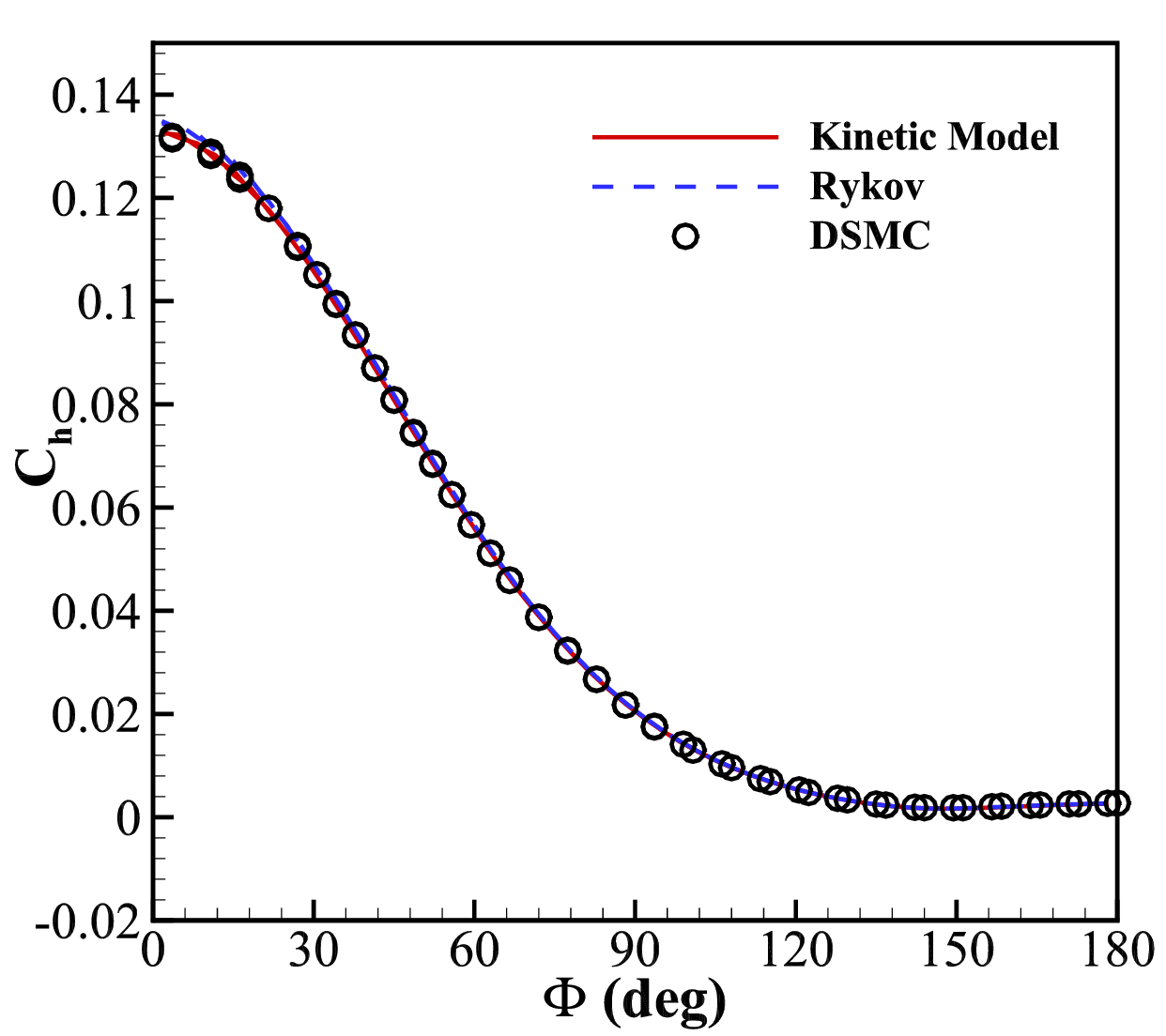}
 		\caption{}
 	\end{subfigure}
 	
 	\caption{Comparison of surface quantities for hypersonic flow past a cylinder at Kn = 0.01. (a) Pressure coefficient, (b) shear stress coefficient, and (c) heat transfer coefficient.}
 	\label{cylinder_Kn0.01_coe}
 \end{figure}

\begin{figure}
 	\centering
 	
 	\begin{subfigure}[b]{0.32\textwidth}
 		\includegraphics[width=\linewidth]{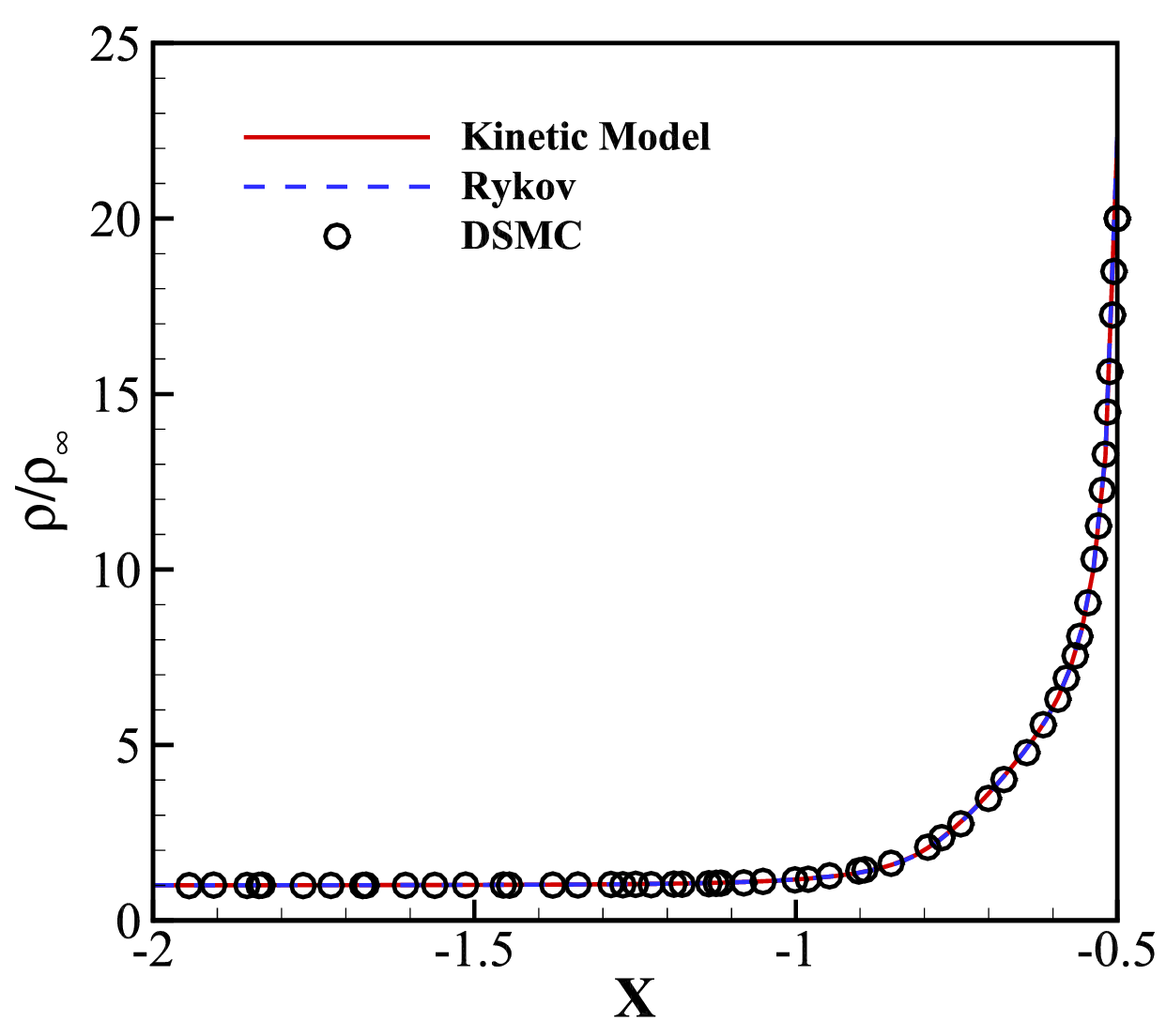}
 		\caption{}
 	\end{subfigure}
 	\hfill
 	\begin{subfigure}[b]{0.32\textwidth}
 		\includegraphics[width=\linewidth]{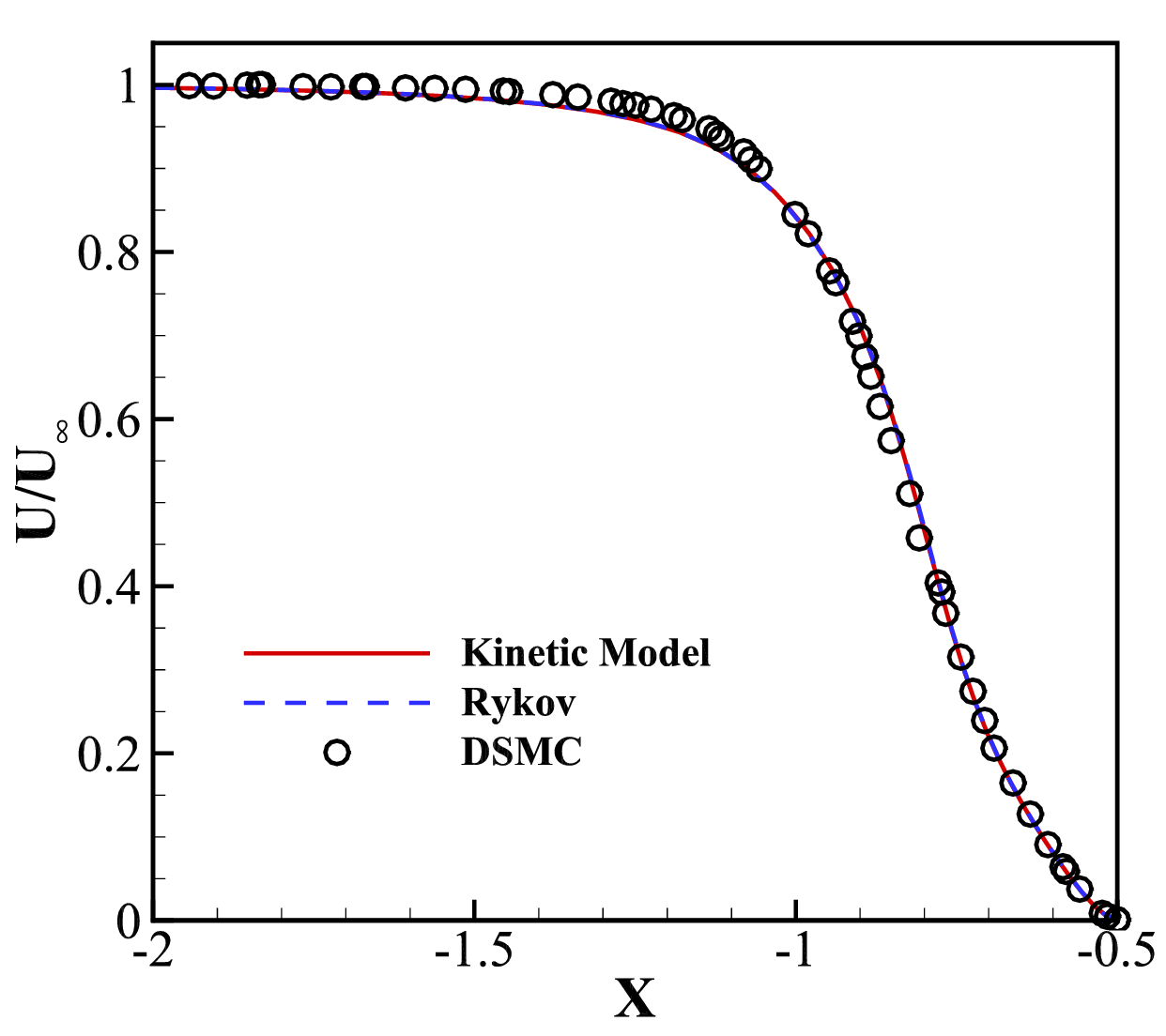}
 		\caption{}
 	\end{subfigure}
 	\hfill
 	\begin{subfigure}[b]{0.32\textwidth}
 		\includegraphics[width=\linewidth]{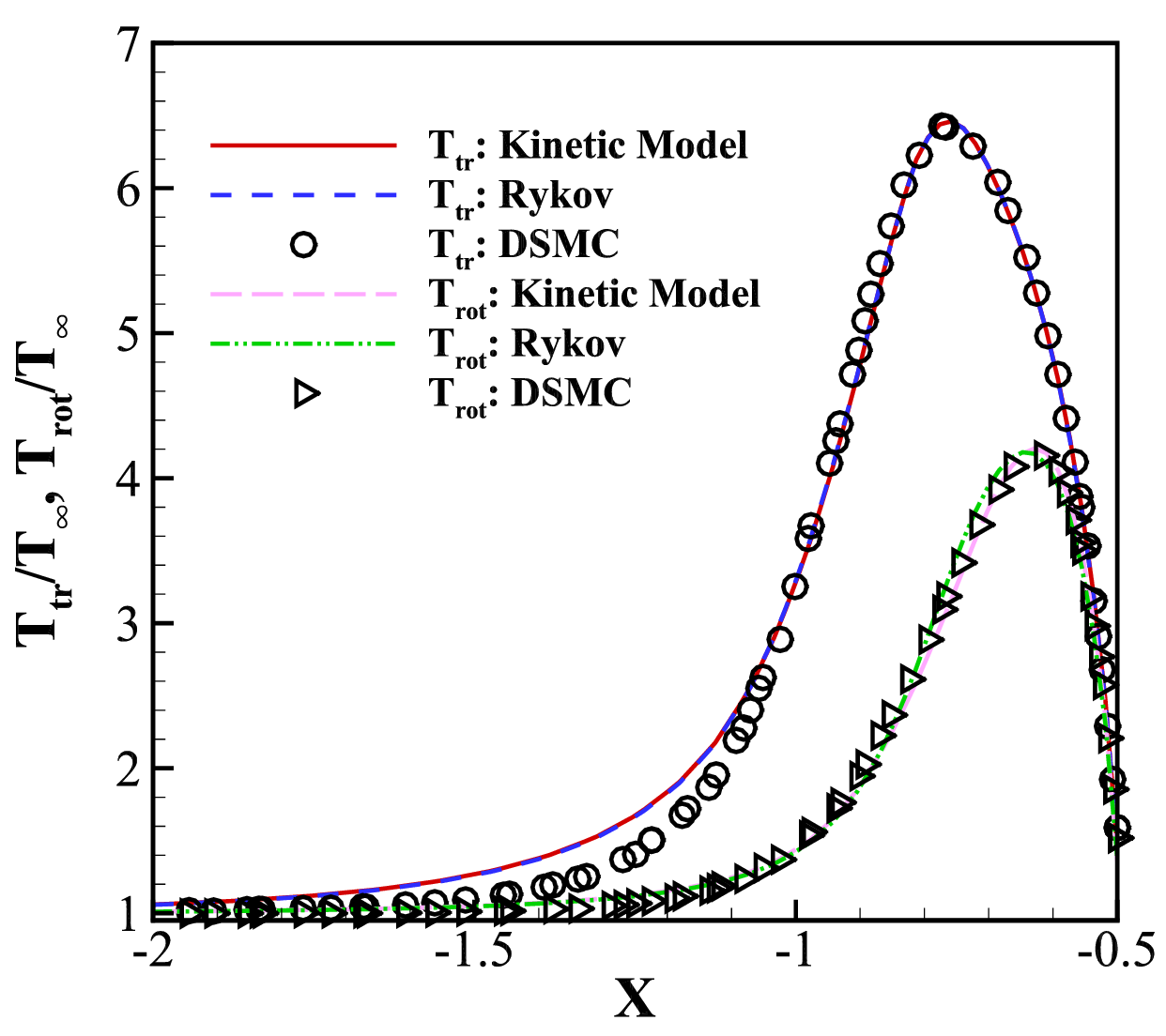}
 		\caption{}
 	\end{subfigure}
 	
 	\caption{Comparison of flow quantities along the stagnation line for hypersonic flow past a cylinder at Kn = 0.1. (a) Density, (b) velocity, and (c) translational and rotational temperatures.}
 	\label{cylinder_Kn0.1_mac}
 \end{figure}

\begin{figure}
 	\centering
 	
 	\begin{subfigure}[b]{0.32\textwidth}
 		\includegraphics[width=\linewidth]{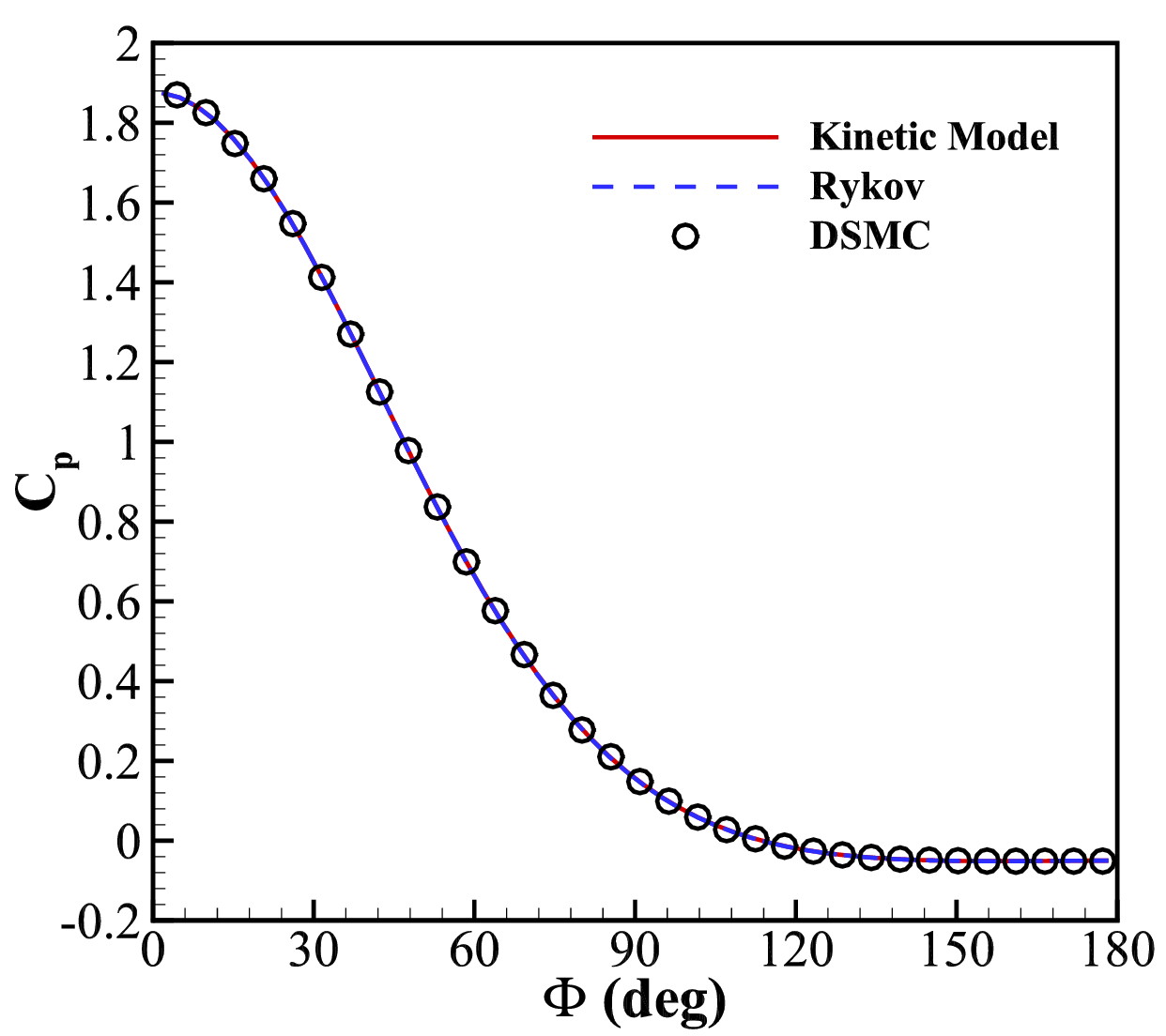}
 		\caption{}
 	\end{subfigure}
 	\hfill
 	\begin{subfigure}[b]{0.32\textwidth}
 		\includegraphics[width=\linewidth]{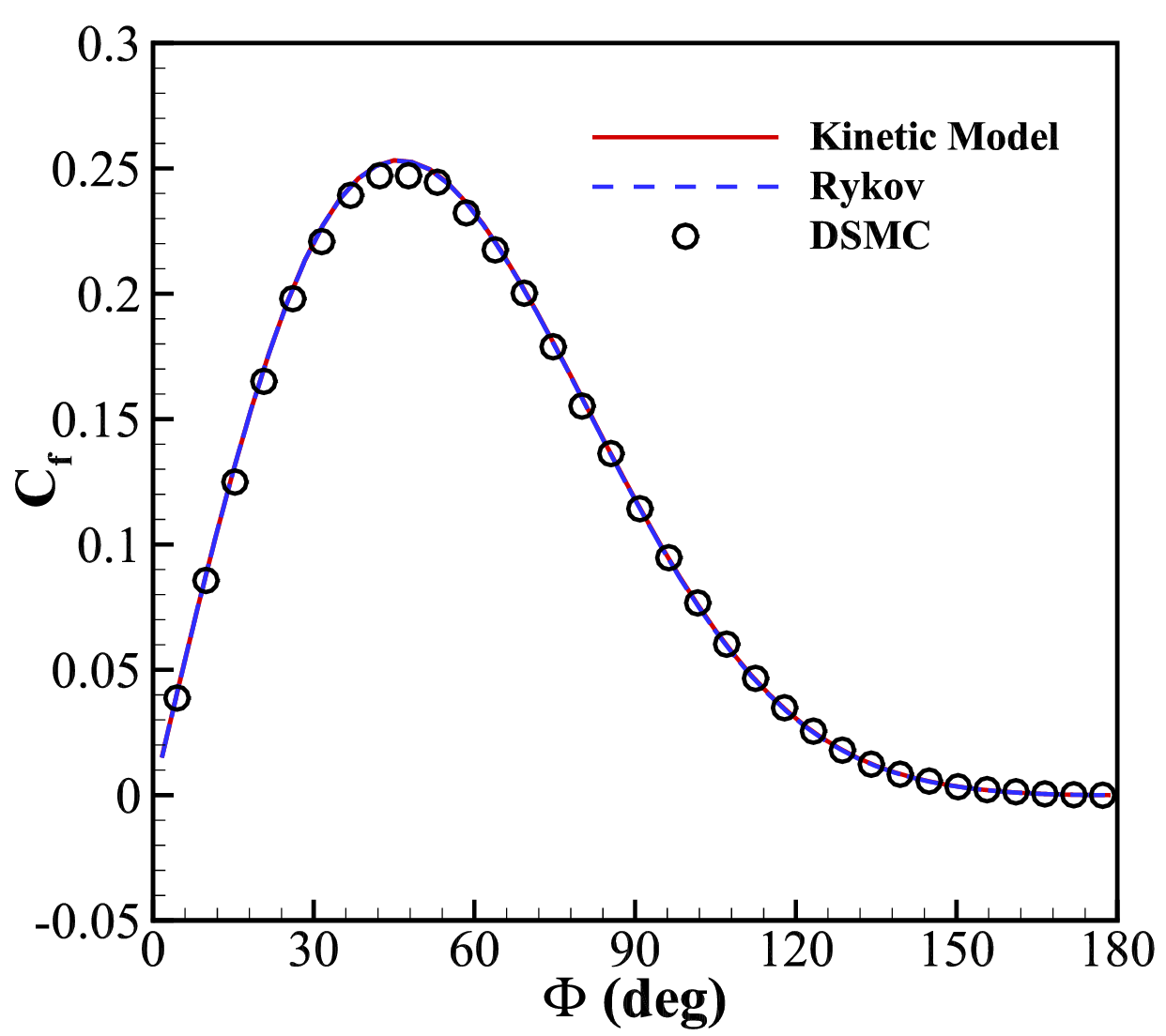}
 		\caption{}
 	\end{subfigure}
 	\hfill
 	\begin{subfigure}[b]{0.32\textwidth}
 		\includegraphics[width=\linewidth]{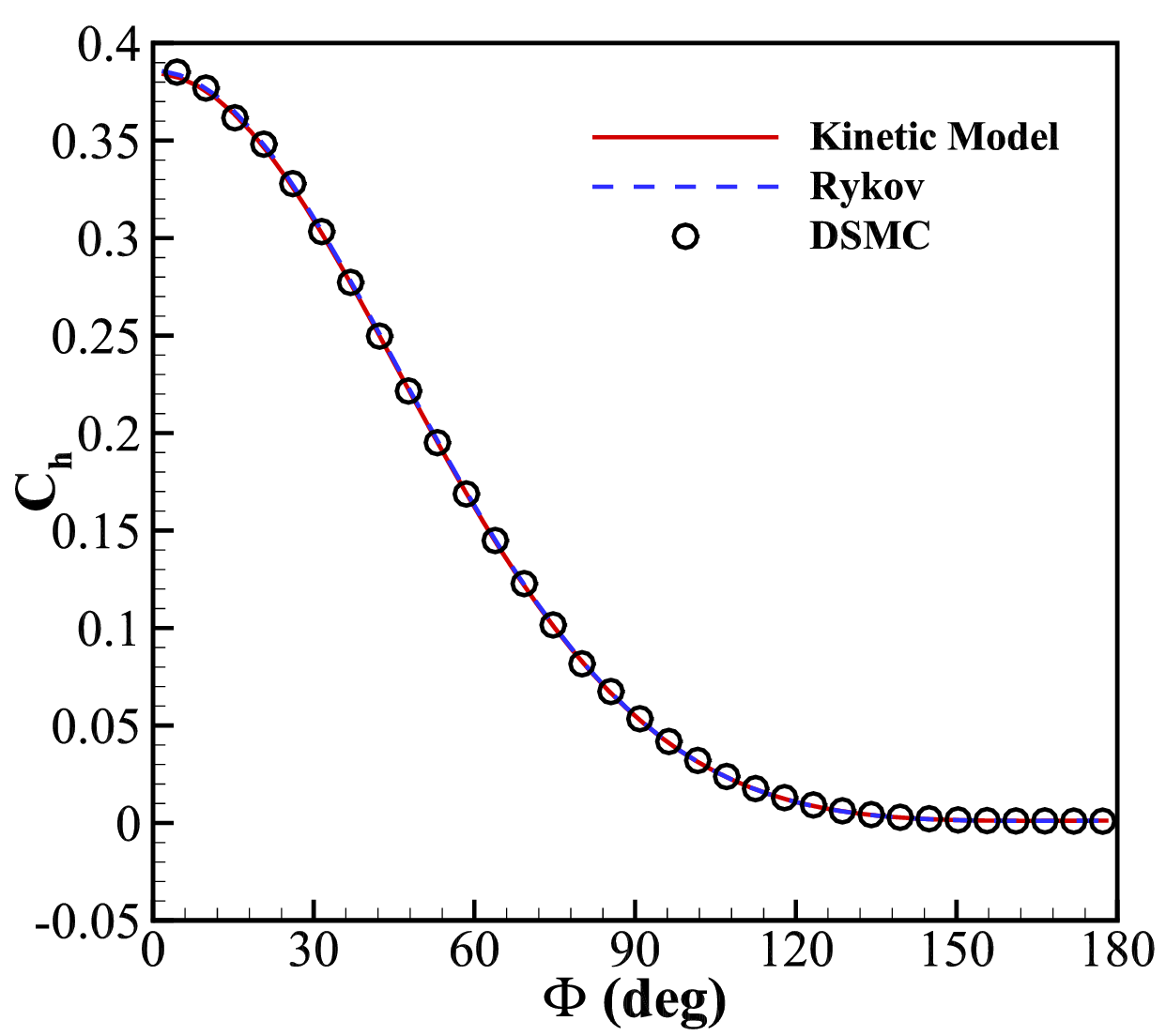}
 		\caption{}
 	\end{subfigure}
 	
 	\caption{Comparison of surface quantities for hypersonic flow past a cylinder at Kn = 0.1. (a) Pressure coefficient, (b) shear stress coefficient, and (c) heat transfer coefficient.}
 	\label{cylinder_Kn0.1_coe}
 \end{figure}

\begin{figure}
 	\centering
 	
 	\begin{subfigure}[b]{0.32\textwidth}
 		\includegraphics[width=\linewidth]{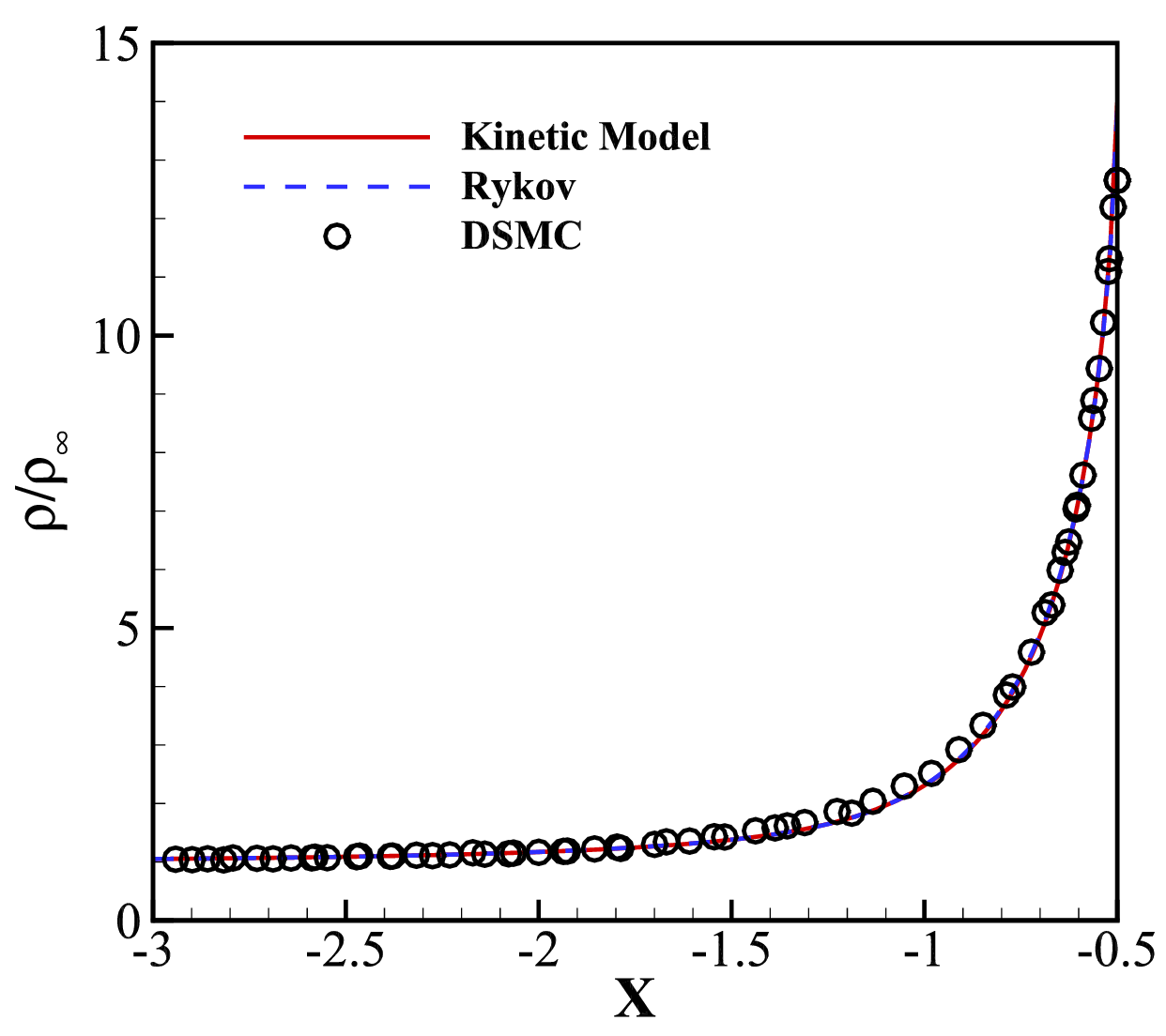}
 		\caption{}
 	\end{subfigure}
 	\hfill
 	\begin{subfigure}[b]{0.32\textwidth}
 		\includegraphics[width=\linewidth]{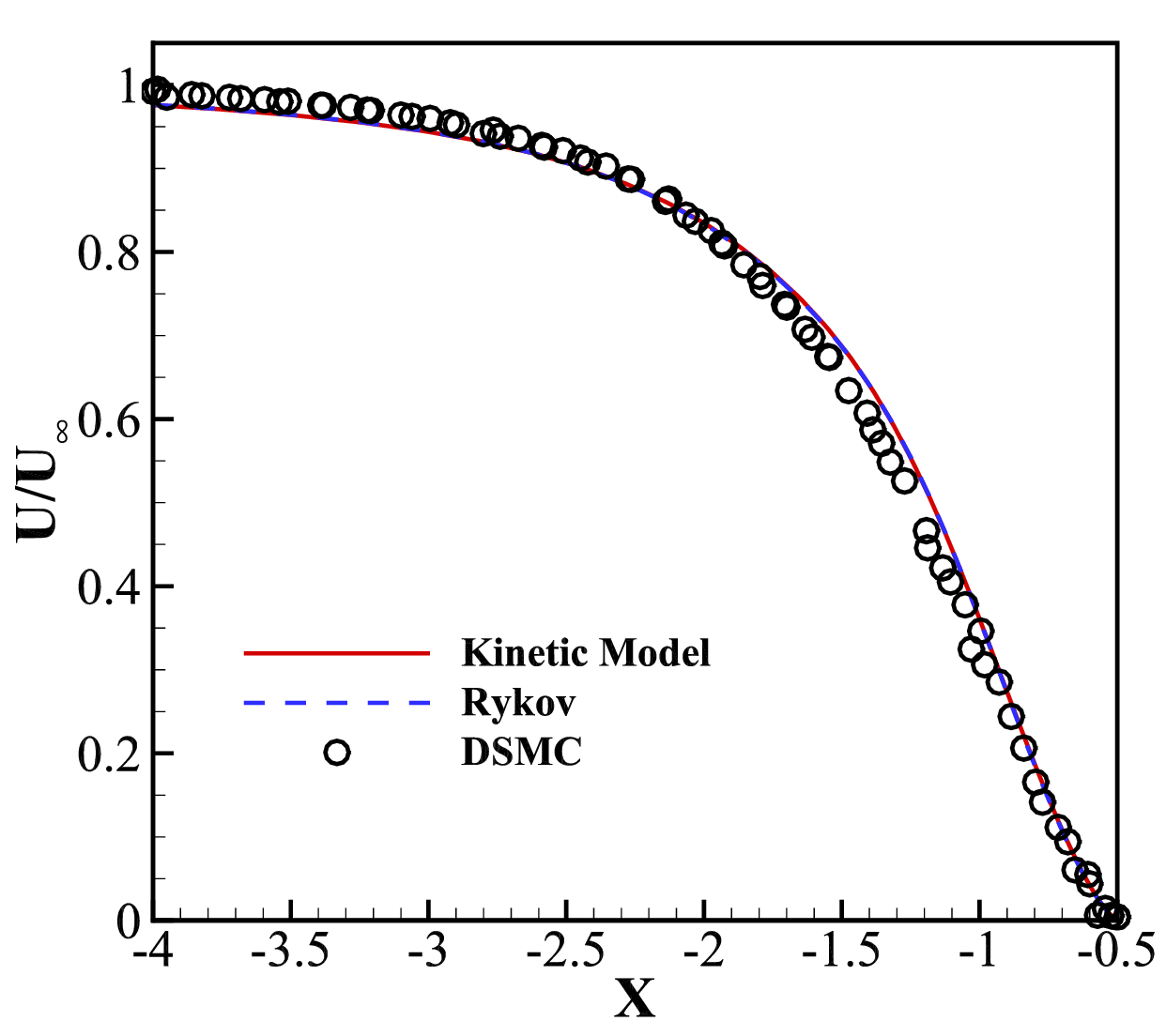}
 		\caption{}
 	\end{subfigure}
 	\hfill
 	\begin{subfigure}[b]{0.32\textwidth}
 		\includegraphics[width=\linewidth]{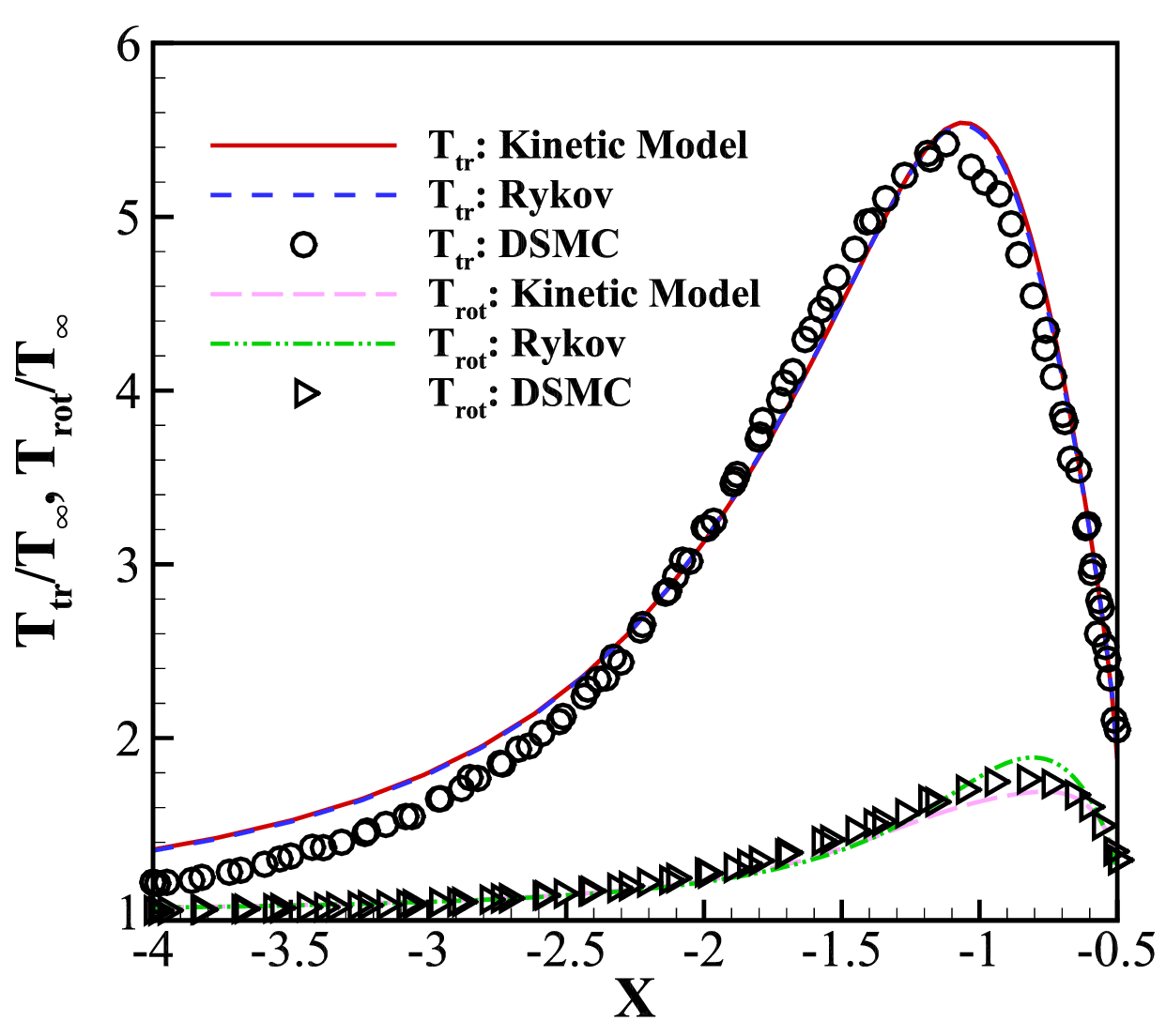}
 		\caption{}
 	\end{subfigure}
 	
 	\caption{Comparison of flow quantities along the stagnation line for hypersonic flow past a cylinder at Kn = 1.0. (a) Density, (b) velocity, and (c) translational and rotational temperatures.}
 	\label{cylinder_Kn1_mac}
 \end{figure}

\begin{figure}
 	\centering
 	
 	\begin{subfigure}[b]{0.32\textwidth}
 		\includegraphics[width=\linewidth]{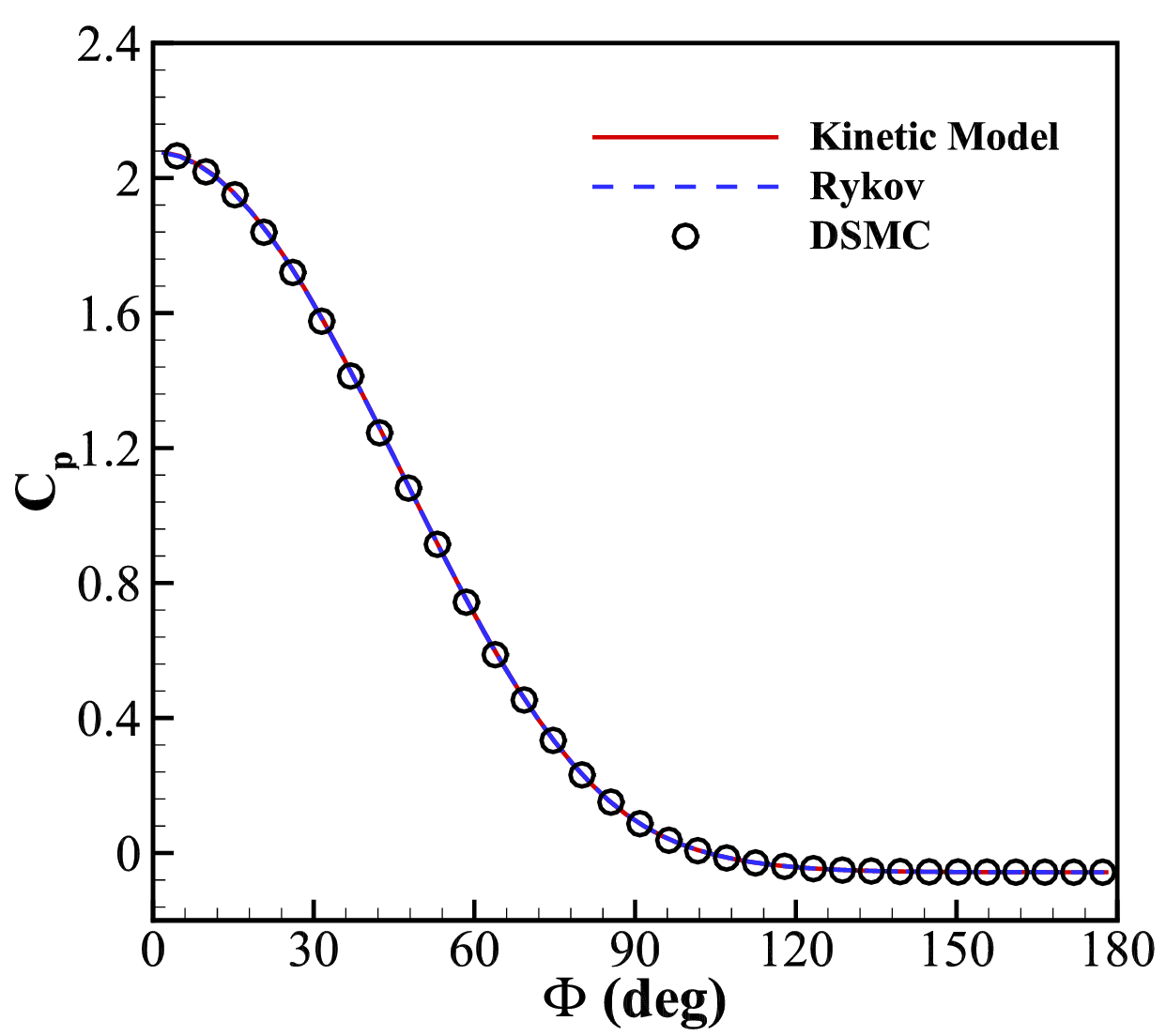}
 		\caption{}
 	\end{subfigure}
 	\hfill
 	\begin{subfigure}[b]{0.32\textwidth}
 		\includegraphics[width=\linewidth]{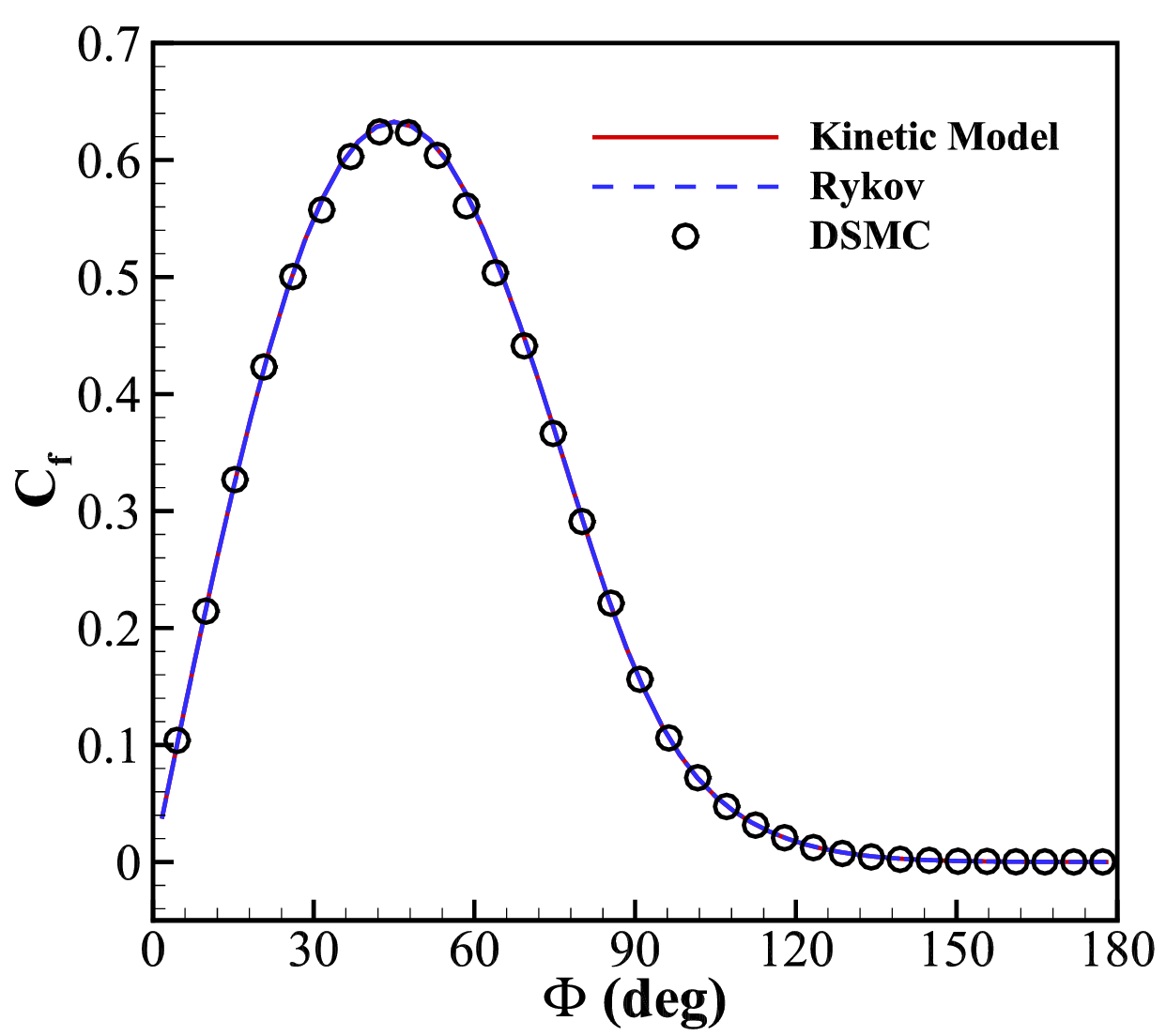}
 		\caption{}
 	\end{subfigure}
 	\hfill
 	\begin{subfigure}[b]{0.32\textwidth}
 		\includegraphics[width=\linewidth]{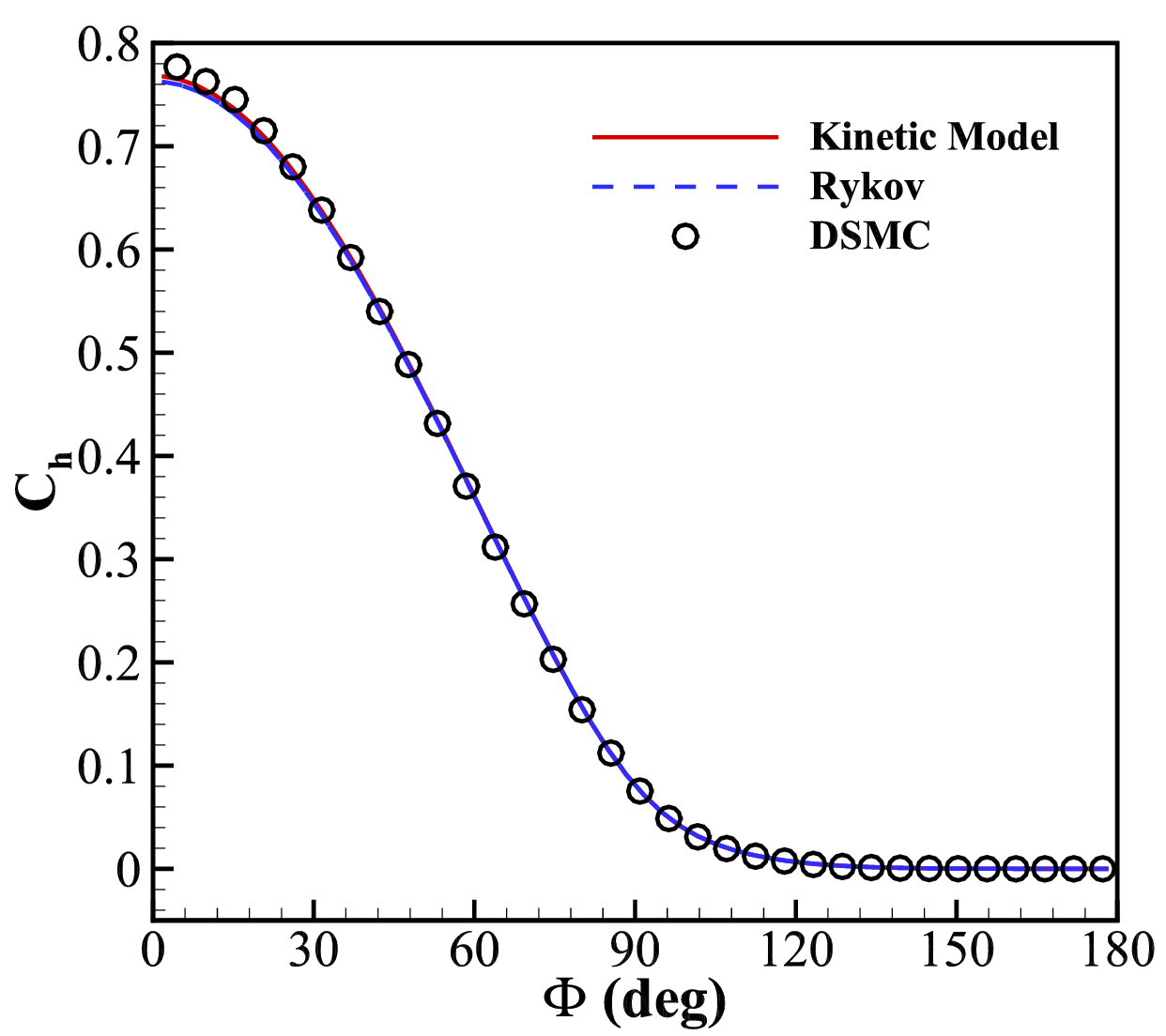}
 		\caption{}
 	\end{subfigure}
 	
 	\caption{Comparison of surface quantities for hypersonic flow past a cylinder at Kn = 1.0. (a) Pressure coefficient, (b) shear stress coefficient, and (c) heat transfer coefficient.}
 	\label{cylinder_Kn1_coe}
 \end{figure}

 Figures~\ref{cylinder_Kn0.01_coe}, \ref{cylinder_Kn0.1_coe}, \ref{cylinder_Kn1_coe} present density, velocity, and temperature distributions along the stagnation streamline for $\rm{Kn = 0.01, 0.1, 1}$. The kinetic model shows essential agreement with Rykov simulations. Compared to DSMC results, density distributions match closely, while velocity profiles exhibit minor deviations. Temperature distributions display an early rise upstream of the shock, which is primarily attributed to insufficient relaxation of backstreaming high-energy particles from downstream. This limitation can be addressed by modifying the relaxation times for energetic particles \citep{xu2021modeling}. Figures~\ref{cylinder_Kn0.01_mac}, \ref{cylinder_Kn0.1_mac}, \ref{cylinder_Kn1_mac} show distributions of normal stress, shear stress, and heat transfer coefficients on the cylinder surface. The kinetic model demonstrates good agreement with both Rykov and DSMC simulations, confirming its capability to accurately predict surface forces and heat transfer in high-speed flows. 
  
\begin{figure}
 	\centering
 	
 	\begin{subfigure}[b]{0.49\textwidth}
 		\includegraphics[width=\linewidth]{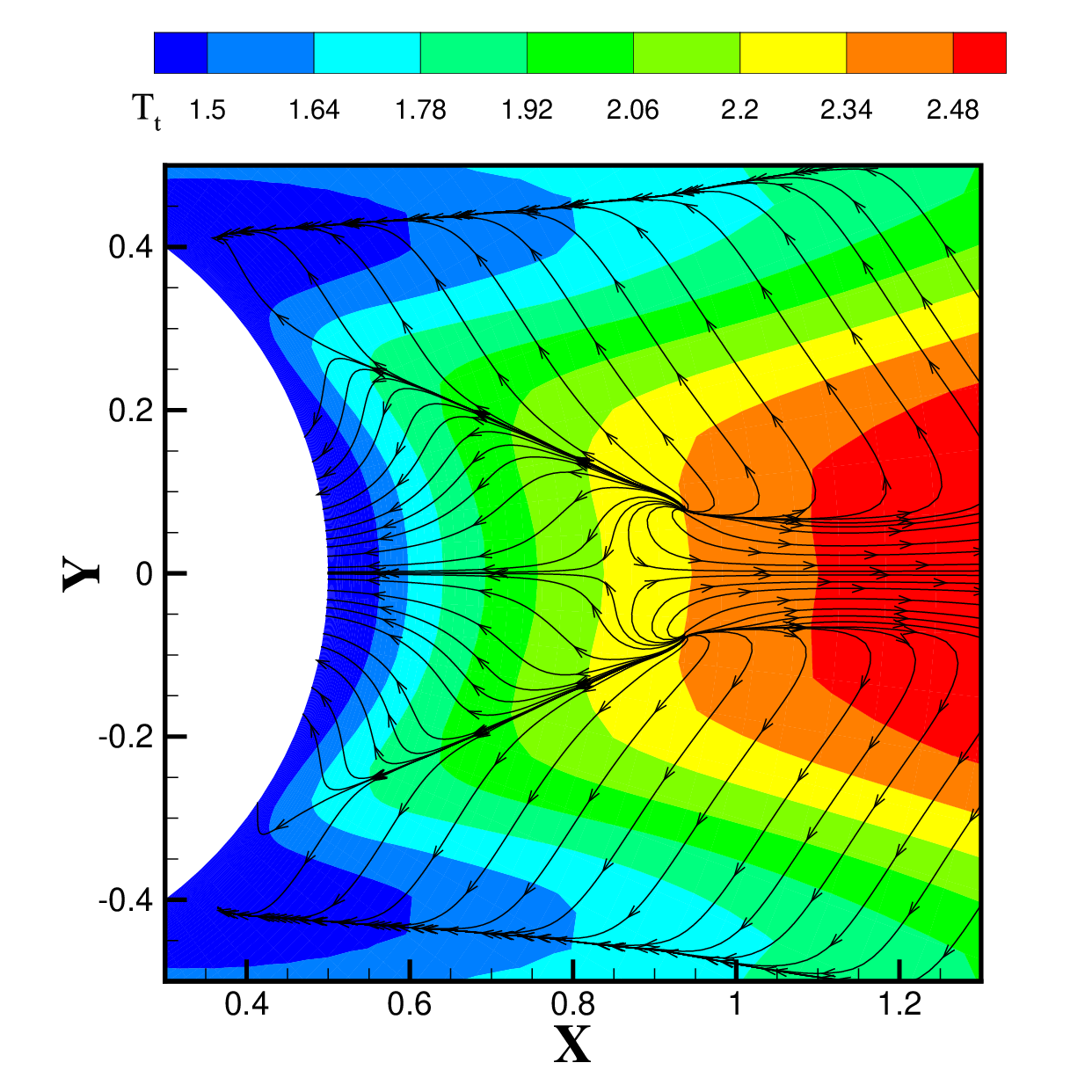}
 		\caption{$T_{\rm{t}}$ and $\mathbf{q}_{\rm{t}}$ by new kinetic model}
 	\end{subfigure} 
 	\hfill
 	\begin{subfigure}[b]{0.49\textwidth}
 		\includegraphics[width=\linewidth]{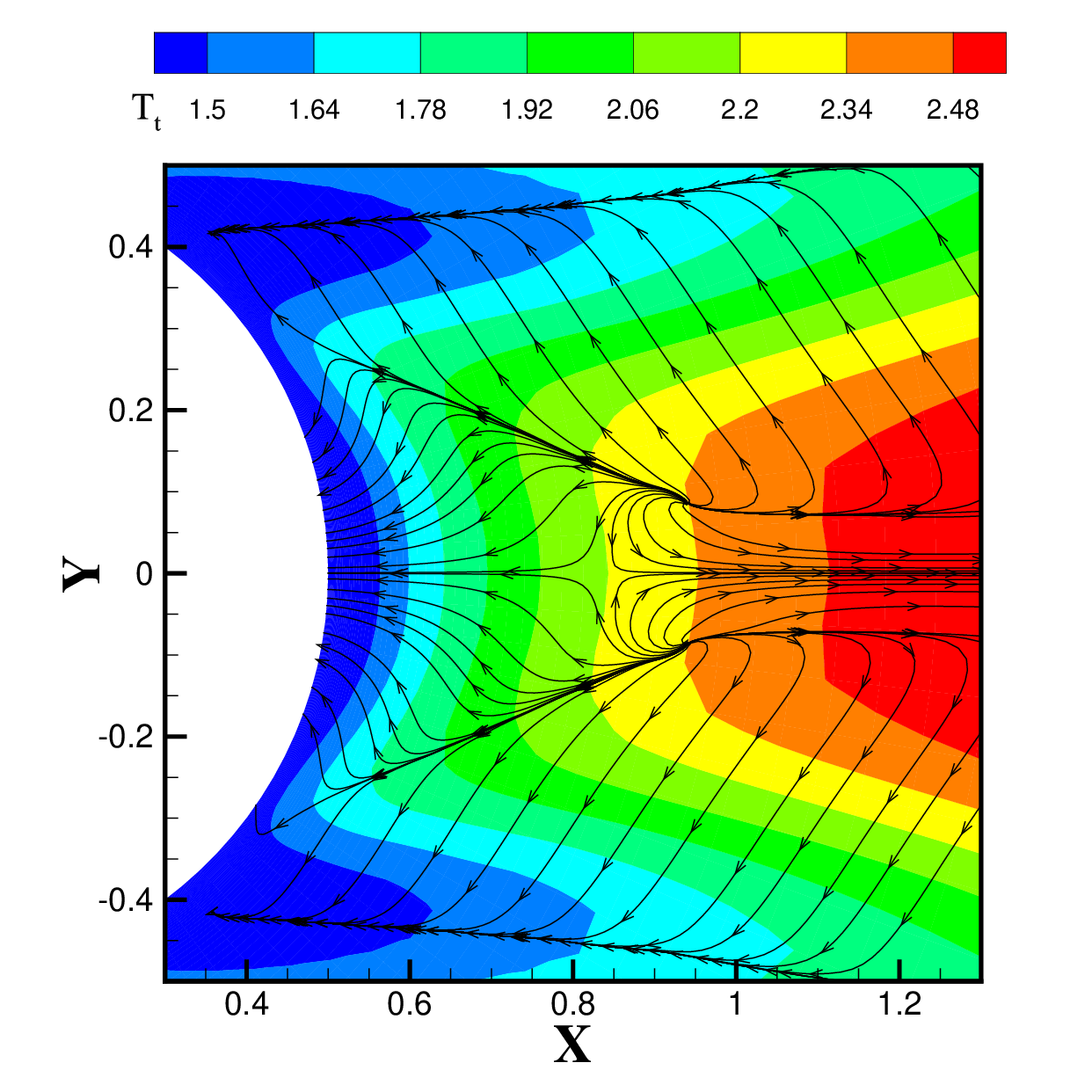}
 		\caption{$T_{\rm{t}}$ and $\mathbf{q}_{\rm{t}}$ by Rykov model}
 	\end{subfigure}
 	
 	\begin{subfigure}[b]{0.49\textwidth}
 		\includegraphics[width=\linewidth]{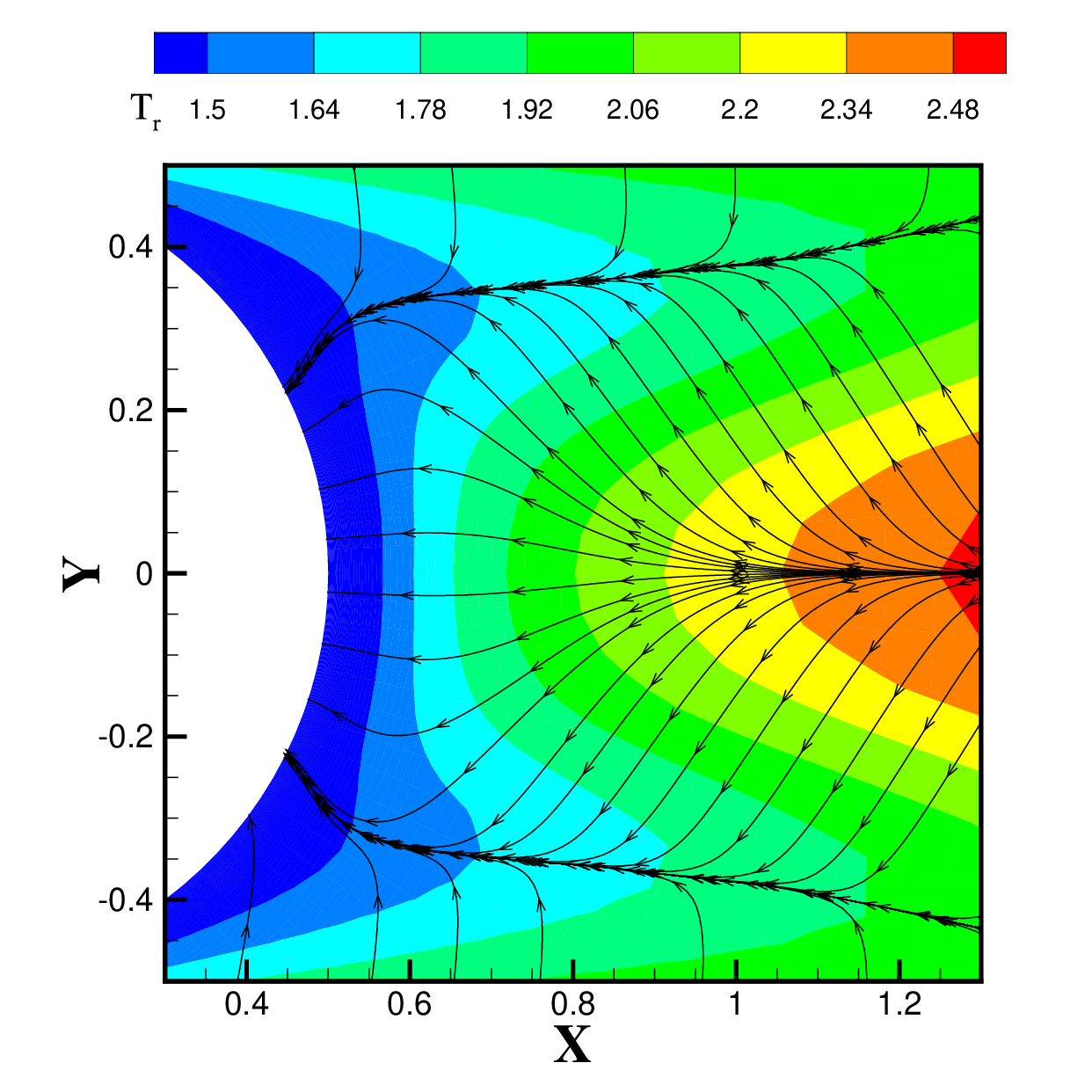}
 		\caption{$T_{\rm{r}}$ and $\mathbf{q}_{\rm{r}}$ by new kinetic model}
 	\end{subfigure}
 	\hfill
 	\begin{subfigure}[b]{0.49\textwidth}
 		\includegraphics[width=\linewidth]{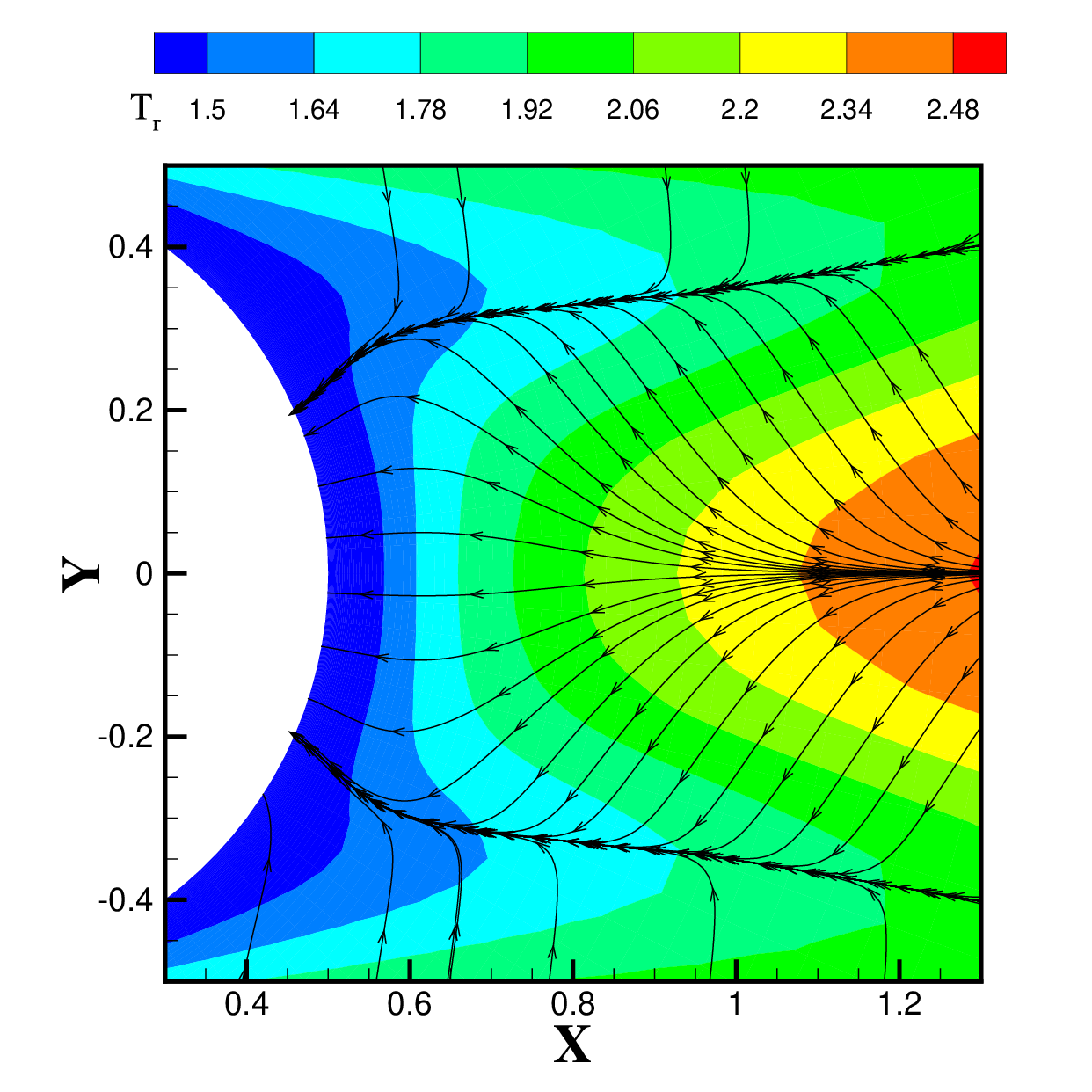}
 		\caption{$T_{\rm{r}}$ and $\mathbf{q}_{\rm{r}}$ by Rykov model}
 	\end{subfigure}
 	
 	\caption{Comparison of the temperature fields (colored fields) and heat flux vectors (black solid lines) between the kinetic model and the Rykov model in the wake region of flow past cylinder at Kn = 0.01. (a)-(b) Translational temperature and heat flux. (c)-(d) rotational temperature and heat flux.}
 	\label{cylinder_Kn0.01_q}
 \end{figure}

\begin{figure}
 	\centering
 	
 	\begin{subfigure}[b]{0.49\textwidth}
 		\includegraphics[width=\linewidth]{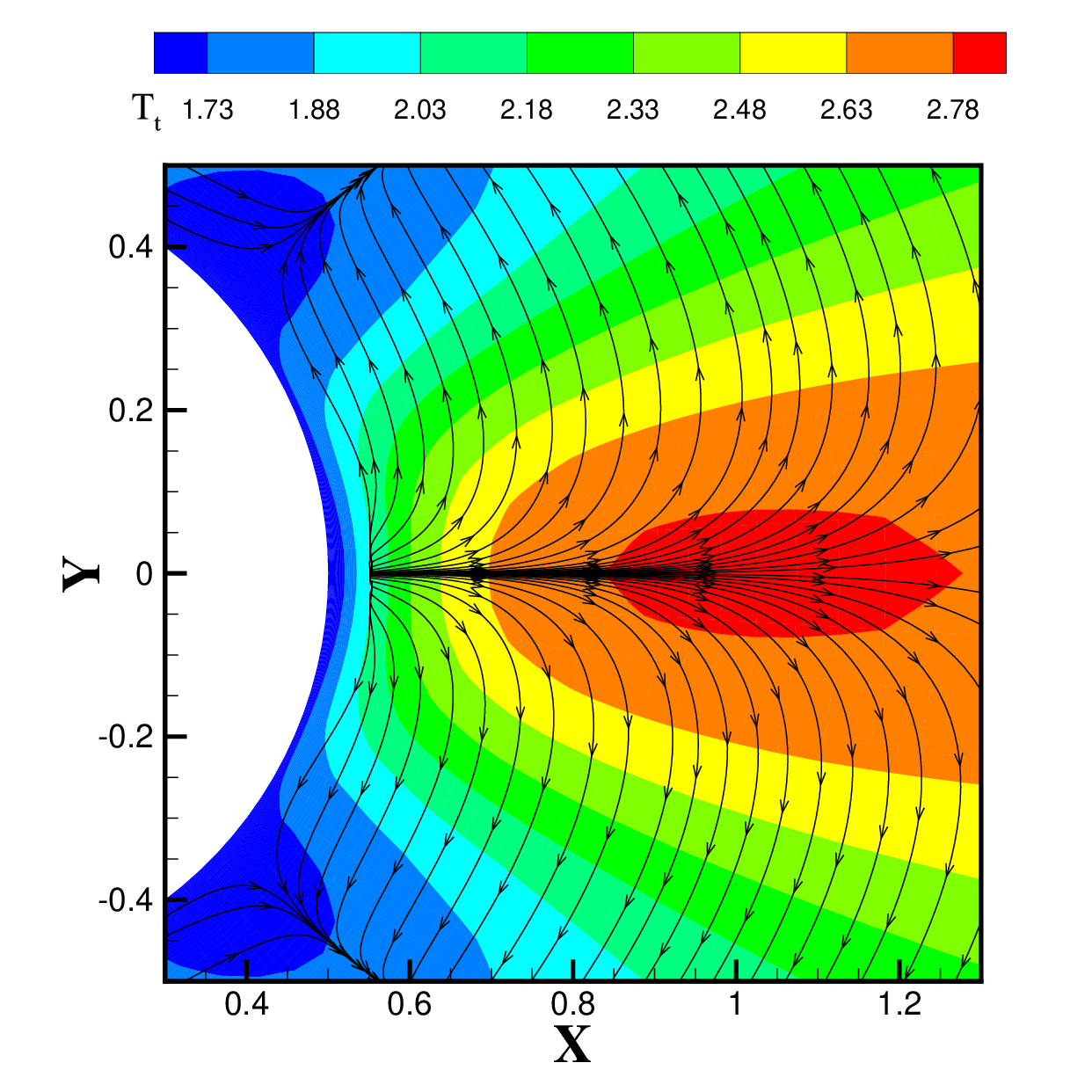}
 		\caption{$T_{\rm{t}}$ and $\mathbf{q}_{\rm{t}}$ by new kinetic model}
 	\end{subfigure}
 	\hfill
 	\begin{subfigure}[b]{0.49\textwidth}
 		\includegraphics[width=\linewidth]{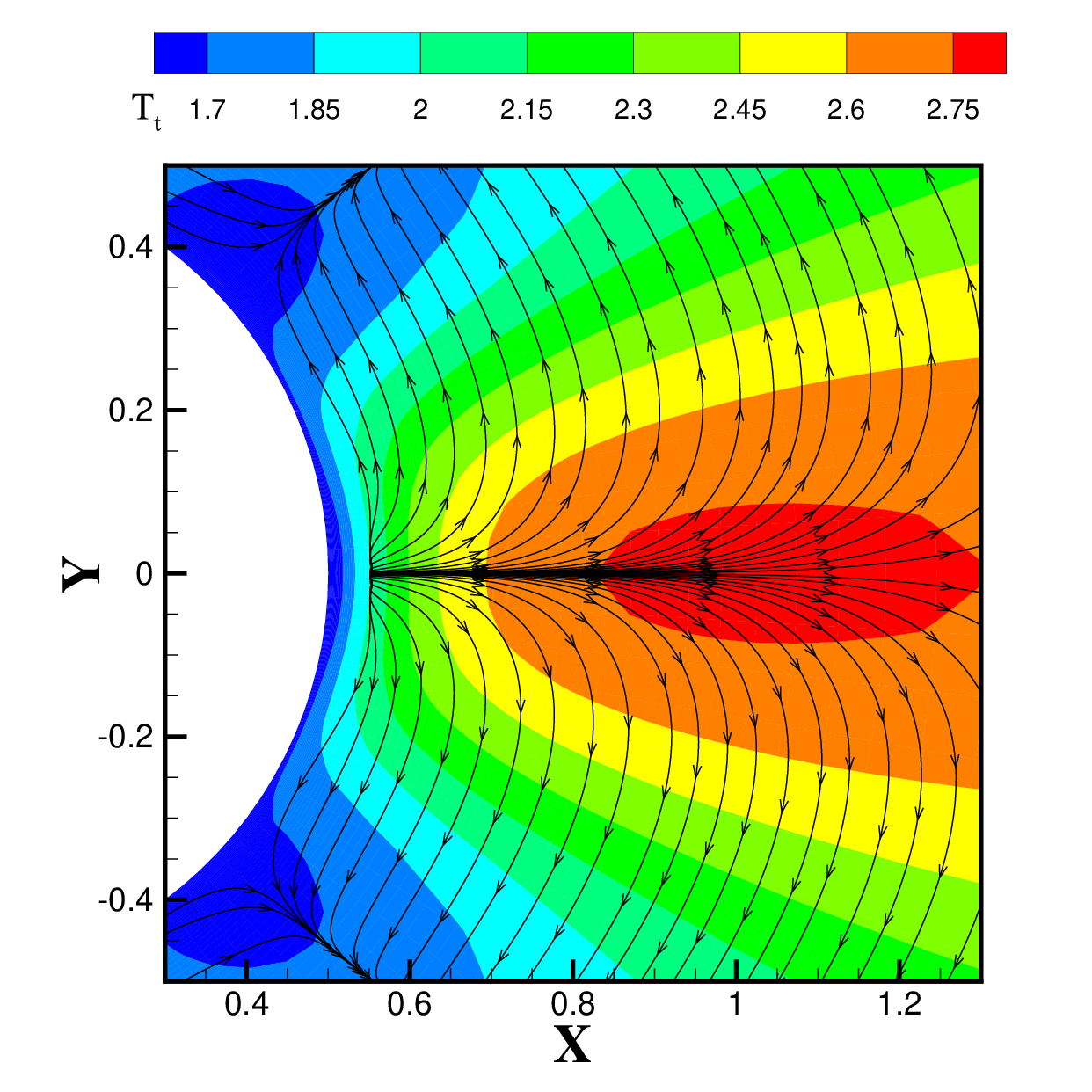}
 		\caption{$T_{\rm{t}}$ and $\mathbf{q}_{\rm{t}}$ by Rykov model}
 	\end{subfigure}
 	
 	\begin{subfigure}[b]{0.49\textwidth}
 		\includegraphics[width=\linewidth]{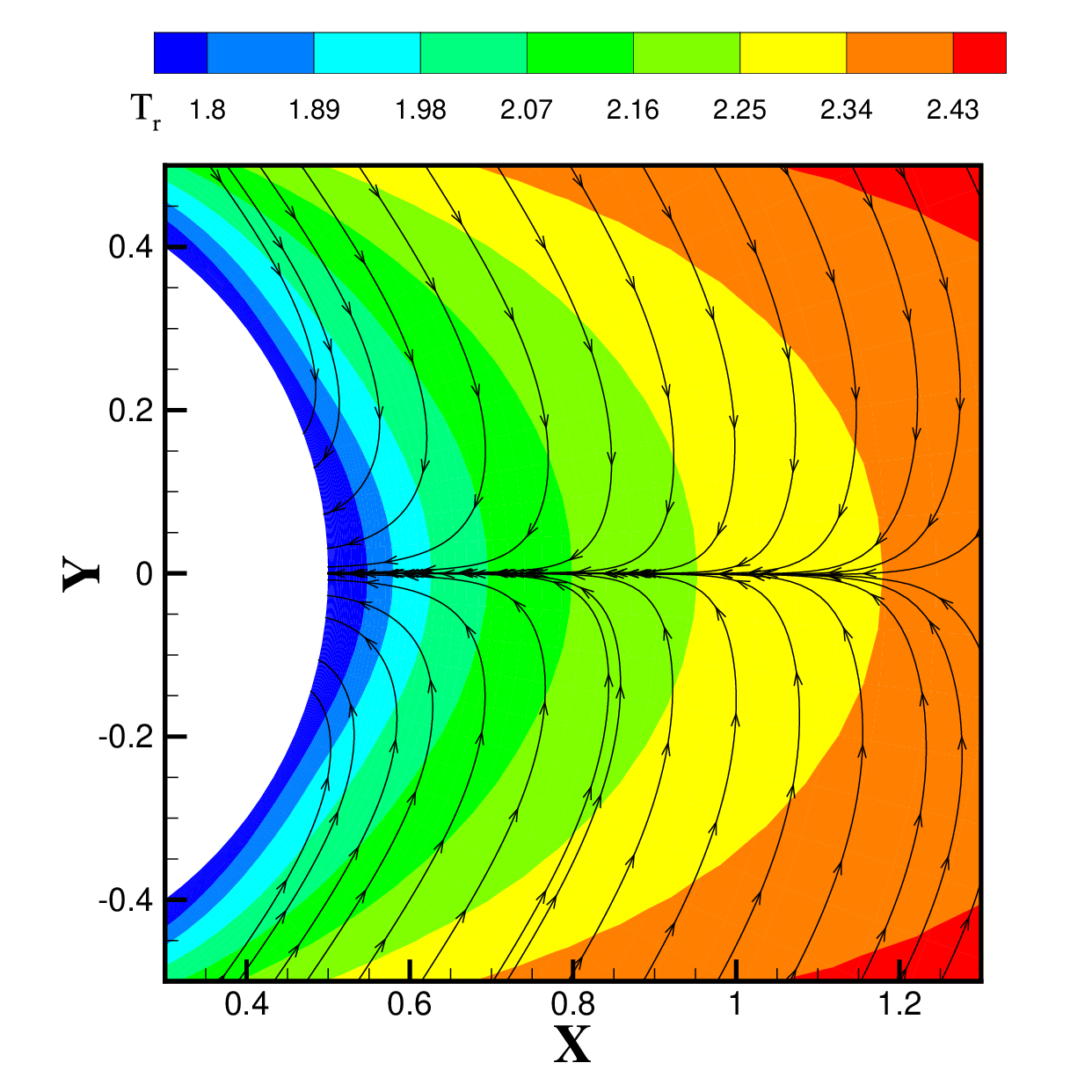}
 		\caption{$T_{\rm{r}}$ and $\mathbf{q}_{\rm{r}}$ by new kinetic model}
 	\end{subfigure}
 	\hfill
 	\begin{subfigure}[b]{0.49\textwidth}
 		\includegraphics[width=\linewidth]{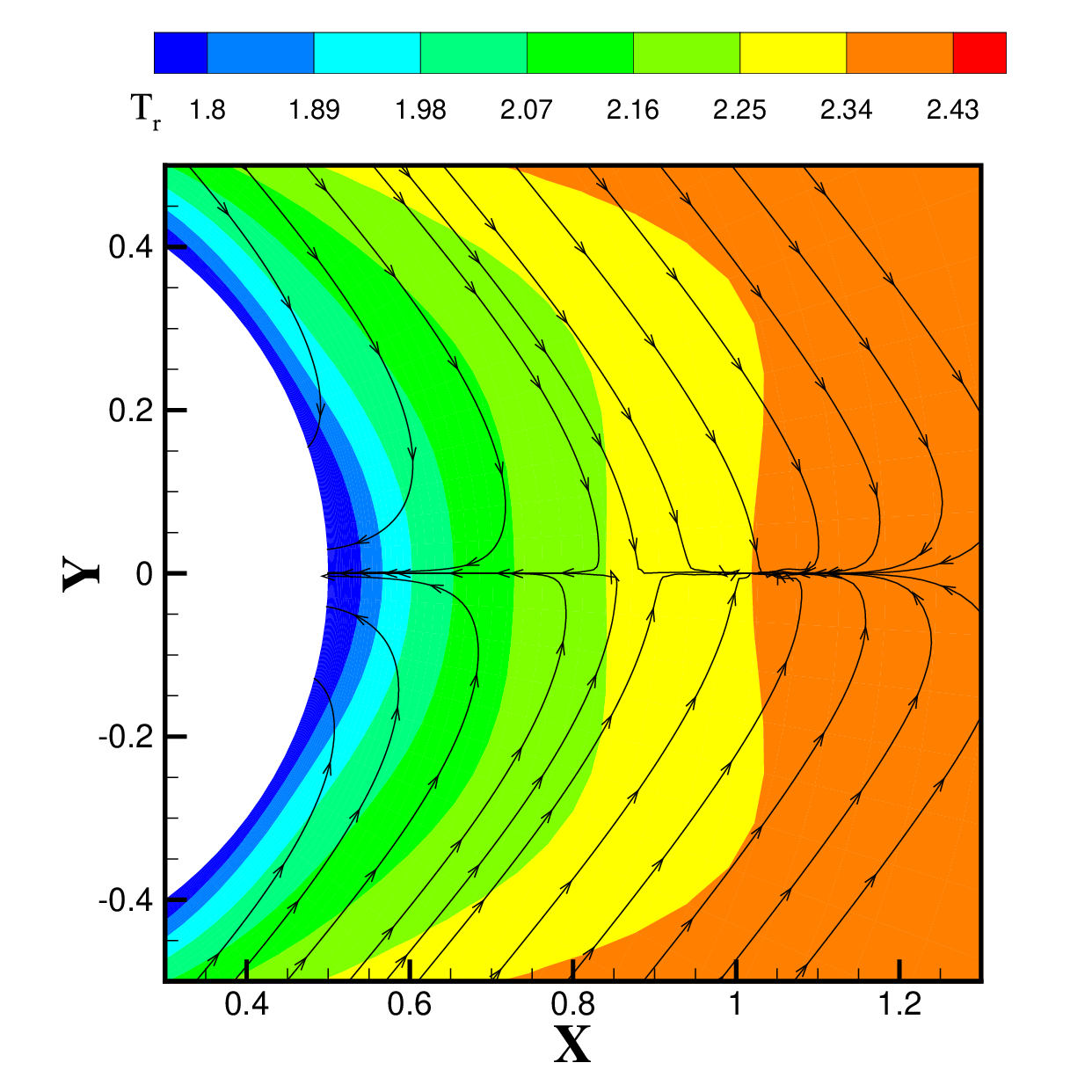}
 		\caption{$T_{\rm{r}}$ and $\mathbf{q}_{\rm{r}}$ by Rykov model}
 	\end{subfigure}
 	
 	\caption{Comparison of the temperature fields (colored fields) and heat flux vectors (black solid lines) between the kinetic model and the Rykov model in the wake region of flow past cylinder at Kn = 0.1. (a)-(b) Translational temperature and heat flux. (c)-(d) rotational temperature and heat flux.}
 	\label{cylinder_Kn0.1_q}
 \end{figure}

\begin{figure}
 	\centering
 	
 	\begin{subfigure}[b]{0.49\textwidth}
 		\includegraphics[width=\linewidth]{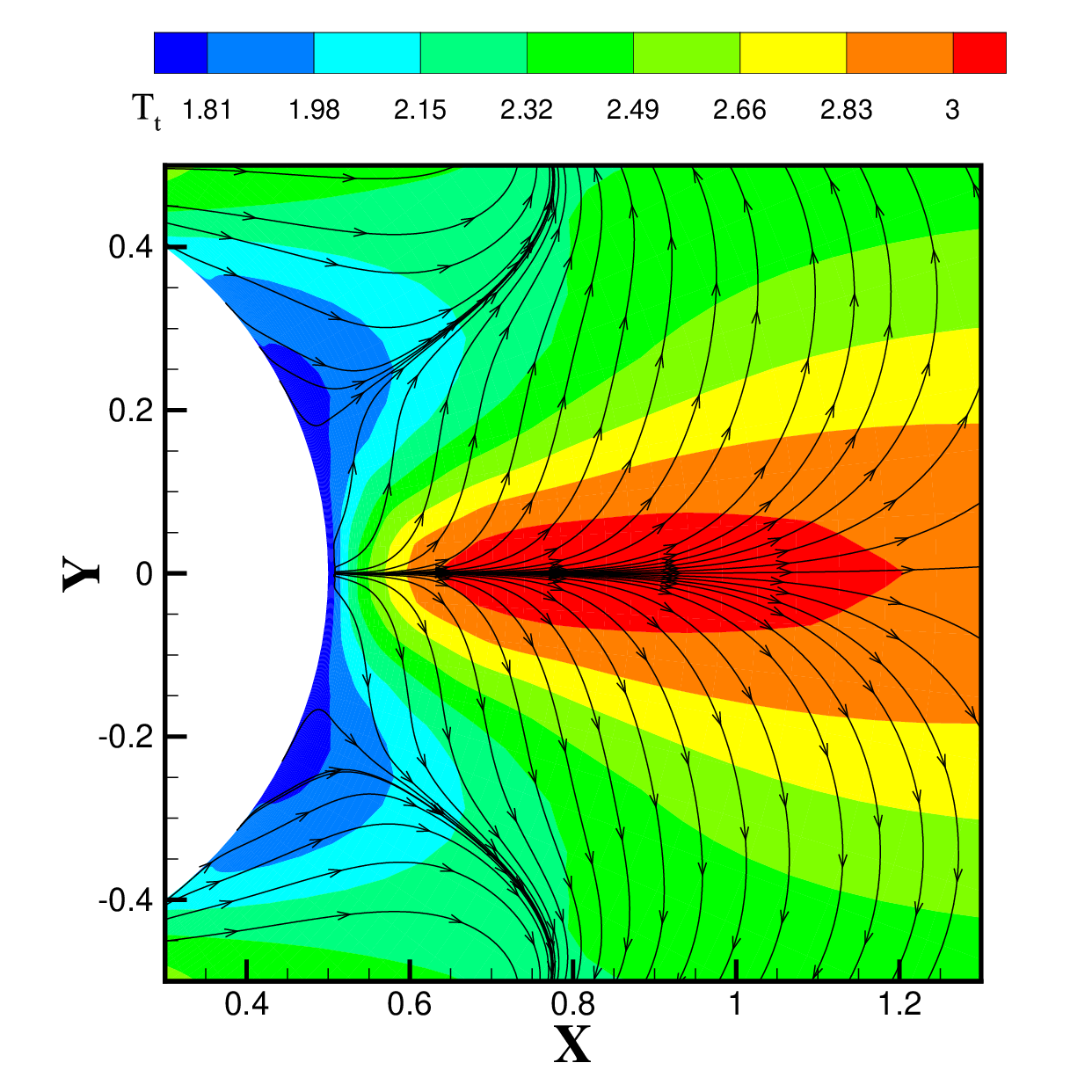}
 		\caption{$T_{\rm{t}}$ and $\mathbf{q}_{\rm{t}}$ by new kinetic model}
 	\end{subfigure}
 	\hfill
 	\begin{subfigure}[b]{0.49\textwidth}
 		\includegraphics[width=\linewidth]{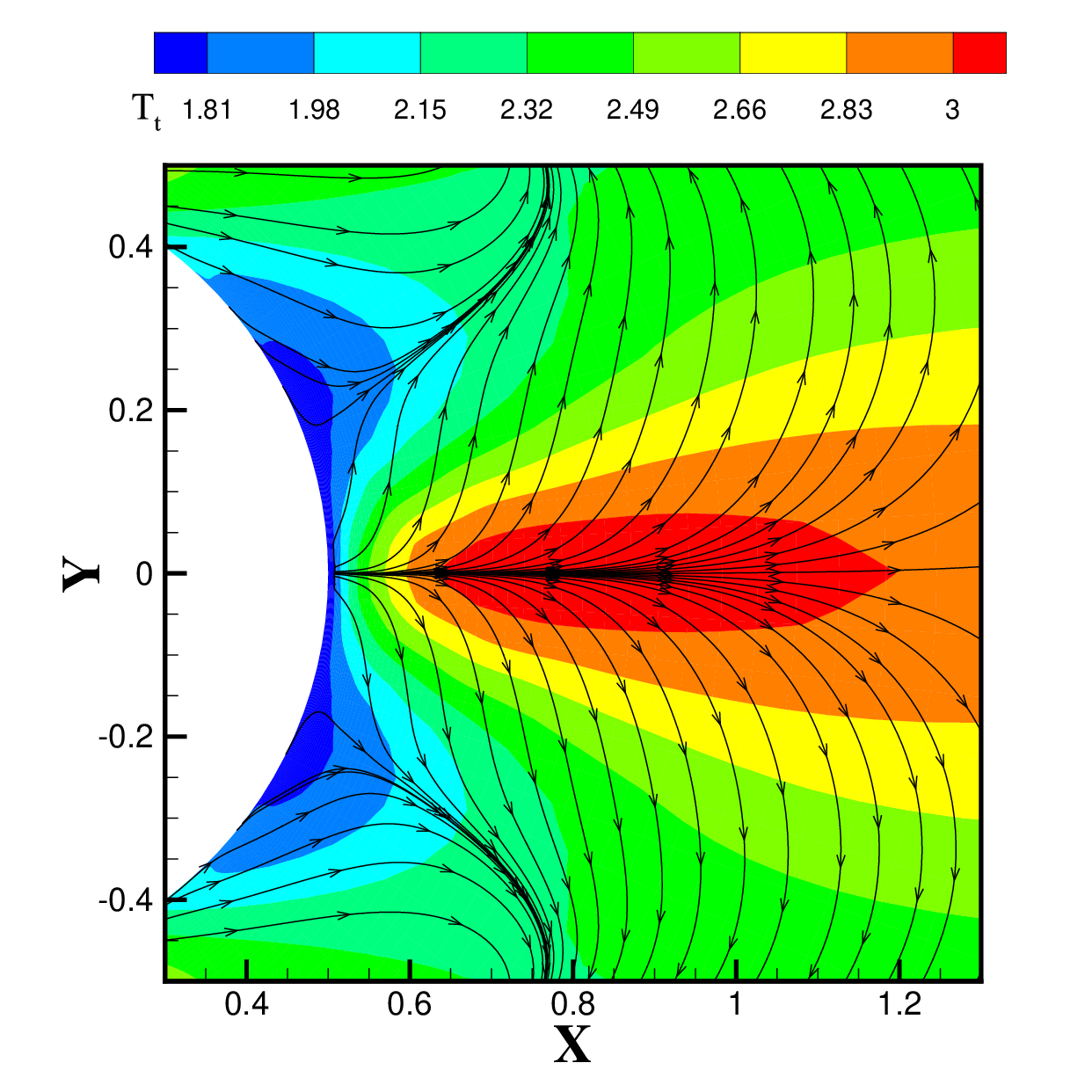}
 		\caption{$T_{\rm{t}}$ and $\mathbf{q}_{\rm{t}}$ by Rykov model}
 	\end{subfigure}
 	
 	\begin{subfigure}[b]{0.49\textwidth}
 		\includegraphics[width=\linewidth]{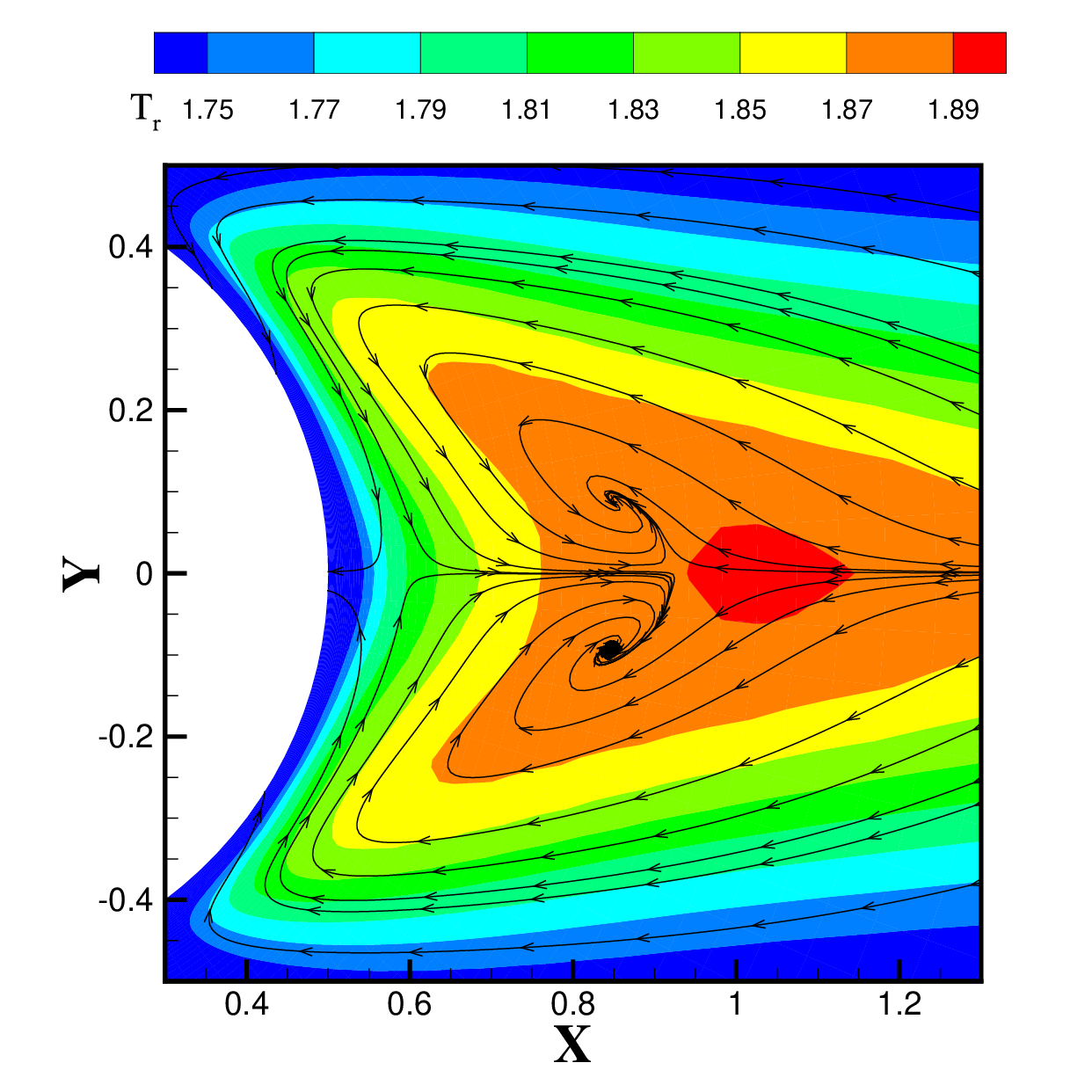}
 		\caption{$T_{\rm{r}}$ and $\mathbf{q}_{\rm{r}}$ by new kinetic model}
 	\end{subfigure}
 	\hfill
 	\begin{subfigure}[b]{0.49\textwidth}
 		\includegraphics[width=\linewidth]{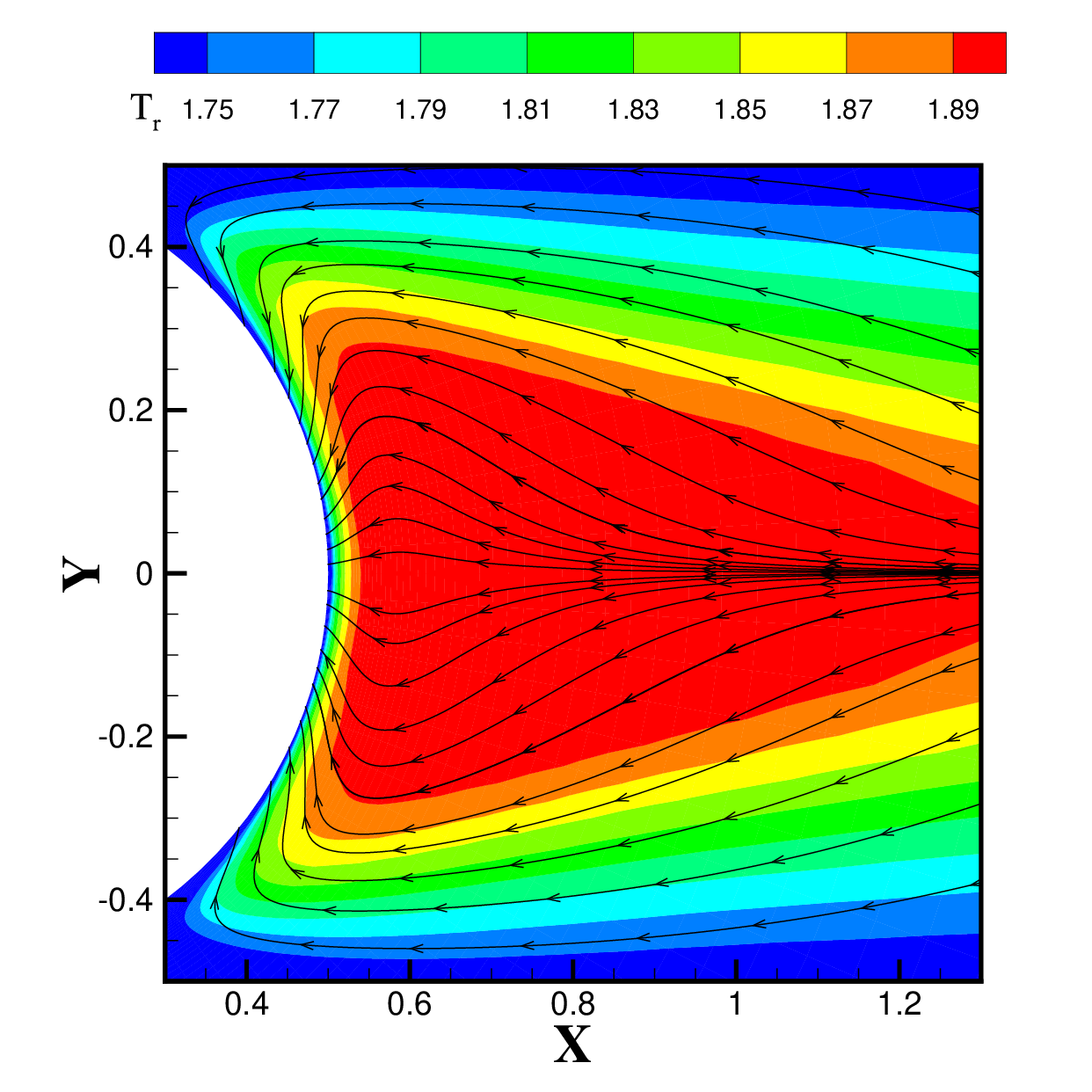}
 		\caption{$T_{\rm{r}}$ and $\mathbf{q}_{\rm{r}}$ by Rykov model}
 	\end{subfigure}
 	
 	\caption{Comparison of the temperature fields (colored fields) and heat flux vectors (black solid lines) between the kinetic model and the Rykov model in the wake region of flow past cylinder at Kn = 1.0. (a)-(b) Translational temperature and heat flux. (c)-(d) rotational temperature and heat flux.}
 	\label{cylinder_Kn1.00_q}
 \end{figure}

 In compressive flow regions, sufficient collisions make the coupling effect in heat flux relaxation less evident in the flow field. However, in wake flows with insufficient collisions, the situation is markedly different. Figures~\ref{cylinder_Kn0.01_q}, \ref{cylinder_Kn0.1_q}, \ref{cylinder_Kn1.00_q} present the distributions of heat flux behind a cylinder at Knudsen numbers of 0.01, 0.1, and 1.0, respectively. The predictions of translational temperature and translational heat flux show good agreement between the kinetic and Rykov models, whereas discrepancies of varying degrees are observed for the rotational temperature and rotational heat flux. These differences in the rotational distributions become more pronounced as the degree of rarefaction increases. It is noteworthy that for the case of Kn = 1.0, the kinetic model exhibits a significant phenomenon wherein the rotational heat flux becomes aligned with the temperature gradient in parts of the flow field. For heat flux distributions in high Knudsen number flows, accurate description cannot be achieved through constitutive relations alone, as it requires simultaneous consideration of heat flux generation and transport. However, results in figure~\ref{cylinder_Kn1.00_q} shows that the heat flux is nearly parallel to the isotherms of the temperature field, indicating that the heat flux in the wake region is primarily governed by thermal relaxation (generation) rather than thermal transport. Consequently, our analysis focuses on heat flux generation, i.e., the relaxation driven by molecular collisions. Similar to the Lid-driven cavity flow case, the differences in rotational heat flux may be attributed to the kinetic model's more comprehensive consideration of the coupling mechanism between the relaxation processes of translational and rotational heat fluxes, compared to the Rykov model. In the central wake region where differences arise, the dimensionless translational heat flux is on the order of $10^{-1}$ to $10^{-2}$, while the rotational heat flux is on the order of $10^{-3}$ to $10^{-4}$. As indicated by the relaxation formula Eqs.~(\ref{relax_Qt_VHS}) and (\ref{relax_Qr_VHS}), the translational heat flux exerts a strong influence on the rotational component here, while the effect of the rotational on the translational component can be ignored. This coupling in the relaxation process ultimately leads to heat flux reversal within the wake. In summary, in rarefied flows where thermal relaxation mechanisms dominate, the influence of the translational-rotational heat flux relaxation coupling becomes more pronounced. Therefore, the kinetic model, established upon more accurate relaxation rates, is expected to yield superior results in thermal predictions for such flow conditions.
 
 \newpage
 \section{Conclusions}\label{chapter5}
 In summary, we adopt the Pullin model with an integrable collision kernel that abides by the detailed balance principle as the extended Boltzmann equation, and derive the analytical expressions of the temporal relaxation rates for key macroscopic quantities via theoretical derivation for the first time. These quantities include the stress tensor, translational and rotational temperatures, as well as translational and rotational heat fluxes. Meanwhile, based on the C-E expansion framework with the same fundamental integrals for the collision term, the corresponding transport coefficients are derived, thereby establishing the fundamental physical connection between non-equilibrium relaxation rates and transport coefficients. The primary results revealed from the theoretical analysis are that translational and rotational heat fluxes relax in a coupled manner, and that both the heat-flux relaxation process and the thermal conductivity are influenced by the non-equilibrium temperature ratio $T_{\rm{t}}/T_{\rm{r}}$. We quantify and evaluate the influence of non-equilibrium temperatures on thermal conductivity, and the results demonstrate that extreme non-equilibrium temperatures can induce drastic variations in thermal conductivity. Additionally, the derived results are compared separately with the theoretical and experimental results under the more common equilibrium temperature condition ($T_{\rm{t}}=T_{\rm{r}}$). The Eucken factor derived in this work reduces to the classical result obtained by Mason and Monchick from the WCU equation under equilibrium temperature condition. On the other hand, comparisons with the viscosity and thermal conductivity data of nitrogen from the NIST experimental database show that the results derived in this work exhibit excellent agreement with the experimental values.
 
 Based on the analytically derived relaxation rates, we have proposed a tractable relaxation-type kinetic model. The kinetic model can not only recover transport coefficients including shear viscosity, bulk viscosity and thermal conductivity, but also accurately capture the relaxation processes of non-equilibrium macroscopic quantities. The accuracy of the proposed model is evaluated by comparisons with DSMC simulation results. The kinetic model demonstrates excellent accuracy in zero-dimensional homogeneous relaxation, planar Couette flow, and hypersonic flow past a cylinder, while it exhibits certain deviations in the normal shock waves. This issue can be resolved by modifying the relaxation time of high-energy particles during the solution process.
 
 Then, the influence of the coupled relaxation mechanism between translational and rotational heat fluxes and the role of the non-equilibrium temperature ratio $T_{\rm{t}}/T_{\rm{r}}$ on the flow are assessed by comparing solutions of the proposed kinetic model with those of the widely used Rykov model, which neglects these mechanisms. In simulations of 0-dimensional homogeneous relaxation and Planar Couette flow, the proposed kinetic model yields more accurate predictions of heat fluxes and temperatures than the Rykov model. Furthermore, under rarefied conditions, significant differences in rotational temperature and heat flux distributions are observed in lid-driven cavity flows and in the wake region of hypersonic flow past a cylinder. This demonstrates that it is essential to account for the coupled relaxation of heat fluxes and the effect of the non-equilibrium temperature ratio $T_{\rm{t}}/T_{\rm{r}}$ in the thermal prediction of rarefied flows.
 
 Although the derivation of relaxation rates and kinetic modeling in this study considers only rotational excitation, the extension to include vibrational relaxation rates and modeling can be achieved within our general framework. Furthermore, the increasingly complete state-to-state database can provide a foundation for achieving higher-precision modeling within this theoretically derived framework. In future work, these elements will be incorporated into the establishment of relaxation-rate models that account for vibrational excitation, to study high-temperature rarefied gas flows with strong thermal non-equilibrium.
 
  \backsection[Supplementary data]{}
 
 \backsection[Acknowledgements]{The authors thank Prof. Quanhua Sun and Dr. Qizhen Hong at the Institute of Mechanics, Chinese Academy of Sciences, and Prof. Lei Wu at Southern University of Science and Technology for discussions on the extended Boltzmann equation. The authors also thank Prof. Kun Xu at Hong Kong University of Science and Technology and Prof. Zhaoli Guo at Huazhong University of Science and Technology for discussions on unified methods.}
 
 \backsection[Funding]{This work is supported by the National Natural Science Foundation of China (Grants No. 12172301) and the 111 Project of China (No. B17037).}
 
 \backsection[Declaration of interests]{The authors report no conflict of interest.}
 
 \backsection[Data availability statement]{The data that support the findings of this study are available from the corresponding author upon reasonable request.}
 
\backsection[Author ORCIDs]{Sha Liu, \url{https://orcid.org/0000-0002-8974-7455}; Ningchao Ding, \url{https://orcid.org/0009-0006-8950-3588}}
 
 \backsection[Author contributions]{Sha Liu: Conceptualization, Methodology, Formal analysis, Validation, Writing– review \& editing, Funding acquisition. Ningchao Ding: Conceptualization, Investigation, Methodology, Formal analysis, Validation, Writing– original draft. Ming Fang: Writing– review \& editing. Hao Jin: Validation. Rui Zhang: Validation. Chengwen Zhong: Funding acquisition. Congshan Zhuo: Writing– review \& editing.
 }
 
 \appendix
 \section{Chapman-Enskog expansion of Pullin equation}\label{appendixA}
 The macroscopic transport coefficients of the Pullin model can be derived through C-E expansion \citep{chapman1990mathematical} and related to macroscopic relaxation rates.Expanding the distribution function and the time derivative operator $\mathcal{D}\equiv \frac{\partial }{\partial t}+{{U}_{i}}\frac{\partial }{\partial {{x}_{i}}}$ as power series in ${{\varepsilon }_{\rm{ce}}}$, we retain terms up to second-order approximation.
 \begin{equation}
 	\mathcal{D}={{\mathcal{D}}^{\left( 0 \right)}}+\varepsilon_{\rm{ce}} {{\mathcal{D}}^{\left( 1 \right)}}
 	\label{expanded_D}
 \end{equation}
 
 \begin{equation}
 	f={{f}^{\left( 0 \right)}}+{{\varepsilon }_{\rm{ce}}}{{f}^{\left( 1 \right)}}={{f}^{\left( 0 \right)}}+{{\varepsilon }_{\rm{ce}}}{{f}^{\left( 0 \right)}}{{\phi }_{\rm{ce}}}
 	\label{expanded_f}
 \end{equation}
 where
 \begin{equation}
 	f_{1}^{\left( 0 \right)}=n{{(\frac{m}{2\pi k{T}_{\rm{t}}})}^{\frac{3}{2}}}\exp \left( -\frac{m{{C}^{2}}}{2k{{T}_{\rm{t}}}} \right)\frac{{{\epsilon }^{\frac{{\nu}}{2}-1}}}{\Gamma \left( \frac{\nu }{2} \right){{\left( k{T}_{\rm{r}} \right)}^{\frac{\nu }{2}}}}\exp \left( -\frac{\epsilon }{k{T}_{\rm{r}}} \right)
 \end{equation}
 The parameter ${{\varepsilon }_{\rm{ce}}}$ is introduced into the collision term of the Pullin equation, resulting in:
 \begin{equation}
 	\mathcal{D}f_{1}^{\left( 0 \right)}+{{\mathbf{C}}_{1}}\frac{\partial f_{1}^{\left( 0 \right)}}{\partial \mathbf{x}}=\frac{1}{{{\varepsilon }_{\rm{ce}}}}Q(f,f)
 	\label{pullin equation_ce}
 \end{equation}
 Substituting the expanded distribution function Eq.~(\ref{expanded_f}) and time derivative Eq.~(\ref{expanded_D}) into the above Eq.~(\ref{pullin equation_ce}), then classifying the resulting equations by order of ${{\varepsilon }_{\rm{ce}}}$,   yields:
 \begin{equation}
 	\mathcal{D}f_{1}^{\left( 0 \right)}+{{\mathbf{C}}_{1}}\frac{\partial f_{1}^{\left( 0 \right)}}{\partial \mathbf{x}}=0
 	\label{pullin_ce1}
 \end{equation}
 
 \begin{equation}
 	\mathcal{D}f_{1}^{\left( 0 \right)}+{{\mathbf{C}}_{1}}\frac{\partial f_{1}^{\left( 0 \right)}}{\partial \mathbf{x}}=\int{f_{1}^{\left( 0 \right)}f_{2}^{\left( 0 \right)}\left( {{{{\phi }}}_{1}'}+{{{{\phi }}}_{2}'}-{{\phi }_{1}}-{{\phi }_{2}} \right)g\sigma h\left( \mathbf{s} \right)d\mathbf{s}d\Omega d{{\mathbf{c}}_{2}}d{{\epsilon }_{2}}}=\mathcal{I}\left[ \phi_{\rm{ce}}  \right]
 	\label{pullin_ce2}
 \end{equation}
 Taking moments of the first-order approximate Eq.~(\ref{pullin_ce1}) with respect to collisional invariants yields the Euler-type equations:
 \begin{equation}
 	\mathcal{D}\rho +\rho \frac{\partial {{U}_{i}}}{\partial {{x}_{i}}}=0
 \end{equation}
 \begin{equation}
 	\rho \mathcal{D}{{U}_{i}}+\frac{\partial {p}_{\rm{t}}}{\partial {{x}_{i}}}=0
 \end{equation}
 \begin{equation}
 	\left( \frac{3}{2}nk\mathcal{D}{{T}_{\rm{t}}}+nk{{T}_{\rm{t}}}\frac{\partial {{U}_{i}}}{\partial {{x}_{i}}} \right)+\frac{{\nu}}{2}nk\mathcal{D}{{T}_{\rm{r}}}=0
 \end{equation}
decomposing the energy equation gives:
 \begin{equation}
 	\left( \frac{3}{2}nk\mathcal{D}{{T}_{\rm{t}}}+nk{{T}_{\rm{t}}}\frac{\partial {{U}_{i}}}{\partial {{x}_{i}}} \right)={{\mathcal{E}}_{\rm{t}}}=\frac{3{\nu}nk}{2\left( 3+{\nu} \right)}\frac{{{T}_{\rm{r}}}-{{T}_{\rm{t}}}}{{{Z}_{\rm{rot}}}\tau_{\rm{t}} }
 	\label{energy_equation1}
 \end{equation}
 \begin{equation}
 	\frac{{\nu}}{2}nk\mathcal{D}{{T}_{\rm{r}}}={{\mathcal{E}}_{\rm{r}}}=\frac{3{\nu}nk}{2\left( 3+{\nu} \right)}\frac{{{T}_{\rm{t}}}-{{T}_{\rm{r}}}}{{{Z}_{\rm{rot}}}\tau_{\rm{t}} }
 	\label{energy_equation2}
 \end{equation}
 where ${{\mathcal{E}}_{\rm{t}}}$ and ${{\mathcal{E}}_{\rm{r}}}$ denote the translational and rotational energy relaxation rates, respectively. From the decomposed energy equations, we derive:
 \begin{equation}
 	\mathcal{D}\left( {{T}_{\rm{t}}}-{{T}_{\rm{r}}} \right)=-\frac{2}{3}{{T}_{\rm{t}}}\frac{\partial {{U}_{i}}}{\partial {{x}_{i}}}-\frac{2\left( 3+{\nu} \right)}{3{\nu}}{{\mathcal{E}}_{\rm{r}}}=-\frac{2}{3}{{T}_{\rm{t}}}\frac{\partial {{U}_{i}}}{\partial {{x}_{i}}}-\frac{{{T}_{\rm{t}}}-{{T}_{\rm{r}}}}{{{Z}_{\rm{rot}}}\tau_{\rm{t}} }
 \end{equation}
 Expanding ${{T}_{\rm{t}}}-{{T}_{\rm{r}}}$ in powers of $\tau_{\rm{t}}$ as ${{T}_{\rm{t}}}-{{T}_{\rm{r}}}={{T}_{\rm{t}}}\tau_{\rm{t}} +\text{O}\left( \tau _{\rm{t}}^{2} \right)$ reveals that the left-hand time derivative is negligible compared to the right-hand terms, yielding:
 \begin{equation}
 	{{T}_{\rm{t}}}-{{T}_{\rm{r}}}\approx -\frac{2}{3}{{Z}_{\rm{rot}}}\tau_{\rm{t}} {{T}_{\rm{t}}}\frac{\partial {{U}_{i}}}{\partial {{x}_{i}}}\approx -\frac{2}{3}{{Z}_{\rm{rot}}}\tau_{\rm{t}} T\frac{\partial {{U}_{i}}}{\partial {{x}_{i}}}
 \end{equation}
 Internal energy relaxation transforms the original pressure ${{p}}=nkT$ into translational pressure ${{p}_{\rm{t}}}=nk{{T}_{\rm{t}}}$. Thus, the translational pressure is expressed as:
 \begin{equation}
 	{{p}_{\rm{t}}}=nk{{T}_{\rm{t}}}=nkT\left( 1+\frac{{\nu}}{3+{\nu}}\left( {{T}_{\rm{t}}}-{{T}_{\rm{r}}} \right) \right)=nkT\left( 1-\frac{2{\nu}}{3\left( 3+{\nu} \right)}{{Z}_{\rm{rot}}}\tau_{\rm{t}} \frac{\partial {{U}_{i}}}{\partial {{x}_{i}}} \right)
 	\label{Pullin_miub}
 \end{equation}
 
 On the other hand, by reformulating the left-hand side of the second-order approximate Eq.~(\ref{pullin_ce2}) as:
 \begin{equation}
 	\left( 
 	\begin{aligned}
 	 & \left( \frac{\partial {{f}^{\left( 0 \right)}}}{\partial \rho }\mathcal{D}\rho +\frac{\partial {{f}^{\left( 0 \right)}}}{\partial {{v}_{i}}}\mathcal{D}{{v}_{i}}+\frac{\partial {{f}^{\left( 0 \right)}}}{\partial {{T}_{\rm{t}}}}\mathcal{D}{{T}_{\rm{t}}}+\frac{\partial {{f}^{\left( 0 \right)}}}{\partial {{T}_{\rm{r}}}}\mathcal{D}{{T}_{\rm{r}}} \right) + \\
 	 & {{C}_{i}}\left( \frac{\partial {{f}^{\left( 0 \right)}}}{\partial \rho }\frac{\partial \rho }{\partial {{x}_{i}}}+\frac{\partial {{f}^{\left( 0 \right)}}}{\partial {{v}_{j}}}\frac{\partial {{v}_{j}}}{\partial {{x}_{i}}}+\frac{\partial {{f}^{\left( 0 \right)}}}{\partial {{T}_{\rm{t}}}}\frac{\partial {{T}_{\rm{t}}}}{\partial {{x}_{i}}}+\frac{\partial {{f}^{\left( 0 \right)}}}{\partial {{T}_{\rm{r}}}}\frac{\partial {{T}_{\rm{r}}}}{\partial {{x}_{i}}} \right) 
 	\end{aligned}
 	 \right)
 	 \label{total derivative}
 \end{equation}
 the function $\phi_{\rm{ce}}$ can be expressed as:
 \begin{equation}
 	\begin{aligned}
 		& \phi ={{\mathbf{A}}_{1}}\left( -\frac{3}{2}+\frac{m}{2k{{T}_{\rm{t}}}}{{C}^{2}} \right)\frac{2}{3nk{{T}_{\rm{t}}}}{{\mathcal{E}}_{\rm{t}}}+{{\mathbf{A}}_{2}}\left( -\frac{{\nu}}{2}+\frac{1}{k{{T}_{\rm{r}}}}{{\epsilon }_{r}} \right)\frac{2}{{\nu}nk{{T}_{\rm{r}}}}{{\mathcal{E}}_{\rm{r}}}+\mathbf{B}\frac{m}{k{{T}_{\rm{t}}}}{{C}_{i}}{{C}_{j}}\frac{\partial {{U}_{\langle j}}}{\partial {{x}_{i\rangle }}}+ \\ 
 		& {{\mathbf{D}}_{1}}{{C}_{i}}\frac{1}{{{T}_{\rm{t}}}}\left( -\frac{5}{2}+\frac{m}{2k{{T}_{\rm{t}}}}{{C}^{2}} \right)\frac{\partial {{T}_{\rm{t}}}}{\partial {{x}_{i}}}+{{\mathbf{D}}_{2}}{{C}_{i}}\frac{1}{{{T}_{\rm{r}}}}\left( -\frac{{\nu}}{2}+\frac{1}{k{{T}_{\rm{r}}}}{{\epsilon }_{r}} \right)\frac{\partial {{T}_{\rm{r}}}}{\partial {{x}_{i}}} \\ 
 	\end{aligned}
 	\label{pullin_ce_solve}
 \end{equation}
 Taking moments of Eq.~(\ref{pullin_ce_solve}) yields:
 \begin{equation}
 	{{\mathbf{A}}_{1}}=0,{{\mathbf{A}}_{2}}=0
 \end{equation}
 \begin{equation}
 	\mathbf{B}=\tau_{\rm{t}} 
 \end{equation}
 \begin{equation}
 	\left[ \begin{matrix}
 		\frac{5}{2}{{\mathbf{D}}_{1}}  \\
 		\frac{{\nu}}{2}{{\mathbf{D}}_{2}}  \\
 	\end{matrix} \right]=\frac{n{{k}^{2}}{{T}_{\rm{t}}}}{m}{{\left[ \begin{matrix}
 				{{A}_{\rm{tt}}} & {{A}_{\rm{tr}}}  \\
 				{{A}_{\rm{rt}}} & {{A}_{\rm{rr}}}  \\
 			\end{matrix} \right]}^{-1}}\left[ \begin{matrix}
 		\frac{5}{2}  \\
 		\frac{{\nu}}{2}  \\
 	\end{matrix} \right]
 \end{equation}
 
 Taking moments of the distribution function $f={{f}^{\left( 0 \right)}}\left( 1+{{\phi }_{\rm{ce}}} \right)$ yields:
 \begin{equation}
 	{{p}_{\rm{t},\mathit{\langle ij\rangle }}}=2\mathbf{B}nk{{T}_{\rm{t}}}\frac{\partial {{U}_{\langle i}}}{\partial {{x}_{j\rangle }}}
 	\label{Pullin_miu}
 \end{equation}
 
 \begin{equation}
 	{{q}_{\rm{t},i}}={{\mathbf{D}}_{1}}\frac{5n{{k}^{2}}{{T}_{\rm{t}}}}{2m}\frac{\partial {{T}_{\rm{t}}}}{\partial {{x}_{i}}}
 	\label{Pullin_lambda_t}
 \end{equation}
 
 \begin{equation}
 	{{q}_{\rm{r},i}}={{\mathbf{D}}_{2}}\frac{n{{k}^{2}}{{T}_{\rm{t}}}}{m}\frac{\partial {{T}_{\rm{r}}}}{\partial {{x}_{i}}}
 	\label{Pullin_lambda_r}
 \end{equation}
 From Eq.~(\ref{Pullin_miu}), the shear viscosity is obtained as:
 \begin{equation}
 	\mu =-nk{{T}_{\rm{t}}}\mathbf{B}
 \end{equation}
 From Eq.~(\ref{Pullin_miub}), the bulk viscosity is derived as:
 \begin{equation}
 	{{\mu }_{\rm{b}}}=\frac{2{\nu}}{3\left( 3+{\nu} \right)}{{Z}_{\rm{rot}}}\tau_{\rm{t}} nkT=\frac{2{\nu}}{3\left( 3+{\nu} \right)}{{Z}_{\rm{rot}}}\mu 
 \end{equation}
 From Eq.~(\ref{Pullin_lambda_t}) and Eq.~(\ref{Pullin_lambda_r}), the translational and rotational thermal conductivities are given by:
 \begin{equation}
 	{{\kappa}_{\rm{t}}}=\frac{5n{{k}^{2}}{{T}_{\rm{t}}}}{2m}{{\mathbf{D}}_{1}}
 \end{equation}
 \begin{equation}
 	{{\kappa}_{\rm{r}}}=\frac{n{{k}^{2}}{{T}_{\rm{t}}}}{m}{{\mathbf{D}}_{2}}
 \end{equation}
 
 \section{Chapman-Enskog expansion of kinetic equation}\label{appendixB}
 The macroscopic transport coefficients of the kinetic model can also be derived through C-E expansion(\citep{chapman1990mathematical}). Introducing the parameter ${{\varepsilon }_{\rm{ce}}}$ into the kinetic model equation yields:
 \begin{equation}
 	\mathcal{D}{{f}_{1}}+{{\mathbf{C}}_{1}}\frac{\partial {{f}_{1}}}{\partial \mathbf{x}}=\frac{1}{{{\varepsilon }_{\rm{ce}}}}\left( \frac{{{f}_{\rm{t}}}-{{f}_{1}}}{\tau_{\rm{t}} }+\frac{{{f}_{\rm{r}}}-{{f}_{\rm{t}}}}{{{Z}_{\rm{rot}}}\tau_{\rm{t}} } \right)
 	\label{kinetic_ce}
 \end{equation}
 The expanded distribution function and the time derivative are substituted into the Pullin equation. By truncating to the first-order approximation and setting ${{\varepsilon }_{\rm{ce}}}=1$, we obtain:
 \begin{equation}
 	\mathcal{D}f_{1}^{\left( 0 \right)}+{{\mathbf{C}}_{1}}\frac{\partial f_{1}^{\left( 0 \right)}}{\partial \mathbf{x}}=\frac{{{f}_{\rm{t}}}-f_{1}^{\left( 0 \right)}}{\tau_{\rm{t}} }+\frac{{{f}_{\rm{r}}}-{{f}_{\rm{t}}}}{{{Z}_{\rm{rot}}}\tau_{\rm{t}} }
 	\label{kinetic_ce1}
 \end{equation}
 where
 \begin{equation}
 	f_{1}^{\left( 0 \right)}=n{{(\frac{m}{2\pi k{T}_{\rm{t}}})}^{\frac{3}{2}}}\exp \left( -\frac{m{{C}^{2}}}{2k{{T}_{\rm{t}}}} \right)\frac{{{\epsilon }^{\frac{{\nu}}{2}-1}}}{\Gamma \left( \frac{\nu }{2} \right){{\left( k{T}_{\rm{r}} \right)}^{\frac{\nu }{2}}}}\exp \left( -\frac{\epsilon }{k{T}_{\rm{r}}} \right)
 \end{equation}
 Taking moments of Eq.~(\ref{kinetic_ce1}) yields the Euler-type equations:
  \begin{equation}
 	\mathcal{D}\rho +\rho \frac{\partial {{U}_{i}}}{\partial {{x}_{i}}}=0
 \end{equation}
 \begin{equation}
 	\rho \mathcal{D}{{U}_{i}}+\frac{\partial {p}_{\rm{t}}}{\partial {{x}_{i}}}=0
 \end{equation}
 \begin{equation}
 	\left( \frac{3}{2}nk\mathcal{D}{{T}_{\rm{t}}}+nk{{T}_{\rm{t}}}\frac{\partial {{U}_{i}}}{\partial {{x}_{i}}} \right)+\frac{\nu}{2}nk\mathcal{D}{{T}_{\rm{r}}}=0
 \end{equation}
 decomposing the energy equation gives:
  \begin{equation}
 	\left( \frac{3}{2}nk\mathcal{D}{{T}_{\rm{t}}}+nk{{T}_{\rm{t}}}\frac{\partial {{U}_{i}}}{\partial {{x}_{i}}} \right)=\frac{3{\nu}nk}{2\left( 3+{\nu} \right)}\frac{{{T}_{\rm{r}}}-{{T}_{\rm{t}}}}{{{Z}_{\rm{rot}}}\tau_{\rm{t}} }
 	\label{kinetic_energy_equation1}
 \end{equation}
 \begin{equation}
 	\frac{{\nu}}{2}nk\mathcal{D}{{T}_{\rm{r}}}=\frac{3{\nu}nk}{2\left( 3+{\nu} \right)}\frac{{{T}_{\rm{t}}}-{{T}_{\rm{r}}}}{{{Z}_{\rm{rot}}}\tau_{\rm{t}} }
 	\label{kinetic_energy_equation2}
 \end{equation}
 From the decomposed energy equations, we derive:
 \begin{equation}
 	\mathcal{D}\left( {{T}_{\rm{t}}}-{{T}_{\rm{r}}} \right)=-\frac{2}{3}{{T}_{\rm{t}}}\frac{\partial {{U}_{i}}}{\partial {{x}_{i}}}-\frac{2\left( 3+{\nu} \right)}{3{\nu}}{{\mathcal{E}}_{\rm{r}}}=-\frac{2}{3}{{T}_{\rm{t}}}\frac{\partial {{U}_{i}}}{\partial {{x}_{i}}}-\frac{{{T}_{\rm{t}}}-{{T}_{\rm{r}}}}{{{Z}_{\rm{rot}}}\tau_{\rm{t}} }
 \end{equation}
 Expanding ${{T}_{\rm{t}}}-{{T}_{\rm{r}}}$ in powers of $\tau_{\rm{t}}$ as ${{T}_{\rm{t}}}-{{T}_{\rm{r}}}={{T}_{\rm{t}}}\tau_{\rm{t}} +\text{O}\left( \tau _{t}^{2} \right)$ reveals that the left-hand time derivative is negligible compared to the right-hand terms, yielding:
 \begin{equation}
 	{{T}_{\rm{t}}}-{{T}_{\rm{r}}}\approx -\frac{2}{3}{{Z}_{\rm{rot}}}\tau {{T}_{\rm{t}}}\frac{\partial {{U}_{i}}}{\partial {{x}_{i}}}\approx -\frac{2}{3}{{Z}_{\rm{rot}}}\tau_{\rm{t}} T\frac{\partial {{U}_{i}}}{\partial {{x}_{i}}}
 \end{equation}
 Internal energy relaxation transforms the original pressure ${{p}_{0}}=nkT$ into translational pressure ${{p}_{\rm{t}}}=nk{{T}_{\rm{t}}}$. Thus, the translational pressure is expressed as:
 \begin{equation}
 	{{p}_{\rm{t}}}=nk{{T}_{\rm{t}}}=nkT\left( 1+\frac{{\nu}}{3+{\nu}}\left( {{T}_{\rm{t}}}-{{T}_{\rm{r}}} \right) \right)=nkT\left( 1-\frac{2{\nu}}{3\left( 3+{\nu} \right)}{{Z}_{\rm{rot}}}\tau_{\rm{t}} \frac{\partial {{U}_{i}}}{\partial {{x}_{i}}} \right)
 	\label{kinetic_miub}
 \end{equation}
, the bulk viscosity is derived as:
 \begin{equation}
 	{{\mu }_{\rm{b}}}=\frac{2{\nu}}{3\left( 3+{\nu} \right)}{{Z}_{\rm{rot}}}\tau_{\rm{t}} nkT=\frac{2{\nu}}{3\left( 3+{\nu} \right)}{{Z}_{\rm{rot}}}\mu 
 \end{equation}
 
 By truncating to the second-order approximation and setting ${{\varepsilon }_{\rm{ce}}}=1$, we obtain:
 \begin{equation}
 	{{\mathcal{D}}^{\left( 0 \right)}}f_{1}^{\left( 0 \right)}+{{\mathbf{C}}_{1}}\frac{\partial f_{1}^{\left( 0 \right)}}{\partial \mathbf{x}}=\frac{{{f}_{\rm{t}}}-\left( f_{1}^{\left( 0 \right)}+f_{1}^{\left( 1 \right)} \right)}{\tau_{\rm{t}} }+\frac{{{f}_{\rm{r}}}-{{f}_{\rm{t}}}}{{{Z}_{\rm{rot}}}\tau_{\rm{t}} }
 	\label{kinetic_ce2}
 \end{equation}
 By transforming the left-hand side of the equation into the form of Eq.~(\ref{total derivative}), the second-order approximate part of the distribution function can be solved from Eq.~(\ref{kinetic_ce2}):
 \begin{equation}
 	f_{1}^{\left( 1 \right)}=\frac{{{f}_{t}}-f_{1}^{\left( 0 \right)}}{{{\tau}_{\rm{t}}}}+\frac{{{f}_{r}}-{{f}_{t}}}{{{Z}_{\rm{rot}}}{{\tau}_{\rm{t}}}}-{F}_{\rm{ce}}
 \end{equation}
 where
 \begin{equation}
 	{{F}_{\rm{ce}}}=f_{1}^{\left( 0 \right)}\left( \begin{aligned}
 		& \left( \left( -\frac{3}{2}+\frac{m}{2k{{T}_{\rm{t}}}}{{C}^{2}} \right)\frac{1}{{{T}_{\rm{t}}}}\frac{T-{{T}_{\rm{t}}}}{{{Z}_{\rm{rot}}}\tau_{\rm{t}} }+\left( -\frac{{\nu}}{2}+\frac{1}{k{{T}_{\rm{r}}}}{{\epsilon }_{r}} \right)\frac{1}{{{T}_{\rm{r}}}}\frac{T-{{T}_{\rm{r}}}}{{{Z}_{\rm{rot}}}\tau_{\rm{t}} } \right)+ \\ 
 		& {{C}_{i}}\left( \frac{m}{k{{T}_{\rm{t}}}}{{C}_{j}}\frac{\partial {{U}_{\langle j}}}{\partial {{x}_{i\rangle }}}+\frac{1}{{{T}_{\rm{t}}}}\left( -\frac{5}{2}+\frac{m}{2k{{T}_{\rm{t}}}}{{C}^{2}} \right)\frac{\partial {{T}_{\rm{t}}}}{\partial {{x}_{i}}}+\frac{1}{{{T}_{\rm{r}}}}\left( -\frac{{\nu}}{2}+\frac{1}{k{{T}_{\rm{r}}}}{{\epsilon }_{r}} \right)\frac{\partial {{T}_{\rm{r}}}}{\partial {{x}_{i}}} \right) \\ 
 	\end{aligned} \right)
 \end{equation}
 
 Taking moments of the distribution function $f={{f}^{\left( 0 \right)}}+{{f}^{\left( 1 \right)}}$ yields:
 \begin{equation}
 	{{p}_{{\rm{t}},\mathit{\langle ij\rangle} }}=2\mu \frac{\partial {{U}_{\langle i}}}{\partial {{x}_{j\rangle }}}=2\tau_{\rm{t}} nk{{T}_{\rm{t}}}\frac{\partial {{U}_{\langle i}}}{\partial {{x}_{j\rangle }}}
 \end{equation}
 \begin{equation}
 	{{q}_{i}}={{q}_{\rm{t},i}}+{{q}_{\rm{r},i}}
 \end{equation}
 \begin{equation}
 	{{q}_{\rm{t},i}}={{\kappa}_{\rm{t}}}\frac{\partial {{T}_{\rm{t}}}}{\partial {{x}_{i}}}
 \end{equation}
 \begin{equation}
 	{{q}_{\rm{r},i}}={{\kappa}_{\rm{r}}}\frac{\partial {{T}_{\rm{r}}}}{\partial {{x}_{i}}}
 \end{equation}
 where
 \begin{equation}
 	\left[ \begin{matrix}
 		{{\kappa}_{\rm{t}}}  \\
 		{{\kappa}_{\rm{r}}}  \\
 	\end{matrix} \right]=\frac{n{{k}^{2}}{{T}_{\rm{t}}}}{m}{{\left[ \begin{matrix}
 				{{A}_{\rm{tt}}} & {{A}_{\rm{tr}}}  \\
 				{{A}_{\rm{rt}}} & {{A}_{\rm{rr}}}  \\
 			\end{matrix} \right]}^{-1}}\left[ \begin{matrix}
 		\frac{5}{2}  \\
 		\frac{{\nu}}{2}  \\
 	\end{matrix} \right]
 \end{equation}
 
 \section{integral operator}\label{appendixC}
 
 In the calculation of relaxation rates, the integral operator ${{\mathcal{I}}_{1}}\left[ A,B \right]$ is used to split and separately compute the integrals of the rates. The definition of ${{\mathcal{I}}_{1}}\left[ A,B \right]$ is as follows:
 \begin{equation}
 	{{\mathcal{I}}_{1}}\left[ A,B \right]={{\left( \frac{k{{T}_{\rm{t}}}}{m} \right)}^{\frac{7}{2}}}{{\left( k{{T}_{\rm{r}}} \right)}^{2}}\int{\left( {A}'-A \right)f_{1}^{\left( 0 \right)}f_{2}^{\left( 0 \right)}\left( {{B}} \right){{g}^{*}}\sigma \text{d}\mathbf{{e}'}h(\mathbf{s})\text{d}\mathbf{s}\text{d}{{\bm{\xi}_{1}}}\text{d}{{\bm{\xi}_{2}}}\text{d}{{\varepsilon }_{2}}\text{d}{{\varepsilon }_{1}}}.
 \end{equation}
 Key intermediate results in the derivation of the relaxation rates are respectively presented by the integral operators ${{\mathcal{I}}_{2}}\left[ A \right]$ and ${{\mathcal{I}}_{3}}\left[ A \right]$, which are defined as follows:
 \begin{equation}
 	\begin{aligned}
 		{{\mathcal{I}}_{2}}\left[ A \right]= \sqrt{\frac{k{{T}_{\rm{t}}}}{m}}
 		\int & A\cdot {{n}^{2}}{{\left( \frac{1}{2\pi } \right)}^{3}}\frac{\varepsilon _{1}^{\frac{\nu }{2}-1}}{\Gamma \left( \frac{\nu }{2} \right)}\frac{\varepsilon _{2}^{\frac{\nu }{2}-1}}{\Gamma \left( \frac{\nu }{2} \right)}{{e}^{-{{G}^{{{*}^{2}}}}}}{{e}^{-\frac{{{g}^{{{*}^{2}}}}}{4}}}{{e}^{-{{\varepsilon }_{1}}}}{{e}^{-{{\varepsilon }_{2}}}} \\
 		& \cdot {{g}^{*}} b\text{d}b\,h(\mathbf{s})\text{d}\mathbf{s}\,\text{d}{{\mathbf{G}}^{*}}\text{d}{{\mathbf{g}}^{*}}\text{d}{{\varepsilon }_{2}}\text{d}{{\varepsilon }_{1}}.
 	\end{aligned}
 \end{equation}
 
 \begin{equation}
 	{{\mathcal{I}}_{3}}\left[ A \right]=\sqrt{\frac{k{{T}_{\rm{t}}}}{m}}\int{A\cdot {{\left( \frac{1}{2\pi } \right)}^{3}}\frac{\varepsilon _{1}^{\frac{\nu }{2}-1}}{\Gamma \left( \frac{\nu }{2} \right)}\frac{\varepsilon _{2}^{\frac{\nu }{2}-1}}{\Gamma \left( \frac{\nu }{2} \right)}{{n}^{2}}{{e}^{-\frac{{{g}^{{{*}^{2}}}}}{4}}}{{e}^{-{{\varepsilon }_{1}}}}{{e}^{-{{\varepsilon }_{2}}}}{{g}^{*}}b\text{d}bh(\mathbf{s})\text{d}\mathbf{s}\text{d}{{g}^{*}}\text{d}{{\varepsilon }_{2}}\text{d}{{\varepsilon }_{1}}}.
 \end{equation}
 The correspondence between the ${{\mathcal{I}}_{1}}\left[ A,B \right]$ and ${{\mathcal{I}}_{2}}\left[ A \right]$ used in the relaxation rate integrals is written as follows:
 \begin{equation}
 	{{\mathcal{I}}_{1}}\left[ \xi _{1}^{2},1 \right]={{\mathcal{I}}_{2}}\left[ \frac{\pi }{2}\left( {{{{g}'}}^{{*}^{2}}}-{{g}^{{{*}^{2}}}} \right) \right]
 \end{equation}
 \begin{equation}
 	{{\mathcal{I}}_{1}}\left[ {{\varepsilon }_{1}},1 \right]={{\mathcal{I}}_{2}}\left[ 2\pi \left( {{\varepsilon }_{1}}^{\prime }-{{\varepsilon }_{1}} \right) \right]
 \end{equation}
 
 \begin{equation}
 	\begin{aligned}
 		& {{\mathcal{I}}_{1}}\left[ {{\xi }_{\langle 1,i}}{{\xi }_{1,j\rangle }},{{\xi }_{\langle 1,k}}{{\xi }_{1,l\rangle }}+{{\xi }_{\langle 2,k}}{{\xi }_{2,l\rangle }} \right]\ \\ 
 		= & {{\mathcal{I}}_{2}}\left[ \pi \left( \frac{1}{2}(3{{\cos }^{2}}\left[\chi\right] -1)\frac{{{\left| {{{{g}'}}^{*}} \right|}^{2}}}{{{\left| {{g}^{*}} \right|}^{2}}}g_{\langle i}^{*}g_{j\rangle }^{*}-g_{\langle i}^{*}g_{j\rangle }^{*} \right)\left( G_{k}^{*}G_{l}^{*}+\frac{1}{4}g_{k}^{*}g_{l}^{*} \right) \right]  
 	\end{aligned}
 \end{equation}
 
 \begin{equation}
 	\begin{aligned}
 		& {{\mathcal{I}}_{1}}\left[ \xi _{1}^{2}{{\xi }_{1,i}},{\xi _{1}^{2}}{{\xi }_{1,k}}+{\xi _{2}^{2}}{{\xi }_{2,k}} \right] \\ 
 		= & {{\mathcal{I}}_{2}}\left[ \begin{aligned}
 			& \left( \frac{\pi }{2}\left( 3{{\cos }^{2}}\left[\chi\right] -1 \right)\frac{{{\left| {{{{g}'}}^{*}} \right|}^{2}}}{{{\left| {{g}^{*}} \right|}^{2}}}g_{i}^{*}g_{j}^{*}G_{j}^{*}+\frac{\pi }{2}\left( {{\sin }^{2}}\left[\chi\right] +1 \right){{{{g}'}}^{{{*}^{2}}}}G_{i}^{*}- \pi g_{i}^{*}g_{j}^{*}G_{j}^{*}+\frac{\pi}{2}{{g}^{{{*}^{2}}}}G_{i}^{*} \right)  \\ 
 			& \cdot 2\left( {{G}^{2}}{{G}_{k}}+\frac{1}{2}{{g}_{k}}{{G}_{j}}{{g}_{j}}+\frac{1}{4}{{g}^{2}}{{G}_{k}} \right) \\ 
 		\end{aligned} \right]  
 	\end{aligned}
 \end{equation}
 
 \begin{equation}
 	\begin{aligned}
 		& {{\mathcal{I}}_{1}}\left[ \xi _{1}^{2}{{\xi }_{1,i}},{{\varepsilon }_{1}} {{\xi }_{1,k}}+{{\varepsilon }_{2}} {{\xi }_{2,k}} \right] \\ 
 		= & {{\mathcal{I}}_{2}}\left[ \left(\begin{aligned} &\frac{\pi }{2}\left( 3{{\cos }^{2}}\left[\chi\right] -1 \right)\frac{{{\left| {{{{g}'}}^{*}} \right|}^{2}}}{{{\left| {{g}^{*}} \right|}^{2}}}g_{i}^{*}g_{j}^{*}G_{j}^{*}+\frac{\pi }{2}\left( {{\sin }^{2}}\left[\chi\right] +1 \right){{{{g}'}}^{{{*}^{2}}}}G_{i}^{*}\\
 			&- \pi g_{i}^{*}g_{j}^{*}G_{j}^{*}+\frac{\pi}{2}{{g}^{{{*}^{2}}}}G_{i}^{*} 		\end{aligned}\right){{G}_{k}}\left( {{\varepsilon }_{1}}+{{\varepsilon }_{2}} \right)\right] \\ 
 		& +{{\mathcal{I}}_{2}}\left[ \pi \left( 1-\cos \left[\chi\right] \frac{\left| {{{{g}'}}^{*}} \right|}{\left| {{g}^{*}} \right|} \right)\left( {{G}^{2}}\frac{1}{2}{{g}_{i}}+{{G}_{i}}{{G}_{j}}{{g}_{j}}+\frac{1}{8}{{g}^{2}}{{g}_{i}} \right) {{g}_{k}}\left( -{{\varepsilon }_{1}}+{{\varepsilon }_{2}} \right) \right]  
 	\end{aligned}
 \end{equation}
 
 \begin{equation}
 	\begin{aligned}
 		& {{\mathcal{I}}_{1}}\left[ \xi _{1}^{2}{{\xi }_{1,i}},{{\xi }_{1,k}}+{{\xi }_{2,k}} \right] \\ 
 		= & {{\mathcal{I}}_{2}}\left[ \left( \frac{\pi }{2}\left( 3{{\cos }^{2}}\left[\chi\right] -1 \right)\frac{{{\left| {{{{g}'}}^{*}} \right|}^{2}}}{{{\left| {{g}^{*}} \right|}^{2}}}g_{i}^{*}g_{j}^{*}G_{j}^{*}+\frac{\pi }{2}\left( {{\sin }^{2}}\left[\chi\right] +1 \right){{{{g}'}}^{{{*}^{2}}}}G_{i}^{*}- \pi g_{i}^{*}g_{j}^{*}G_{j}^{*}+\frac{\pi}{2}{{g}^{{{*}^{2}}}}G_{i}^{*} \right)2{{G}_{k}} \right]  
 	\end{aligned}
 \end{equation}
 
 \begin{equation}
 	\begin{aligned}
 		& {{\mathcal{I}}_{1}}\left[ {{\xi }_{1,i}}{{\varepsilon }_{1}},{\xi _{1}^{2}}{{\xi }_{1,k}}+{\xi _{2}^{2}}{{\xi}_{2,k}}\right] \\ 
 		= & {{\mathcal{I}}_{2}}\left[ 4\pi {{G}_{i}}\left( {{{{\varepsilon }}}_{1}}'-{{\varepsilon }_{1}} \right)\left( {{G}^{2}}{{G}_{k}}+\frac{1}{2}{{g}_{k}}{{G}_{j}}{{g}_{j}}+\frac{1}{4}{{g}^{2}}{{G}_{k}} \right) \right]  
 	\end{aligned}
 \end{equation}
 
 \begin{equation}
 	\begin{aligned}
 		& {{\mathcal{I}}_{1}}\left[ {{\xi }_{1,i}}{{\varepsilon }_{1}},{{\varepsilon }_{1}}{{\xi }_{1,k}}+{{\varepsilon }_{2}} {{\xi }_{2,k}} \right] \\ 
 		= & {{\mathcal{I}}_{2}}\left[ 2\pi {{G}_{i}}\left( {{{{\varepsilon }}}_{1}}'-{{\varepsilon }_{1}} \right){{G}_{k}}\left( {{\varepsilon }_{1}}+{{\varepsilon }_{2}} \right) \right] + \\
 		& {{\mathcal{I}}_{2}}\left[ \frac{\pi }{2}\left( -\cos \left[\chi\right] \frac{\left| {{{{g}'}}^{*}} \right|}{\left| {{g}^{*}} \right|}{{g}_{i}}{{{{\varepsilon }}}_{1}}'+{{g}_{i}}{{\varepsilon }_{1}} \right){{g}_{k}}\left( -{{\varepsilon }_{1}}+{{\varepsilon }_{2}} \right) \right]  
 	\end{aligned}
 \end{equation}
 
 \begin{equation}
 	{{\mathcal{I}}_{1}}\left[ {{\xi }_{1,i}}{{\varepsilon }_{1}},{{\xi }_{1,k}}+{{\xi }_{2,k}} \right] = {{\mathcal{I}}_{2}}\left[ 4\pi {{G}_{i}}\left( {{{{\varepsilon }}}_{1}}'-{{\varepsilon }_{1}} \right) {{G}_{k}} \right]  
 \end{equation}
 Furthermore, the correspondence between the ${{\mathcal{I}}_{1}}\left[ A,B \right]$ and ${{\mathcal{I}}_{3}}\left[ A \right]$ used in the relaxation rate integrals is given by:
 \begin{equation}
 	{{\mathcal{I}}_{1}}\left[ \xi _{1}^{2},1 \right]={{\mathcal{I}}_{3}}\left[ 2{{\pi }^{\frac{7}{2}}}\left( {{{{g}'}}^{{*}^{2}}}-{{g}^{{{*}^{2}}}} \right){{g}^{{{*}^{2}}}} \right]
 	\label{integrate3-Tt}
 \end{equation}
 
 \begin{equation}
 	{{\mathcal{I}}_{1}}\left[ {{\varepsilon }_{1}},1 \right]={{\mathcal{I}}_{3}}\left[ 8{{\pi }^{\frac{7}{2}}}\left( {{\varepsilon }_{1}}^{\prime }-{{\varepsilon }_{1}} \right){{g}^{{{*}^{2}}}} \right]
 \end{equation}
 
 \begin{equation}
 	\begin{aligned}
 		{{\mathcal{I}}_{1}}\left[ {{\xi }_{\langle 1,i}}{{\xi }_{1,j\rangle }},{{\xi }_{\langle 1,k}}{{\xi }_{1,l\rangle }}+{{\xi }_{\langle 2,k}}{{\xi }_{2,l\rangle }} \right] = {{\mathcal{I}}_{3}}\left[ \frac{1}{15}{{\pi }^{\frac{7}{2}}}{{g}^{{{*}^{4}}}}\left( {{{{g}'}}^{{{*}^{2}}}}(3{{\cos }^{2}}\left[\chi\right] -1)-2{{g}^{{{*}^{2}}}} \right) \right]  
 	\end{aligned}
 \end{equation}
 
 \begin{equation}
 	\begin{aligned}
 		{{\mathcal{I}}_{1}}\left[ \xi _{1}^{2}{{\xi }_{1,i}},{\xi _{1}^{2}}{{\xi }_{1,k}}+{\xi _{2}^{2}}{{\xi }_{2,k}} \right]= & {{\mathcal{I}}_{3}}\left[ \frac{1}{3}{{\pi }^{\frac{7}{2}}}{{g}^{{{*}^{2}}}}\left( 10+3{{g}^{{{*}^{2}}}} \right)\left( \frac{1}{2}\left( 3{{\cos }^{2}}\left[\chi\right] -1 \right){{{{g}'}}^{{{*}^{2}}}}-{{g}^{{{*}^{2}}}} \right) \right]\\
 		&+{{\mathcal{I}}_{3}}\left[ \frac{5}{6}{{\pi }^{\frac{7}{2}}}{{g}^{{{*}^{2}}}}\left( 6+{{g}^{{{*}^{2}}}} \right)\left( \left( {{\sin }^{2}}\left[\chi\right] +1 \right){{{{g}'}}^{{{*}^{2}}}}-{{g}^{{{*}^{2}}}} \right) \right]   
 	\end{aligned}
 \end{equation}
 
 \begin{equation}
 	\begin{aligned}
 		{{\mathcal{I}}_{1}}\left[ \xi _{1}^{2}{{\xi }_{1,i}},{{\varepsilon }_{1}} {{\xi }_{1,k}}+{{\varepsilon }_{2}} {{\xi }_{2,k}} \right]	= & {{\mathcal{I}}_{3}}\left[ \frac{2}{3}{{\pi }^{\frac{7}{2}}}{{g}^{{{*}^{2}}}}\left( {{\varepsilon }_{1}}+{{\varepsilon }_{2}} \right)\left( \frac{1}{2}\left( 3{{\cos }^{2}}\left[\chi\right] -1 \right){{{{g}'}}^{{{*}^{2}}}}-{{g}^{{{*}^{2}}}} \right) \right]\\
 		& +{{\mathcal{I}}_{3}}\left[ {{\pi }^{\frac{7}{2}}}{{g}^{{{*}^{2}}}}\left( {{\varepsilon }_{1}}+{{\varepsilon }_{2}} \right)\left( \left( {{\sin }^{2}}\left[\chi\right] +1 \right){{{{g}'}}^{{{*}^{2}}}}-{{g}^{{{*}^{2}}}} \right) \right] \\ 
 		& +{{\mathcal{I}}_{3}}\left[ \frac{1}{6}{{\pi }^{\frac{7}{2}}}{{g}^{{{*}^{2}}}}\left( 10+{{g}^{{{*}^{2}}}} \right)\left( -{{\varepsilon }_{1}}+{{\varepsilon }_{2}} \right)\left( {{g}^{{{*}^{2}}}}-\cos \left[\chi\right] {{{{g}'}}^{*}}{{g}^{{{*}}}} \right) \right]  
 	\end{aligned}
 	\label{I3_Omega+1}
 \end{equation}
 
 \begin{equation}
 	\begin{aligned}
 		{{\mathcal{I}}_{1}}\left[ \xi _{1}^{2}{{\xi }_{1,i}},{{\xi }_{1,k}}+{{\xi }_{2,k}} \right]	= & {{\mathcal{I}}_{3}}\left[ \frac{4}{3}{{\pi }^{\frac{7}{2}}}{{g}^{{{*}^{2}}}}\left( \frac{1}{2}\left( 3{{\cos }^{2}}\left[\chi\right] -1 \right){{{{g}'}}^{{{*}^{2}}}}-{{g}^{{{*}^{2}}}} \right) \right]\\
 		&+{{\mathcal{I}}_{3}}\left[ 2{{\pi }^{\frac{7}{2}}}{{g}^{{{*}^{2}}}}\left( \left( {{\sin }^{2}}\left[\chi\right] +1 \right){{{{g}'}}^{{{*}^{2}}}}-{{g}^{{{*}^{2}}}} \right) \right]  
 	\end{aligned}
 \end{equation}
 
 \begin{equation}
 	\begin{aligned}
 		{{\mathcal{I}}_{1}}\left[ {{\xi }_{1,i}}{{\varepsilon }_{1}},{\xi _{1}^{2}}{{\xi }_{1,k}}+{\xi _{2}^{2}}{{\xi}_{2,k}}\right]= & {{\mathcal{I}}_{3}}\left[ \frac{10}{3}{{\pi }^{\frac{7}{2}}}{{g}^{{{*}^{2}}}}\left( 6+{{g}^{{{*}^{2}}}} \right)\left( {{{{\varepsilon }}}_{1}}'-{{\varepsilon }_{1}} \right) \right]  
 	\end{aligned}
 \end{equation}
 
 \begin{equation}
 	\begin{aligned}
 		{{\mathcal{I}}_{1}}\left[ {{\xi }_{1,i}}{{\varepsilon }_{1}},{{\varepsilon }_{1}} {{\xi }_{1,k}}+{{\varepsilon }_{2}} {{\xi }_{2,k}} \right]=&{{\mathcal{I}}_{3}}\left[ \frac{2}{3}{{\pi }^{\frac{7}{2}}}{{g}^{{{*}^{3}}}}\left( -\cos \left[\chi\right] {{{{g}'}}^{*}}{{{{\varepsilon }}}_{1}}'+{{g}^{*}}{{\varepsilon }_{1}} \right)\left( -{{\varepsilon }_{1}}+{{\varepsilon }_{2}} \right) \right]\\ 
 		&+  {{\mathcal{I}}_{3}}\left[ 4{{\pi }^{\frac{7}{2}}}{{g}^{{{*}^{2}}}}\left( {{{{\varepsilon }}}_{1}}'-{{\varepsilon }_{1}} \right)\left( {{\varepsilon }_{1}}+{{\varepsilon }_{2}} \right) \right]
 	\end{aligned}
 	 	\label{I3_Omega+2}
 \end{equation}
 
 \begin{equation}
 	{{\mathcal{I}}_{1}}\left[ {{\xi }_{1,i}}{{\varepsilon }_{1}},{{\xi }_{1,k}}+{{\xi }_{2,k}} \right] ={{\mathcal{I}}_{3}}\left[ 8{{\pi }^{\frac{7}{2}}}{{g}^{{{*}^{2}}}}\left( {{{{\varepsilon }}}_{1}}'-{{\varepsilon }_{1}} \right) \right]
 	\label{integrate3-Qr3}
 \end{equation}

\bibliographystyle{jfm}
\bibliography{jfm}

\end{document}